%% file: thesis.tex
\documentclass[reqno,12pt,oneside]{report} 

\usepackage{rac}                                       
\usepackage[intlimits]{amsmath}                 
\usepackage{amsxtra}                                
\usepackage{amsthm}
\usepackage{amssymb}
\usepackage{amsfonts}
\usepackage{graphicx}                               
\usepackage{rotating}
\usepackage{color}
\usepackage{epsfig}
\usepackage{subfigure}                             
\usepackage{verbatim}
\usepackage{cite}
\usepackage[printonlyused]{acronym}         
\usepackage{setspace}                              

\usepackage{array}

\usepackage{hyperref}                              
\hypersetup{
    colorlinks=true,
    linkcolor=blue,
    filecolor=magenta,      
    urlcolor=cyan,
}

\usepackage{amsmath}

\usepackage{lineno}                                 

\theoremstyle{plain}

\theoremstyle{definition}

\theoremstyle{remark}

\numberwithin{theorem}{chapter}     

\makeatletter
\def\cleardoublepage{\clearpage\if@twoside \ifodd\c@page\else
\hbox{}
\thispagestyle{empty}
\newpage
\if@twocolumn\hbox{}\newpage\fi\fi\fi}
\makeatother 

\begin{document}

\bibliographystyle{ieeetr} 

\titlepage{Transverse Single-Spin Asymmetries of Midrapidity Direct Photons, Neutral Pions, and Eta Mesons in 200 GeV Polarized Proton-Proton Collisions at PHENIX}{Nicole Lewis}{Doctor of Philosophy}
{Physics}{2020}
{Associate Professor Christine Aidala, Chair\\
 Professor James Liu \\
 Associate Professor Joern Putschke \\
 Associate Professor Thomas Schwarz\\
  Professor Alejandro Uribe }

\initializefrontsections


\copyrightpage{Nicole Lewis}{nialewis@umich.edu}{0000-0003-0146-1565}

\makeatletter
\if@twoside \setcounter{page}{4} \else \setcounter{page}{1} \fi
\makeatother
 
\dedicationpage{To Alan Lin, who supported me through every step of this process }

\startacknowledgementspage
\input{Intro/Acknowledgements}
\label{Acknowledgements}


\tableofcontents     
\listoffigures       
\listoftables        
\listofappendices    

\startabstractpage
\input{Abstract/Abstract}
\label{Abstract}

\startthechapters 

\chapter{Introduction}
\label{Chapter:Intro}
\input{Intro/Intro}

\chapter{Experimental Set Up}
\label{Chapter:ExperimentalSetUp}
\input{Chap2/chap2}

\chapter{Analysis Details}
\label{Chapter:AnalysisDetails}
\input{Chap3/chap3}

\chapter{Results}
\label{Chapter:Results}
\input{Chap4/chap4}

\chapter{Conclusion}
\label{Chapter:Conclusion}
\input{Chap5/chap5}

\startappendices
\appendix{\( \pi^0 \) and \( \eta \) Asymmetry Cross Checks}
\label{Appendix:AsymmetryPloots}
\input{Appendices/pi0_eta_Asymmetry_Ploots}

\appendix{\( \pi^0 \) and \( \eta \) Asymmetry Bunch Shuffling Results}
\label{Appendix:BunchShuffling}
\input{Appendices/Bunch_Shuffling}


\begin{singlespace} 
\bibliography{thesis_references}   
\end{singlespace}


\end{document}

%% file: Intro/Acknowledgements.tex
Obviously, any thesis would not be possible without the graduate student's adviser, but Christine went above and beyond to ensure that I was not only growing as a scientist, but that I felt valued as a person.   I felt comfortable asking for help when I needed it and she modeled how to be secure in my expertise while also being open about what I don't know.   She taught me how to not let the minutiae of an analysis project distract from the big-picture physics that we were trying to measure and not to be afraid of questioning conventional assumptions.  She even had the patience to wade through my unreasonable number of typos. 

Christine also created an environment in which in her students looked after each other instead of competing.  When I started doing research I had a much weaker background in coding than what was ideal and a level of impostor syndrome that made asking for help really difficult.  I was able to overcome both of these with the endless patience of Joe Osborn who answered my numerous coding questions with a friendly smile, made sure that I knew the motivation behind every step, and often checked in with me to make sure I wasn't stuck.  Catherine Ayuso was and will always be my best of conference buddies.  Kara Mattioli and Jordan Roth very generously laughed at some of my terrible jokes and were there to take over when the SiPM project started taking away my ability to do research.  There are so many more people in the Aidala Group that I am grateful for getting the opportunity to work with. 

\pagebreak
There are so many friends in my cohort whose commiseration was instrumental in getting me through my classes and helping me learn how to be comfortable being stuck on research.   Thank you to the Society for Women in Physics for fighting for a more inclusive environment in our department and giving me the opportunity to give back to my community.  I also want to thank the Graduate Employee Organization for increasing the quality of life of graduate students at the university and showing me how to pull a community together to fight for a common cause.  

PHENIX is a wonderful experiment to be on while earning a PhD because of the well-established expertise and a culture of mentorship for its analyzers.  Sasha Bazilevsky was very generous with his time and his knowledge and always gave very detailed answers to all of my questions.  All my analyses were guided by the keen eyes of Sanghwa Park and Ralf Seidl, who made sure that each step was completed and crossed checked and were always available when I needed to ask for help.  Chris Pinkenburg was very patient with my numerous stupid questions on how to access the computing facility.  

Alan Lin has been my rock throughout this whole process from encouraging me when I wanted to switch to a physics major in undergrad, to supporting me through my chaotic grad school application process, to being willing to sit through the day to day minutiae of homework and research in grad school.  He talked me through numerous ``freak outs'' and consistently made me laugh when I felt like I was too stressed out to smile.  He helped me remember that life outside of grad school still existed and he was forced to hear more about the internal structure of the proton than any reasonable person would expect out of life.  

My parents not only ensured that my scientific curiosity could flourish and that I had everything I needed, they taught me how to value my own mental well-being.  They were always encouraging even when listening to my early attempts at explaining my research when my scientific communication skills were pretty dubious.  I also want to take a sentence or two to thank my high school physics teacher Mr. Laderman who was able convey how exciting and fun physics could be even with limited classroom resources.  He taught me that the word “obviously” should always be ignored in the context of physics and instilled in me a very proud love of physics puns.   

%% file: Abstract/Abstract.tex
Experimental observations of strikingly large transverse single-spin asymmetries (TSSAs) opened a window into quark and gluon dynamics present in hadronic collisions, revealing large spin-momentum correlations within nucleons and in the process of forming hadrons.  Though originally measured in lower energy fixed target experiments, they have been found to persist in collisions with momentum transfer well into the perturbative regime of quantum chromodynamics (QCD) and yet their origin remains poorly understood.  The Relativistic Heavy Ion Collider (RHIC) is the only collider in the world that can run polarized proton beams, allowing for these asymmetries to be measured at higher energies, with center of mass energies ranging from \( \sqrt{s} = 60 \) to \(500\) GeV.  TSSA measurements have allowed for the development of both transverse momentum dependent and collinear twist-3 descriptions of nonperturbative spin-momentum correlations for both initial- and final-state effects. 

Results are presented for the TSSAs of direct photons, neutral pions, and eta mesons in the pseudorapidity range \( | \eta | < 0.35 \) from \( p^\uparrow + p \) collisions with \( \sqrt{s} = 200 \) GeV at PHENIX.   As hadrons, \( \pi^0 \) and \( \eta \) mesons are sensitive to both initial- and final-state effects. At midrapidity, \( \pi^0 \) and \( \eta \) measurements are sensitive to the dynamics of gluons along with a mix of quark flavors.  Comparisons of the differences in the \( \pi^0 \) and \( \eta \) TSSAs are sensitive to potential effects from strangeness, isospin, or mass.  These results are a factor of three increase in statistical precision and extend to higher transverse momentum when compared with previous PHENIX measurements in this kinematic region.  Because direct photon production does not include hadronization,  the direct photon TSSA is only sensitive to spin-momentum correlations in the proton. The kinematics of this result in particular make the direct photon TSSA a clean probe of gluon dynamics in the transversely polarized proton.  This is the first time direct photons have been used as a probe of spin-momentum correlations in polarized protons at RHIC.  All three of these asymmetries will help constrain the twist-3 trigluon collinear correlation function as well as the gluon Sivers function, improving our knowledge of spin-dependent gluon dynamics in QCD. 

%% file: Intro/Intro.tex

\section{Quantum Chromodynamics}\label{Section:QCD}
The strong nuclear force is one of four fundamental forces of nature.  It is responsible for both binding protons and neutrons together into an atomic nucleus as well as binding quarks and gluons together to form protons, neutrons, and other strong force bound states, called hadrons.  This process is responsible for 98\% of the mass in the visible universe.  Quantum Chromodynamics (QCD) is the quantum field theory for the strong nuclear force and describes interactions between quarks and gluons, which are collectively referred to as partons.  

The ``chromo'' refers to color charge.  In Quantum Electrodynamics (QED) the charge can either be positive or negative, but in QCD there are three types of charges which are referred to as: red, blue, and green.  They are governed by the SU(3) color symmetry group, which is non-Abelian.  Each quark can be either red, blue, or green and each antiquark can be antired, antiblue, or antigreen. They can change their color charge by exchanging gluons.  The gluon is the force carrier in QCD, similar to the photon in QED, except that photons are not electrically charged and so do not interact with other photons.  Gluons \textit{do} carry color charge and so interact with other gluons.  
There are eight independent gluon color charges which are made up of combinations of quark colors and anticolors.  

Color charge is meant to be an analogy to light where combing red, blue, and green light creates white light.  Equivalently combining a red, a blue, and a green quark creates a color neutral three-quark bound state called a baryon, which includes protons and neutrons.  A quark-antiquark pair can also combine to form a color-neutral bound state referred to as a meson which includes pions and eta mesons.  There are six different types of quarks, or quark flavors, which in order of lightest to heaviest are: up, down, strange, charm, bottom, and top.  Protons consist of two up quarks and one down quark, as wells as quark-antiquark pairs of the lightest three flavors (up, down, and strange) which pop in and out of existence via gluon splitting: \( g \rightarrow q \bar{q} \).  These virtual quark-antiquark pairs are collectively referred to as ``sea quarks''. 

\subsection{Asymptotic Freedom}
\begin{figure}
  \centering
  \includegraphics[width=.65\textwidth]{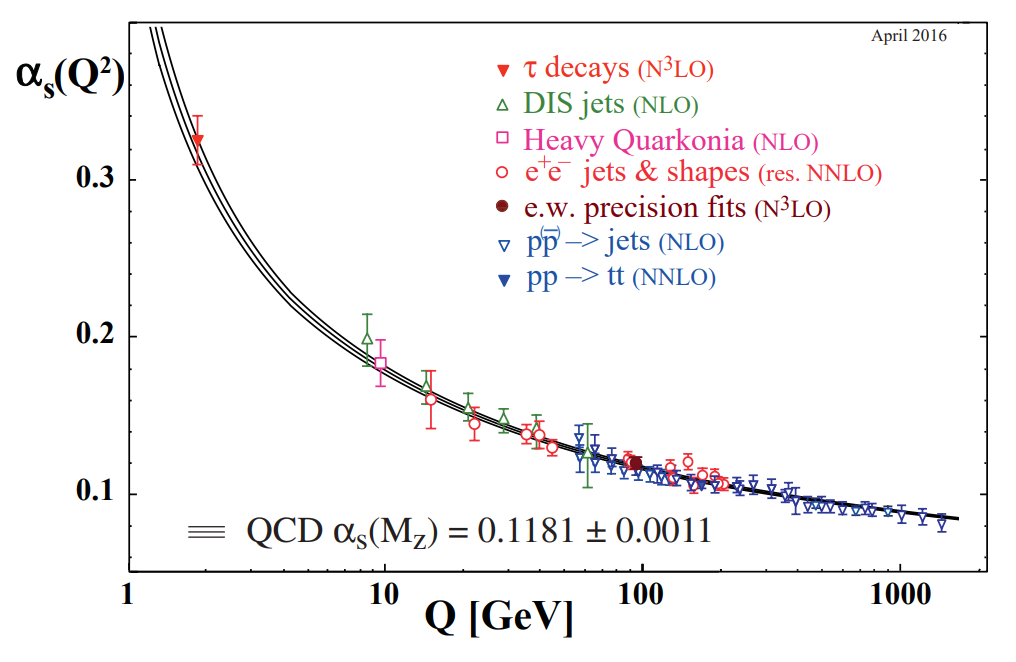}
  \caption[Summary of the measurements of the strong coupling constant, \( \alpha_s \), over a wide range of energy scales.]{ Summary of the measurements of the strong coupling constant, \( \alpha_s \), over a wide range of energy scales.  \cite{particlePhysicsReview}}
  \label{Figure:AsymptoticFreedom}
\end{figure}

The value of the strong force coupling constant, \( \alpha_s \), changes as a function of interaction energy.  This comes from higher order corrections to the coupling constant and is not unique among quantum field theories.  But for QCD the coupling constant decreases as a function of interaction energy unlike in QED where the coupling constant increases.  This concept is referred to as asymptotic freedom: with increasing scattering energy quarks and gluons become asymptotically closer to being free, \cite{asymptoticFreedom1, asymptoticFreedom2} for which the Nobel prize was awarded in 2004.   Figure~\ref{Figure:AsymptoticFreedom} shows a summary of measurements of \( \alpha_s \), displaying how the constant falls off with scattering energy.\cite{particlePhysicsReview} This figure also shows how well these measurements agree with perturbative calculations.

In practice this means that perturbative Quantum Chromodynamics (pQCD) is only able to describe collisions between quarks and gluons at high energies or ``hard''-scale energies.  Fixed order pQCD calculations become less and less accurate with decreasing scattering energy until eventually there is a cut off where the energy of a partonic collision is so low that pQCD can longer describe these strong force interactions.  These are referred to as nonperturbative interactions or ``soft''-scale interactions.  Exactly where this cut off exists depends on the calculation, but perturbative QCD calculations must be done for energies much larger than \( \Lambda_{QCD} \sim 200 \) MeV and in general are only done for collisions with momentum transfers of at least 1-2 GeV.  
Alternatively these hard-scale perturbative interactions can be described as ``short-range'' nuclear interactions, while soft-scale nonperturbative scattering is described as a ``long-range'' nuclear interaction.  In the context of pQCD, the radius of the proton, which is on the order of \( 10^{-15} \) meters, is considered a ``long'' range and so proton structure cannot be calculated perturbatively.  

\subsection{Color Confinement}\label{Subsection:ColorConfinement}
Perturbative QCD is only capable of directly calculating scattering between approximately free partons, but these high energy scatterings between individual quarks and gluons cannot be directly observed.  This is due to a property of the strong force called color confinement (often referred to as just confinement) which states that no color charged object can ever be observed on its own.  So even though the existence of quarks was confirmed over 50 years ago, \cite{ quarkDiscovery1, quarkDiscovery2 } a quark has never directly been observed and if color confinement remains true, never will be.  Unlike asymptotic freedom, color confinement has yet to be analytically proven.  

\begin{figure}
  \centering
  \includegraphics[width=.65\textwidth]{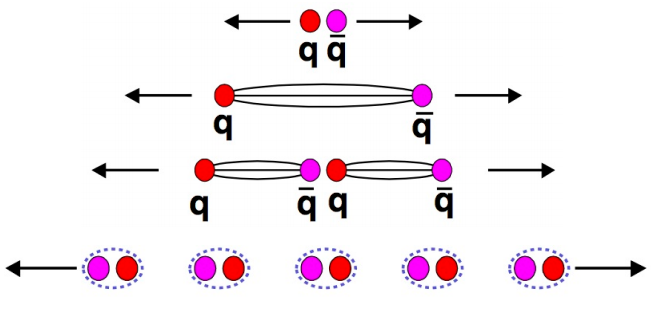}
  \caption[A cartoon explaining color confinement by depicting what would happen if we tried to pull the quark-antiquark pair within a meson apart.]{A cartoon explaining color confinement by depicting what would happen if we tried to pull the quark-antiquark pair within a meson apart.\cite{LukLecture}   }
  \label{Figure:Confinement}
\end{figure}

To build some physical intuition behind this curious property, one can think through the thought experiment of pulling apart the quark-antiquark pair within a meson.  This is depicted in Figure~\ref{Figure:Confinement} where the red quark is pictured as red and the antired antiquark is depicted as magenta.\cite{LukLecture} These quarks are being held together by the strong force and the further they are pulled apart, the more the potential energy in the gluon field between them increases, similar to stretching a rubber band.  Eventually there is enough energy built up, that another quark-anti quark pair appears, similar to a rubber band breaking when it is pulled too hard.  So instead of extracting the quarks from their color-neutral bound state, we are left with two separate mesons and this process repeats as we continue to attempt to pull apart these quark-antiquark bound states.  This is a heuristic description of how color confinement works, but serves to show how closely asymptotic freedom and color confinement are linked.  

\subsection{Cross Sections and Nonperturbative Functions}
Because of asymptotic freedom, perturbative QCD calculations can only be applied to quark and gluon scattering with large enough momentum transfer.  And these high energy partonic interactions cannot be directly observed in the laboratory because due to color confinement, we can only manipulate and observe hadrons.  Thus, in order to use pQCD to interpret the results from high energy collisions we must use nonperturbative functions to parameterize the effects of the bound-state structure of hadrons. Parton distribution functions (PDFs) describe the nonperturbative partonic structure of a proton relevant when probed at high energies 
and fragmentation functions (FFs) are used to capture the nonperturbative process of hadronization.  These functions cannot be directly calculated with pQCD and need to be measured in data.  

Quark PDFs are often denoted as \( f_{q/h}( x, Q^2 ) \) for a quark of flavor \( q \) that is being scattered out of an initial-state hadron \( h \) with squared momentum transfer \( Q^2 \).  Bjorken \( x \), commonly referred to as just \( x \), is the longitudinal fraction of the proton's momentum that this quark was carrying: \( x = p_q / p_h \) and can range from 0 to 1.  At leading order (LO) in the strong coupling constant, \( f_{q/h}( x, Q^2 ) \) can be interpreted as the probability of finding a quark of flavor \( q \) with longitudinal momentum fraction \( x \) when probing a hadron \( h \) at scattering energy \( Q \).  But next to leading order (NLO) pQCD contributions make the physical interpretation of these functions less straight forward.  Most of the early information about parton distribution functions came from high energy collisions between protons and leptons, \( \ell + p \rightarrow \ell + X \), such as electrons, muons, and neutrinos, which do not interact via the strong force.   These collisions are referred to as deep inelastic scattering (DIS) if the energy of the collision is high enough such that the proton breaks apart and the behavior of individual quarks and gluons can be resolved.  The energy and scattering angle of the lepton provides direct access to both \( Q^2 \) and \( x \), where a smaller lepton scattering angle corresponds to the proton being probed at lower \( x \).  Flavor dependent PDFs can be measured with semi-inclusive deep inelastic scattering (SIDIS) where in addition to the scattered lepton at least one final state hadron is measured, \( \ell + p \rightarrow \ell + h + X \).  Drell-Yan, \( p + p \rightarrow \ell^+ + \ell^- + X \), has served as another clean probe of proton structure because there are no strong force interactions in the final-state, and it is also sensitive to the antiquark PDFs via quark-antiquark 
annihilation.  There are also gluon PDFs, often denoted as 
\( G( x, Q^2 ) \), which have proven more difficult to constrain because gluons do not have electric charge and so QED processes like SIDIS and Drell-Yan are not sensitive to gluon dynamics in the proton at leading order.  

\begin{figure}
 \centering
 \includegraphics[width=.9\textwidth]{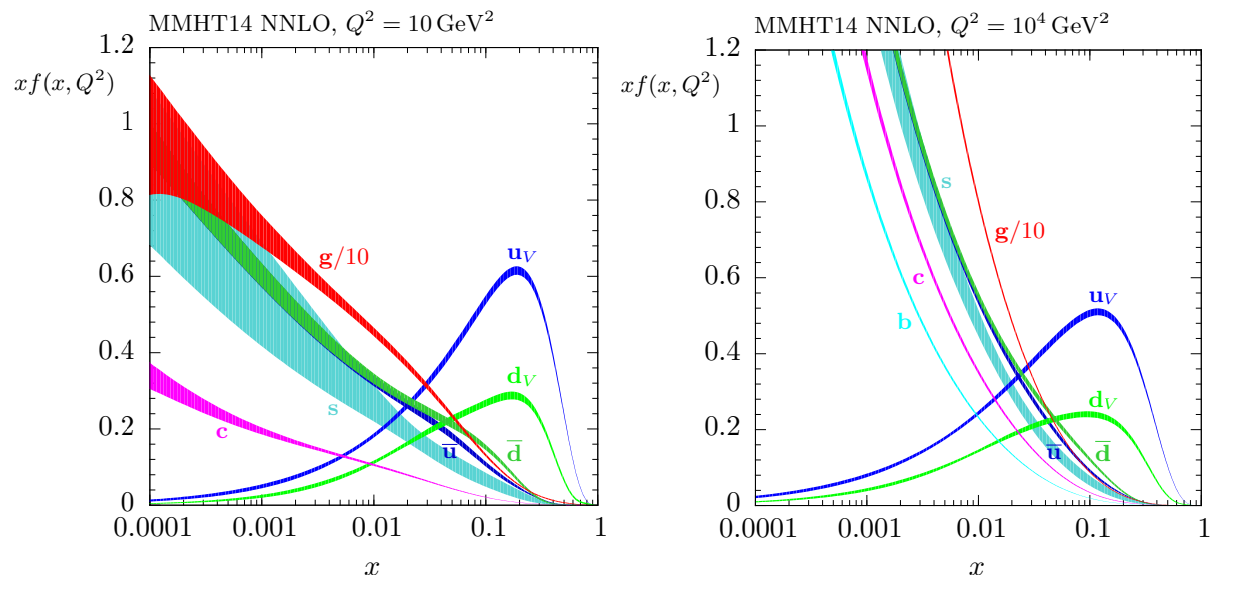}
 \caption[The 2014 MMHT extraction of NNLO parton distribution functions  at \( Q^2 = 10 \) GeV\textsuperscript{2} and \( Q^2 = 10^4 \) GeV\textsuperscript{2}. ]{The 2014 MMHT extraction of NNLO parton distribution functions  at \( Q^2 = 10 \) GeV\textsuperscript{2} and \( Q^2 = 10^4 \) GeV\textsuperscript{2}. \cite{PDFs}}
 \label{Figure:PDFs}
\end{figure}

Today parton distribution functions are extracted from global analyses of available hard scattering data, including data from electron-proton, proton-proton, and proton-antiproton collisions. 
 Figure~\ref{Figure:PDFs} shows the 2014 next-to-next to leading order (NNLO) extraction of proton PDFs from the Martin, Motylinski, Harland-Lang, Thorne theory group (MMHT) for \( Q^2 = 10 \) GeV\textsuperscript{2} and \( Q^2 = 10^4 \) GeV\textsuperscript{2}.\cite{PDFs}  As expected for a proton, the valence up quark PDF (\( u_v \)) is about two times larger than the valence down quark PDF (\( d_v \)).  Thus the valence quark PDFs dominate at higher momentum fractions of about \( 0.1 \) and at lower \( x \) the sea quarks and gluons have a much higher contribution.  This MMHT 2014 extraction was specifically done with the high energy proton-proton collisions at the Large Hadron Collider (LHC) in mind and the fact that they include charm and bottom sea quark PDFs in their extraction is atypical of the field.  It is generally assumed that the contributions of ``intrinsic'' charm and bottom to the proton are negligible, especially at the comparatively lower collision energies of the results that will be presented later in this document.  
 
Note that the gluon PDF plotted Figure~\ref{Figure:PDFs} is divided by 10 such that it can fit on the plot, indicating just how much gluons dominate the proton at lower \( x \).  Lower \( x \) corresponds to a lower parton momentum where it is possible to have more partons.  By definition there cannot be more than one parton with \( x > 0.5 \), but in principle there could be \( 10^4 \) partons with  \( x = 10^{-4} \).  The theory of gluon saturation predicts that at low enough \( x \) there are enough low momentum gluons that gluon splitting becomes just as likely as two gluons recombining into one and the total number of gluons in the proton reaches some kind of equilibrium.  This saturated gluon regime has yet to be unambiguously observed. \cite{gluonSaturationEIC} More data is needed to sufficiently describe the low \( x \) behavior of the proton, but suffice it to say that it is dominated by gluon dynamics.  

Recent progress in lattice QCD provides a potential way forward to calculating PDFs from first principles.  Lattice QCD is an alternative method to pQCD techniques where partons are placed on a three dimensional discrete lattice and their strong force interactions are captured via computationally intensive methods.  Unlike pQCD, lattice QCD can be evaluated at all energy scales and recent innovations in lattice techniques have shown promise towards calculating the full \( x \) dependence of some PDFs.  \cite{latticeQCD}  These studies are still at an early stage, but show tremendous promise in expanding our knowledge of nucleon structure especially as computational power limitations become less of an issue.  


Hadronization is the process by which a quark turns into a color-neutral bound state that can be directly observed in the lab.  This nonperturbative process is described in cross section calculations with fragmentation functions (FFs).  The quark FF \( D_q^h (z, Q^2 ) \) describes the nonperturbative process of a quark \( q \) hadronizing into a particular hadron \( h \) which carries a longitudinal fraction \( z \) of the initial quark's momentum: \( z = p_h / p_q \).  Just like \( x \), the longitudinal momentum fraction \( z \) ranges from 0 to 1.  At LO, \( D_q^h (z, Q^2 ) \) has its own probabilistic interpretation as the probability that a quark \( q \) exiting a high energy scattering event with momentum transfer \( Q \) will produce hadron \( h \) that carries \( z \) of the quark's momentum before it hadronized.  But again this physical interpretation becomes more complicated when we consider NLO effects from gluon radiation.  FFs have been constrained with data from SIDIS and also from \( e^+ e ^- \) annihilation, which does not include effects from initial-state nucleon structure.  

As indicated by Figure~\ref{Figure:PDFs}, PDFs change with the energy of the scattering event just like \( \alpha_s \).  Increasing the energy of an interaction can be thought of as shortening the length scale at which the proton is being probed, allowing an experiment to resolve smaller distances and so see higher contributions from gluons and sea quarks.  Fragmentation functions also depend on the scattering energy of the interaction.  Also, the probability that a quark or gluon will radiate another gluon depends on the amount of energy that is available, which in turn will affect the momentum distribution of the partons.  This effect is captured in the Dokshitzer–Gribov–Lipatov–Altarelli–Parisi (DGLAP) evolution equations, which allow for PDFs and FFs that are measured at one energy scale to be evolved and applied at a different energy scale. \cite{ DGLAP1, DGLAP2, DGLAP3 }  These equations allow us to extract information about the gluon PDFs even from DIS data, where gluons do not interact directly with the scattering lepton. 

A full cross section calculation of a proton-proton collision to a single hadron will require both parton distribution functions and fragmentation functions, in addition to the pQCD calculation of all of the relevant parton 2-to-2 scattering processes.  The fact that all three of these processes: the initial-state nonperturbative effects, the perturbative hard scattering, and the nonperturbative final-state hadronization, are described with separate expressions is referred to as factorization.  Which nonperturbative functions need to be included in a cross section calculation depends heavily on the process.  For example, the Drell-Yan (\( p + p \rightarrow \ell^+ + \ell^- + X \)) cross section calculation does not include any FFs because Drell-Yan itself does not include hadronization.  In contrast, \( e^+ \) \( e^- \) annihilation does not include any effects from initial-state nucleon structure and so \( e^+ + e^- \rightarrow h + X \) cross sections do not include PDFs.  Factorization is generally assumed for all cross section calculations, but it has only been rigorously proven in the following processes: \( \ell^+ + \ell^- \rightarrow h + X \), \( \ell + p \rightarrow \ell + X \) and \( p + p \rightarrow \ell^+ + \ell^- + X \). \cite{CollinsTextbook} Tightly coupled with the assumption of factorization is the assumption of universality.  This assumes that all of these nonperturbative functions will remain the same regardless of the interaction that they are describing.  So the PDFs that were measured by the high energy electron-proton collider HERA, can be used to analyze proton-proton data at the LHC.   Universality also allows PDF extractions like the one in Figure~\ref{Figure:PDFs} to fit to data sets from multiple different collision systems.  By assuming both factorization and universality, perturbative QCD has been able to successfully interpret high energy collisions involving hadrons for a wide variety of colliding systems.

\section{Transverse Single-Spin Asymmetries}
\begin{figure}
  \centering
  \includegraphics[width=.8\textwidth]{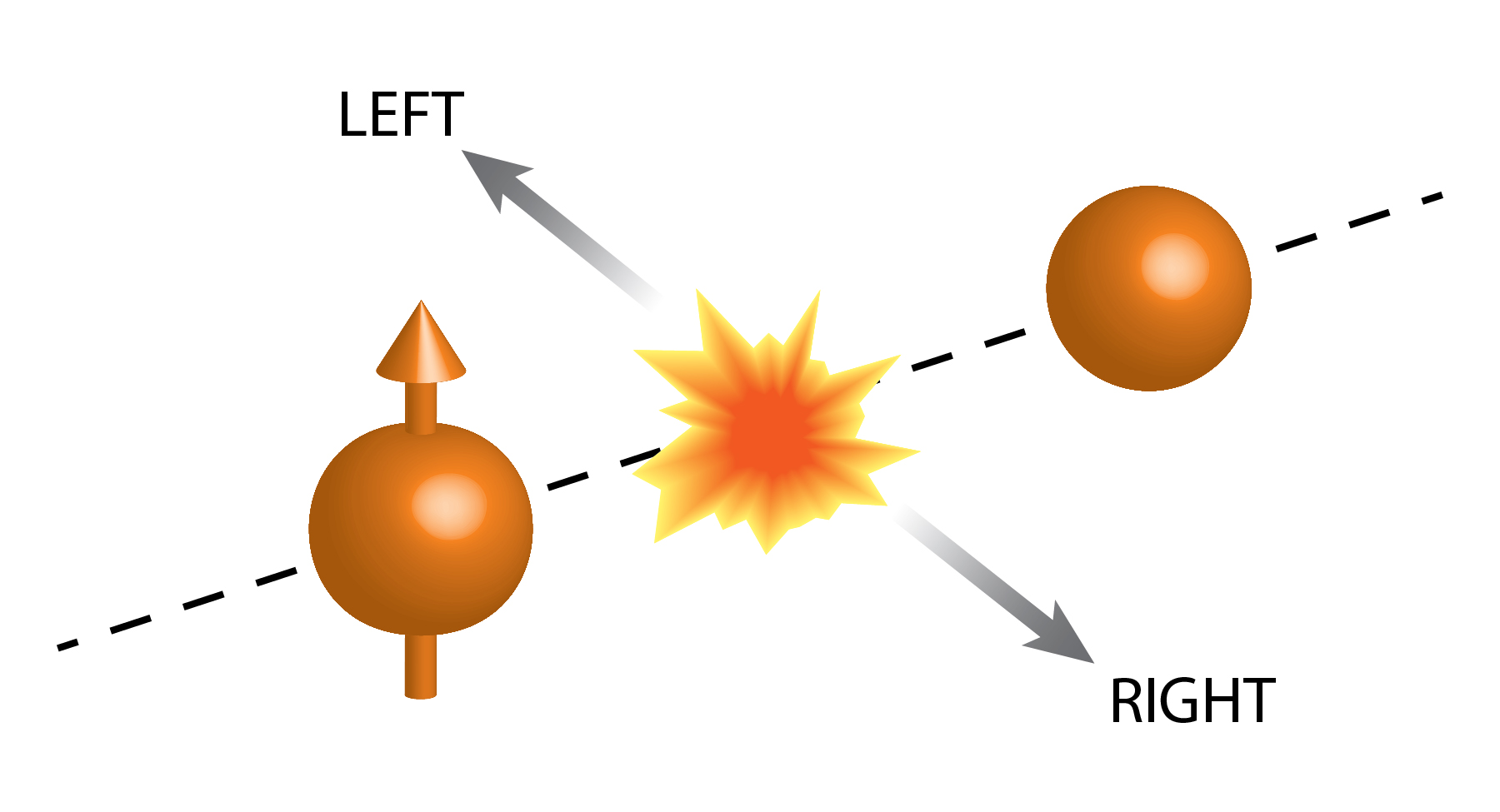}
  \caption[Diagram of the transverse single-spin asymmetry in proton-proton collisions. ]{Diagram of the transverse single-spin asymmetry in proton-proton collisions. TSSAs are measured in collisions between one transversely polarized proton and one unpolarized proton.  They measure the difference in yields of particles that travel to the left versus the right of the polarized-proton-going direction. \cite{PrettyPrettyTSSAPicture} }
  \label{Figure:TSSAprettypretty}
\end{figure}

In 1976 the spontaneous polarization of the \( \Lambda^0 \) baryon was measured for the first time in collisions between an unpolarized proton beam on a beryllium target.   It was found to be about 30\% \cite{LambdaPolarization}  even though pQCD calculations had found that the spin-momentum correlations from NLO quark and gluon scattering are small and fall off with increasing collision energies.\cite{GordyPaper} Around the same time the left-right asymmetry of charged pions was measured to be up to about 40\% in collisions between a transversely polarized proton beam and an unpolarized hydrogen target. \cite{sqrts4.9GeV}  This observable would later come to be known as a transverse single-spin asymmetry (TSSA), which is a spin-momentum correlation that is measured in hadronic collisions between one transversely polarized particle and one unpolarized one.  As depicted in Figure~\ref{Figure:TSSAprettypretty}, it measures the asymmetry in yields of particles that travel to the left versus the right with respect to the direction that the transversely polarized particle is traveling:
\begin{equation}
A_N = \frac{ \sigma_L - \sigma_R}{\sigma_L + \sigma_R}
\end{equation}

\begin{figure}
  \centering
  \includegraphics[width=\textwidth]{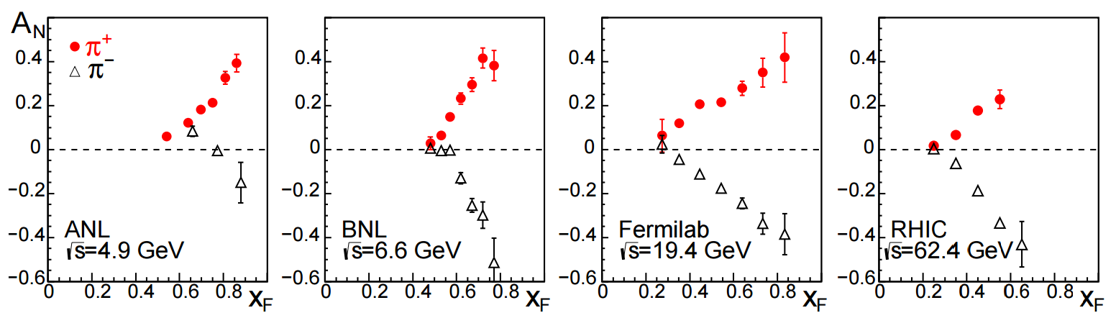}
  \caption[The charged pion TSSA has been measured to have a large \( x _F \) dependence over a wide range of collision energies.]{The charged pion TSSA has been measured to have a large \( x _F \) dependence over a wide range of collision energies. \cite{NonzeroAsymmetry}}
  \label{Figure:NonzeroTSSAs}
\end{figure}

Following the NLO pQCD results from Ref. \cite{GordyPaper}, if the sole source of these asymmetries was hard partonic scattering, these asymmetries would be very small (less than 1\%) and be proportional to \( A_N \sim \alpha_s \, m_q / p_T \), where \( m_q \) is the mass of the scattering quark and \( p_T \) is the measured hadron momentum transverse to the beam.  But large nonzero TSSAs have been measured for a wide range of collision energies as shown in Figure~\ref{Figure:NonzeroTSSAs}. \cite{sqrts4.9GeV, sqrts6.6GeV, sqrts19.4GeV, sqrts62.4GeV}  The asymmetries in this plot are evaluated as a function of Feynman \( x \), \( x_F = 2 p_z / \sqrt{s} \) where \( p_z \) is the pion's momentum parallel to the beam, which 
allows for the comparison of results from hadronic collisions with multiple center of mass energies.  Nonzero forward \( \pi^0 \) TSSA have also been measured  at collision energies up to \( \sqrt{s} = 500 \) GeV with \( p_T = 7 \) GeV/c \cite{STAR500GeVpreliminary}, well into the perturbative regime of QCD.   Because the perturbative part of the calculation cannot account for the large spin-momentum correlations that have been measured, we must reexamine our nonperturbative functions.  This led to the development of two theoretical frameworks: transverse momentum dependent functions and twist-3 collinear correlation functions which describe spin-momentum correlations within the nucleon and in the process of hadronization.  



\subsection{Transverse Momentum Dependent Functions}
Traditional PDFs and FFs are collinear, meaning they only depend on the longitudinal momentum fractions and integrate over the nonperturbative dynamics of partons within the proton and in hadronization.  Transverse momentum dependent functions (TMDs), as the name implies, explicitly depend on relative parton transverse momentum.  In order for TMD factorization to apply, this parton transverse momentum needs to be nonperturbative and much smaller than the hard scattering energy.  Thus, it is most straightforward to extract a TMD function from a two-scale process that is sensitive to both a soft- and a hard-momentum scale.  This includes SIDIS where the energy and scattering angle of the electron can be used to directly measure the hard-scale energy and the transverse momentum of the hadron can serve as the soft-scale.  The Collins-Soper-Sterman (CSS) evolution equations describe how these TMD functions evolve with the hard scattering energies, \cite{CSS1, CSS2} similar to the DGLAP evolution equation for the collinear functions.  But unlike the DGLAP evolution equations, CSS evolution depends on nonperturbative contributions \cite{CSS3} which are still in the process of being constrained by data. \cite{CSS2}

The TMD formalism began with the Sivers TMD PDF, \(  f_{1T}^{\perp } (x,  k_T^2 ) \),  which was first proposed by Dennis Sivers in 1990 as a way of explaining the large TSSA measurements in Figure~\ref{Figure:NonzeroTSSAs}. \cite{SiversPaper1, SiversPaper2} The Sivers function describes the correlation between the transverse spin of the proton and the nonperturbative transverse momentum, \( k_T \), of the parton within it.  It has a QED analogue in the hyperfine structure: the shifting of atomic spectral lines due to the coupling between the orbital angular momentum of the electron and the spin of the nucleus.  This spin-momentum correlation was shown to be able to generate these large TSSAs in \( p^\uparrow + p \) collisions, a phenomenon which is known today as the ``Sivers Effect''.  A few years later, John Collins published a paper claiming that the Sivers function had to be zero because it violated time reversal invariance.  As an alternative explanation for these large TSSA, he proposed a transverse momentum dependent fragmentation function which came to be known as the Collins function, \( H_1^{\perp} (z, j_T^2 ) \). \cite{CollinsPaper}  The Collins function describes the spin-momentum correlation between the transverse spin of a quark and the soft-scale relative transverse momentum, \( j_T \), of the unpolarized hadron it produces.   The Collins function was shown to be able to generate these large TSSAs in \( p^\uparrow + p \) collisions with what came to be known as the ``Collins effect'': the Collins function is convolved with the transversity function, a (in this case) collinear PDF  that describes the spin-spin correlation between the transverse spin of the proton and the transverse spin of the scattering quark.


Nearly a decade later, Brodsky, Hwang, and Schmidt pointed out that Collins’ Sivers function calculation was not fully gauge invariant in the TMD framework.\cite{SiversSignChange1}  This concept was later dubbed “naive T-odd,” or more accurately parity and time (PT) reversal odd, when Collins revised his argument to show that the Sivers function could be nonzero if the scattered quark engaged in a soft gluon exchange with a left-over proton fragment.  This concept led to the prediction of modified universality for the Sivers function: there should be a relative sign between the Sivers function for the Drell-Yan (\( p + p \rightarrow \ell^+ + \ell^- + X \)) and SIDIS (\( \ell + p \rightarrow \ell + h + X \)) processes. \cite{SiversSignChange2, SiversSignChange3}  In SIDIS, this soft gluon exchange happens after the QED hard scattering event, between the scattered quark and the left over proton fragment.  Since this quark and proton fragment were once combined into a color-neutral bound state, they have opposite color charge, which means that this is an attractive interaction.  In Drell–Yan, a quark is scattered out of a proton and exchanges a soft gluon with the fragment of the other proton.  This gluon exchange happens before the QED hard scattering event such that the quark and fragment of the other proton have the same color charge, making this a repulsive interaction.    The HERMES collaboration \cite{SiversNonzeroHeremes} and later the COMPASS collaboration \cite{SiversNonzeroCompass} confirmed that the Sivers function was nonzero by extracting it from transversely polarized SIDIS collisions.  
Recent measurements of the Sivers function in Drell-Yan \cite{SiversDYCompass} and Drell-Yan like processes \cite{SiversDYSTAR} favor the modified universality prediction, but because these results are statistically limited, more measurements will be needed to draw a firmer conclusion.  The fact that the Sivers function is PT-odd becomes even more interesting when considering the scattering process of proton-proton to hadrons where soft gluon exchanges can happen both before and after the hard partonic scattering event.  This idea led to the prediction of color entanglement effects and factorization breaking in the \( p + p \rightarrow h_1 + h_2 + X \) scattering process. \cite{ObligatoryFactorizationCitation}  Because most information about the Sivers function has come from SIDIS, the gluon Sivers function has remained poorly constrained in comparison to the quark Sivers functions. 

The Collins function has also been measured to be nonzero in both in SIDIS \cite{SiversNonzeroHeremes} and in \(e^+e^-\) annihilation.\cite{CollinsBelle, CollinsBABAR}  Because it is chiral odd, the Collins function needs to be measured in conjunction with another chiral odd function to make the overall hadronic scattering process chiral even.  This is why the Collins effect involved the convolution between the Collins function and the chiral odd transversity function.  In the case of \(e^+e^-\) annihilation two final state hadrons are measured such that one Collins function is convolved with another.  In unpolarized SIDIS the Collins function can be measured with the Boer-Mulders function, \(  h_{1}^{\perp } (x,  k_T^2 ) \). \cite{COMPASSBoerMuldersCollins}  This is another PT-odd TMD PDF which describes the spin-momentum correlation between a quark's transverse spin and its own soft-scale transverse momentum \( k_T \).  \cite{BoerMulders}

Because leptons are not composite particles, QED processes like SIDIS and Drell-Yan are able to provide direct access to the kinematics of the scattering event.  This is not possible when studying hadronic collisions that only reconstruct a single final-state particle.  Not only must the hard scale be approximated with e.g. the transverse momentum of that final state particle, \( p_T \), there is no way to access a soft momentum scale.  In order to do so one would need to measure more particles in the event, such as multiparticle angular correlations \cite{ PPG095, PPG195, PPG217} or particle-in-jet asymmetry measurements. \cite{STARJetTSSA500GeV, STAR500GeVpreliminary}  But inclusive TSSA measurements have the advantage of a higher statistical precision when compared to multiparticle correlations or particle-in-jet asymmetries and do not include the systematic uncertainties associated with measuring jets.  While both the Sivers function and Collins function were proposed as a way of explaining the strikingly large TSSAs present in Figure~\ref{Figure:NonzeroTSSAs}, they need to be integrated over their soft-scale transverse momentum dependence in order to be applied to those measurements. These calculations do indicate that the Collins effect \cite{CollinsEffect} may have a smaller contribution than the Sivers effect \cite{SiversEffect} for TSSAs in \( p^\uparrow + p \rightarrow h + X \) collisions, which is supported by the small Collins asymmetry that was measured in a forward rapidity \( \pi^0 \)-in-jet measurement. \cite{STAR500GeVpreliminary}  But both the Sivers and Collins effects are needed for a full phenomenological study of hadron TSSA in proton-proton collisions, because measuring the asymmetry as a function of \( p_T \) provides no clear way of separating out initial- from final-state effects.  


\subsection{Twist-3 Collinear Correlation Functions}
The ``twist'' of a nonperturbative function is defined as the mass dimension minus spin of the operator within a matrix element in the Operator Product Expansion. \cite{CollinsTextbook}  Traditional PDFs and FFs are leading twist, or twist-2, and only consider the scattering of one parton at a time.  The next term in this expansion is not scattering off of two partons, but the quantum mechanical interference between scattering off of one parton versus scattering off of two. These are referred to as twist-3 correlation functions and are split into two types: the quark-gluon-quark functions (qgq) and the trigluon (or three-gluon) functions (ggg).  When considering initial-state effects from proton structure, the qgq functions describe the quantum mechanical interference from scattering off of one quark versus scattering off of a gluon and a quark of the same flavor, while the ggg functions capture the quantum mechanical interference between scattering off of one gluon versus scattering off of two.  There are also related twist-3 correlation functions for final-state hadronization effects which describe the quantum mechanical interference between two partons undergoing hadronization together versus a single parton hadronizing on its own.

These multiparton correlation functions can be used to describe spin-momentum correlations from both initial-state and final-state effects.  They have the added benefit that they do not depend on a soft momentum scale and so are uniquely suited to describe TSSA in proton-proton collisions where only one final state particle is measured.   Rather than representing entirely new nonperturbative parton dynamics, many of these twist-3 correlation functions are related to the \( k_T \) moments of twist-2  TMD PDF and FFs.  \cite{Twist3toTMD1, Twist3toTMD2}  The fact that these twist-3 functions are nonzero reflects that scattering partons do in fact interact with their surrounding color fields.  

\pagebreak
In the twist-3 collinear QCD factorization scheme, the polarized cross section for a general \( A^\uparrow + B \rightarrow C + X \) process is written as:
\begin{equation}\label{Equation:Twist3}
\begin{split}
d\sigma( S_T) = H \otimes f_{a/A (3)} \otimes f_{b/B (2)} \otimes D_{C/c (2)} \\
	                 + H' \otimes f_{a/A (2)} \otimes f_{b/B (3)} \otimes D_{C/c (2)} \\
	                 + H'' \otimes f_{a/A (2)} \otimes f_{b/B (2)} \otimes D_{C/c (3)}
\end{split}
\end{equation}
\noindent  where \( f_{a/A (t)} \) is the PDF associated with the parton \( a \) in the transversely polarized hadron \( A^\uparrow \), \( f_{b/B (t)} \) is the PDF associated with the parton \( b \) in the unpolarized hadron \( B \), and \( D_{C/c (t)} \) is the FF associated with parton \( c \) producing the final-state (unpolarized) hadron \( C \).  The number in the subscripts of the nonperturbative functions corresponds to the twist of that function.  (Technically both PDFs and FFs are defined as twist-2 objects, but phrases like ``twist-3 PDFs'' and ``twist-3 FFs'' are common in the literature.)  The \( H \), \( H' \), and \( H'' \) are the hard pQCD parts of the calculation and the \( \otimes \) symbol denotes the convolution of each term across the appropriate momentum fractions.  Evaluating this expression for a given scattering process will include summing over all appropriate scattering channels and parton flavors.\cite{PitonyakPaper} 

The first term of Equation~\ref{Equation:Twist3} is often described as the ``Sivers-like'' term and contains spin-momentum correlations from partons in the polarized hadron.  Depending on the process, \( f_{a/A (3)} \) can include contributions from both qgq and ggg twist-3 correlation functions.  The second term is referred to as the ``Boer-Mulders-like'' term and contains spin-momentum correlations of a transversely polarized parton within the unpolarized hadron.  \( f_{b/B (3)} \) only includes quark-gluon-quark functions and not trigluon correlation functions because gluons are massless and so cannot be transversely polarized.  
(A twist-3 linearly polarized gluon correlation function is not generally included with phenomenological discussions of TSSA at this point in time.)
The third term in Equation~\ref{Equation:Twist3} is often described as ``Collins-like,'' where the source of the spin-momentum correlation comes from the process of hadronization.  Unlike twist-2 parton distribution functions, twist-3 initial-state correlation functions do not have a probabilistic interpretation so there are less constraints placed on them, which poses a challenge to extracting these collinear twist-3 functions from data.  

The initial-state quark-gluon-quark correlation functions are often split into two different terms: the soft-gluon pole (SGP) and the soft-fermion pole (SFP).  These terms arise because TSSA are PT-odd which causes a pole in the complex plane to appear in the hard scattering part of the calculation.  This pole will cause either the gluon or the quark in the multiparton correlator to vanish, which leads to either a SGP or a SFP in the qgq correlator, respectively.  \cite{PitonyakPaper}
The Qiu-Sterman function, \( G_F( x, x ) \), is the SGP of the quark-gluon-quark function in the polarized proton.\cite{ QiuSterman1, QiuSterman2 }  (Note there are a few  common notations for the Qiu-Sterman function in the literature which also include \( F_{FT}(x ,x ) \) and \( T_F( x ,x ) \).)  It is related to the Sivers TMD PDF, \(  f_{1T}^{\perp } (x,  k_T^2 ) \), by: 
\begin{equation}\label{Equation:QuiStermanToSivers}
G_F( x, x ) = \mp \frac{1}{ \pi M_N^2 } \int \mathrm{d}^2 k_T \, k_T^2 \, f_{1T}^{\perp (\pm) } (x,  k_T^2 ).
\end{equation}
\noindent where the \( \pm \) superscript takes into account the predicted Sivers sign change and \( M_N \) is the nucleon mass.\cite{Twist3toTMD2}  While both TMD and twist-3 collinear correlation functions provide a way forward into including a more complete three dimensional picture of the proton, there are a few subtleties hiding in the simplicity of these types of expressions. The first is that \(  f_{1T}^{\perp } (x,  k_T^2 ) \) is only defined for a very specific range of \( k_T \) and this relation will depend on what range of \( k_T \) it integrates over. The second is that while \(  f_{1T}^{\perp } (x,  k_T^2 ) \) and  \( G_F( x, x ) \) probably both contain some of the same color force dynamics present inside of the proton, this expression converts between the NLO term of two \textit{different} types of perturbative expansions.  But while expressions like Equation~\ref{Equation:QuiStermanToSivers} might contain some implicit caveats, they have allowed twist-3 correlation function phenomenology to take advantage the comparatively well-established TMD formalism, allowing them to describe TSSAs in proton-proton collisions.  Just like the polarized proton, the unpolarized proton has an equivalent SGP term for its own quark-gluon-quark function, \( E_F( x, x ) \), which is related to the Boer-Mulders function, \(  h_{1}^{\perp } (x,  k_T^2 ) \), by a similar \( k_T \) moment:
\begin{equation}\label{Equation:Twist3ToBoerMulders}
E_F( x, x ) = \pm \frac{1}{ \pi M_N^2 } \int \mathrm{d}^2 k_T \, k_T^2 \, h_{1}^{\perp (\pm) } (x,  k_T^2 ).
\end{equation}
\noindent where the \( \pm \) superscript takes into account that the Boer-Mulders functions is also PT-odd.  \cite{Twist3toTMD2} \( E_F( x, x ) \) is chiral odd and needs to be convolved with another chiral odd function, like the collinear twist-2 transversity function, to make the overall term chiral even.  The qgq correlation functions also include SFP terms for both the polarized and unpolarized protons.

The trigluon correlation function is only present in the first term of Equation~\ref{Equation:Twist3}.  
While the Qiu-Sterman function is able to take advantage of the previously extracted quark Sivers function, the trigluon correlation function has no equivalent to Equation~\ref{Equation:QuiStermanToSivers} because the gluon Sivers function remains comparatively poorly constrained.  
The trigluon correlation function was first proposed in Ref. \cite{Origggg} but was later clarified to have two independent trigluon correlation functions due to the difference in the contraction of color indices.  \cite{gggTwoFunctions1, gggTwoFunctions2 } These complex functions are often denoted as \( N(x_1, x_2) \) and \( O(x_1, x_2) \), where \( x_ 1 \) and \( x_ 2 \) are the linear momentum fractions of the single gluon and the gluon pair.  The overall cross section becomes real when  \( x_1 \) is set to \( x_2 \), creating a soft gluon pole (SGP) in these ggg function terms.    Current predictions for the trigluon correlation function's contribution to various TSSAs involve parameterizing in terms of the twist-2 unpolarized gluon PDF, \( G(x) \).

Heavy flavor TSSA measurements at the Relativistic Heavy Ion Collider (RHIC) are uniquely sensitive to the trigluon correlation function because most heavy flavor quarks that are produced at RHIC collision energies are created through gluon-gluon fusion.  This means that only the first term in Equation~\ref{Equation:Twist3} is nonzero, because the collinear twist-2 transversity function (\( f_{a/A (2)} \) Equation~\ref{Equation:Twist3}'s notation) is zero for gluons since they
cannot be transversely polarized.  Based off the kinematics of the twist-3 cross section for \( p^\uparrow + p \rightarrow D + X \), where D denotes a heavy flavor D meson which carries a charm quark, an estimate \cite{OpenHeavyFlavorTSSAggg} of the ggg function's contribution to the D meson TSSA assumed \( O(x, x) =O(x, 0) =  N(x, x) = - N(x, 0) \) and chose two different models for \( O( x, x) \):  
\begin{equation}
\mathrm{Model \, 1:} \, O(x, x) =  0.004 \, x \, G(x) 
\end{equation}
\begin{equation}
\mathrm{Model \, 2:} \,  O(x, x) = 0.001 \, \sqrt{x} \, G(x) 
\end{equation}
The coefficients 0.004 and 0.001 were determined such that the calculated \( A_N^D \) did not exceed the RHIC preliminary data for \( A_N^D \).  These two model ansatz were chosen in order to study the effect of the three-gluon correlations in comparison with the gluon density and study the sensitivity of the TSSA to small-x behavior. \cite{OpenHeavyFlavorTSSAggg}  The PHENIX forward open heavy flavor TSSA was found to be consistent with zero and also consistent with this trigluon correlation function prediction.\cite{PPG196} The TSSA of forward production of J/Psi, a charm-anticharm bound state, was also measured to be consistent with zero. \cite{PPG211} 

The forward pion TSSA in proton-proton collisions has been measured to be nonzero both at STAR\cite{sqrts200GeV, STARForwardPi0AndEta}, BRAHMS\cite{sqrts62.4GeV} and PHENIX \cite{PPG135}.  Forward is in relation to the polarized proton going direction and so samples a higher \( x \) region of the polarized proton when compared to other collision kinematics.  This means that light hadron production in the forward region is dominated by valence quarks in the polarized proton.   For many years it was assumed that the Qiu-Sterman function was the dominant source of these large forward pion \( A_N \) \cite{QiuSterman2, QiuStermanOnlyForwardPion1,  QiuStermanOnlyForwardPion2}. From Equation~\ref{Equation:QuiStermanToSivers} it follows that there are two ways to extract the Qiu-Sterman function: directly from the \(p^\uparrow p \rightarrow \pi X \) TSSA \cite{QiuStermanOnlyForwardPion1, QiuStermanOnlyForwardPion2} and by taking the \( k_T \) moment of the Sivers function that has already been extracted  from SIDIS data. \cite{SiversFunctionExtraction}  It was discovered however that these two approaches yielded different results which disagreed by a sign. \cite{SignMismatch}  The contribution from the unpolarized proton (the second term in Equation~\ref{Equation:Twist3}) had previously been found to be negligible. \cite{ForwardPionUnpolarized1, ForwardPionUnpolarized2}  Using the models that were developed for the open heavy flavor TSSA in Ref. \cite{OpenHeavyFlavorTSSAggg}, the ggg correlation function's contribution to light hadron TSSAs was found to be too small to account for this ``sign-mismatch'' problem. \cite{ForwardPionggg}  And the same went for the SFP term of the qgq correlation function in the polarized proton. \cite{ForwardPionSFP}

Eventually the there was enough progress on the hadronization contribution to TSSA, \cite{weCanDoTwist3FF} that the third term of Equation~\ref{Equation:Twist3} could be included in descriptions of the forward pion asymmetries.  Ref. \cite{forwardPi0qgqWithHadronization} found that it could simultaneously describe forward \( A_N^{\pi^0} \) and \( A_N^{\pi^\pm} \) results when they included these twist-3 effects from hadronization.  This result demonstrated progress towards resolving the ``sign-mismatch'' discrepancy since the Qiu-Sterman function that was used in this calculation came from applying Equation~\ref{Equation:QuiStermanToSivers} to two different extractions of the Sivers function from SIDIS data.  Part of the twist-3 final-state effects were described by a quark-gluon-quark correlation term that is related to the \( k_T \) moment of the Collins function.  These fits used an additional collinear twist-3 final-state correlation function which corresponds to the imaginary part of the qgq matrix element.  It was parameterized in terms of the standard twist-2 collinear unpolarized FF.  This term was found to be the dominant contribution to these large forward pion asymmetries, while the contribution from the \( k_T \) moment of the Collins function term was small.  Further measurements of these twist-3 fragmentation effects is needed before these forward pion asymmetries can be fully understood. But the fact that they found that the Collins effect is small agrees with both the TMD formalism \cite{CollinsEffect, SiversEffect} and a forward \( \pi^0 \)-in-jet Collins asymmetry measurement.  \cite{STAR500GeVpreliminary}

Since their inception, there has been theoretical evidence that the collinear twist-3 and TMD factorization pictures could combine to form a unified picture of TSSAs in hard processes.   This concept was recently put to the test with the first  simultaneous global analysis of TSSAs in SIDIS, Drell-Yan, \( e^+ e^- \) annihilation, and proton-proton collisions.  \cite{2020EverythingFits} This study used quark TMD PDFs and FFs to describe the asymmetries in processes that are sensitive to the soft scale momentum, i.e. SIDIS, Drell-Yan, and \( e^+ e^- \) annihilation.  To describe the forward pion asymmetries that were measured at RHIC, they used twist-3 qgq correlation functions that were calculated by taking the \( k_T \) moments of the same TMD functions that were used to describe TSSAs in the QED processes above.  They concluded that their simultaneous description of these TSSAs across multiple collision species indicated that all TSSAs had a common origin related to the quantum mechanical interference from multiparton interactions.  

\section{\( \pi^0 \) and \( \eta \) Mesons}
\begin{figure}
\centering
\includegraphics[width=.8\textwidth]{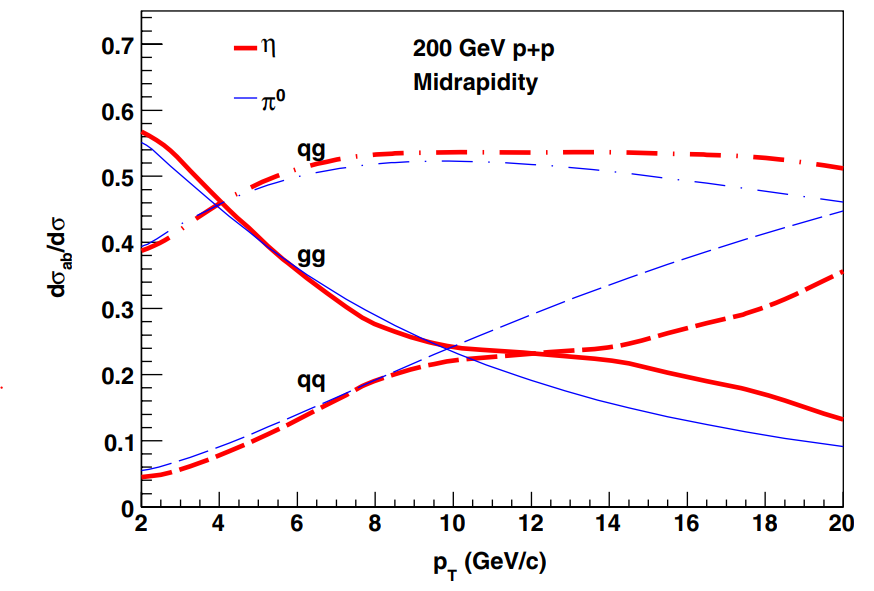}
\caption[The fractional contribution of parton scattering processes to \(\pi^0 \) and \( \eta \) meson at midrapidity in proton-proton collisions with \( \sqrt{s} = 200 \) GeV. ]{The fractional contribution of parton scattering processes to \(\pi^0 \) and \( \eta \) meson at midrapidity in proton-proton collisions with \( \sqrt{s} = 200 \) GeV. \cite{PPG107} }
\label{Figure:pi0AndEtaPartonScattering}
\end{figure}

At forward rapidity, the polarized proton is being probed at relatively high \( x \) and so light hadron TSSAs are dominated by valence quark spin-momentum correlations.  The forward \( \pi^0 \) asymmetry has been used to constrain quark-gluon-quark correlation functions both from the polarized proton and the process of hadronization.  \cite{QiuStermanOnlyForwardPion1, QiuStermanOnlyForwardPion2, forwardPi0qgqWithHadronization, 2020EverythingFits}  In contrast, midrapidity \( \pi^0 \) and \( \eta \) measurements probe the polarized proton at comparatively moderate \( x \) and so are sensitive to both quark and gluon dynamics at leading order.  This can be seen in Figure~\ref{Figure:pi0AndEtaPartonScattering} which shows the fractional contributions of different parton scattering processes to midrapidity \( \pi^0 \) and \( \eta \) production.\cite{PPG107}  Since \( \pi^0 \) and \( \eta \) mesons are both hadrons, their TSSAs are sensitive to quark and gluon spin-momentum correlations both in the proton and also in the process of hadronization.  

Midrapidity \( \pi^0 \) TSSA results from PHENIX have already been used to constrain the gluon Sivers function, \cite{ConstrainGluonsSivers1, ConstrainGluonsSivers2} which continues to have large uncertainties compared to the quark Sivers functions because  Sivers asymmetries measured with SIDIS are not sensitive to gluon dynamics at leading order.  These TMD calculations are done in what is often referred to as the generalized parton model (GPM) which takes the \( k_T \) moment of TMD functions such that they can be applied to single-scale measurements and also conditionally assumes that TMD PDFs and FFs are universal by not including NLO interactions with the proton fragments.  The color gauge invariant generalized parton model (CGI-GPM) relaxes this assumption of universality by allowing for initial- and final-state interactions through the one-gluon exchange approximation.  The CGI-GPM has been shown to reproduce the quark Sivers function's predicted sign change between the SIDIS and Drell-Yan processes.  The midrapidity \( \pi^0 \) TSSA has been shown to have the potential of distinguishing between the GPM and CGI-GPM frameworks.  \cite{CGI-GPM} Similarly, midrapidity light hadron TSSAs have also been also shown to be sensitive to the poorly constrained trigluon correlation function in the transversely polarized proton.  \cite{ForwardPionggg}


There are enough similarities between \( \pi^0 \) and \( \eta \) mesons to make differences in their results interesting.  They are both flavorless pseudoscalar mesons  which are made up of light quarks, so they are produced under more or less similar circumstances.  But because their quark content differs, \( \pi^0 = \frac{1}{ \sqrt{2} }( u\bar{u} - d\bar{d} ) \) and \( \eta \approx \frac{1}{ \sqrt{6} }( u\bar{u} + d\bar{d} - 2s\bar{s} ) \), differences in \( A_N^{\pi^0} \) and \( A_N^\eta \) results are sensitive to the potential effects that isospin and strangeness have on spin-momentum correlations.  Also since the \( \eta \) meson is about four times heavier than the \( \pi^ 0 \),  differences in their TSSA results could also be caused by hadron mass.  At midrapidity, previous measurements of both asymmetries have been consistent with zero, but at forward rapidity there has been some indication that the \( \eta \) meson asymmetry is slightly larger than the \( \pi^0 \). \cite{PPG170, STARForwardPi0AndEta} However, more data is needed in order to make a more definitive statement.  

The updated midrapidity \( \pi^0 \) and \( \eta \) asymmetries presented in this document are a factor of 3 increase in precision from previously published results and extend to higher \( p_T \).  \cite{PPG135}  Since these analyses use PHENIX's last polarized proton data set, these will be the final midrapidity \( \pi^0 \) and \( \eta \) TSSA results from PHENIX.  

\section{Direct Photons}
\begin{figure}
\centering
\includegraphics[width=.8\textwidth]{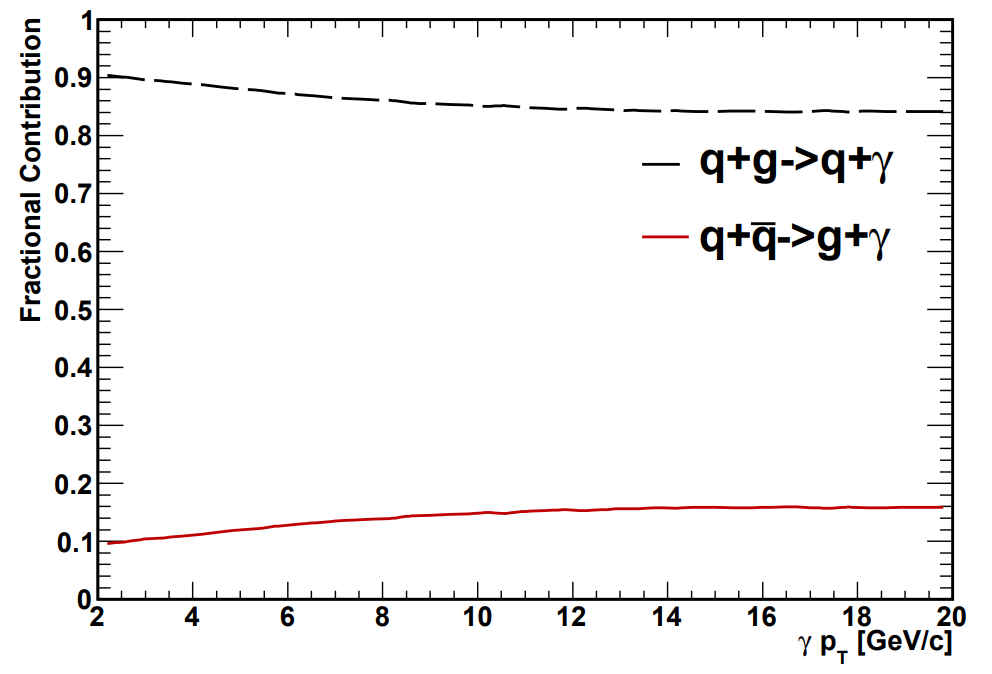}
\caption[The fractional contribution of different partonic processes to direct photon production at midrapidity in proton-proton collisions with \( \sqrt{s} = 200 \) GeV. ]{The fractional contribution of different partonic processes to direct photon production at midrapidity in proton-proton collisions with \( \sqrt{s} = 200 \) GeV. \cite{PPG095} }
\label{Figure:dpPartonScattering}
\end{figure}

In contrast to \( \pi^0 \) and \( \eta \) mesons, direct photons do not undergo hadronization and so are only sensitive to initial-state effects, i.e. proton structure.  At leading order in pQCD, direct photons are produced \textit{directly} from the hard scattering of partons.  At large transverse momentum they are predominately produced in QCD 2-to-2 hard scattering subprocesses: quark-gluon Compton scattering (\( g + q \rightarrow \gamma + q \)) and quark-antiquark annihilation (\( \bar{q} + q \rightarrow \gamma + g \)).  As Figure ~\ref{Figure:dpPartonScattering} shows, Compton scattering dominates over quark-antiquark annihilation at midrapidity \cite{PPG095} because the gluon PDF is much larger than the antiquark PDFs even at smaller \( x \).  This means that direct photon production is sensitive at leading order to both quark and gluon distributions in the proton.  So quark PDFs that have already been previously measured in cleaner QED processes like DIS can be used as inputs to direct photon phenomenological calculations such that the gluon PDF can be cleanly extracted.  

At NLO, photons can also be produced by a quark emitting bremsstrahlung radiation.   Results that include these fragmentation photons not only depend on final-state nonperturbative effects, but analyzing them requires using parton-to-photon fragmentation functions which are poorly constrained compared to their parton-to-hadron counterparts.  \cite{PhotonFragementationFunction}  An isolation cut is used to suppress the contribution from these fragmentation photons (in addition to photons from hadronic decays) by requiring that the energy of the photon be much greater than the energy of all of the surrounding event activity.  \cite{PPG136}  
The remaining contribution of fragmentation photons to the isolated direct photon sample has been estimated to be less than 15\% \cite{PPG095} and results using these isolated direct photons can then be compared with NLO calculations which apply the same isolation criteria.  Isolated direct photons at PHENIX have been used to measure nonperturbative transverse momentum effects in azimuthal correlations \cite{PPG095, PPG195, PPG217} and used to probe the dynamics present in different heavy ion collision systems. \cite{PPG210}

Direct photons do not interact via the strong force and so do not undergo hadronization.  Thus, the direct photon TSSA has been proposed as a clean probe of both quark and gluon spin-momentum correlations in the proton.  At forward rapidity, direct photon production is dominated by valence quarks in the polarized proton and so the  forward direct photon TSSA has been found to be sensitive to the quark-gluon-quark correlation functions.  Ref. \cite{dpTSSAqgq} calculated the full quark-gluon-quark correlation functions' contribution to the direct photon TSSA, which included the SGP and SFP terms for both the polarized and unpolarized proton.  
The SGP term for the unpolarized proton was calculated for the first time using the Boer-Mulders function and Equation~\ref{Equation:Twist3ToBoerMulders}.  The SFP contribution from the unpolarized proton disappeared after summing over all possible Feynman diagrams. After combining all contributions, they found as expected that the qgq contribution would be negligible at backward rapidity.  At forward rapidity, they found that the Qiu-Sterman function dominated to produce an overall negative asymmetry.  Because the trigluon correlation function's contribution is small at forward rapidity, the forward direct photon TSSA has been proposed as a method of cleanly extracting the Qiu-Sterman function and even having the potential for providing insight into the Sivers sign change.  The Qiu-Sterman function's contribution to the direct photon TSSA has also been studied in the context of color entanglement in the twist-3 factorization framework.  It was found that this contribution changed when one considered the quantum mechanical effects of simultaneous initial- and final-state soft gluon exchanges.  This means that the forward direct photon TSSA could also serve as a potential probe of nontrivial gauge links in collinear factorization.  \cite{dpTwist3ColorEntanglement}  A calculation of the quark Sivers function's contribution to the forward direct photon TSSA predicted the opposite sign for the asymmetry. \cite{SiversEffect} So it could also be used as a way of comparing the collinear twist-3 and TMD approaches and evaluating the validity of parton model identities like those in Equations ~\ref{Equation:QuiStermanToSivers} and \ref{Equation:Twist3ToBoerMulders}.

At backward rapidity, the transversely polarized proton is being probed at much lower \( x \) and so the direct photon TSSA is dominated by gluon dynamics in the polarized proton.  Calculations of the ggg function's contribution to the direct photon TSSA have shown that the backward rapidity asymmetry is sensitive to the magnitude of the trigluon correlation function.  \cite{dpTSSAggg} This calculation used the same parameterizations to make the open heavy flavor TSSA prediction in Ref. \cite{OpenHeavyFlavorTSSAggg} and found that for \( x_F < 0 \) (backward production) the direct photon asymmetry is sensitive to the relative sign between the \( O( x_1, x_2 ) \) and \( N( x_1,  x_2 ) \) functions.  The direct photon TSSA at RHIC energies has also been found to be sensitive to the gluon Sivers function in the TMD factorization framework, both at midrapidity and backward rapidity.  Ref. \cite{dpConstrainGluon} found that at backward rapidity where the contributions from quark spin-momentum correlations are the most suppressed, the direct photon TSSA could be as large as 10\% with current constraints to the gluon Sivers function and that this potentially large asymmetry was suppressed with the inclusion of soft gluon exchanges.  

At midrapidity the direct photon TSSA is sensitive to both quark and gluon dynamics in the proton and it is not affected by hadronization.  
The only previous direct photon TSSA measurement was published in 1995 by the E704 experiment at Fermilab.  It was measured with a 200 GeV/c polarized proton beam on a unpolarized proton target and found to be consistent with zero with large error bars. \cite{FNALE704dpTSSA}  This will be the first direct photon TSSA published at RHIC, which will help constrain gluon spin-momentum correlations in the transversely polarized proton.  

Transverse single-spin asymmetries are spin-momentum correlations which probe parton dynamics present in the proton and the process of hadronization.  They can be analyzed through the TMD framework which explicitly keeps information on nonperturbative parton transverse momentum.  Alternatively, collinear twist-3 functions describe the quantum mechanical interference between interacting with two partons versus interacting with one and have been shown to be able to generate large TSSAs.  The TMD and collinear twist-3 frameworks both encode spin-momentum correlations that come from partons interacting with their surrounding color fields.  The results presented in this dissertation measure the TSSA of a single particle produced in a proton-proton collision event and calculate the asymmetry as a function of this particle's \( p_T \).  Thus in order to analyze these TSSA measurements in the TMD framework the \( k_T \) moment of TMD functions must be taken, while collinear twist-3 functions can be applied to these TSSA measurements directly.  This dissertation describes the direct photon TSSA which is a clean probe of proton structure and the TSSAs of \( \pi^0 \) and \( \eta \) mesons which are sensitive to both initial- and final-state effects.  These results will constrain the trigluon collinear twist-3 function as well as the gluon Sivers function.

%% file: Chap2/chap2.tex
\section{Relativistic Heavy Ion Collider}\label{Section:RHIC}
The Relativistic Heavy Ion Collider (RHIC) is located on Long Island, New York at Brookhaven National Lab (BNL).  It is one of two high-energy hadronic colliders in the world that are currently operating and the only collider that is able to run polarized proton beams.  This allows for studies of spin-spin and spin-momentum correlations at much higher energies  compared to polarized fixed target experiments. RHIC is also unique in that it is able to run a wide variety of light and heavy ion beams including gold, deuterium, copper, and uranium.  
 
\begin{figure}
  \centering
  \includegraphics[width=1.0\textwidth]{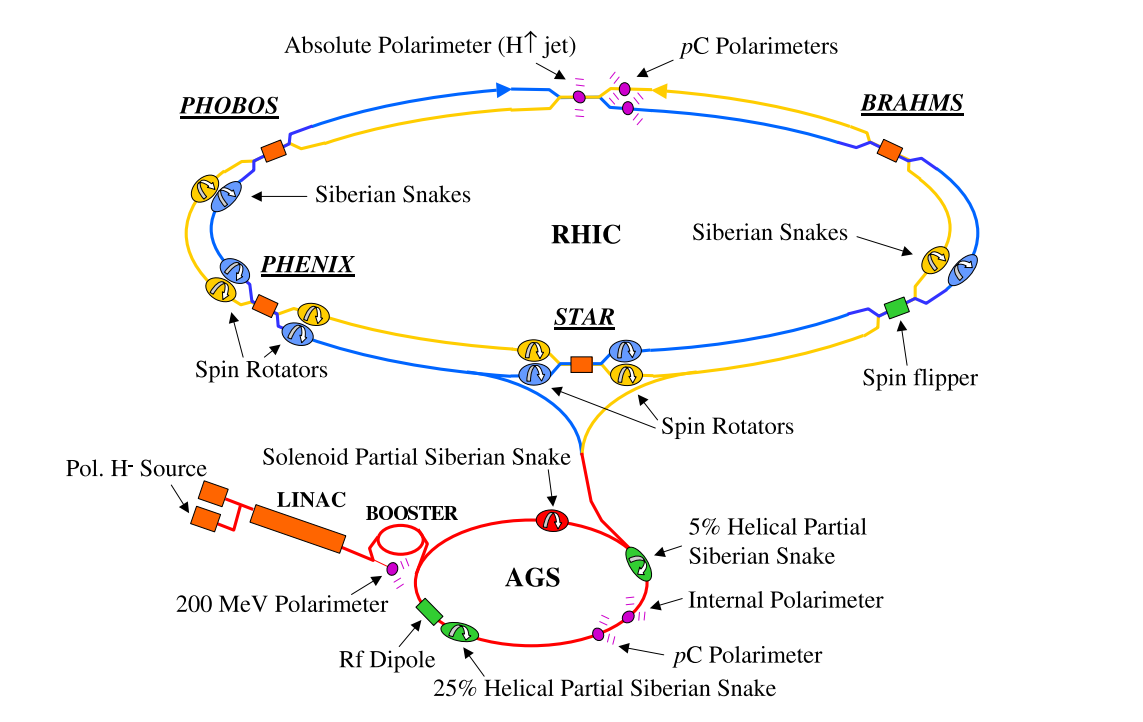}
  \caption[Schematic of the RHIC-AGS equipment required to run polarized proton beam]{Schematic of the RHIC-AGS equipment required to run polarized proton beam}
  \label{Figure:RHIC}
\end{figure}

RHIC is 3.8 km in circumference and made up of two separate rings that are capable of running both heavy ion and polarized proton beams.  The beam that travels counterclockwise is referred to as the yellow beam and the clockwise beam is called the blue beam, as illustrated in Figure~\ref{Figure:RHIC}, named for the colored stripes painted on their respective magnet systems.  Each beam has 120 separate bunches that collide in 106 ns intervals at up to six interaction points.  Originally there were four major detectors taking data at the RHIC complex: PHENIX, STAR, BRAHMS, and PHOBOS, of which STAR is the only experiment that is still currently taking data.  STAR will be joined by a follow up experiment to PHENIX called sPHENIX, which is scheduled to start taking data in 2023.\cite{RHIC1,RHIC2}

A number of technological developments were required to make RHIC not only polarize the proton beam but maintain the polarization as the beam is steered around the RHIC ring by strong magnetic fields.  Polarized proton bunches are injected into RHIC one bunch at a time, which allows the polarization direction of each bunch to be selected independently.    Not only does this reduce the systematics of polarized measurements, it also allows for polarization-averaged analyses to be done with the same proton-proton data sets.  The polarization direction of each bunch is set at the beginning of the fill and care is taken to ensure that there are nearly the same number of bunches that are polarized up as polarized down.  The fact that the polarization direction changes can also be used as a tool in spin-spin and spin-momentum correlation measurements.  

\subsection{Polarized Source and Injection}
The source for polarized proton beams is optically polarized \( H^- \) ions and is referred to as the Optically-Pumped Polarized Ion Source (OPPIS).\cite{RHIC3}  It starts with unpolarized protons picking up electrons that were polarized in an optically pumped rubidium vapor that was placed in a high magnetic field, making hydrogen atoms with a polarized electron.  The electron's polarization is then transferred over to the proton nucleus with static magnetic fields.  These atoms then pick up a second unpolarized electron from a sodium vapor negative-ionizer cell, forming \(H^-\) ions with polarized proton nuclei.  

A pulse of polarized \(H^-\) ions is then transferred to the Radio Frequency Quadrupole (RFQ) cavities and then the Linear Accelerator (LINAC) where is it accelerated to 200 MeV.   As it is injected into the Booster, the \(H^-\) ions are stripped of their electrons producing a bunch of polarized protons that the Booster then accelerates to 1.5 GeV.  This bunch is then transferred to the Alternating Gradient Synchrotron (AGS) and accelerated to 25 GeV before being injected into RHIC where it can be accelerated to a maximum of 255 GeV.  Each bunch is accelerated in the AGS and injected into RHIC separately, allowing for the polarization direction to differ bunch to bunch.  

For Run-15 \( p + p\) data, taken in 2015, each polarized proton beam was accelerated to 100 GeV for a total center of mass energy of \( \sqrt{s} = 200 \) GeV.  Of the 120 bunches, 109 bunches were filled with protons and nine consecutive bunches were left empty to give the abort kicker magnet enough time to raise its current to the level needed to dump the beam.  Two additional bunches were left empty to serve as a crosscheck for the bunch patterns.  These empty bunches are called the abort gap. A collision between two bunches in separate beams is called a crossing.  The beam luminosity for Run-15 was about \(10^{29} \) to \(10^{32}\) cm\textsuperscript{-2}s\textsuperscript{-1} and the average beam polarization was 55\%.  The integrated luminosity delivered to the PHENIX interaction point was 196.7 pb\textsuperscript{-1}.  Each cycle of beam storage is referred to as a fill and identified by a fill number.  Each fill typically runs for about 8 hours, though it can be dumped earlier because of nonideal beam conditions such as low intensity or polarization or due to equipment problems.  

\subsection{Maintaining the Polarization of the Beam}
Once the polarized proton bunches are injected into RHIC, the next challenge is maintaining the polarization.  Over the course of a fill some loss of polarization is unavoidable because of the magnetic fields used to steer the beam.  But certain depolarization effects can be additive due to the procession of the protons around the ring.  The precession of the spin vector \( \vec{ P } \) of a relativistic proton traveling in a circle is given by the Thomas-BMT equation: \cite{SpinPrecession1, SpinPrecession2}
\begin{equation}
\frac{ d \vec{P} }{dt} = - \Big( \frac{e}{\gamma M} \Big) \Big( G \gamma \vec{ B_\perp } + (1 + G) \vec{ B_\parallel} \Big) \times \vec{ P }
\label{Equation:ProtonPrecession}
\end{equation}
\noindent where \( e \) is the electric charge of the proton, \( M \) is the proton mass, \( \gamma \) is the relativistic Lorentz factor and \( G = 1.7928 \) is the anomalous magnetic moment of the proton.  \( \vec{ B_\perp }\) is the component of the magnetic field that is perpendicular to the plane of the accelerator and  \( \vec{ B_\parallel }\) is the longitudinal component. For this equation, \( \vec{ P } \) is measured in the rest frame of the proton.  As the speed of the proton increases, so does \( \gamma \) which means the \( \vec{ B_\perp }\) term will dominate.  The magnetic field is of course necessary to steer the protons around the ring and the equation of motion for a charged particle orbiting in a magnetic field is:
\begin{equation}
\frac{ d \vec{v} }{dt} =  - \left( \frac{e}{\gamma M} \right) \vec{ B_\perp } \times \vec{ v }
\label{Equation:ProtonVelocity}
\end{equation}

\noindent By comparing Equations~\ref{Equation:ProtonPrecession} and ~\ref{Equation:ProtonVelocity}, we can see that for each revolution around the RHIC ring the spin of each proton precesses \( G \gamma \) times.  \( \nu_{sp}  = G \gamma \) is referred to as the spin tune.  

\pagebreak
Looking at Equation~\ref{Equation:ProtonPrecession} we can see that the polarization direction would be stable if \(\vec{ B_\parallel} \) were negligible and if \( \vec{ P }  \) and \( \vec{ B_\perp } \) pointed in the same direction.  For this reason, each proton bunch is polarized perpendicular to the accelerator plane for the majority of time that it travels through RHIC. 

But magnetic fields that are parallel to the beam axis are unavoidable and can cause additive depolarization effects that occur when the spin precession frequency is such that the polarization is pointing in the exact same direction each time the bunch encounters a particular depolarizing field.  These resonance effects are categorized into two different types: imperfection and intrinsic resonances.  Imperfection resonances are caused by small errors in the magnetic currents and alignments which occur more or less randomly around the ring.  These imperfections cause resonant depolarization when the spin tune is an integer, \( \nu_{sp}  = G \gamma  = n \).  Intrinsic resonances are caused by the beam focusing quadrupole magnets.  Intrinsic resonances occur when \( \nu_{sp}  = G \gamma  = kP \pm \nu_y \), where \( k \) is an integer and \( P \) is the superperiodicity or the regularity of the focusing-defocusing lattice; at RHIC \( P = 12 \).  \( \nu_y \) is the vertical component of the betatron tune or the number of oscillations per beam revolution possible during a stable beam in the vertical plane; at RHIC \( \nu_y \approx 8.8 \).  The closer \( \nu_{sp} \) gets to these resonant frequencies, the faster the beam polarization is lost.  

In order to avoid both imperfection and intrinsic resonances, a series of spin-rotating helical dipoles called Siberian snakes are installed around both the RHIC and AGS rings.  Each RHIC ring has two Siberian snakes \cite{SiberianSnake} at diametrically opposite points along the rings, which flip each bunch's spin direction 180\( ^\circ \) without distorting the trajectory of the beam.  This makes the spin tune a half integer and causes these additive effects from the RHIC magnets to cancel out.  The AGS does not have enough space for a full snake, but has two partial snakes that rotate the spin vector by less than 180\( ^\circ \) to keep the spin tune away from integer values.

\pagebreak
Even though transversely polarized proton beams are the most stable, there is still interesting physics to be measured with a longitudinally polarized proton beam.  In particular, double longitudinal-spin asymmetries are  sensitive to the helicity distributions of partons within the proton.  To achieve these collisions, spin rotators are located outside of the interaction regions of both PHENIX and STAR which can rotate each bunch's polarization from transverse to longitudinal.  After the crossing, the longitudinally polarized bunch is then returned to a transverse polarization by another spin rotator as it leaves either PHENIX or STAR.  This way each experiment can independently choose the polarization direction for each of their data sets.

\subsection{Polarimeters}\label{Section:Polarimeters}
Knowing the polarization of the beam is not only important for monitoring the performance of the beam, but the absolute polarization is needed as a correction in spin analyses.  The RHIC polarimeters are located at the 12 o’clock position on the ring and use two separate polarimeters to measure the beam polarization in the vertical direction.  The basic idea behind measuring the polarization is to measure an already known physics asymmetry, \( A_N \).   The measurement of this asymmetry, \( \epsilon_N  \), will then be diluted by the beam polarization, \( P^{beam} \):
\begin{equation}
A_N = \frac{1}{ P^{beam} }\epsilon_N
\end{equation}
\noindent If the beam were 100\% polarized then \( \epsilon_N  = A_N \), but since it is not, the measured asymmetry can then be compared to the physics asymmetry and used to calculate the polarization of the beam: \( P^{beam} = \epsilon_N / A_N \).  Thus, the usefulness of the polarimetry measurement method is determined both by how precisely the raw asymmetry can be measured and by how well the physics asymmetry is known.  

\pagebreak
The proton-carbon (\( p \)C) polarimeter \cite{pCPolarimeter} studies the elastic scattering of polarized protons on a carbon target using an array of silicon detectors.  This asymmetry is caused by Coulomb-nuclear interference (CNI) or the interference of the electromagnetic and hadronic elastic scattering amplitudes. The \( p \)C  polarimeter measures the left-right asymmetry of elastically scattered protons off of a carbon target, which is known to be \(  A_N^{p\textrm{C}} \approx 0.01 \):
\begin{equation}
  P^{beam} = \frac{1}{ A_N^{p\textrm{C}} }
    \frac
    {\sqrt{N^\uparrow_L N^\downarrow_R } - \sqrt{N^\downarrow_L N^\uparrow_R}}
    {\sqrt{N^\uparrow_L N^\downarrow_R } + \sqrt{N^\downarrow_L N^\uparrow_R}}  
\end{equation}

\noindent where \( N^\uparrow \) and \( N^\downarrow \) refer to counts where the polarization of the beam was pointing up or down respectively, and \( N_L \) and \( N_R \) refer to the number of carbon atoms that recoil to the left or right of the polarized beam going direction.  (This equation can be compared to Equation~\ref{Equation:SquareRootFormula} in Section~\ref{Section:SquareRootFormula}.) This is a fast measurement that can get up to 2 or 3\% statistical precision within a few minutes of data taking and so this measurement can be performed multiple times in a fill to monitor the depolarization of the beam.  The physics asymmetry for this process, however, is not known to very high precision, so the absolute asymmetry needs to be measured using a different method.  

The hydrogen jet target (H-jet) polarimeter \cite{HJetPolarimeter} uses a similar method to the \( p \)C but with a polarized Atomic Beam Source (ABS) target which flips the polarization direction every 10 minutes. The absolute polarization of the ABS is measured very precisely by a Breit-Rabi polarimeter and is typically about 92\%.  The protons from the RHIC beam scatter off of the ABS target and the left-right asymmetry of elastically scattered protons due to the CNI process is measured. Even though the physics asymmetry of this process is not precisely known, the asymmetry can be measured twice: once when taking into account the polarization of the target and averaging over the spin state of the beam (\( \epsilon_N^{target} \)) and again by using the polarization direction of the beam and averaging over the polarization of the target (\( \epsilon_N^{beam} \)).  Since the target polarization is precisely known, it can be used calculate absolute polarization of the beam:

\begin{equation}
A_N = \frac{ \epsilon_N^{target} }{ P^{target} } = \frac{ \epsilon_N^{beam} }{ P^{beam} } \Longrightarrow P^{beam} = P^{target} \frac{ \epsilon_N^{beam} }{ \epsilon_N^{target} }   
\end{equation}

\noindent Low target density however means that this measurement needs to be taken over a long period of time.  Even after collecting data for an entire fill, the asymmetries are measured with a statistical uncertainty of about 5\%.  In contrast, the \( p \)C can take a precise measurement in less than 10 seconds and can get to 2-3\% precision within a few minutes.  Therefore the \( p \)C is used to monitor the polarization of the beam as it changes across the fill and also how the polarization changes between fills and the H-jet polarization measurements are averaged over multiple fills and used to normalize the \( p \)C results.

\section{The PHENIX Experiment}\label{Section:PHENIX}
The \textbf{P}ioneering \textbf{H}igh \textbf{E}nergy \textbf{N}uclear \textbf{I}nteraction e\textbf{X}periment (PHENIX) is located at the 8 o'clock position along the RHIC ring.  PHENIX was designed to measure a wide variety of probes to study both cold and hot nuclear matter in \(p + p\), \( p/d/h + A\) and \( A + A \) systems.  Figure~\ref{Figure:PHENIX} shows the two central arms of the PHENIX detector, which are nearly back-to-back and each cover \(\Delta \phi = \pi / 2 \) in azimuth and \( | \eta | = 0.35 \) in pseudorapidity.  The PHENIX detector design sacrificed acceptance for the ability to measure rare processes with a combination of high energy and spatial resolution, high rate capability, and advanced trigger systems.  Additionally, there are two forward arms which both cover full azimuth and \( 1.2 < |\eta | < 2.4 \) with spectrometers designed to measure muons and decays from heavy flavor.   The very far forward Zero Degree Calorimeter (ZDC) is a hadronic calorimeter designed to detect very far forward neutrons, which can be used to identify diffractive events and calculate centrality in heavy ion collisions,  a proxy for the impact parameter between two nuclei in a relativistic collision.  In 2012 and 2013 two silicon detectors were installed to provide secondary vertex measurement capabilities for heavy flavor decays: the forward FVTX located at \( 1 < |\eta| < 3 \) and the VTX covering the barrel region around the interaction vertex.    The Muon Piston Calorimeters (MPCs) are two forward electromagnetic calorimeters that are placed inside of the piston holes of the Muon Magnets.  Their primary goal was to identify and measure \( \pi^0 \) and \( \eta \) mesons and they cover \( -3.7 < \eta < -3.1 \) on the south arm and \( 3.1 < \eta < 3.9 \) on the north arm.  An overview of the PHENIX detector can be found at \cite{PHENIXoverview}.

\subsection{Beam-Beam Counters}
Global event information like the timing, vertexing, and luminosity are determined by the Beam-Beam Counters (BBC).\cite{InnerPHENIX}  They consist of an array of  quartz Cherenkov radiators that surround the beam pipe and are placed \( \pm 144 \) cm away from the nominal collision point.  The BBC cover full azimuth and \( 3.0 < |\eta| < 3.9 \) in pseudorapidity and are designed to detect charged particles with velocity \( \beta > 0.7 \).  The z-vertex of inelastic collisions is calculated using the difference in the  average hit times between the north and south side of the BBC.  Thus, it is critical that the BBC have excellent timing resolution.  Each element in the BBC has an intrinsic resolution of roughly 50 ps, which translates to a vertex resolution of roughly two cm in \( p + p \) collisions. This timing information is also used to calculate the time zero of each event which is crucial for measuring time of flight and the PHENIX Level1 trigger system.  

The minimum bias trigger fires on crossings where at least one charged particle is measured in both the north and south sides of the BBC.  This trigger detects about 50\% of all inelastic \( p + p \) collisions with \( \sqrt{s}  = 200 \) GeV.  The standard BBC level 1 trigger has an online vertex requirement of \( \pm 30 \) cm.  The amount of times each crossing fires this trigger is used to calculate the relative luminosity between polarization configurations which will be described in Section~\ref{Section:RelativeLuminosity} and used to calculate the TSSA.

\subsection{Central Arms}
\begin{figure}
  \centering
  \includegraphics[width=.65\textwidth]{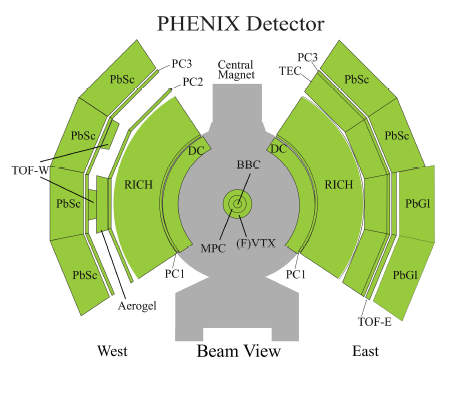}
  \caption[A diagram of the subsystems of the PHENIX central arms after the 2012 running period. ]{A diagram of the subsystems of the PHENIX central arms after the 2012 running period.  In this diagram the beam pipe extends in and out of the page.}
  \label{Figure:PHENIX}
\end{figure}

PHENIX central detector coverage consists of two arms referred to as west and east and shown in Figure~\ref{Figure:PHENIX}.  Each arm covers roughly \( \pi / 2 \) in azimuth and \( | \eta | < 0.35 \) in pseudorapidity.  These arms are not exactly back-to-back but slightly offset such that instead of having an angle of \( \pi / 2 \) between them, the angle at the top is \( 3\pi / 8 \) and at the bottom is \( 5\pi / 8 \).  The sub detectors in the central arms are highly segmented leading to excellent spatial resolution.  One of the primary subdetectors is the electromagnetic calorimeter (EMCal) which measures the position and energy of charged particles and photons. Charged tracks are measured using hits from the Drift Chamber (DC) and the Pad Chamber (PC).  The Ring Imaging Cherenkov (RICH) detector is used for charged particle identification and combined with signals from the EMCal as part of the EMCal RICH Trigger (ERT) which triggers on rare high \( p_T \) processes.  

Cylindrical coordinates are generally used to describe the geometry of the PHENIX detector with the z-axis and \( \theta = 0 \) pointing north and along the beam axis.   The y axis and \( \phi = \pi/2 \) points directly up from the ground, which leaves the x axis and \( \phi = 0 \) pointing towards the west arm.

\subsubsection{EMCal}
The central detector furthest from the beamline is the EMCal\cite{EMCal} whose primary purpose is to measure the spatial position and energy of photons and electrons.  This energy information can be used for particle identification and triggering on high \( p_T \) events.  There are a total of 24,768 individual towers split between eight sectors, six of which are made of lead scintillating (PbSc) sampling calorimeters and two of which are made of lead glass (PbGl) Cherenkov calorimeters.  The PbSc and PbGl sectors have different properties and different strengths and weaknesses, so they often have to be treated differently during calibration and analysis.  The two different technologies for the EMCal allow independent cross checks of results within the same experiment.

\begin{figure}
  \centering
  \includegraphics[width=.7\textwidth]{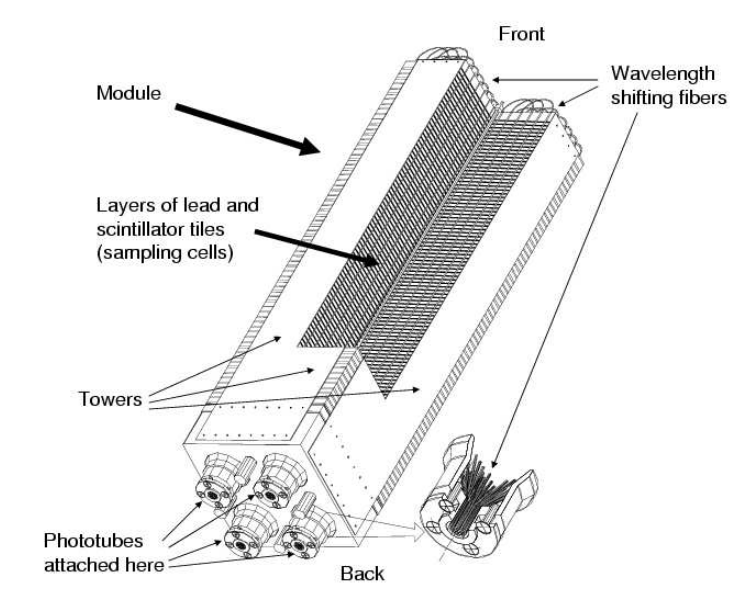}
  \caption[Interior view of the PbSc module showing the alternating scintillator and lead plates and the wavelength shifting fiber readout.]{Interior view of the PbSc module showing the alternating scintillator and lead plates and the wavelength shifting fiber readout. \cite{EMCal}}
  \label{Figure:PbScTower}
\end{figure}

The PbSc sectors are referred to as a shashlik type sampling calorimeter because each tower is made of alternating plates of lead and scintillator, as shown in Figure~\ref{Figure:PbScTower}.  The PbSc has a nominal energy resolution of \( 8.1\% / \sqrt{E} \text{[GeV]} \oplus 2.1\% \), a radiation length of 18 \( X_0 \), and an intrinsic timing resolution better than 200 ps for electromagnetic showers. 

There are a total of 15,552 PbSc towers which collectively take up approximately 48 square meters.   These towers produce a pseudorapidity (\( \Delta \eta \)) and azimuthal (\( \Delta \phi \)) resolution of \( \Delta \eta \times \Delta \phi \approx 0.011 \times 0.011 \).  Each tower contains 66 layers of alternating lead and scintillator plates which are connected with wavelength-shifting, fiber-optic cables that penetrate longitudinally through the tower to connect to PhotoMultiplier Tubes (PMTs) at the back which read out the light produced by the electromagnetic shower.  The edges of each tower are plated with aluminum and then they are mechanically attached into groups of four which are called modules.  Thirty-six of these modules are grouped together into what are called supermodules that are held together by welded stainless steel skins on the outside to form a rigid structure.  Eighteen of these supermodules are held together in a two-meters-by-four-meters steel frame to form a sector.  The PbSc sectors also have better energy linearity and faster timing when compared to the PbGl.  

\begin{figure}
  \centering
  \includegraphics[width=.7\textwidth]{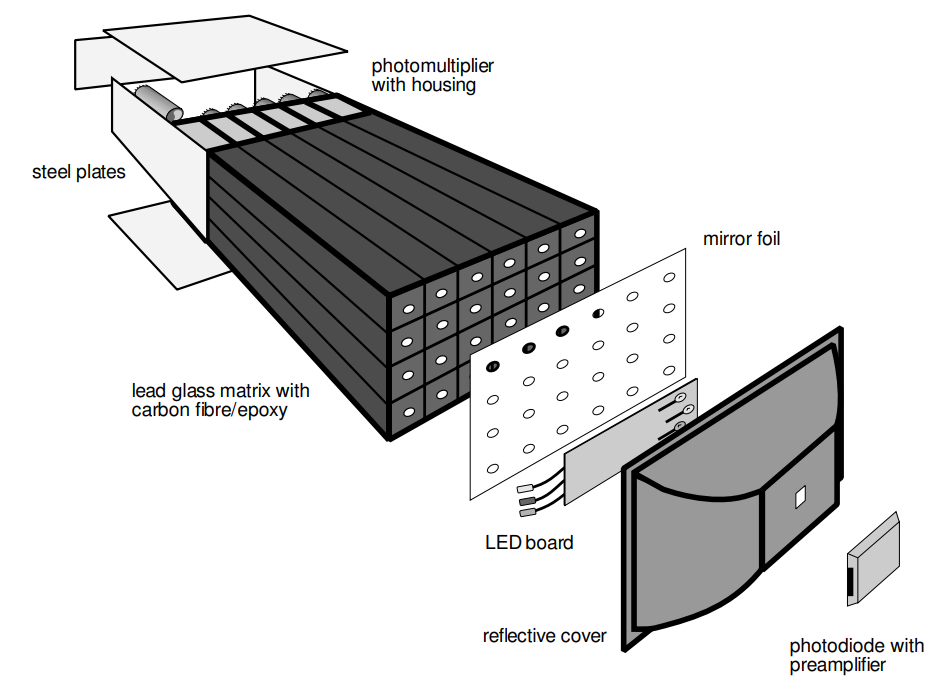}
  \caption[Interior view of a lead PbGl supermodule where the front part of this image is the part of the supermodule closest to the interaction point.]{Interior view of a lead PbGl supermodule where the front part of this image is the part of the supermodule closest to the interaction point. \cite{EMCal}}
  \label{Figure:PbGlSupermodule}
\end{figure}

The PbGl sectors however have a higher energy resolution of \( 5.9\% / \sqrt{E} \text{[GeV]} \oplus 0.8\% \) and finer spatial resolution with \( \Delta \eta \times \Delta \phi \approx 0.008 \times 0.008 \).   The PbGl also has a lower radiation length of about 14 \( X_0 \) when compared to the PbSc.  There are a total of 9216 PbGl towers which occupy the lower two EMCal sectors in the east arm.  These sectors were previously used in the WA98 CERN experiment.  Each PbGl sector is comprised of 192 supermodules in an array of 16 supermodules wide and 12 high.  Each supermodule, shown in Figure~\ref{Figure:PbGlSupermodule}, is made up of 24 lead glass towers that are arranged in an array of 6 towers wide and 4 towers high.  These 24 towers are individually wrapped in aluminized mylar and shrink tube and then glued together with carbon fiber and epoxy resin.  A single PMT is attached to the end of each tower behind the lead glass matrix to read out the signal of each electromagnetic shower.  

\subsubsection{DC and PC}
The primary subsystem used for PHENIX central tracking is the DC \cite{CentralTracking} which measures the trajectory of charged particles in the \( r - \phi \) plane.  The curvature of these charged tracks in the PHENIX magnetic field can then be used to calculate the \( p_T \) of the charged particle.  The DC is located on both of the PHENIX central arms as shown in Figure~\ref{Figure:PHENIX} and the cylindrical frames are filled with a gas mixture of 50\% argon and 50\% ethane.  When a charged particle passes through the DC, the gas mixture ionizes.  Wires within the chamber are held at a high enough voltage such that the resulting charged particles drift towards them.  Which wire detects these charged particles can then be used to reconstruct track position information using the drift time of the electrons.  Each frame has six types of wires that are stacked radially and run the length of the DC in the direction parallel to the beam.  The wires themselves do not run parallel to each other, but instead are attached at an angle such that signals from multiple wires can be used to reconstruct the full \( r \), \( \phi \), and \( z \) information.  

The PC is used to improve the track \( z \) and \( p_T \) resolution and also reduces combinatorial background from tracks in the DC.  Each of the components of the PC contains a  plane of anode wires surrounded by a gas chamber.  This anode plane is sandwiched between two cathode planes,  one of which is finely segmented into an array of pixels.  When a charged particle passes through the gas volume, it starts an avalanche on an anode wire which induces charge on this cathode plane which is then read out by electronics specially designed for the PC.  There are three separate PC multiwire chambers that are labeled PC1, PC2, and PC3 which can be seen in Figure~\ref{Figure:PHENIX}.  The PC1 is the innermost pad chamber plane, located between the DC and the RICH.  It measures the z coordinate of a track as it exits the DC.  The PC2 is located only on the west arm just behind the RICH and the PC3 is mounted on both arms just in front of the EMCal.  In photon analyses the track position information from the PC3 is used to eliminate EMCal clusters that are associated with charged tracks.

\subsection{EMCal-RICH Trigger}
The ERT is designed to identify rare processes in PHENIX by firing on high \( p_T \) photons and electrons.  EMCal towers are grouped into what are called tiles over which the deposited energy is summed.  If this total tile energy is above some preset threshold, then this event is recorded as firing this trigger.  The energy threshold of the triggers is set depending on the center of mass energy of the collision system.  The electron ERT trigger uses non-overlapping 2 x 2-tower tiles in conjunction with the RICH.  The high \( p_T \) photon trigger uses overlapping 4 x 4-tower tiles and there are three different types of this trigger: the ERTA, ERTB, and ERTC.  Each has a  different energy threshold and the trigger with the highest energy threshold is the ERTB while the trigger with the lowest energy threshold is the ERTC.

%% file: Chap3/chap3.tex
\section{Data Selection}
These transverse single-spin asymmetry results were calculated using the 2015 \( p + p \) data set with \( \sqrt{s} = 200 \) GeV.  The total recorded luminosity of transversely polarized collisions was 60 pb\textsuperscript{-1} and the average polarization of the yellow beam was 59\% while the average blue beam polarization was 57\%.  Because both of the proton beams are transversely polarized and the polarization direction changes bunch to bunch, the same TSSA can be calculated twice using the same data set.  Once by keeping track of the polarization directions for only the yellow beam and effectively averaging over the polarization directions of the blue beam; we shall refer to this as the yellow beam asymmetry.  Then the asymmetry can be calculated for a second time by considering the polarization directions of the blue beam and effectively averaging over the yellow, which we will call the blue beam asymmetry.  These two asymmetry measurements are completely statistically independent and are averaged together to find the final result.  

\subsection{Run Quality Assurance}\label{Section:RunQA}
All data taken at PHENIX is segmented into what are called runs which correspond to one cycle of DAQ data-taking which typically lasts about an hour.  Each run is identified by its own run number and required to pass certain criteria to be included in these analyses.  Eight runs are eliminated because of either a low vertex distribution or a low trigger efficiency.  
 
The beam spin information, like the polarization directions of each crossing and the bunch-to-bunch luminosity, is taken from PHENIX's spin database.  All information put into this spin database has already undergone its own quality assurance to ensure all of its information is as accurate as possible and any runs that were flagged as bad by the spin data base quality assurance are eliminated from the data sample.  Additionally, seven runs are eliminated because the recorded bunch by bunch luminosity was either small or zero.  Another nine runs are removed from the sample because their overall fill relative luminosity was significantly different from the other calculated relative luminosities, see Section~\ref{Section:RelativeLuminosity}.  After run quality assurance, there are a total of 797 runs remaining.  

\subsection{Event Selection}\label{Section:EventSelection}
In high energy physics, an event refers to the set of outgoing particles produced in a collision between two incoming particles.  Like most high energy experiments, all of PHENIX's detectors are read out for each triggered bunch crossing such that they collect as much information about the event as possible.  Each event used in the TSSA analyses is required to fire at least one of ERT 4 x 4 photon triggers and pass a vertex cut of \( | z_{vtx} | \leq 30 \) cm.  Photons that came from events that occurred during empty crossings are also eliminated.  

\subsection{Photon Selection}\label{Section:PhotonSelection}
The direct photon, \( \pi^0 \), and \( \eta \) meson analyses all use photons that are measured as clusters in the EMCal, which measures the photon's position and energy information.  The corrected cluster energy that was calculated during data calibration is referred to as \( E_{core} \).  In the PbSc sectors this means that the energy is corrected for fiber attenuation, energy leakage, and incident angle.  Clusters in the PbGl have had their energy corrected for the incident angle and nonlinear detector effects.  These corrections assume that the cluster is formed by an electromagnetic shower and the ``core'' subscript refers to the assumption that the majority of the cluster's energy should be concentrated at its center.  Because photons have zero charge, their paths do not curve as they travel in the magnetic field.  Since they are massless, the position and energy information from the EMCal can be used to directly calculate the photon's momentum: \( \vec{ p } = E_{core} \vec{x} \).  

These clusters are required to pass certain fiducial cuts which include an energy cut of \( 0.5 < E_{core} < 20 \) GeV.  There are also shower shape cuts designed to eliminate clusters from charged hadrons which tend to have wider particle showers.  These cuts look for narrower clusters with a Gaussian shape and also perform a \( \chi^2 < 3 \) on clusters in PbSc sectors and a dispersion cut on clusters in PbGl sectors.  Hot and dead map cuts eliminate clusters whose central tower either is or is next to an EMCal tower that had previously been identified as problematic.  The edge tower cut eliminates clusters whose central tower is either on the edge or one over from the edge of the sector.  The time of flight cut requires that \( |TOF| < 5 \) ns in order to remove additional detector noise.  

\begin{figure}
\centering
\subfigure[Sector 3]{ \includegraphics[scale = 0.35]{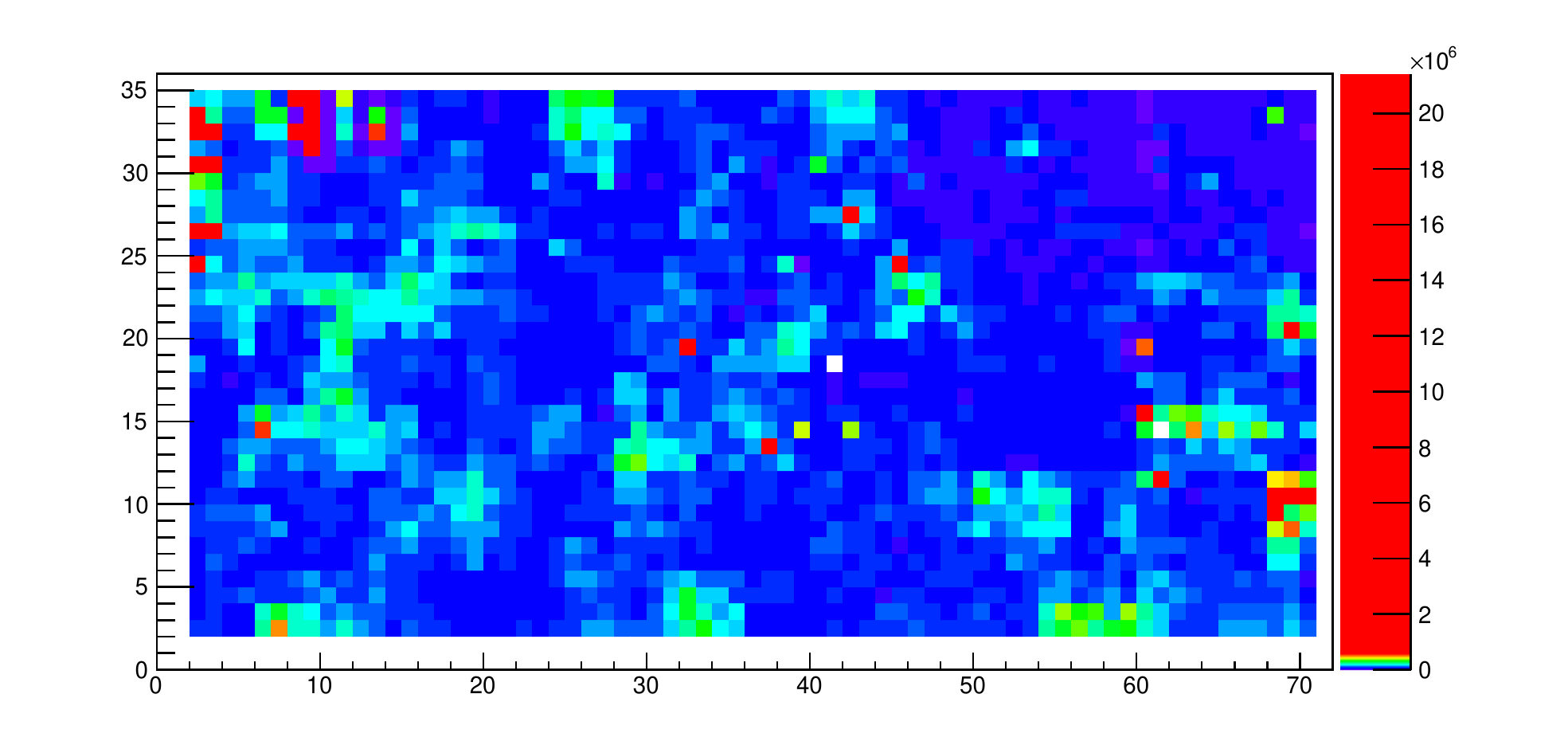} } 
\subfigure[Sector 4]{ \includegraphics[scale = 0.35]{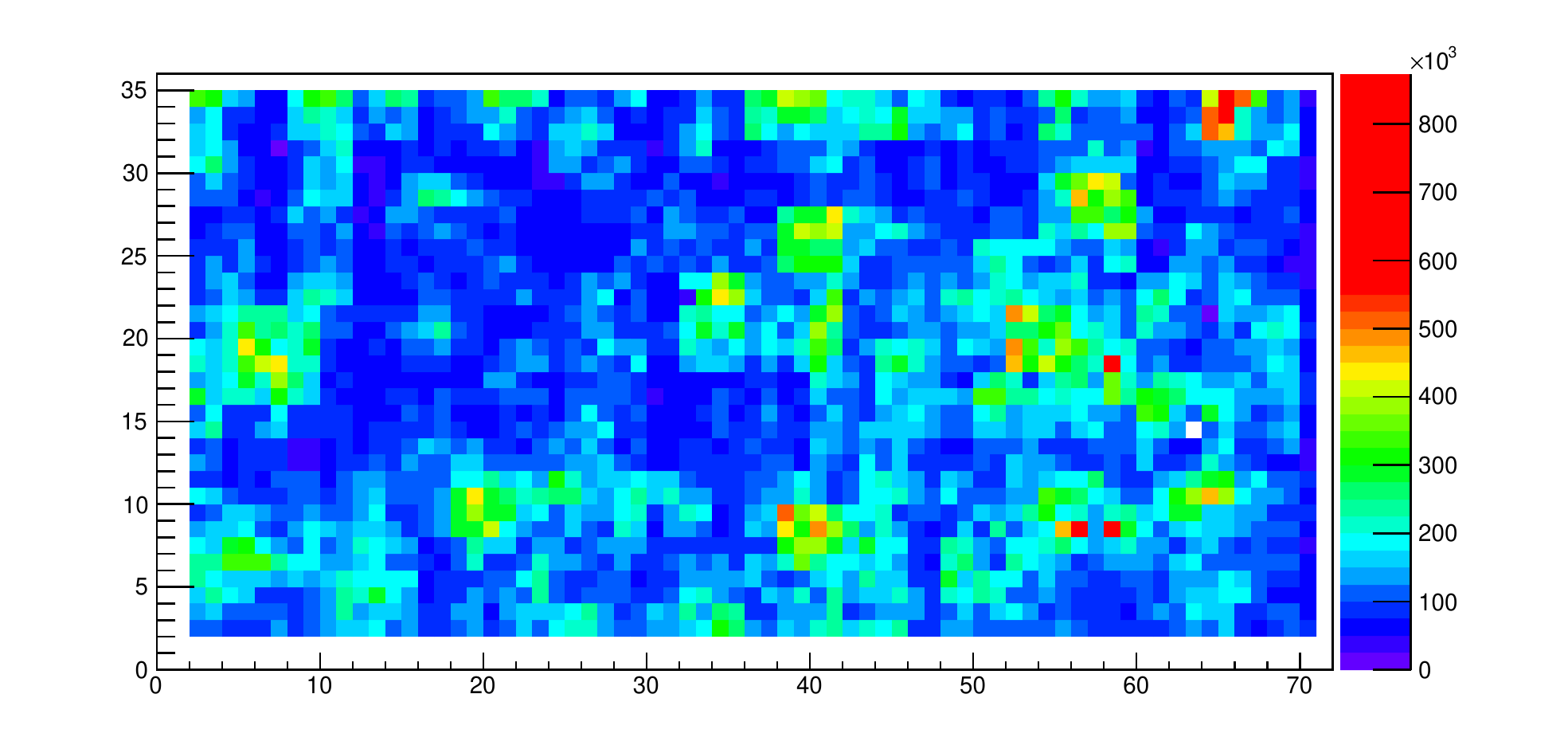} }\\
\subfigure[Sector 2]{ \includegraphics[scale = 0.35]{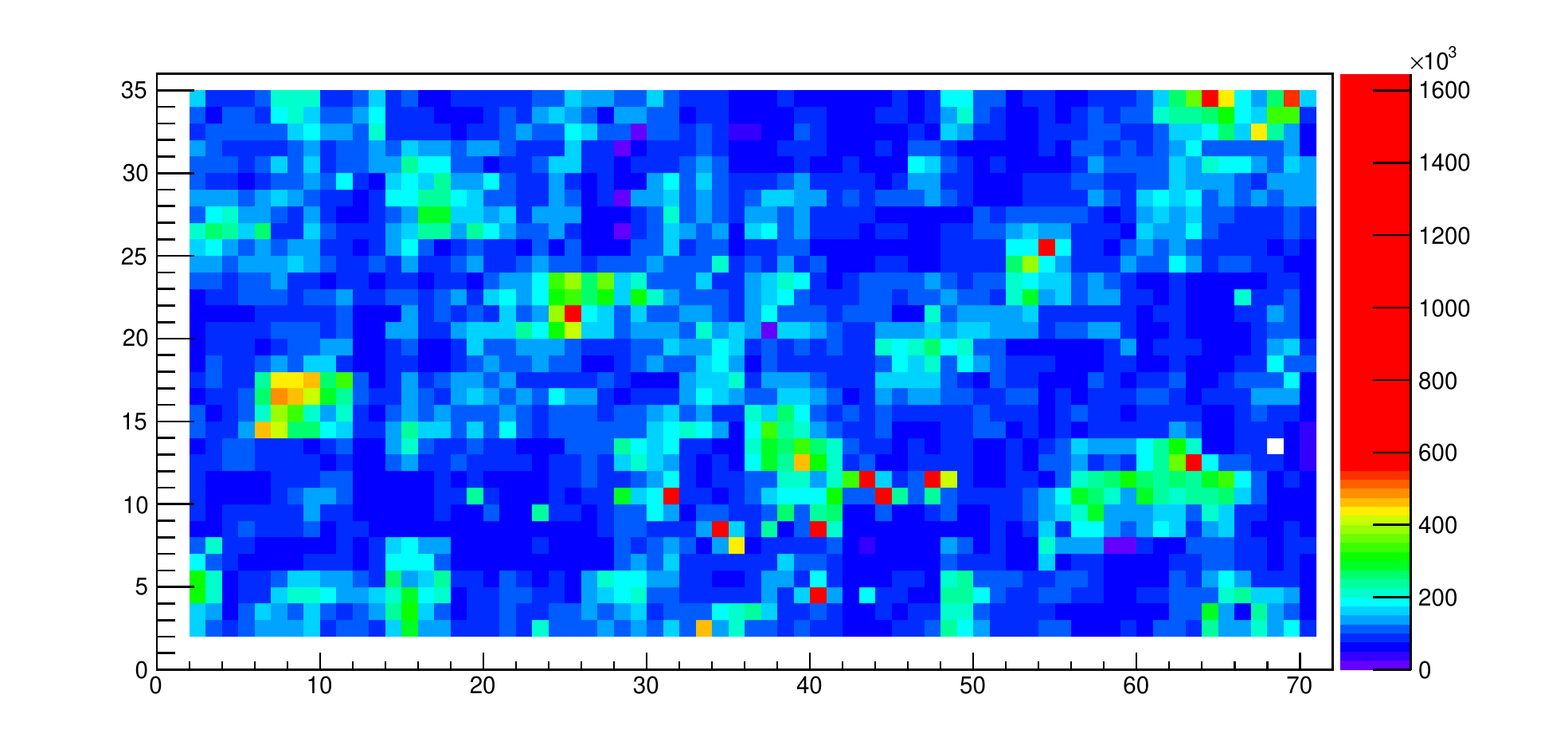} } 
\subfigure[Sector 5]{ \includegraphics[scale = 0.35]{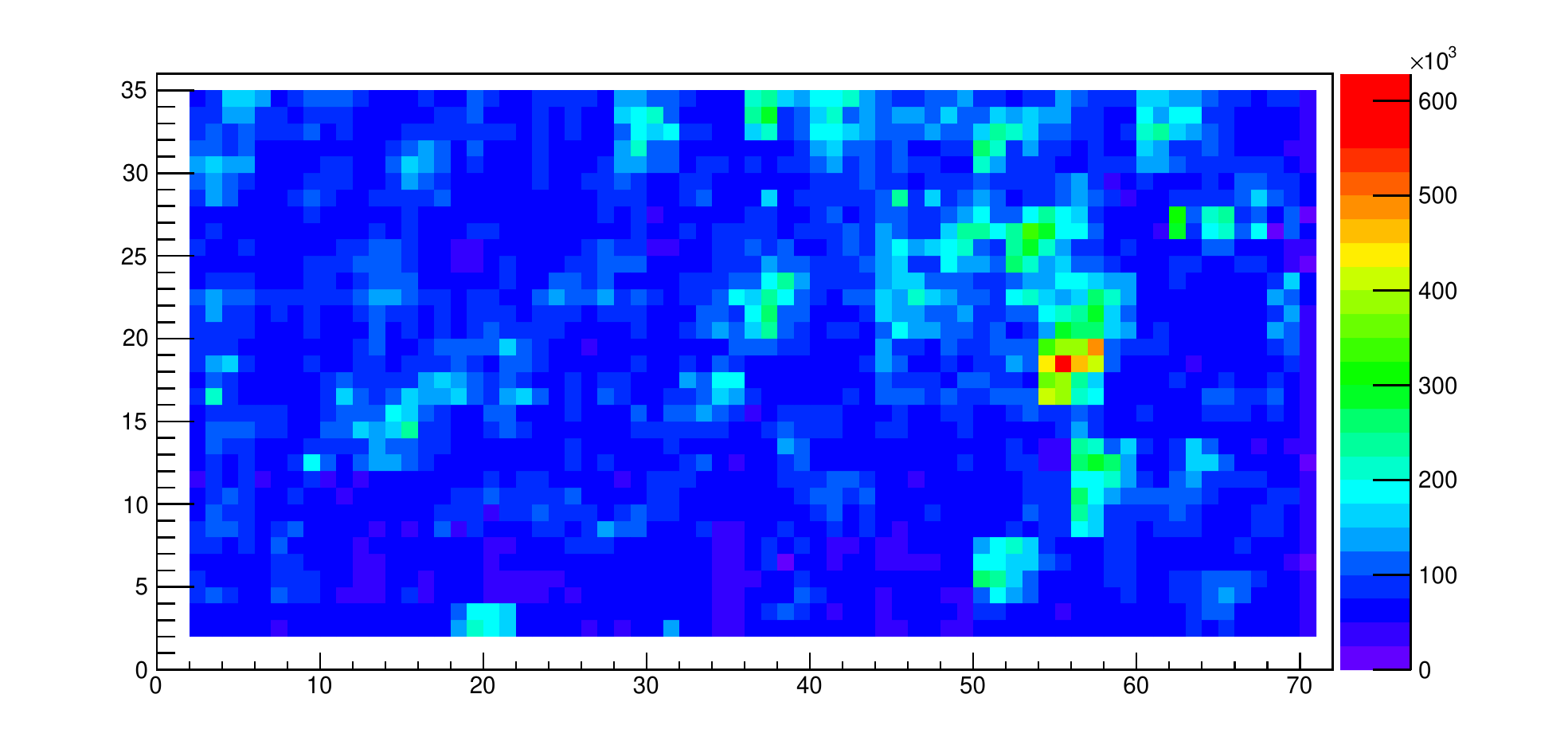} }\\
\subfigure[Sector 1]{ \includegraphics[scale = 0.35]{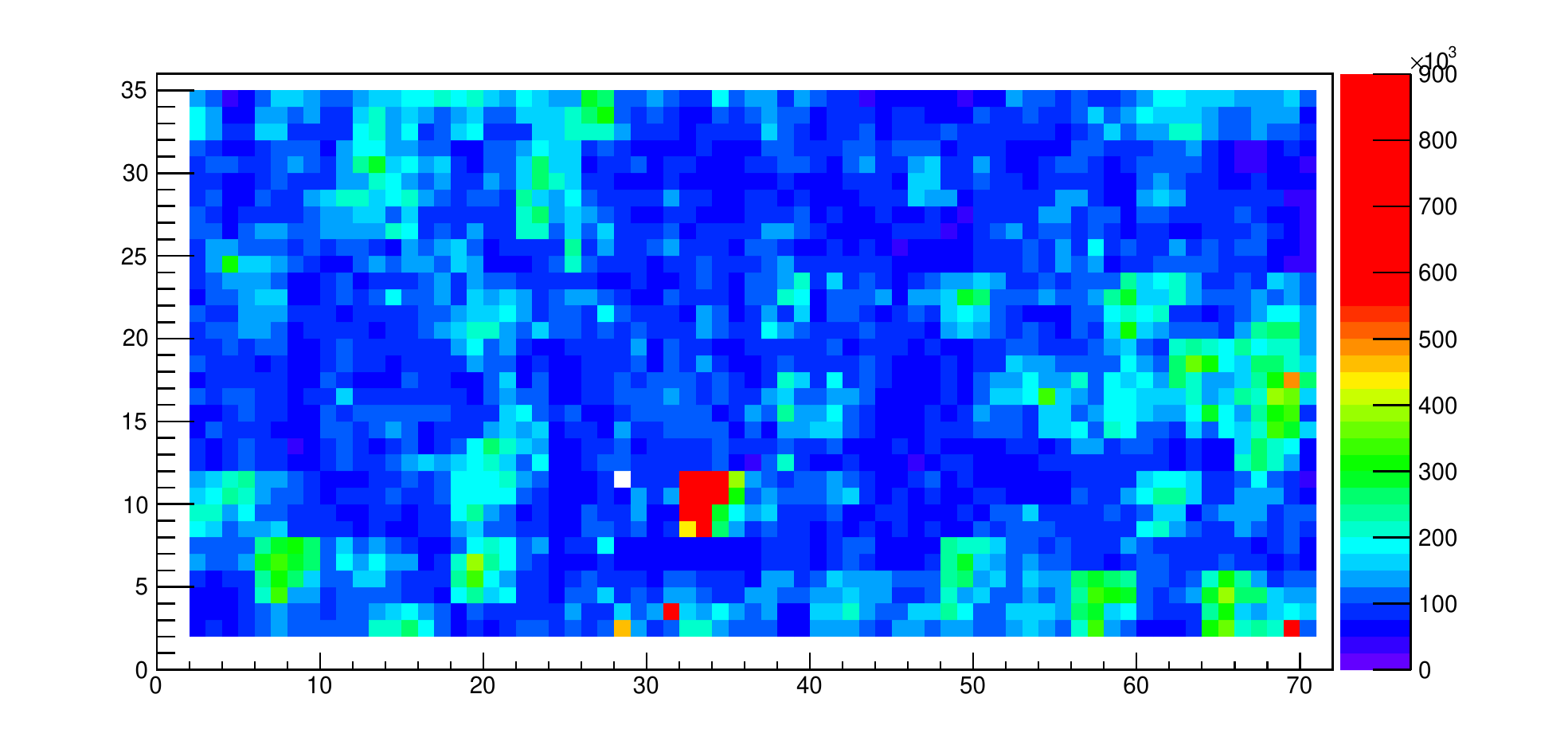} } 
\subfigure[Sector 6]{ \includegraphics[scale = 0.35]{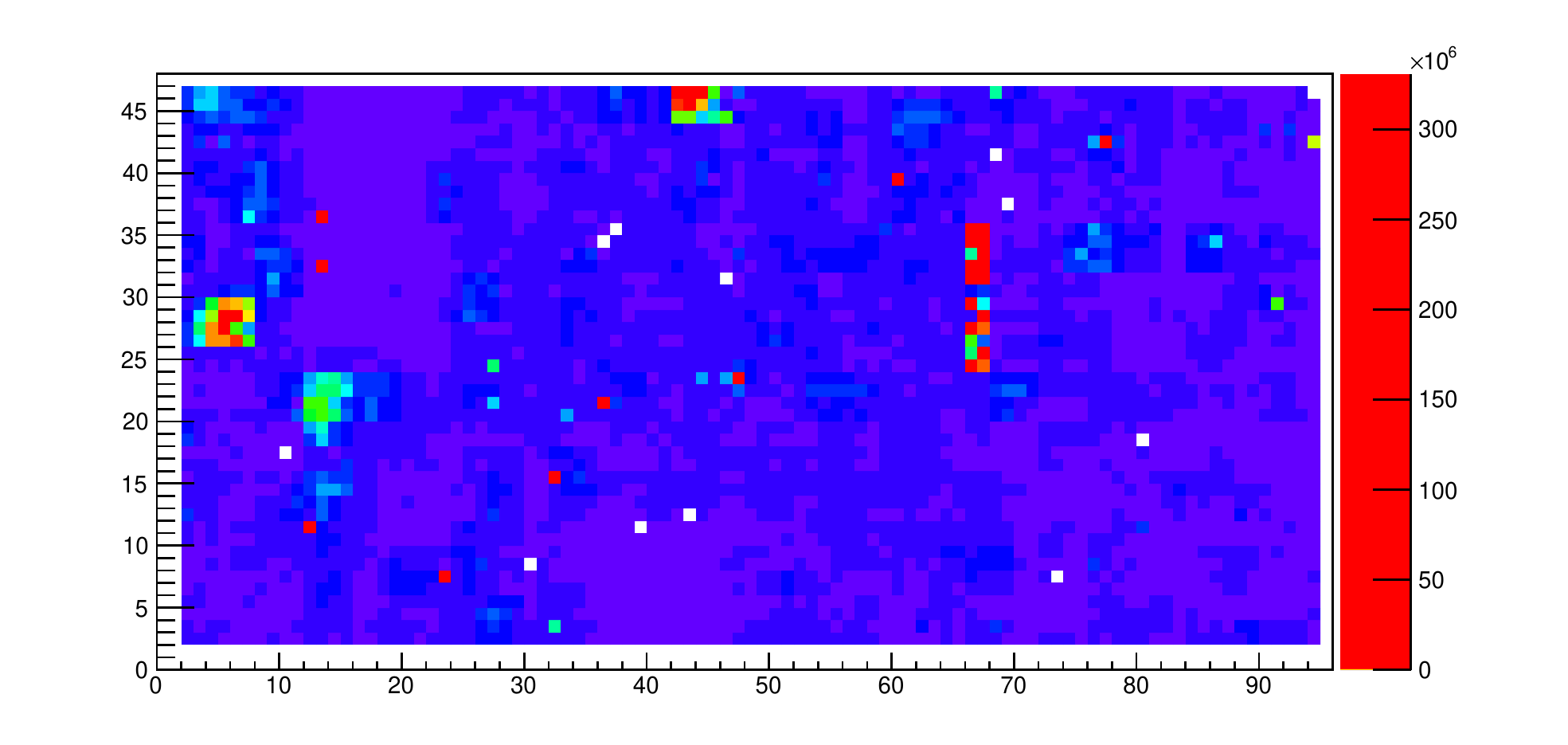} }\\
\subfigure[Sector 0]{ \includegraphics[scale = 0.35]{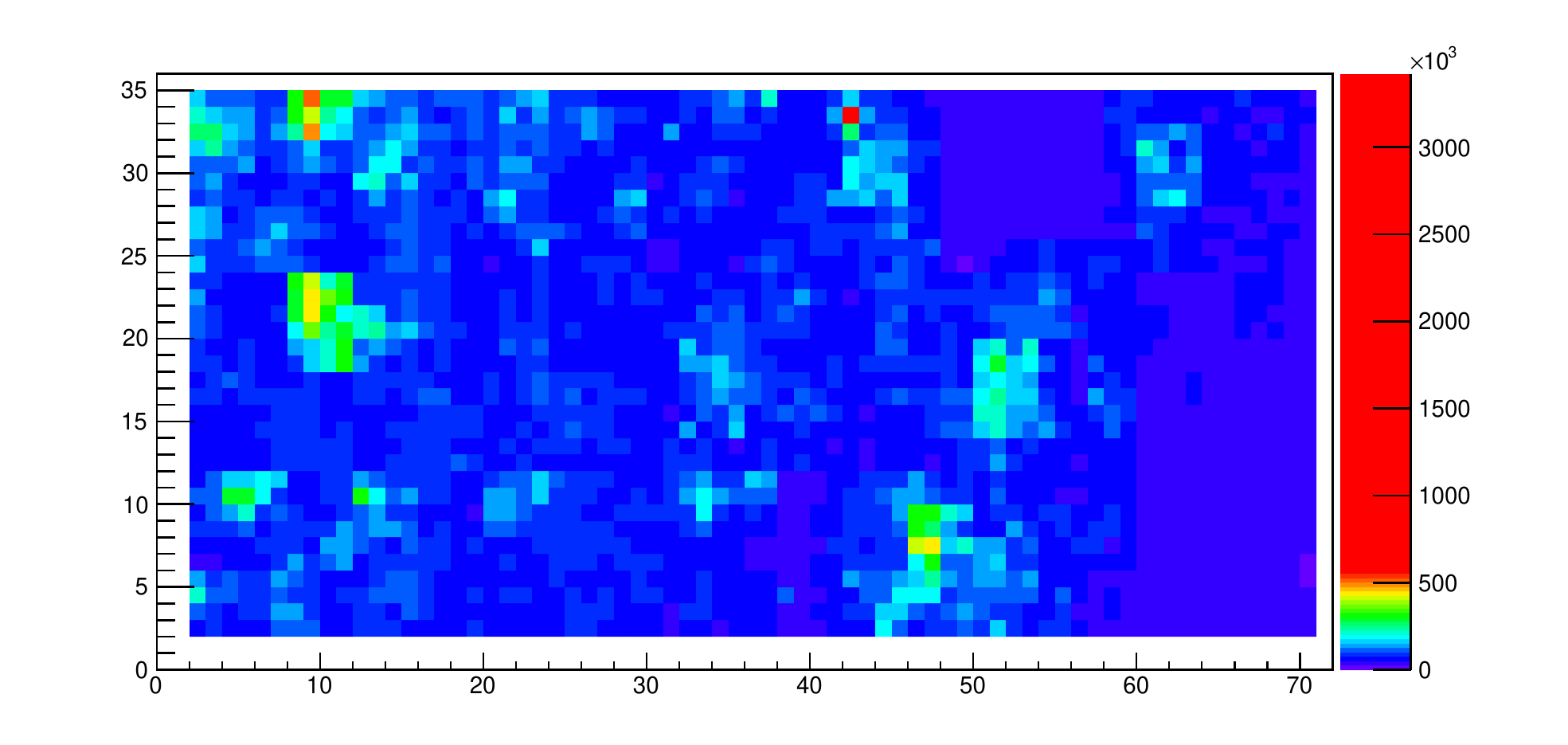} }
\subfigure[Sector 7]{ \includegraphics[scale = 0.35]{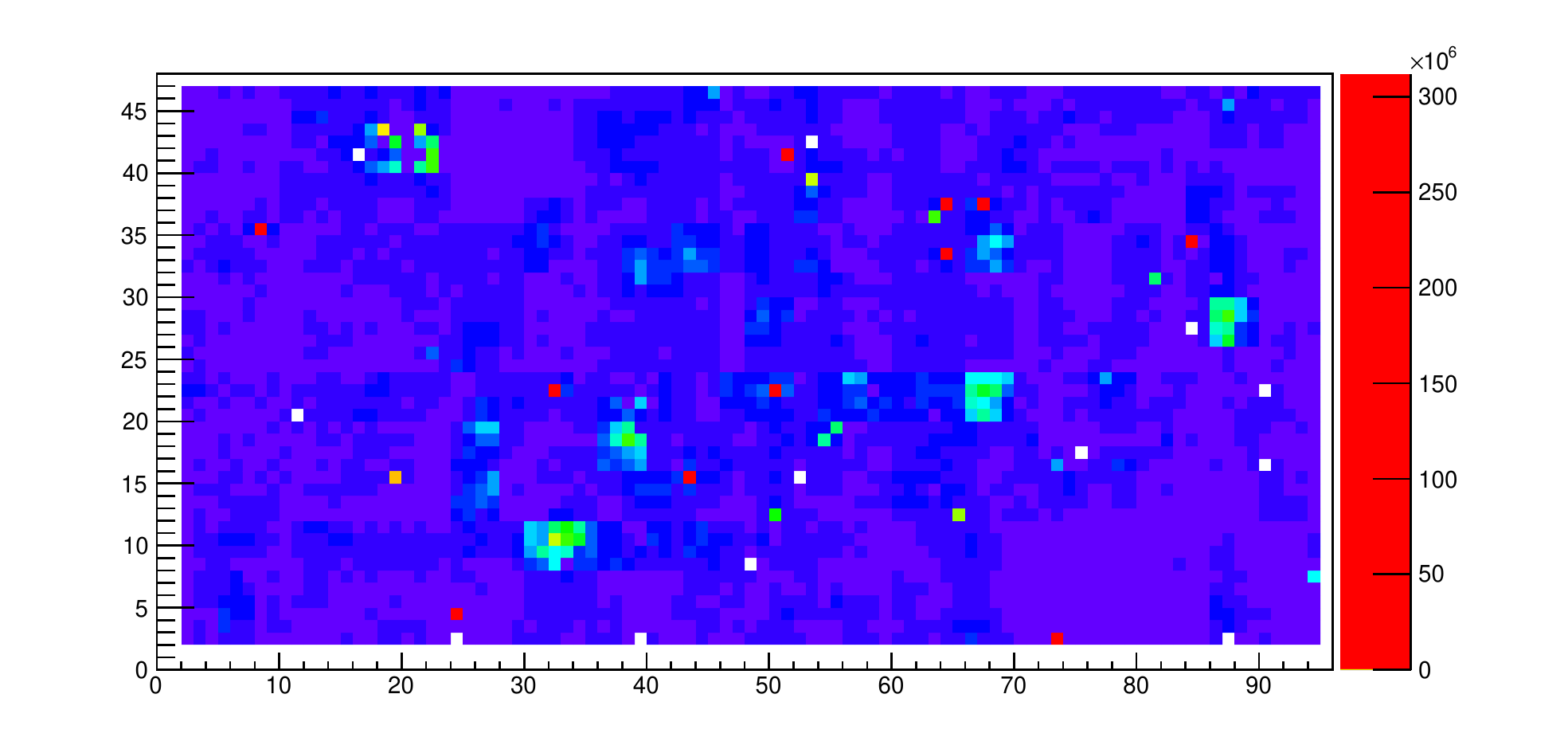} } 
\caption[Example of a hot tower map: a sector by sector comparison hits per tower used to identify problematic towers in Run-15.]{Example of a hot tower map: a sector by sector comparison hits per tower used to identify problematic towers in Run-15. The x-axis corresponds to the tower's position along the sector's z coordinate and the y-axis is the tower's y coordinate in the sector.  This example is for clusters with \( 0.5 < E_{core} < 5 \) GeV.  The contour levels of each histogram are set to be the same to showcase the towers that have the largest difference in hits from the rest of the sector.  }
\label{Figure:HotMap}
\end{figure}

An additional list of hot towers is also created for this analysis.  These are EMCal towers that due to electronic noise, fire many more times than all of the other towers in the sector.  In order to identify hot towers, hot maps like the one in Figure~\ref{Figure:HotMap} are created.  (Here the sectors are numbered clockwise from the west arm such that sector 0 is the PbSc sector at the bottom of the west arm and sector 7 is the PbGl sector at the bottom of the east arm.)  These are sector-by-sector two dimensional histograms of the central tower for all clusters in Run-15 that have passed the cuts listed in the previous paragraph.  Towers are considered ``hot" if the total number of hits is larger than 6 times the RMS of hits for that sector.  This RMS should not include any values from towers that have been previously flagged as hot. And so, this becomes an iterative process: hot towers are flagged, the RMS is recalculated without using the counts from these flagged towers, more towers are now flagged as having counts that are 6 times higher than this new lower RMS value and so on. This process is repeated until no new hot towers are found.   There are two different lists created for different bins in energy: one for clusters with energies between \( 0.5 \) and \(5 \) GeV (Figure~\ref{Figure:HotMap}) and another for energies between \( 5 \) and \( 20 \) GeV.  This is to ensure that higher energy clusters are not eliminated based on how their towers behaved at lower energies, since there are four times as many towers in the lower energy list than the high energy one.  Since clusters consist of multiple adjacent towers, the hot tower cut not only eliminates clusters whose central tower is on the hot tower list, but also clusters whose central tower is adjacent to a flagged tower.  

In order to be added to the photon sample, clusters are also required to pass a charged track veto cut.  This helps eliminate electrons as well as charged hadrons that have not been eliminated by the shower shape cut, by checking that all photons candidates do not have a matching angle with a charged track as determined by the PC3.  This is an elliptical cut of 12 cm in  \( z \) and 8 cm in \( \phi \).  This cut does not require that the area of the PC3 in front of the photon cluster be live, just that no matching charged track is found.  

\subsubsection{Photon pair selection}\label{Section:PhotonPairSelection}
\begin{figure}
\centering
\includegraphics[width=1\textwidth]{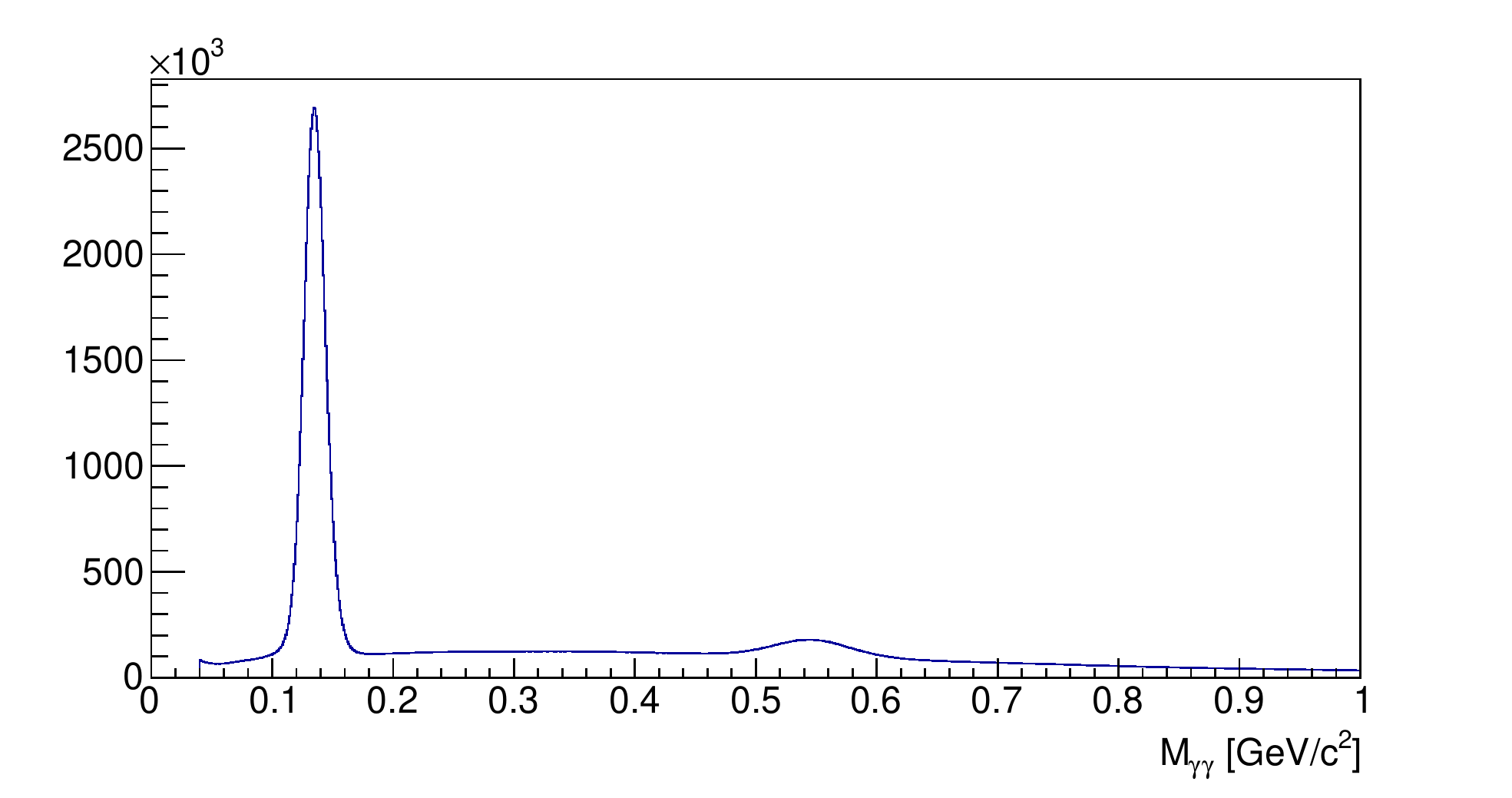}
\caption[Invariant mass distribution of all diphoton pairs in Run-15.]{Invariant mass distribution of all diphoton pairs in Run-15.}\label{Figure:invmass}
\end{figure}

\( \pi^0 \) and \( \eta \) mesons are measured via their diphoton decay channel which has a branching fraction of about 99\% for \( \pi^0 \)s and 39\% for \( \eta \)s.  All of the photons that are included in these \( h \rightarrow \gamma \gamma \) pairs are required to pass all of the fiducial photon cuts described earlier in this section in addition to some photon pair cuts.  First the pairs are required to be in the same arm of the PHENIX detector and be at least 8 cm apart from each other in the EMCal.  All diphoton pairs with invariant mass less than 1 GeV/c\textsuperscript{2} that pass these cuts are shown in Figure~\ref{Figure:invmass}.  The \( \pi^0 \) invariant mass peak can clearly be seen around 0.13 GeV/c\textsuperscript{2} and the \( \eta \) peak can be seen somewhat less clearly around 0.55 GeV/c\textsuperscript{2}.  There are less \( \eta \rightarrow \gamma \gamma \) photon pairs measured than \( \pi^0 \rightarrow \gamma \gamma \) partially because of the different branching fractions and partially because there are about twice as many \( \pi^0 \)s produced because of their lighter mass.  Photon pairs used in \( \pi^0 \) and \( \eta \) analyses are also required to pass an energy asymmetry cut of \( \alpha = \mid E_1 - E_2 \mid / (E_1 + E_2) < 0.8 \).  This cut is designed to eliminate asymmetric decays which are more difficult to reconstruct partially because of PHENIX's limited central acceptance and partially because there is a higher contribution from detector noise at lower energies.  

The higher energy photons in the diphoton pairs are also required to pass additional cuts.  They must have \( p_T \geq 1.5 \) GeV/c such that their energy is high enough for the ERT efficiency to be constant.  
The higher energy photon also needs to pass the ERT check by being in the same supermodule that fired one of the three 4 x 4 ERTs.  This ensures that the photon is at least in the same region of the detector that fired the ERT trigger since we cannot verify what particle actually fired the trigger.  The higher energy photon also needs to be the photon with the highest energy in the event.  This is to avoid random benefit: when a lower energy photon just happens to be in the supermodule that fired the ERT, so it passes the ERT check but did not actually fire the trigger itself.  

Photon pairs are considered to be within the \( \pi^0 \) invariant mass peak if they have \( 112 < M_{\gamma\gamma} < 162 \) MeV/c\textsuperscript{2} and the \( \pi^0 \) combinatorial background is studied with photon pairs with \( 47 < M_{\gamma\gamma} < 97 \) MeV/c\textsuperscript{2} or \( 177 < M_{\gamma\gamma} < 227 \) MeV/c\textsuperscript{2}.  The signal range used for the \( \eta \) invariant mass peak is 480 to 620 MeV/c\textsuperscript{2} and the combinatorial background is studied with the invariant mass ranges 300 to 400 MeV/c\textsuperscript{2} and 700 to 800 MeV/c\textsuperscript{2}, both of which are the same ranges that were used in Ref. \cite{PPG135}.  

\subsubsection{Direct Photon Selection} \label{Section:DirectPhotonCuts}
In addition to the previously listed photon requirements, clusters are required to pass additional cuts to be included in the direct photon sample.  This includes the ERT check that was described in the previous section.  Tagging cuts are used to eliminate photons that are tagged as coming from either a \( \pi^0 \) or \( \eta \) decay.  The photon in question is matched with another photon in the same event that has passed all of the photon cuts listed at the beginning of Section~\ref{Section:PhotonSelection}.  If these matched photons are in the same arm, have more than 8 cm between them in the EMCal, and have an invariant mass of either \( 105 < M_{\gamma\gamma} < 165 \) MeV/c\textsuperscript{2} for \( \pi^0 \) tagging or \( 480 < M_{\gamma\gamma} < 620 \) MeV/c\textsuperscript{2} for \( \eta \) tagging, then these photons are eliminated by the tagging cut.  The invariant mass range used for the \( \pi^0 \) tagging cut is looser than the signal region used in the \( \pi^0 \) TSSA analysis described in the previous paragraph which ensures that the tagging cut eliminates as many decay photons as possible.  

Isolation cuts have a been standard practice for direct photon analyses at PHENIX.\cite{PPG095, PPG136, PPG195, PPG217}   These are designed to suppress the background contribution from decay photons as well as NLO fragmentation photons.  Since direct photons come directly from the hard scattering, they should have an energy that is much higher than the energy of the surrounding event:
\begin{equation}
E_\gamma \cdot 10\% > E_{cone}
\label{Equation:PhotonIsolationCut}
\end{equation}

\noindent Here \( E_\gamma \) is the energy of the photon in question and it is required to be 10 times larger than \( E_{cone} \): the sum of the energies of all of the surrounding clusters and the momenta of all of the surrounding tracks that are within \( r = \sqrt{ \Delta\phi^2 + \Delta\eta^2 } < 0.4 \) radians of this photon.  The clusters that are included in the cone sum have to pass much less stringent cuts than what is needed to be included in a photon sample since they are only being used to characterize the surrounding event.  Similarly, the tracks whose momenta are included in \( E_{cone} \) are also only required to pass a bare minimum of quality cuts.  The charged hadron veto cut is still implemented to ensure charged particles are not being double counted by the energy that they deposit in the EMCal and the momentum from a reconstructed charged track, but no shower shape cuts are used.  This ensures that neutral hadrons and charged hadrons that were not reconstructed as charged tracks, can still contribute to \( E_{cone} \) with energy that they deposit in the EMCal.  

The main source of background for direct photons comes from hadronic diphoton decays where one of the photons is missed.  There are three things of note that can happen when a hadron decays to two photons: (1) both photons are measured, (2) only one of the photons is measured, and (3) the two photons are so close together that they are reconstructed as a single cluster, also known as merging.  When both photons from a \( h \rightarrow \gamma \gamma \) decay are measured, these photons are eliminated from the direct photon sample by the tagging cut.  Missing the second photon is especially a concern with PHENIX's limited central acceptance, but it can also happen if the other photon hits a dead area of the detector or its energy is too low to pass the minimum energy requirement of 0.5 GeV.  In contrast, merging is a much smaller effect at the \( p_T \) range of this analysis and is further mitigated with the shower shape cuts, this will be discussed in more detail in Section~\ref{Section:DirectPhotonBackgroundFraction}.  

In order to be included in the direct photon sample, a photon from a \( h \rightarrow \gamma \gamma \) decay would need to have its sister photon missed and also pass the isolation cut.  In order to study this isolated background, a photon \textit{pair} isolation cut is used:

\begin{equation}
E_\gamma \cdot 10\% > E_{cone} - E_{partner}
\label{Equation:PhotonPairIsolationCut}
\end{equation}
\noindent  In this equation the photon with energy \( E_\gamma \) has been matched into a photon pair with a second photon of energy \( E_{partner} \).  This pair isolation cut is slightly more lenient than the photon isolation cut from Equation~\ref{Equation:PhotonIsolationCut} with the idea being that if the energy from second the photon, \( E_{partner} \), had not been measured, then this photon with energy \( E_\gamma \) would have passed the photon isolation cut and been added to the direct photon sample. These tagged photons that are in an isolated pair will be used to estimate the background contribution to the direct photon sample in Section~\ref{Section:DirectPhotonBackgroundFraction}.  

\section{Corrections to the Asymmetry}
For TSSAs measured at RHIC, the raw asymmetry measures difference in yields of particles that travel to the left versus the right of the polarized proton going direction.  These asymmetries need to be corrected for the relative luminosity between different beam polarization configurations, the absolute polarization of the beam, and for being measured across a wide range in azimuth.    

\subsection{Acceptance Correction}\label{Section:AcceptanceCorrection}

\begin{figure}
\centering
\includegraphics[width=0.8\linewidth]{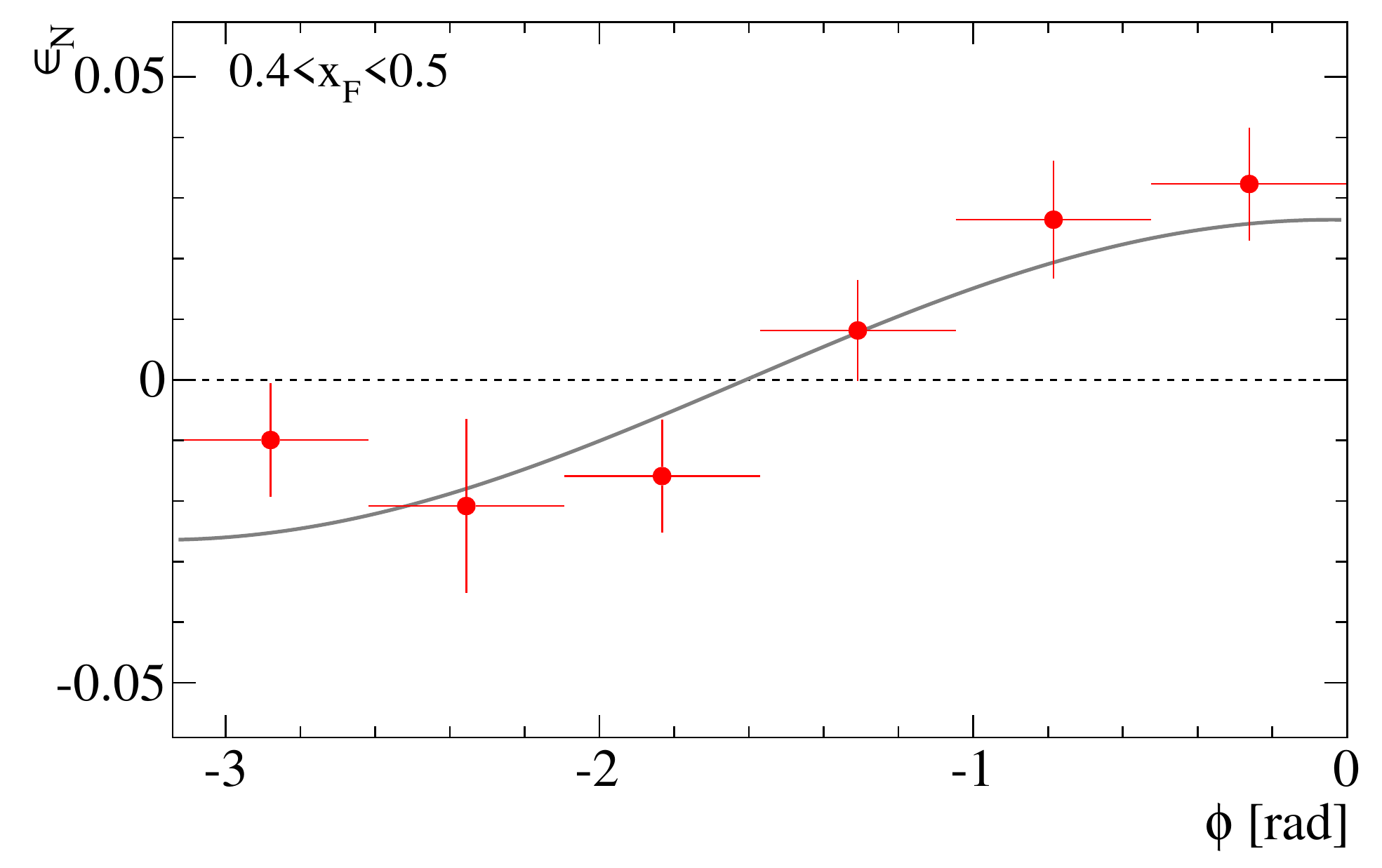}
\caption[An example of extracting the TSSA from a sinusoidal fit function]{An example of extracting the TSSA from a sinusoidal fit function.  The forward \( \eta \) TSSA \cite{PPG170} had full azimuthal acceptance and a relatively large asymmetry.  So the analyzers were able to measure the raw asymmetry as a function of \( \phi \) and extract \( A_N \) as the magnitude of a fit that used a sinusoidal function.  Here is an example for a single \( x_F \) bin for data from the south arm of the MPC.}\label{Figure:ForwardEtaTSSAFit}
\end{figure}

The TSSA describes the azimuthal-angular dependence of particle production relative to the transverse-spin direction of the polarized proton.  An azimuthally dependent polarized cross section can be parameterized as:
\begin{equation}
\frac{d\sigma}{d\Omega} = \bigg( \frac{d\sigma}{d\Omega} \bigg)_0 (1 + P \cdot A_N \cdot \cos{\phi} )
\end{equation}

\noindent where \( \bigg( \frac{d\sigma}{d\Omega} \bigg)_0 \) is the unpolarized differential cross section and \( P \) is the beam polarization which will be discussed in the next section.  Here, \( \phi \) is the azimuthal angle around the beam axis where \( \phi = 0 \) is defined as an angle of \( \pi/2 \) to the left of the spin direction such that \( \phi = 0 \) and \( \phi = \pi \) are where the cross section is maximal.  The TSSA can then be extracted by measuring \( \epsilon_N(\phi) \), the raw asymmetry as a function of \( \phi \), and fitting it to a sinusoid to extract the amplitude \( A_N \):  
\begin{equation}
P \cdot A_N \cdot \cos{ \phi } = \epsilon_N(\phi) 
\end{equation}

\noindent This was the method that was used for the PHENIX forward \( \eta \) meson TSSA which measured photons with the MPC which has full azimuthal acceptance.\cite{PPG170}  Figure~\ref{Figure:ForwardEtaTSSAFit} shows an example of a sinusoidal fit to the raw asymmetry to extract the \( A_N \) for a single bin in \( x_F \). 

At central rapidity, however, this is less practical.  Not only do the central arms only cover \( 180^\circ \) in azimuth, these midrapidity asymmetries tend to be consistent with zero and statistically limited.  So while it is possible to extract an amplitude that is consistent with zero from a sinusoidal fit function (Section~\ref{Section:SinPhi} will explain how this is used as a cross check), it is much more straightforward to integrate over the full \( \phi \) range of each arm.  But now this means that the asymmetry is being diluted across a wide range in \( \phi \) and needs to be corrected by \( \langle \mid \cos{ \phi }\mid \rangle \), the azimuthal acceptance correction:
\begin{equation}
    A_N = \frac{1}{P} \frac{1}{\langle \mid \cos{ \phi } \mid \rangle } A_N^{raw}
\end{equation} 

Assuming uniform azimuthal coverage across each PHENIX arm, the acceptance correction could be calculated as: 
\begin{equation}
    \langle \mid \cos { \phi  }\mid \rangle = \frac{\int \mid \cos{ \phi } \mid d\phi}{\int d\phi} 
\end{equation}

\begin{figure}
\centering
\subfigure[West Arm\label{Figure:dpWestPhi}]{ \includegraphics[scale = 0.36]{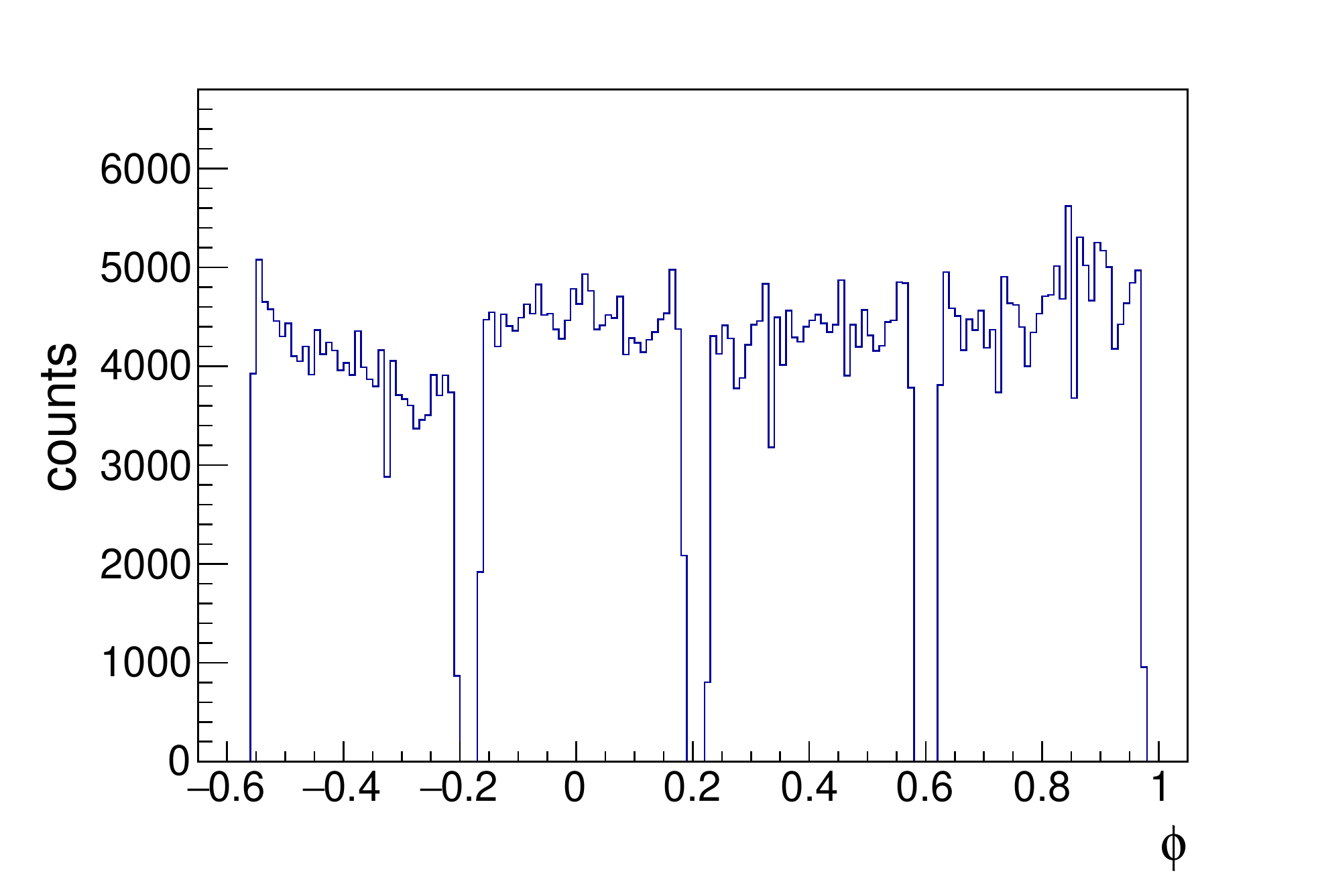} } 
\subfigure[East Arm\label{Figure:dpEastPhi}] { \includegraphics[scale = 0.36]{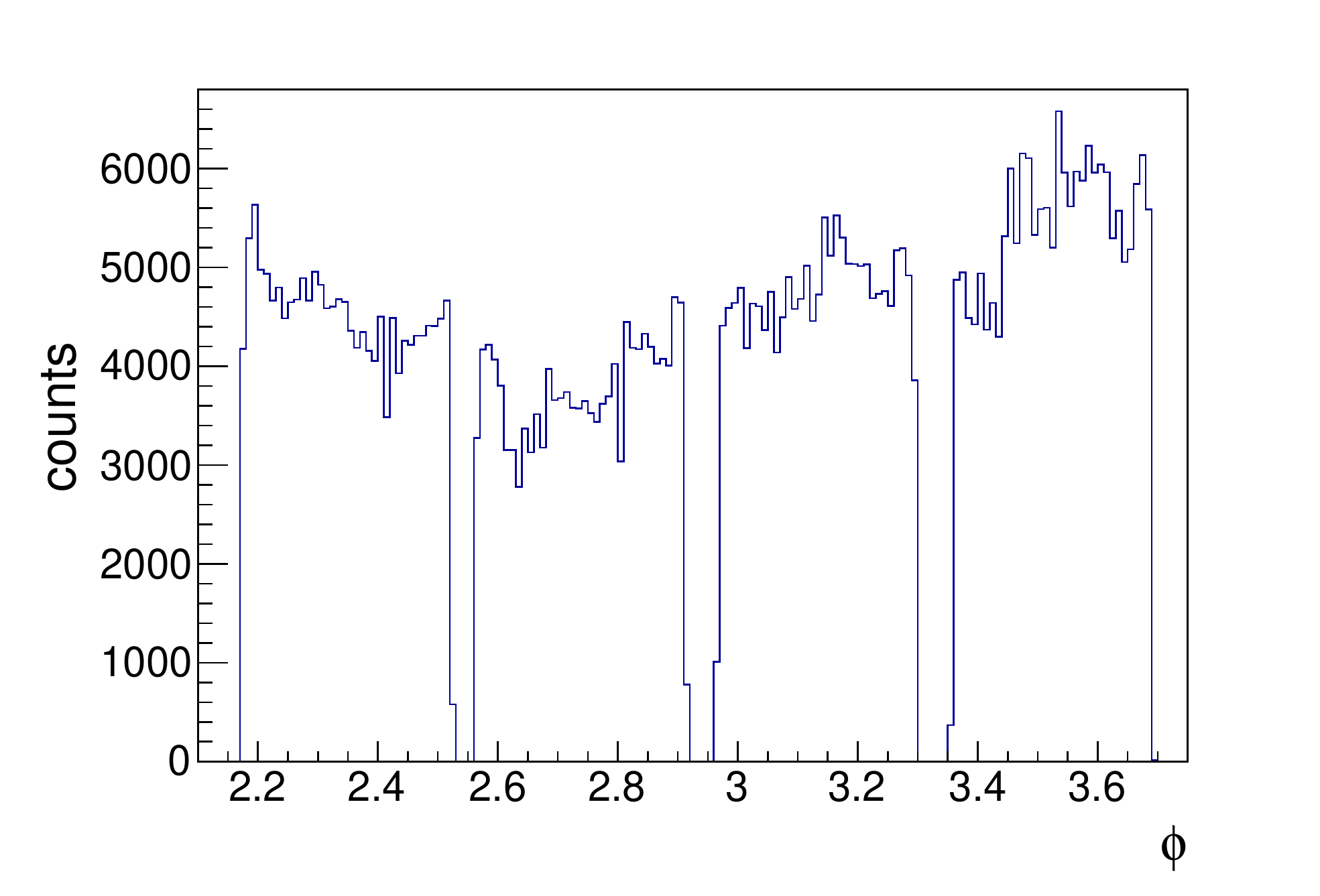} }
\caption[The isolated direct photon \( \phi \) distribution is not uniform across both arms.]{The isolated direct photon \( \phi \) distribution is not uniform across both arms.}
\label{Figure:dpPhi}
\end{figure}

\begin{figure}
\centering
\subfigure[\( \pi^0 \) pairs: \( 2 < p_T^{\pi^0} < 3 \) GeV/c\label{Figure:pi0WestPhi2to3}]{ \includegraphics[scale = 0.36]{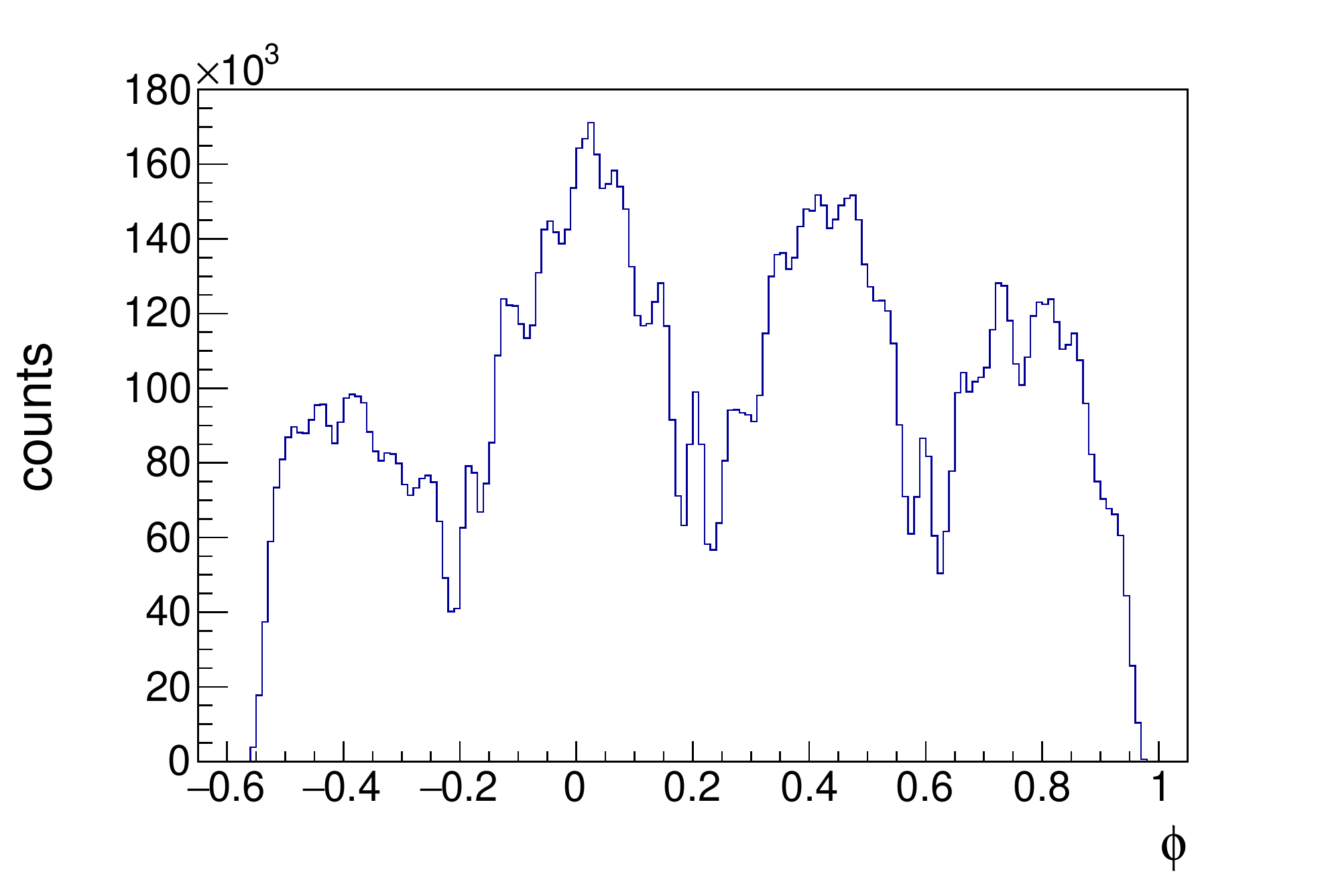} }
\subfigure[\( \eta \) pairs: \( 2 < p_T^\eta < 3 \) GeV/c\label{Figure:etaWestPhi2to3}]{ \includegraphics[scale = 0.36]{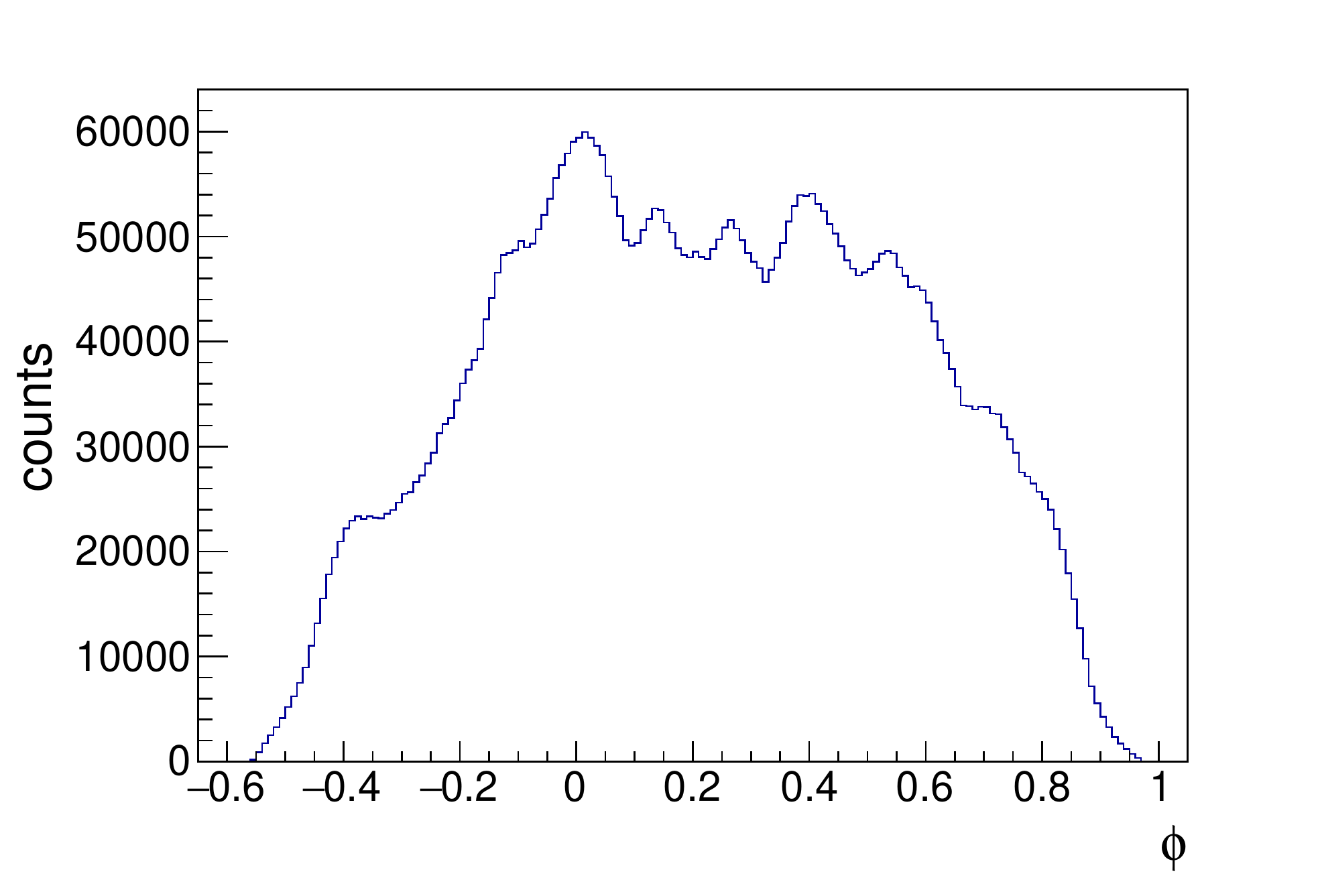} } \\
\subfigure[\( \pi^0 \) pairs: \( 5 < p_T^{\pi^0} < 6 \) GeV/c\label{Figure:pi0WestPhi5to6}]{ \includegraphics[scale = 0.36]{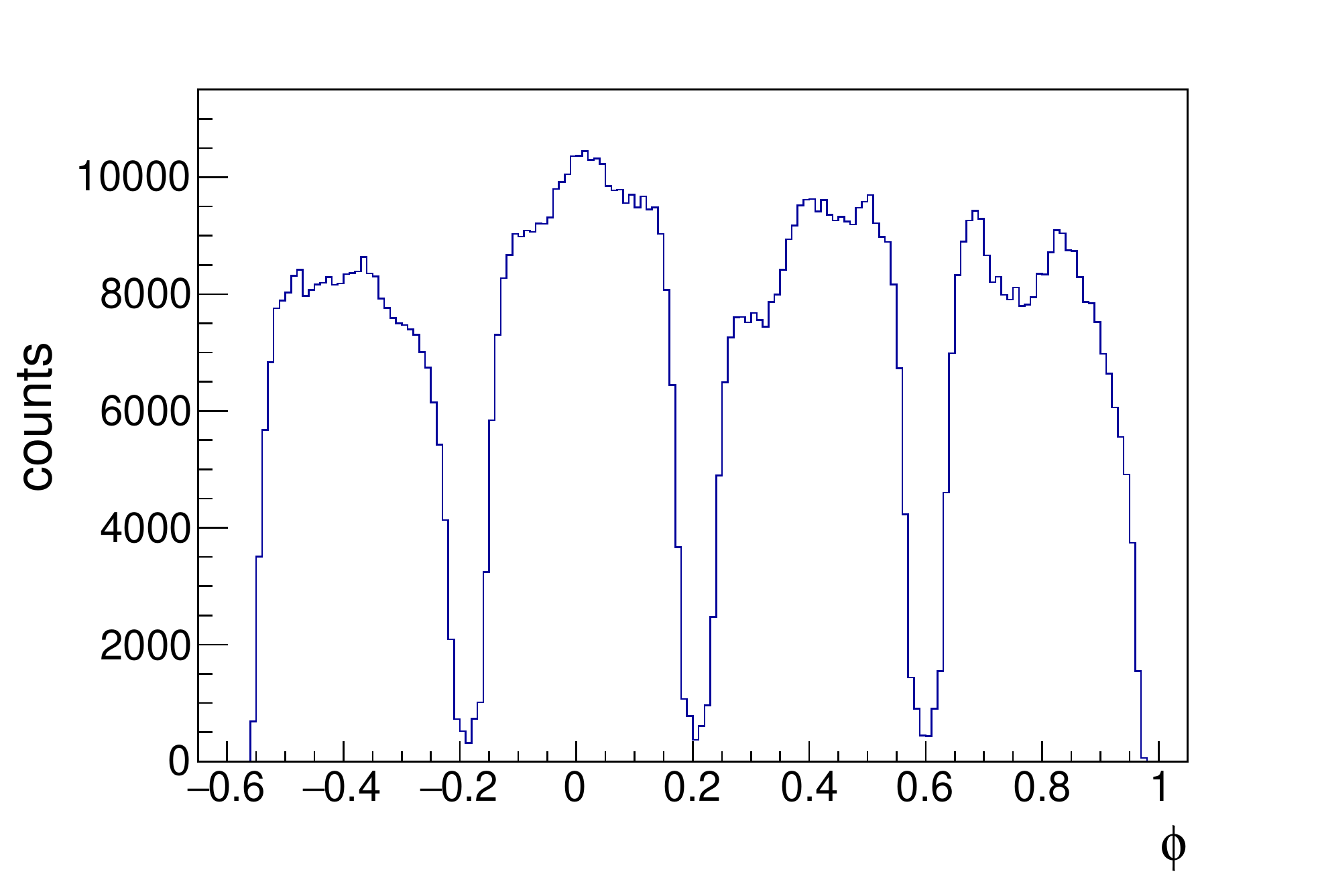} }
\subfigure[\( \eta \)  pairs: \( 5 < p_T^\eta < 6 \) GeV/c\label{Figure:etaWestPhi5to6}]{ \includegraphics[scale = 0.36]{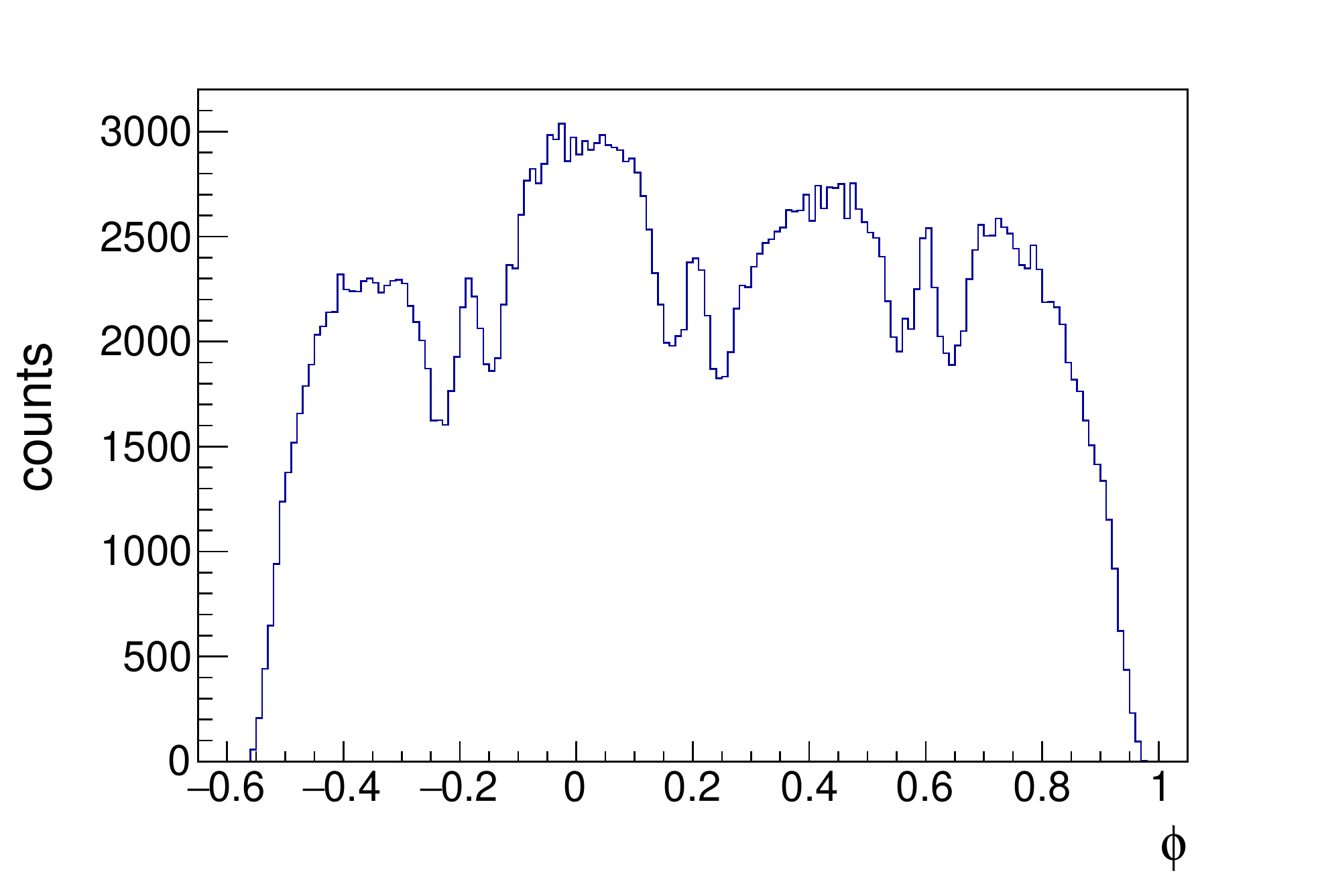} } \\
\subfigure[\( \pi^0 \) pairs: \( 12 < p_T^{\pi^0}< 20 \) GeV/c\label{Figure:pi0WestPhi12to20}]{ \includegraphics[scale = 0.36]{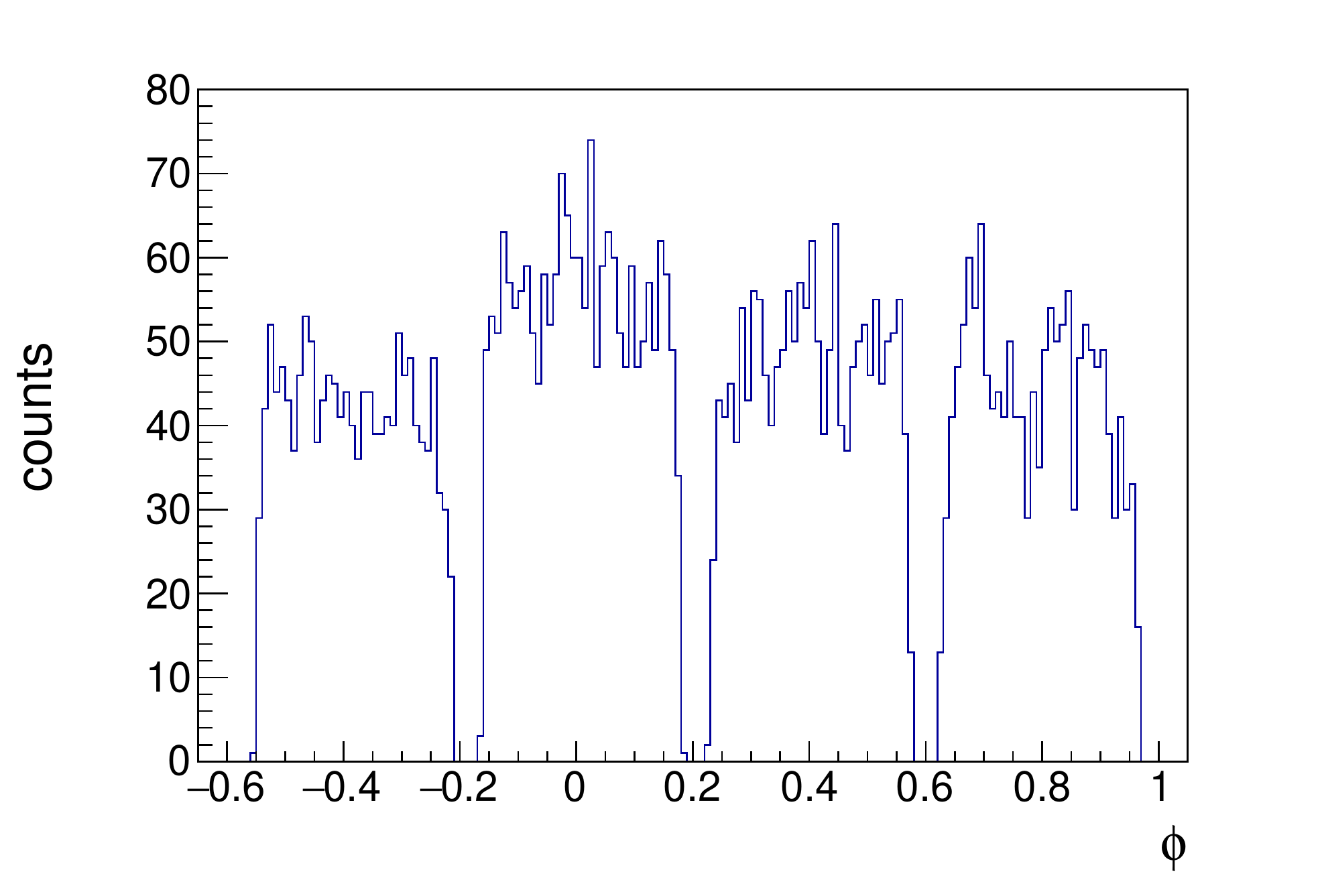} }
\subfigure[\( \eta \) pairs: \( 10 < p_T^\eta < 20 \) GeV/c\label{Figure:etaWestPhi10to20}]{ \includegraphics[scale = 0.36]{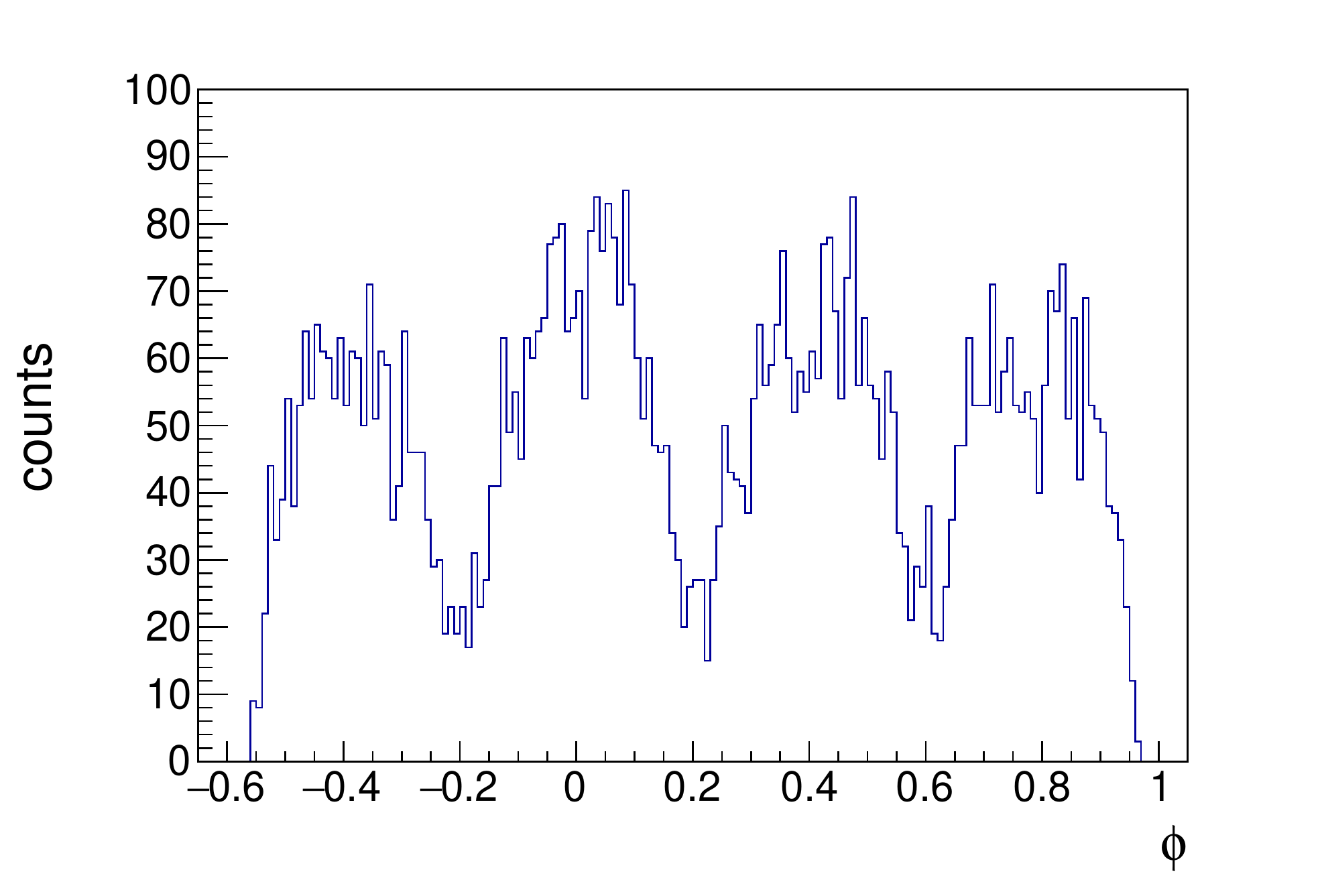} } \\
\caption[Example plots of the \( \phi \) distribution photon pairs in the west arm with invariant mass in either the \( \pi^0 \) or \( \eta \) peak. These plots clearly show that the azimuthal acceptance of diphoton decays changes a function of \( p_T \).]{Example plots of the \( \phi \) distribution photon pairs in the west arm with invariant mass in either the \( \pi^0 \) or \( \eta \) peak .  These plots clearly show that the azimuthal acceptance of diphoton decays changes a function of \( p_T \).}
\label{Figure:pairPhi}
\end{figure}

\noindent However, the \( \phi \) distributions in Figures~\ref{Figure:dpPhi}  and ~\ref{Figure:pairPhi} show that this is not the case.   Figure~\ref{Figure:dpPhi} shows the \( \phi \) distribution for all photons in the direct photon sample. Not only are there gaps between the sectors because of the edge tower cuts, but there are also some fluctuations in yields across the sectors due to dead areas and the difference in the behaviors of the PbSc and PbGl calorimeters. These plots use PHENIX coordinates such that the block on the left side of Figure~\ref{Figure:dpWestPhi} corresponds to the sector at the bottom of the west arm and the block on the right side of Figure~\ref{Figure:dpEastPhi} corresponds to the PbGl sector at the bottom of the east arm. Because the \( \phi \) distribution is not flat across both arms, the azimuthal acceptance correction is instead calculated by taking the average \( \mid \cos{ \phi } \mid  \) value for all direct photon candidates:  
\begin{equation}\label{Equation:cosSum}
\langle \mid \cos{ \phi } \mid \rangle = \frac{\sum_{i=1}^{N} \mid \cos{ \phi_i } \mid }{N}
\end{equation}
\noindent For the direct photon analysis this \( \langle \mid \cos{ \phi }\mid \rangle \) is found to be 0.878 in the west arm and 0.882 in the east arm.  

The acceptance correction for diphoton pairs is slightly more complicated because their azimuthal acceptance changes as a function of \( p_T \), which is shown with example plots in Figure~\ref{Figure:pairPhi}.  In the rest frame of the hadron, the photons from a \( h \rightarrow \gamma \gamma \) decay travel back to back to conserve momentum.  In the lab frame, they travel with some decay angle between them and the faster that this hadron was traveling, the smaller this decay angle tends to be.  At lower \( p_T \), the decay angle tends to be wider and \( \eta \rightarrow \gamma \gamma \) decays tend to have wider decay angles than \( \pi^0 \rightarrow \gamma \gamma \) decays at the same momentum.  This is because \( \eta \) mesons are about four times heavier than \( \pi^0 \)s and so their decays have access to a wider range of phase space.  So in Figure~\ref{Figure:etaWestPhi2to3}, the structure of the separate EMCal sectors from the single photon spectrum in Figure~\ref{Figure:dpPhi} is almost completely washed out in this lower \( p_T \) \( \eta \) photon pair azimuthal spectrum and the number of detected pairs falls off towards the edges of the detector arm.  For photon pairs with invariant mass near the \( \pi^0 \) peak, the lower \( p_T \) azimuthal distribution in Figure~\ref{Figure:pi0WestPhi2to3} shows some of the separate EMCal structure from the single photon spectrum because of the smaller decay angle.  The smaller peaks in between the wider sector peaks represent pairs where the photons are detected in adjacent sectors.  Again, the higher the \( p_T \) of the photon pair, the smaller the decay angle, so in Figures~\ref{Figure:pi0WestPhi5to6} and ~\ref{Figure:pi0WestPhi12to20} the \( \pi^0 \) azimuthal distribution even more closely resembles the single photon spectrum.  In Figures~\ref{Figure:etaWestPhi5to6} and \ref{Figure:etaWestPhi10to20} the \( \eta \) azimuthal distribution starts to resemble the lower \( p_T \) \( \pi^0 \) \( \phi \) spectrum in Figure~\ref{Figure:pi0WestPhi2to3}.  Thus, it is crucial that the acceptance correction from Equation~\ref{Equation:cosSum} is calculated as a function of \( p_T \) for the photon pair TSSAs.  The \( \pi^0 \) and \( \eta \)  azimuthal acceptance corrections range in the west arm from about 0.95 at low \( p_T \) to about 0.89 at high \( p_T \).  In the east arm there is far more dead area, so the acceptance correction stays roughly constant with \(p_T \) at about 0.88.  

\subsection{Beam Polarization} \label{Section:Polarization} 
\begin{figure}
\centering
\includegraphics[width=0.9\linewidth]{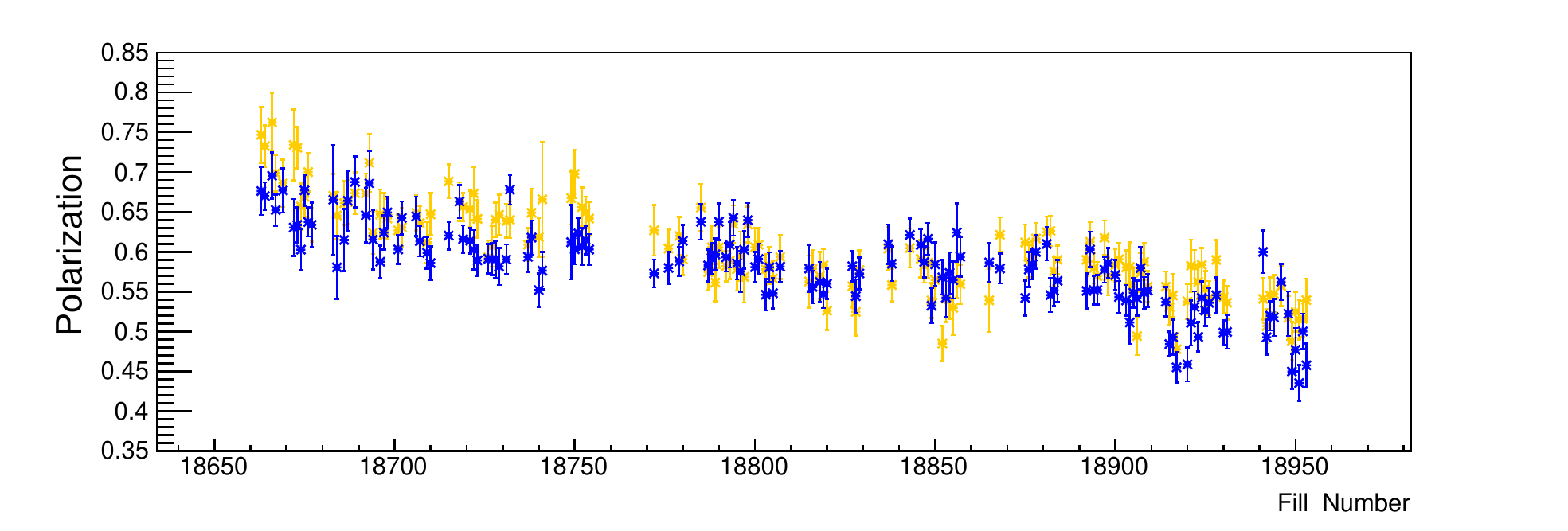}
\caption[The beam polarization as a function of fill number for both the blue and yellow beams.]{The beam polarization as a function of fill number for both the blue and yellow beams.}\label{Figure:polarization}
\end{figure}

Collisions from unpolarized protons dilute the TSSA measurement, causing the measured asymmetry to be smaller than it would have been had the beam had a higher level of polarization.  Thus, the asymmetry needs to be corrected by dividing by the average polarization of the beam.  The fill by fill polarization values for both the blue and yellow beams are measured by the CNI polarimetry group and plotted in Figure \ref{Figure:polarization}.  A description of how the absolute beam polarization is measured can be found in Section~\ref{Section:Polarimeters}.

\subsection{Relative Luminosity}\label{Section:RelativeLuminosity}
The relative luminosity formula is a method of calculating the TSSA that uses the fact that the beam can be polarized either up or down and will be described in more detail in Section~\ref{Section:RelativeLuminosityFormula}.  One has to be careful to avoid the potential systematic effect from having more \( p^\uparrow + p \) collisions than \(p^\downarrow + p \) collisions or vice versa.  To correct for such effects, this formula includes the relative luminosity between when the beam is polarized up versus when the beam is polarized down: \( \mathcal{R} =  \mathcal{L}^\uparrow /  \mathcal{L}^\downarrow \). 

\begin{figure} 
 \centering
 \includegraphics[width=0.9\linewidth]{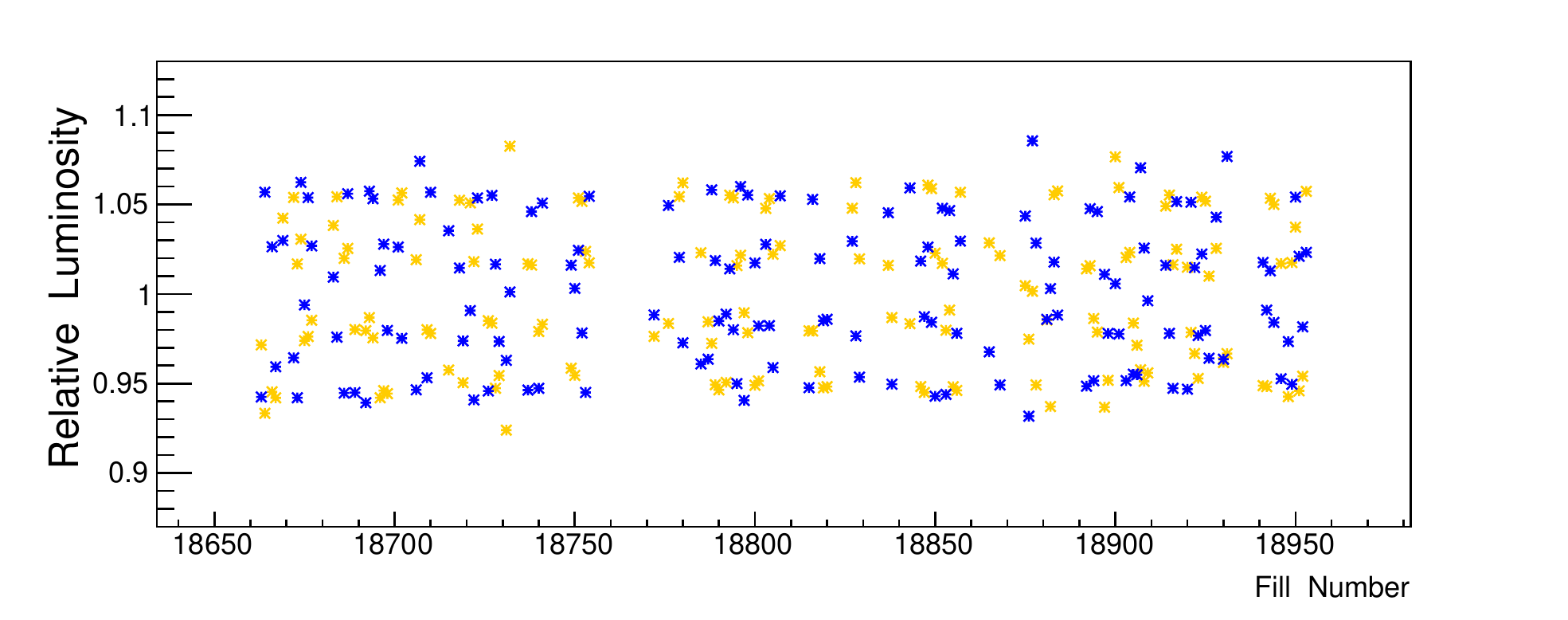}
\caption[The relative luminosity as a function of fill number for the both blue and yellow beams.]{The relative luminosity as a function of fill number for the both blue and yellow beams.}\label{Figure:relativeLuminosity}
\end{figure}

These crossing-dependent luminosities are calculated by summing the GL1P scalers.  These are numbers taken from the spin database that keep track of how many times each crossing fires a particular Global Level 1 (GL1) trigger over the course of a run.  These TSSA analyses used the GL1P scaler for the minimum bias BBC Local Level-1 (BBCLL1) trigger which incorporates a 30 cm vertex cut.  The relative luminosity is calculated as the sum of these scalers for when the crossings were spin up for a particular beam divided by the sum for when the crossings were spin down for that same beam.  This has to be done separately for the blue and yellow beams since their spin configurations are independent from each other.  The relative luminosity values that are used in these analyses are between 0.9 and 1.1 and are shown as a function of fill number in Figure~\ref{Figure:relativeLuminosity}.  

\section{Calculating the Asymmetry}\label{Section:CalculatingTheAsymmetry}
\begin{figure} 
\centering
\includegraphics[scale = 0.7]{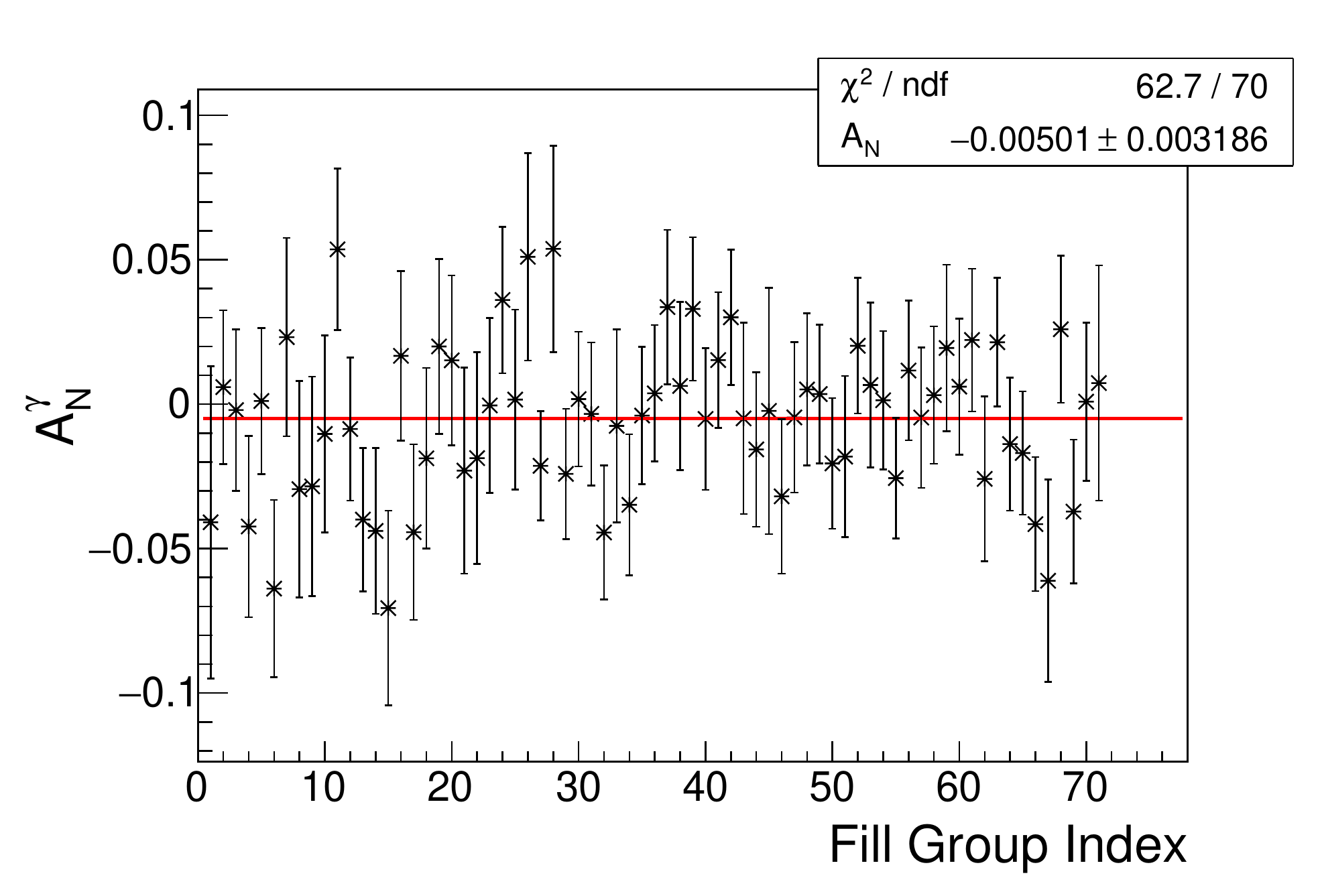}
\caption[An example plot of the asymmetry as function of fill group for the direct photon blue beam, left relative luminosity asymmetry for \( 5 < p_T^\gamma < 6 \) GeV/c. ]{An example plot of the asymmetry as function of fill group for the direct photon blue beam, left relative luminosity asymmetry for \( 5 < p_T^\gamma < 6 \) GeV/c. }
\label{Figure:dpAsymmetryVSFill}
\end{figure}

The data is broken into different fill groups with a separate asymmetry calculated for each data group that is corrected by the corresponding polarization and relative luminosity.  
This ensures that the trigger and reconstruction efficiencies for that specific data group cancel out in the ratio and also makes the polarization correction more accurate since the polarization of both beams decreased over the length of Run-15 (see Figure~\ref{Figure:polarization}).  The statistical uncertainty for the asymmetry is calculated using standard error propagation which assumes Poissonian statistics, so each count that is plugged into every fill group's asymmetry is required to be 10 or higher.  If any of the counts are less than 10, then the data from this fill group is thrown out for that particular \( p_T \) bin.  In order to increase the number of \( p_T \) bins possible, these asymmetries combine data from two separate fills together instead of calculating the asymmetry fill by fill.  The fills are grouped together chronologically such that the first two fills of Run-15 become the first group, the third and fourth fills are put into the second group, and so on. 
These asymmetries are then plotted as a function of fill group to ensure that there are no systematic effects that would cause the asymmetry to change over time,
 an example of which is shown in Figure~\ref{Figure:dpAsymmetryVSFill} for the direct photon asymmetry.  
The polarization values used for these fill groups are the average polarization of the two fills weighted by each fill's luminosity. These plots are then fit to a constant which effectively takes the average of the asymmetry over all fill groups, weighted by the statistical error. 



\subsection{Relative Luminosity Formula}\label{Section:RelativeLuminosityFormula}
Even though TSSAs are usually described as ``left-right'' asymmetries, comparing particle yields from different parts of a detector involves correcting for the differences in the detector reconstruction efficiencies.  This is especially important for the PHENIX EMCal where the entire west arm is made of all PbSc sectors and the east arm is half PbSc and half PbGl sectors.  The relative luminosity formula combines data in such a way that these detector effects cancel out by taking advantage of the fact that the proton beam can either be polarized up or polarized down.    So instead of comparing the number of particles to the left when the beam is polarized up, \( N^\uparrow_L \), to the counts to the right when the beam polarization points in the same direction, \( N^\uparrow _R \),  the \( N^\uparrow_L \) counts are compared to the number of particles of the left when the beam was polarized \textit{down}, \( N^\downarrow _L \):
\begin{equation}
    A_N = \frac{1}{P} \bigg( \frac{\sigma^\uparrow - \sigma^\downarrow}{\sigma^\uparrow + \sigma^\downarrow} \bigg)
    =\frac{1}{P} \frac{1}{\langle \mid \cos{ \phi } \mid \rangle } \frac{N^\uparrow_L - \mathcal{R} \cdot N^\downarrow _L}{N^\uparrow_L +  \mathcal{R} \cdot N^\downarrow _L}
\end{equation}

\noindent  The \( N^\downarrow _L \) counts are equivalent to the \( N^\uparrow _R \) ones but do not need to be corrected for any detector effects.  This is an exact expression of the asymmetry and because it takes the ratio of counts that are all found in the same side of the detector, effects from detector acceptance and efficiency cancel out.  However, this formula does need to be corrected by \( \mathcal{R} \), the relative luminosity between the different beam configurations, which was discussed is Section~\ref{Section:RelativeLuminosity}.  There is an equivalent formula for the right-side counts, where the signs in the numerator are flipped to preserve the left-right asymmetry convention.   

\begin{figure}
\centering
\subfigure[Yellow Beam]{ \includegraphics[scale = 0.32]{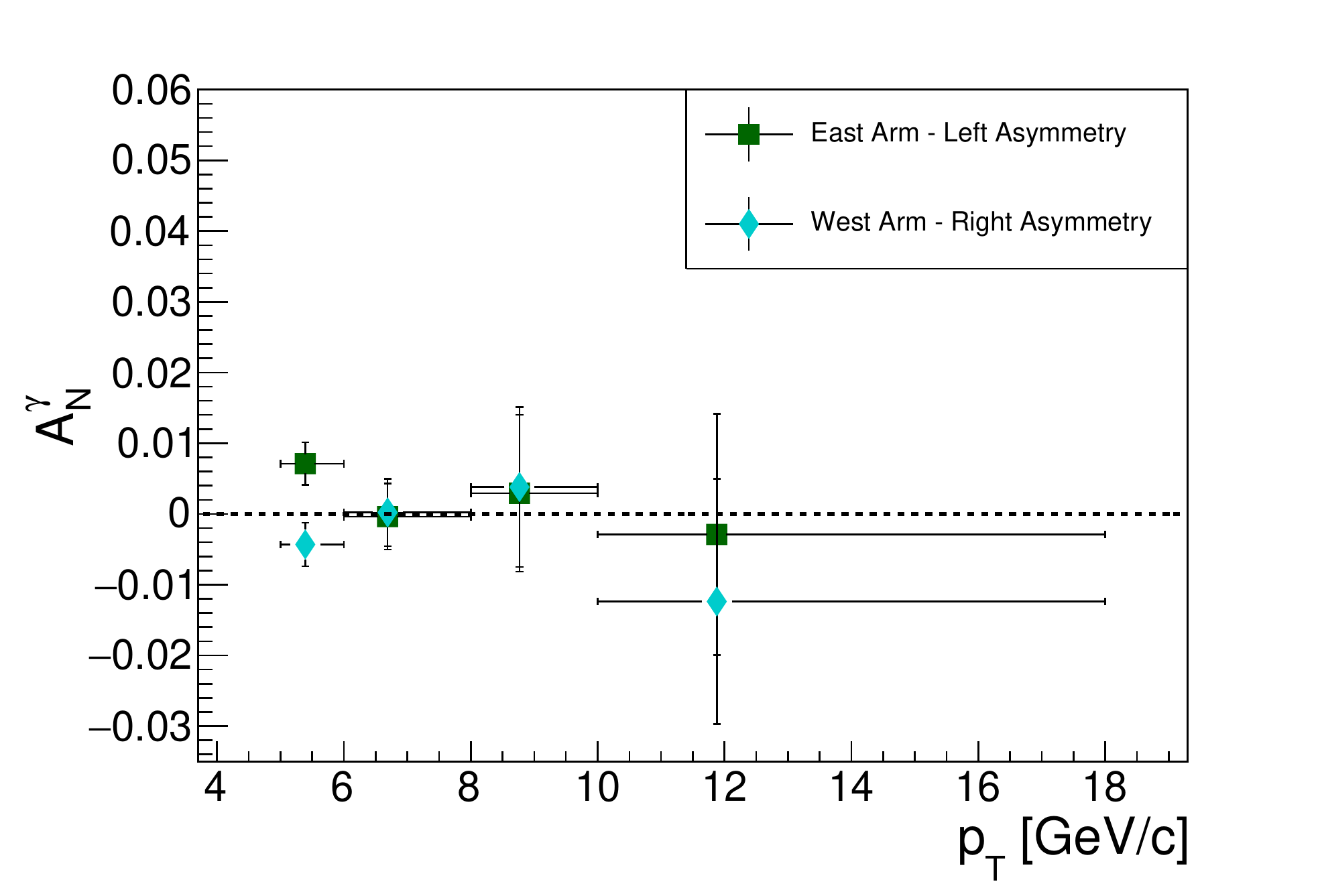} }
\subfigure[T test of the yellow beam left and right asymmetry results]{ \includegraphics[scale = 0.32]{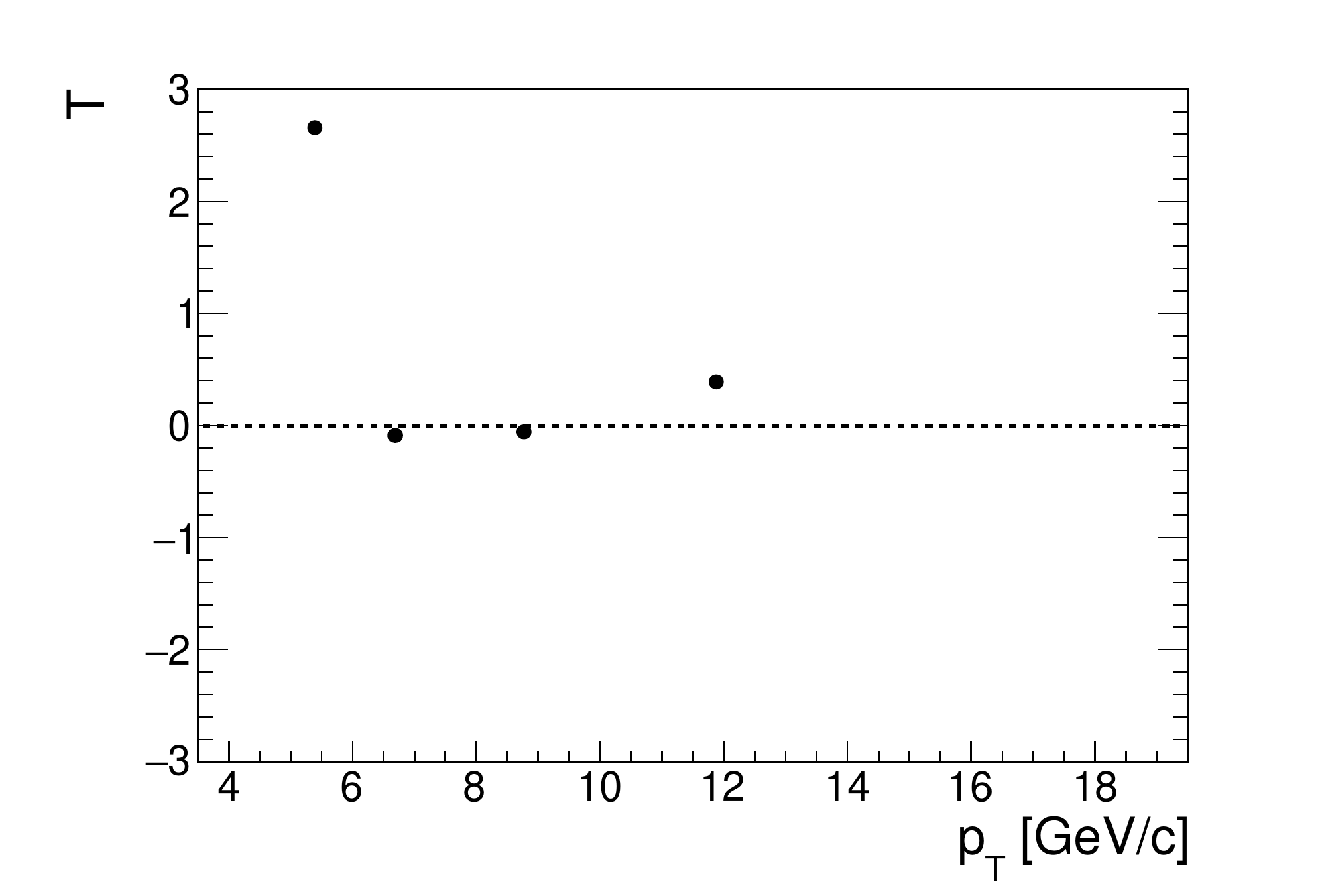} }\\
\subfigure[Blue Beam]{ \includegraphics[scale = 0.32]{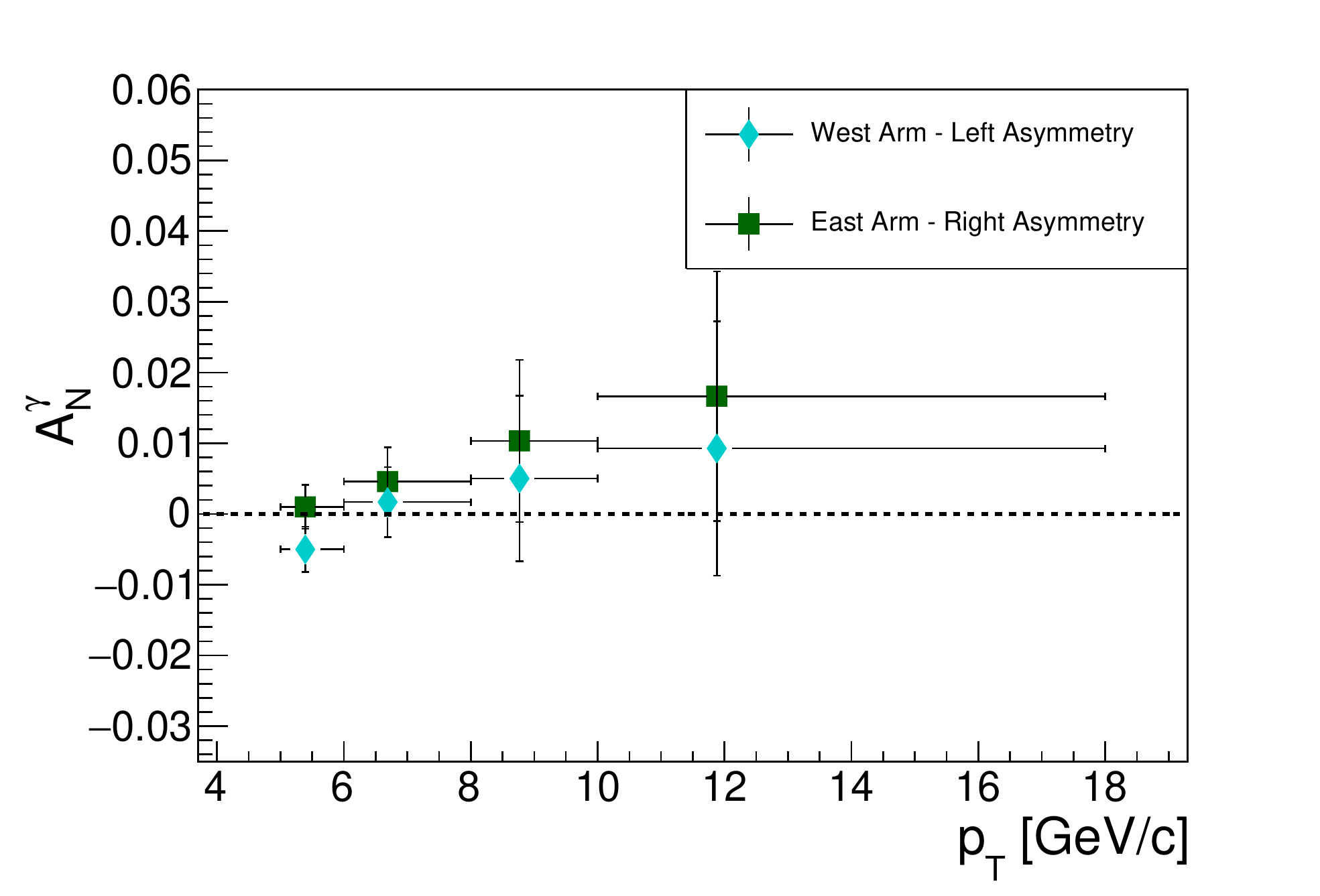} } 
\subfigure[T test of the blue beam left and right asymmetry results]{ \includegraphics[scale = 0.32]{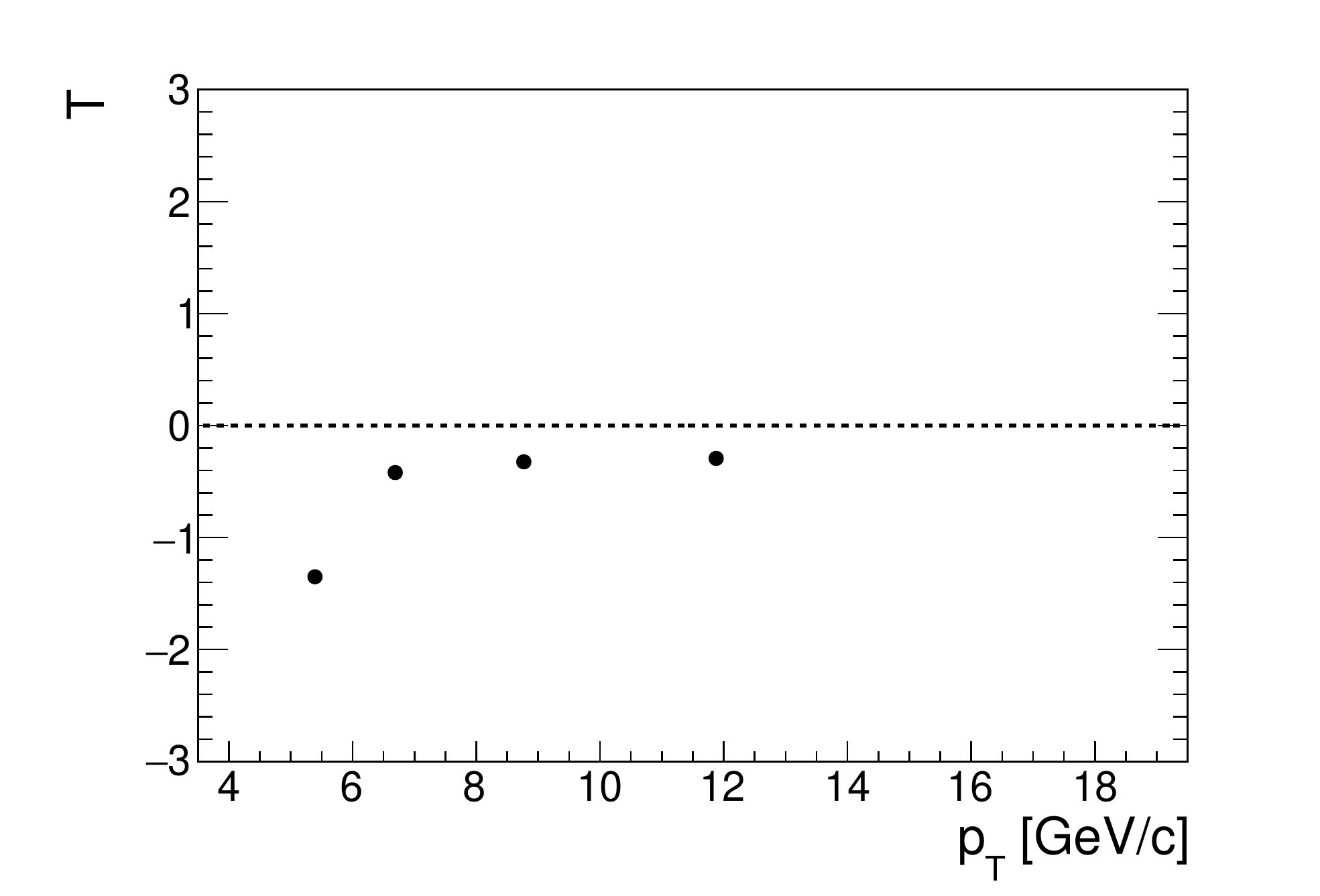} } 
\caption[The direct photon asymmetry left and right asymmetry results calculated using the relative luminosity formula.]{The direct photon asymmetry left and right asymmetry results calculated using the relative luminosity formula. These results are for the direct photon asymmetry before the background correction that will be discussed in Section~\ref{Section:BackgroundSubraction}.}
\label{Figure:dpLumiLeftRight}
\end{figure}

There are four different relative luminosity formula results, two for each beam, which are all shown for the direct photon asymmetry in Figure~\ref{Figure:dpLumiLeftRight}.  These are all statistically independent from each other and will be averaged together to calculate the final asymmetry after correcting for background, discussed in Section~\ref{Section:BackgroundSubraction}.  Before averaging the results, we can also check their consistency with each other.  A T test provides a way of quantifying the differences between the left and right asymmetries:
\begin{equation} \label{Equation:TTestLeftRight}
T(p_T) = \frac{ A_N^{Left} - A_N^{Right} }{ \sqrt{ (\sigma^{Left})^2 + (\sigma^{Right})^2 } }
\end{equation}
\noindent To demonstrate that the differences between these results are not statistically significant, we would expect the T values to follow a normal distribution: an even split between positive and negative values, with approximately 68\% of them having a magnitude that is less than 1, and about 95\% with a magnitude less than 2.  But as Figure~\ref{Figure:dpLumiLeftRight} shows, it is difficult to apply Gaussian statistics to only eight points.  Still, six out of eight of the T values have a magnitude that is less than 1 and only a single T value has a magnitude greater than 2.  And two out of the eight points are positive which is two points away from an even four-four split.  Thus, this T test shows no evidence that the results are not consistent with one another.

\begin{figure}
\centering
\subfigure[Yellow and Blue Beam Asymmetries]{\includegraphics[scale = 0.32]{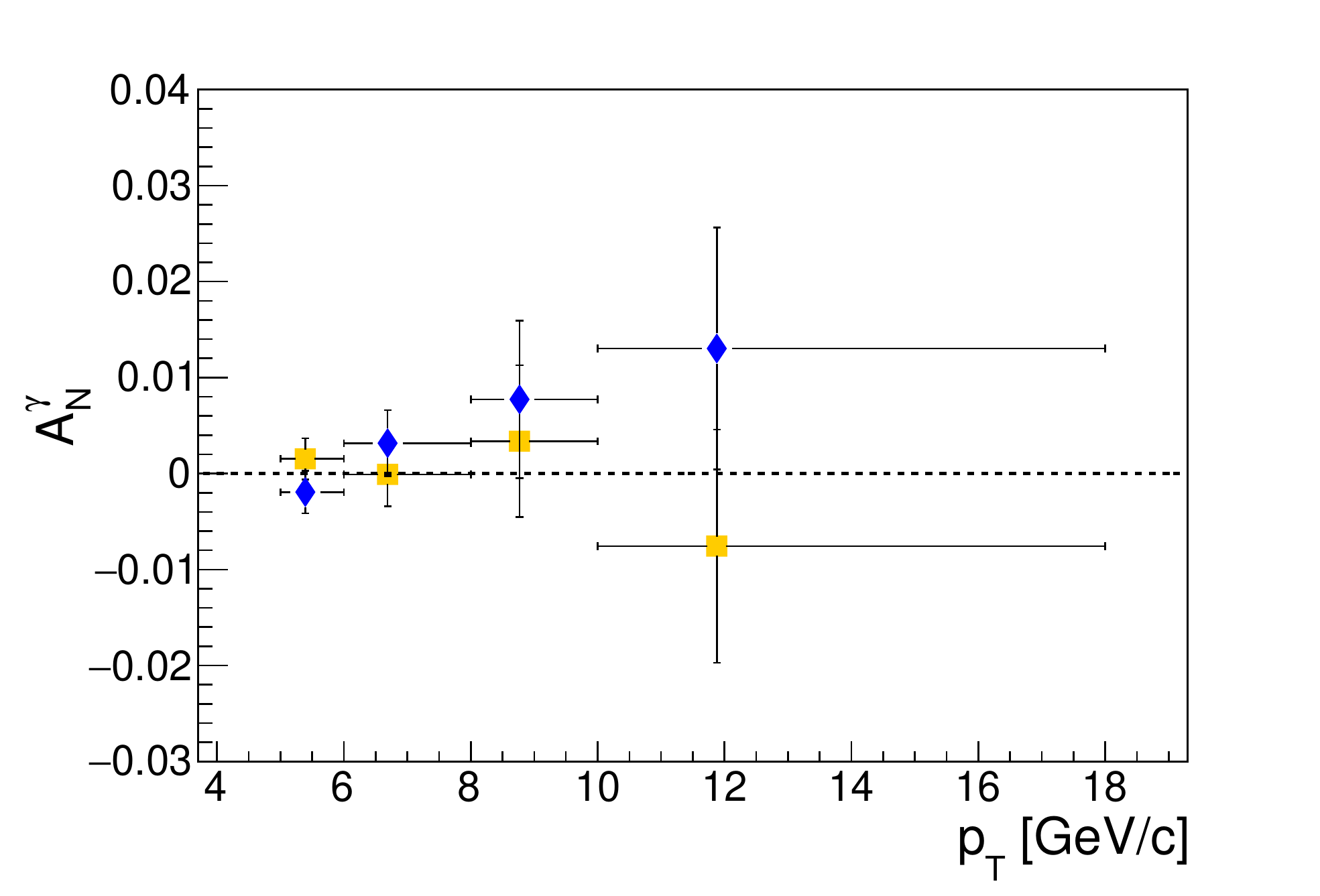} } 
\subfigure[T test comparing the result of the blue and yellow beam asymmetries\label{dpLumiYellowBlueTTest}]{\includegraphics[scale = 0.32]{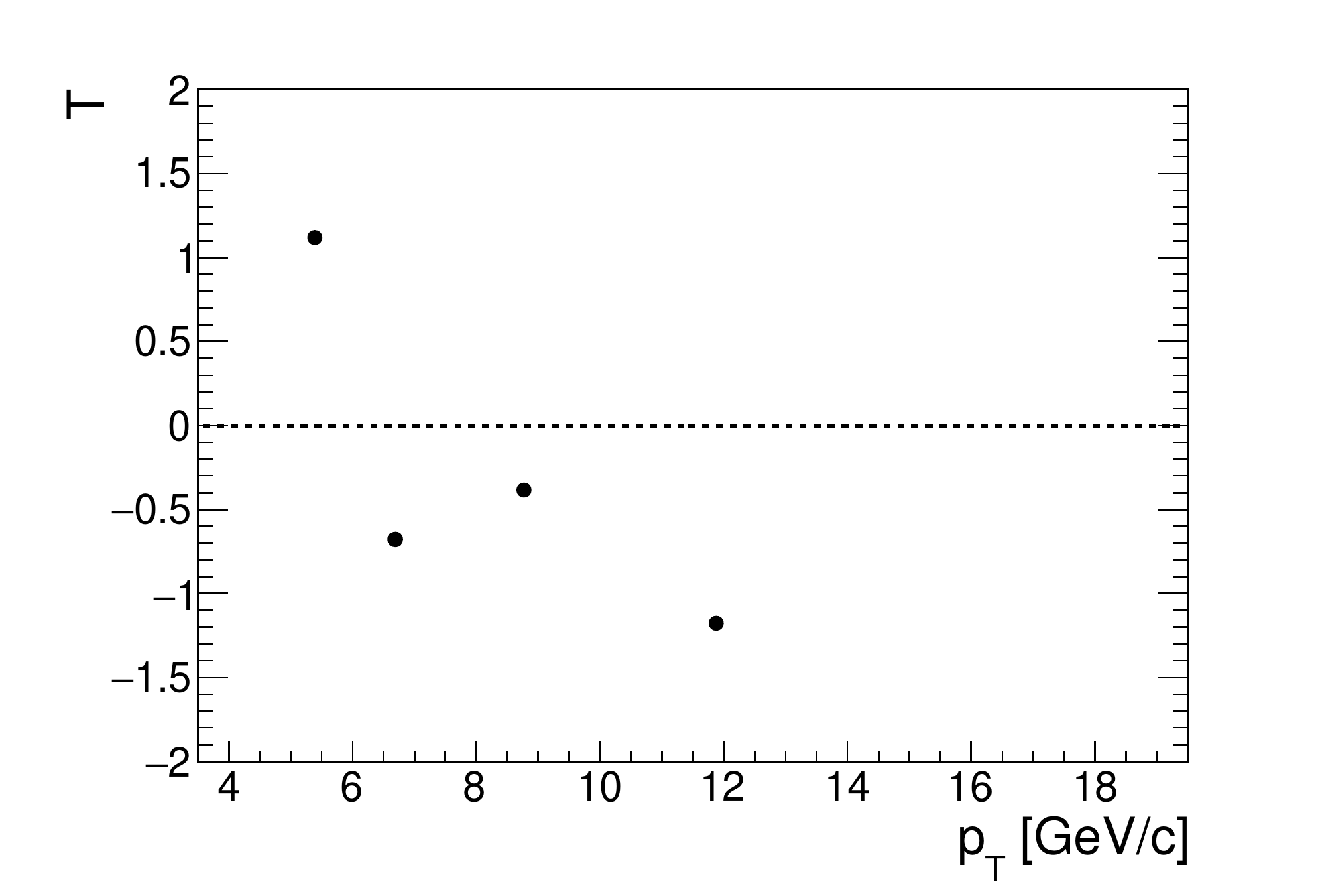} }
\caption[Relative luminosity yellow and blue beam asymmetries for the direct photon asymmetry before the background correction]{Relative luminosity yellow and blue beam asymmetries for the direct photon asymmetry before the background correction, which will be explained in Section~\ref{Section:BackgroundSubraction}. }
\label{Figure:dpLumiYellowBlue}
\end{figure}

\begin{figure}
\centering
\subfigure[Yellow Beam]{ \includegraphics[scale = 0.32]{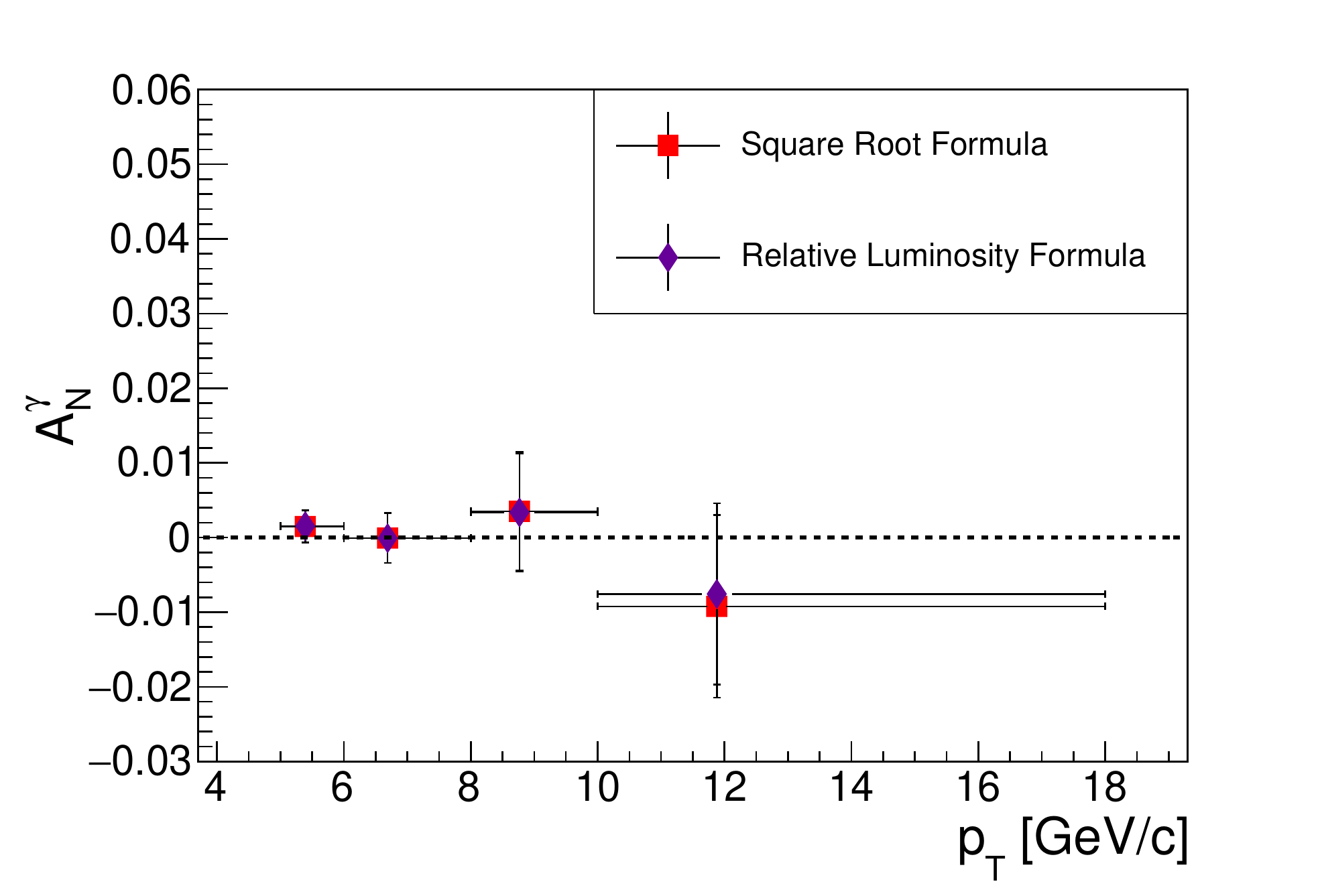} }
\subfigure[T test comparing the yellow beam square root and relative luminosity formula results]{ \includegraphics[scale = 0.32]{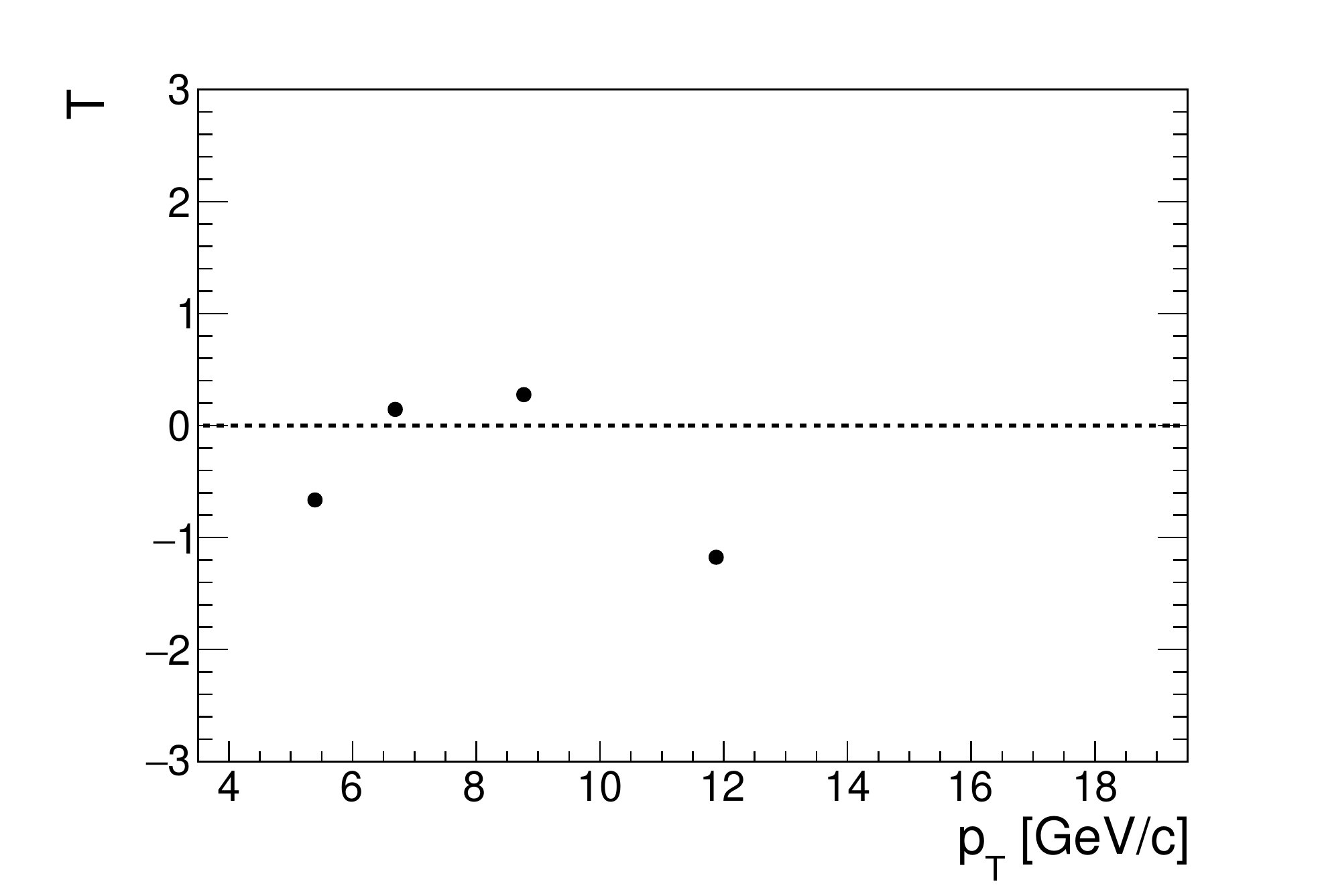} }\\
\subfigure[Blue Beam]{ \includegraphics[scale = 0.32]{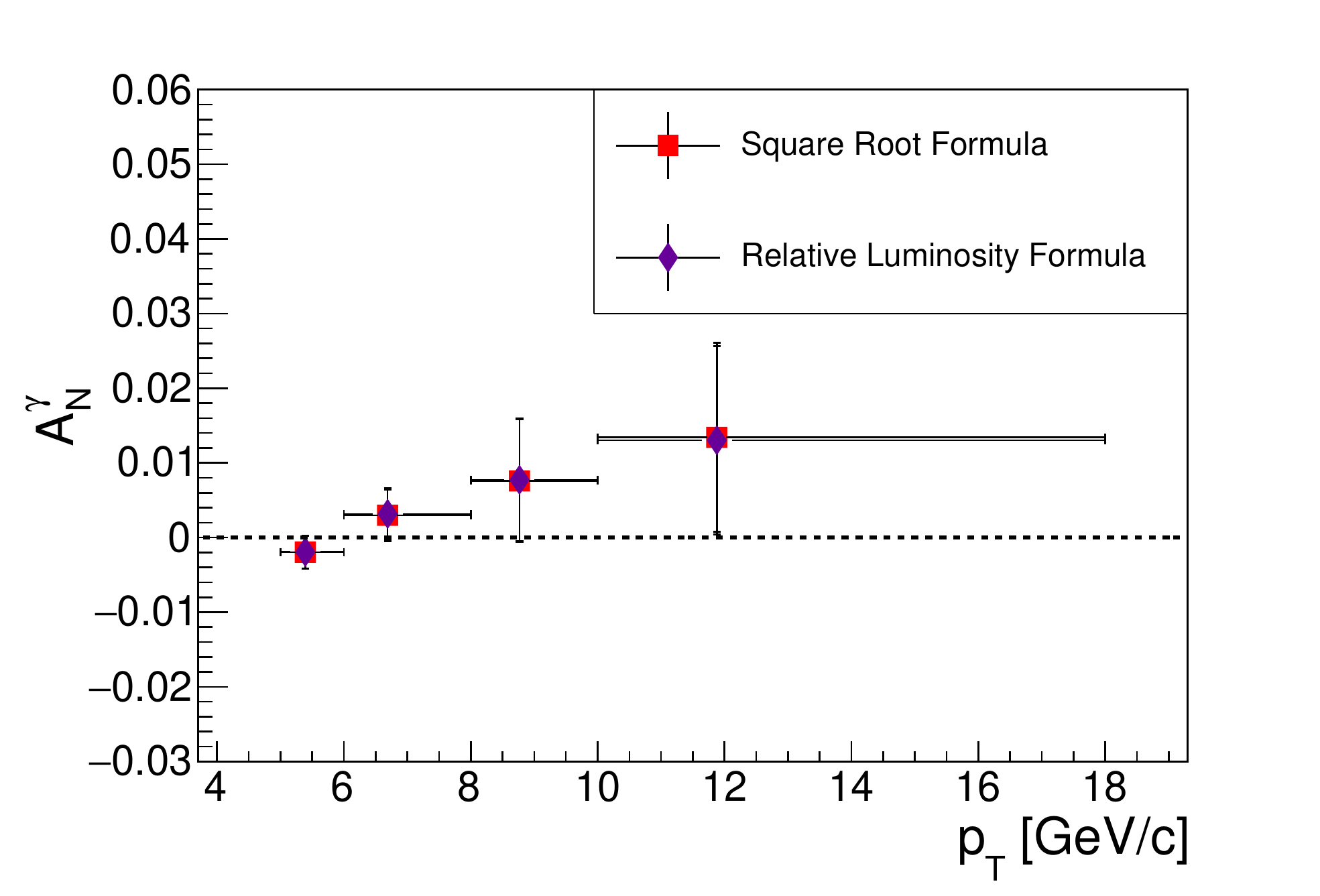} } 
\subfigure[T test comparing the blue beam square root and relative luminosity formula results]{ \includegraphics[scale = 0.32]{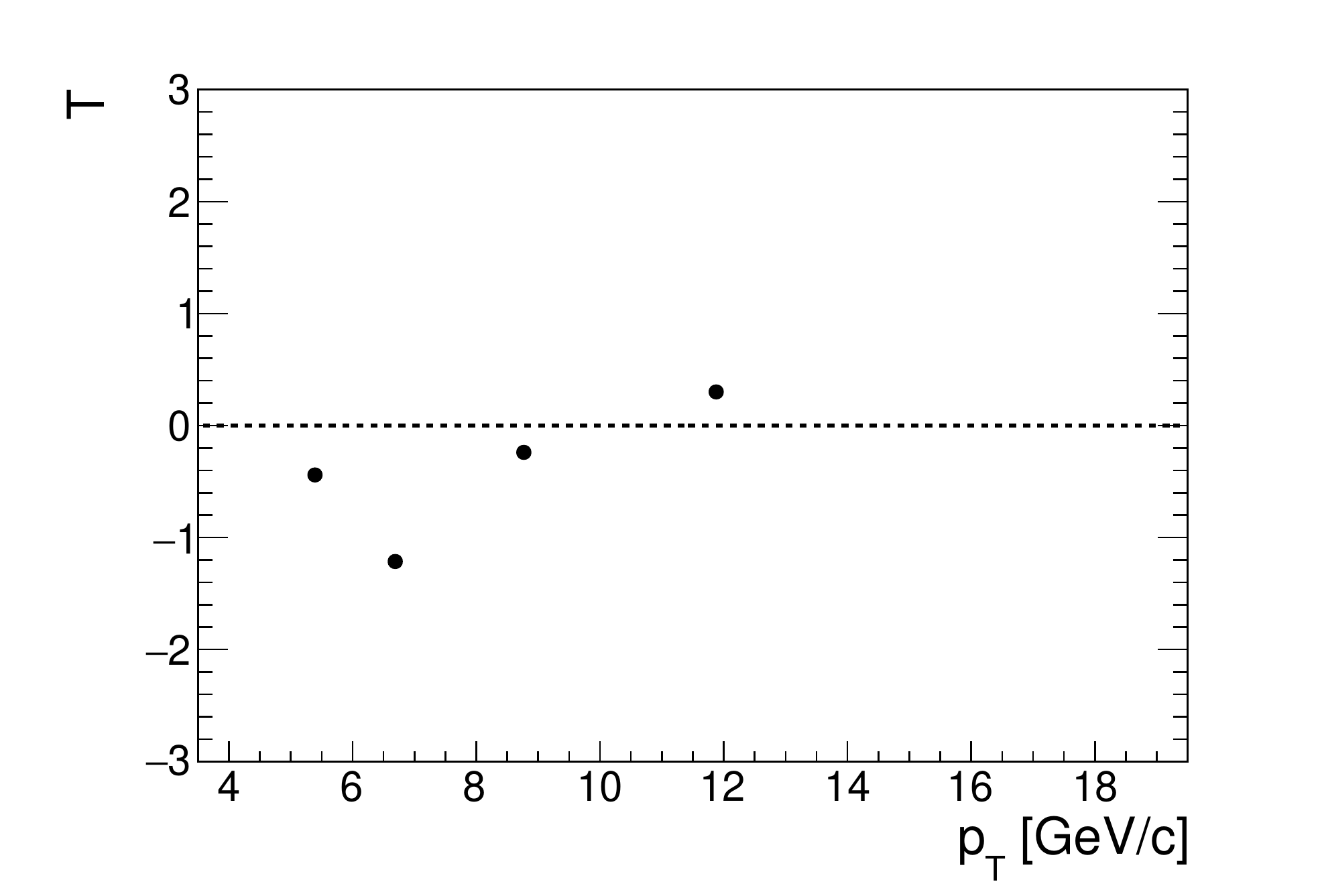} } \\
\subfigure[Final Averaged Asymmetry\label{Figure:dpAveCompare}]{ \includegraphics[scale = 0.32]{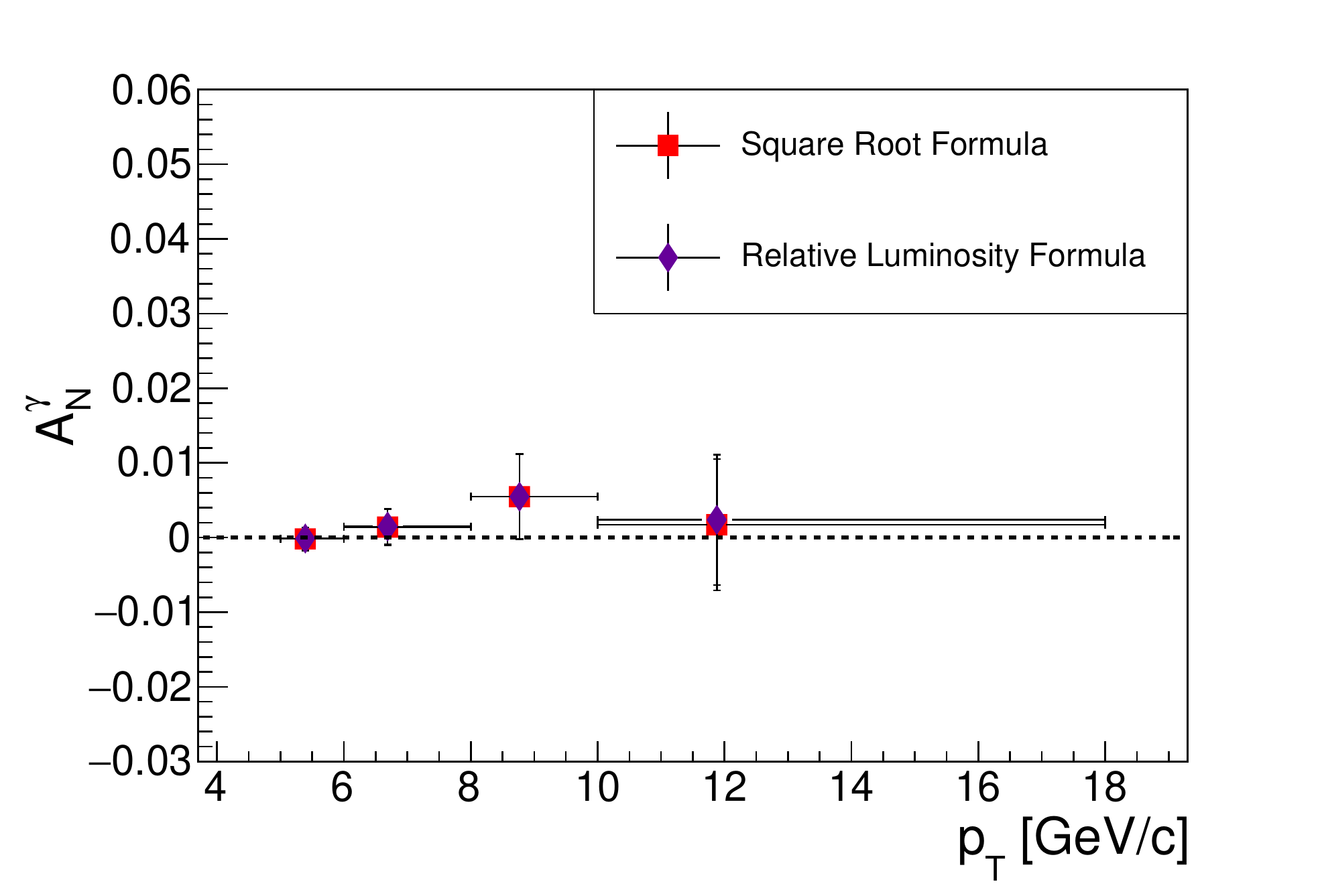} } 
\subfigure[T test comparing the averaged asymmetry results for the square root and relative luminosity formulas]{ \includegraphics[scale = 0.32]{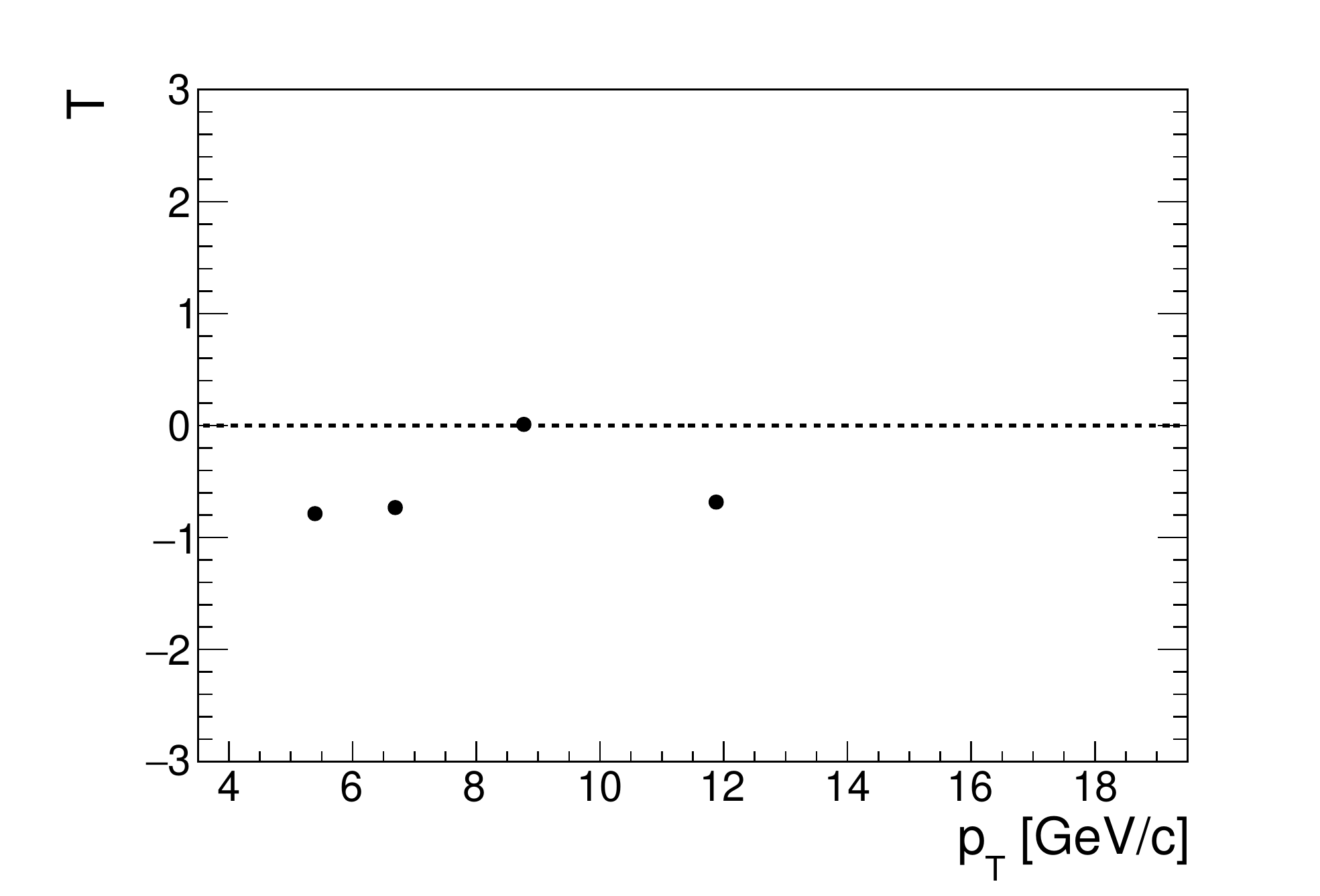} } 
\caption[Comparing the results of the relative luminosity and square root formulas for the direct photon asymmetry before the background correction]{Comparing the results of the relative luminosity and square root formulas for the direct photon asymmetry before the background correction, which will be discussed in detail in Section~\ref{Section:BackgroundSubraction}.}
\label{Figure:dpCompare}
\end{figure}

To get the yellow and blue beam results, the weighted average of the left and right asymmetries are taken, which is shown for the direct photon asymmetry in Figure~\ref{Figure:dpLumiYellowBlue}. The difference between these results can also be quantified with a similar T test: 
\begin{equation}\label{Equation:TTestYellowBlue}
T(p_T) = \frac{ A_N^{Yellow} - A_N^{Blue} } { \sqrt{ (\sigma^{Yellow})^2 + (\sigma^{Blue})^2 } }
\end{equation}
\noindent Again it is somewhat impractical to apply Gaussian statistics to the four points that are present in Figure~\ref{dpLumiYellowBlueTTest}, but all of the T values have a magnitude that is either less than or close to 1 and are not systematically positive or negative.  

\subsection{Square Root Formula Cross Check}\label{Section:SquareRootFormula}
The square root formula is an alternative way of calculating the TSSA using a geometric mean and will be used as a cross check for the relative luminosity formula result.  The square root formula is not an exact expression for the asymmetry, but to first order effects from both detector acceptance and the relative luminosity cancel out: 
\begin{equation}
A_N =\frac{1}{P} \frac{1}{\langle \mid \cos{ \phi } \mid \rangle } \frac{\sqrt{N^\uparrow_L N^\downarrow_R }  - \sqrt{N^\downarrow_L N^\uparrow_R}}
                                                                                                     {\sqrt{N^\uparrow_L N^\downarrow_R } + \sqrt{N^\downarrow_L N^\uparrow_R}}  
\label{Equation:SquareRootFormula}
\end{equation}

The relative luminosity and square root formulas calculate the asymmetry with a slightly different method, but the results are still 100\% correlated because they are using the exact same data set.  Thus, the T test formula used to quantify the differences between the formula results has a minus sign in the denominator: 
\begin{equation}\label{Equation:TTestSqrtLumi}
T(p_T) = \frac{ A_N^{Sqrt} - A_N^{Lumi} } { \sqrt{ | (\sigma^{Sqrt})^2 - (\sigma^{Lumi})^2 } | }
\end{equation}
\noindent An example of this is shown in Figure~\ref{Figure:dpCompare} for the direct photon asymmetry, where Figure~\ref{Figure:dpAveCompare} shows the  averaged asymmetry values for the different beam asymmetries.  The different formula results already show a excellent agreement, but this will be further improved once the asymmetry is corrected for background, which can be seen in Figure~\ref{Figure:dp_correctedCompare}.

\subsection{\( \sin{\phi_{s}} \) Modulation Cross Check} \label{Section:SinPhi}
Section~\ref{Section:AcceptanceCorrection} describes how the TSSA is generally extracted by measuring the raw asymmetry as a function of \( \phi \) and fitting to a sinusoid to measure the amplitude.  At midrapidity this becomes more difficult not only because the PHENIX central detectors do not have full azimuthal coverage, but also because these asymmetries are both consistent with zero and statistically limited.  Thus, this method is only used as another cross check.  A modified version of the relative luminosity formula is used:
\begin{equation}
A_N \cdot \sin{ \phi_s } = \frac{1}{P} \epsilon_N(\phi_s) = 
                    \frac{1}{P} \frac{ N^\uparrow( \phi_s ) - \mathcal{R} \cdot N^\downarrow( \phi_s ) }
                                           { N^\uparrow( \phi_s ) + \mathcal{R} \cdot N^\downarrow( \phi_s ) }
\end{equation}

\noindent Unlike the azimuthal angle used in Section~\ref{Section:AcceptanceCorrection},  \( \phi_s \) is the angle from the spin up direction (y = 0 in PHENIX coordinates).  It increases to the left of the polarized proton beam going direction, preserving the left-right asymmetry convention.  \( P \) is the beam polarization from Section~\ref{Section:Polarization} and \(\mathcal{R} \) is the same relative luminosity explained  Section~\ref{Section:RelativeLuminosity}.  No azimuthal correction is needed because we are no longer integrating over \( \phi \).  

\begin{figure}
\centering
\subfigure[Yellow Beam\label{Figure:dpSinPhiYellow5to6}]{ \includegraphics[scale = 0.36]{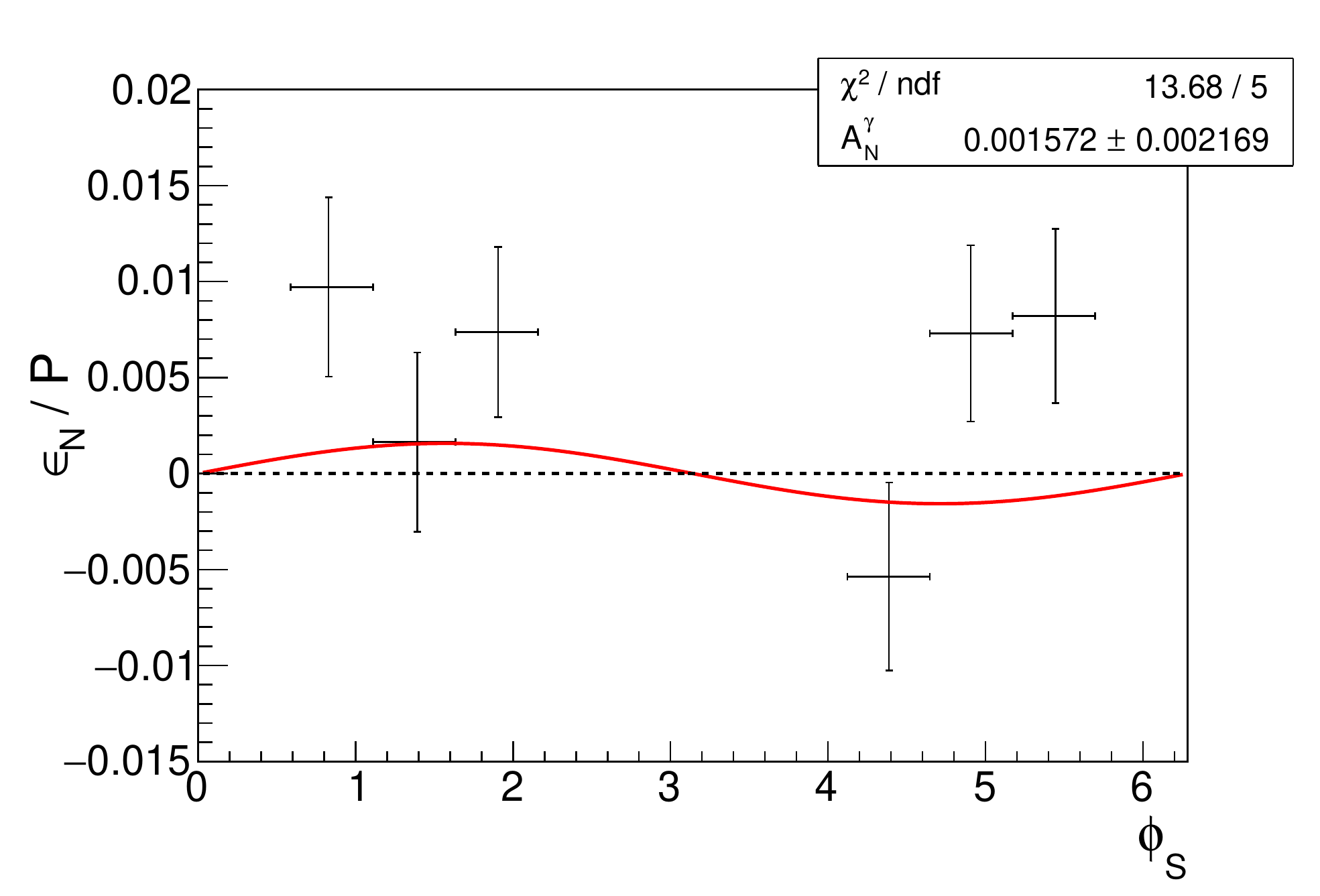} } 
\subfigure[Blue Beam\label{Figure:dpSinPhiBlue5to6}] { \includegraphics[scale = 0.36]{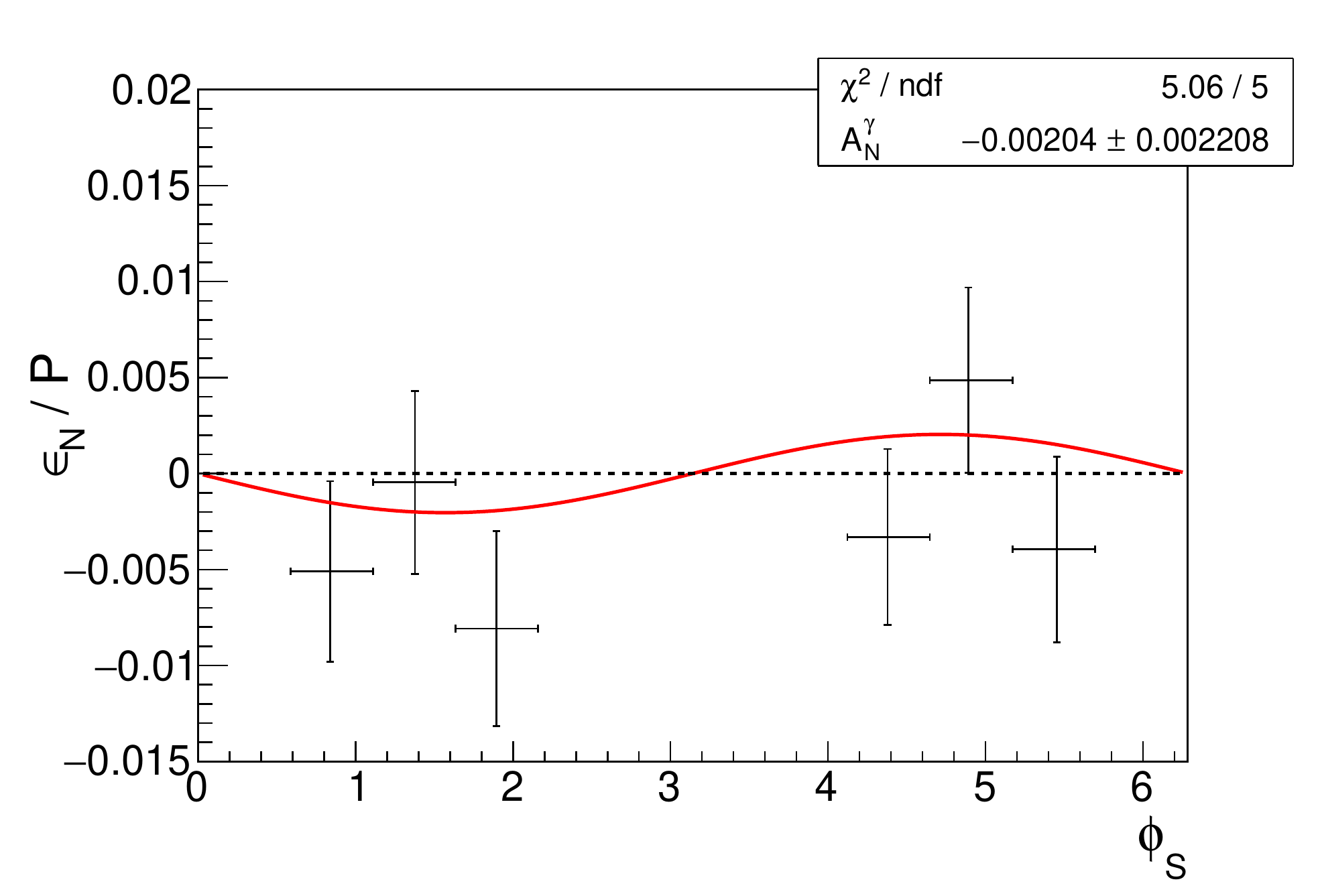} }
\caption[Example plots of the \( A_N \sin{\phi_s} \) fits for the direct photon asymmetry]{Example plots of the \( A_N \sin{\phi_s} \) fits for the direct photon asymmetry. The raw direct photon asymmetry for \( 5 < p_T^\gamma < 6 \) GeV/c measured as a function of \( \phi_s \), divided by the beam polarization, and fit with the function \( A_N \sin{\phi_s} \).}
\label{Figure:dpSinPhi5to6}
\end{figure}

\begin{figure} 
 \centering
 \includegraphics[scale = 0.7]{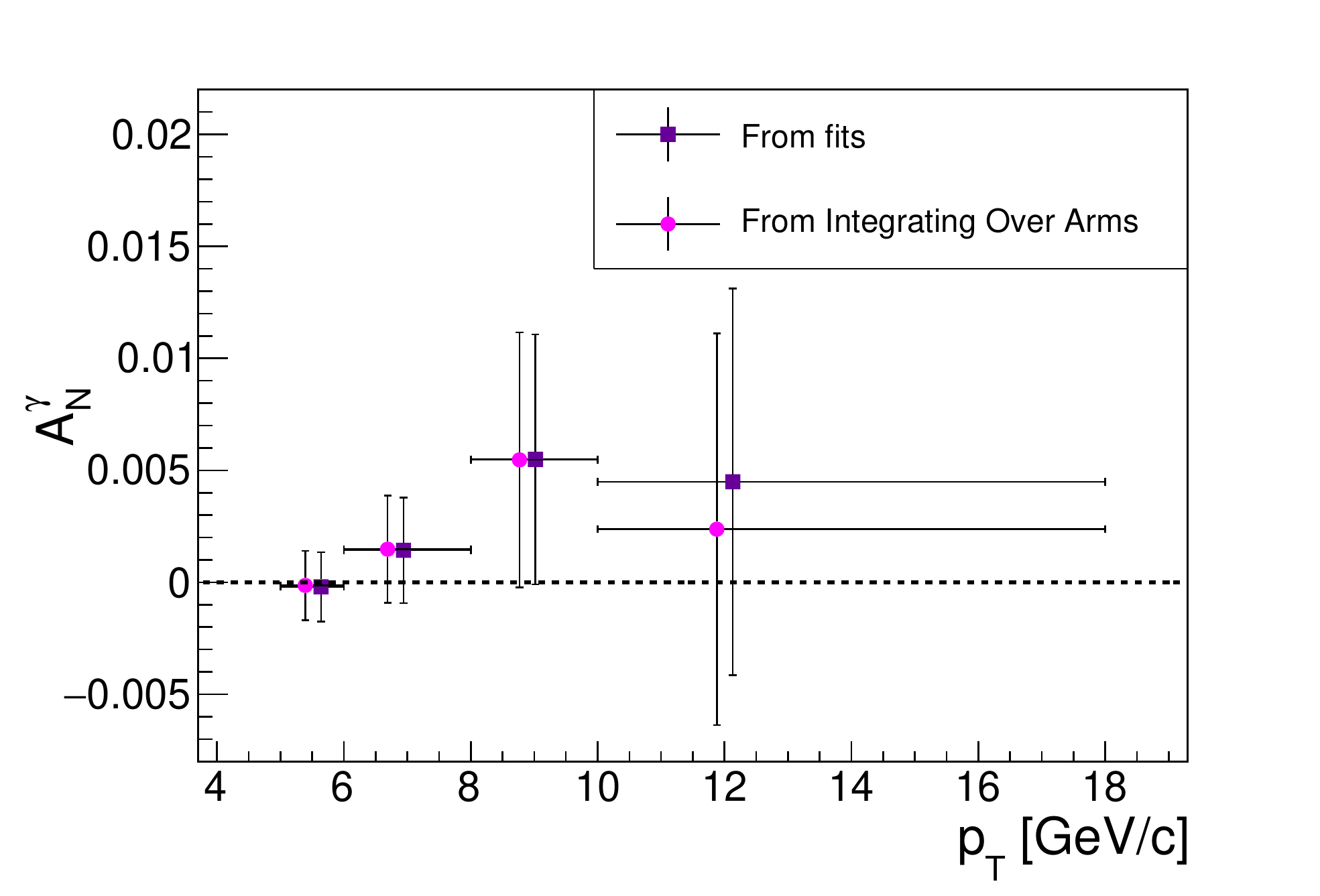}
\caption[Comparison of the direct photon asymmetry from sinusoidal fits to the asymmetry from integrating over the \( \phi \) ranges of the arms. ]{Comparison of the direct photon asymmetry from sinusoidal fits to the asymmetry from integrating over the \( \phi \) ranges of the arms.  These points have been slightly offset in \( p_T \) such that they are more visible.}\label{Figure:dpCompareSinPhi}
\end{figure}

Each arm is split into three bins in \( \phi_s \), as shown in Figure~\ref{Figure:dpSinPhi5to6}.  This is the direct photon asymmetry for the blue and yellow beam results for the lowest \( p_T \) bin.  Even though this bin has the highest number of counts, these asymmetries become statistically limited when they are split across the six \( \phi_s \) bins.  
Even though there is no clear sinusoidal distribution like what is shown in Figure~\ref{Figure:ForwardEtaTSSAFit}, these asymmetries are still fit to a \( A_N \sin{ \phi_s } \) function to extract the amplitude, as shown in the red curves in Figure~\ref{Figure:dpSinPhi5to6}.  These fits are done separately for the yellow and blue beam results and then averaged together.  Figure~\ref{Figure:dpCompareSinPhi} shows an example of the final results from these fits for all \( p_T \) bins plotted with the results calculated by integrating over the \( \phi \) range of the arms explained in Section~\ref{Section:RelativeLuminosityFormula}.  This process is repeated for the \( \pi^0 \) and \( \eta \) asymmetries, which also showed a excellent level of agreement.  This indicates that the acceptance correction from Section~\ref{Section:AcceptanceCorrection} is being calculated correctly.  

\section{Background Correction}\label{Section:BackgroundSubraction}
Correcting for background in TSSA analyses is not as simple as subtracting off the estimated contribution from background like what would be done to the yields used in a cross section analysis.  This is because all of the yields used in TSSA calculations have beam polarization information associated with them.  Thus, we must instead correct for the effects from background by subtracting off the background asymmetry: 
\begin{equation} 
    A_N^S = \frac{A_N^{S + B} - r A_N^B}{1 - r}
\end{equation}
\noindent where \( S \) stands for signal, \( B \) stands for background, and \( r = N^B/(N^B + N^S) \) is the background fraction. The background subtraction formula is applied separately for the different beam and arm results, i.e. the yellow left background asymmetry, \( A_N^B \) is subtracted off of the yellow left sample asymmetry, \( A_N^{S + B} \), using a background fraction that is specifically calculated for east arm, the arm to the left of the direction that the yellow beam is traveling.   This background correction procedure has been used for all previously published PHENIX asymmetries including \cite{ PPG135, PPG170}

\subsection{Background Asymmetries}
By definition, the background within a sample cannot be directly measured, because otherwise it would have been eliminated with a cut.  Thus, the background asymmetries must be calculated using some kind of proxy.  

\begin{figure}
\centering
\subfigure[\( \pi^0 \) pairs: \( 2 < p_T^{\pi^0} < 3 \) GeV/c\label{Figure:pi0Inv2to3}]{ \includegraphics[scale = 0.36]{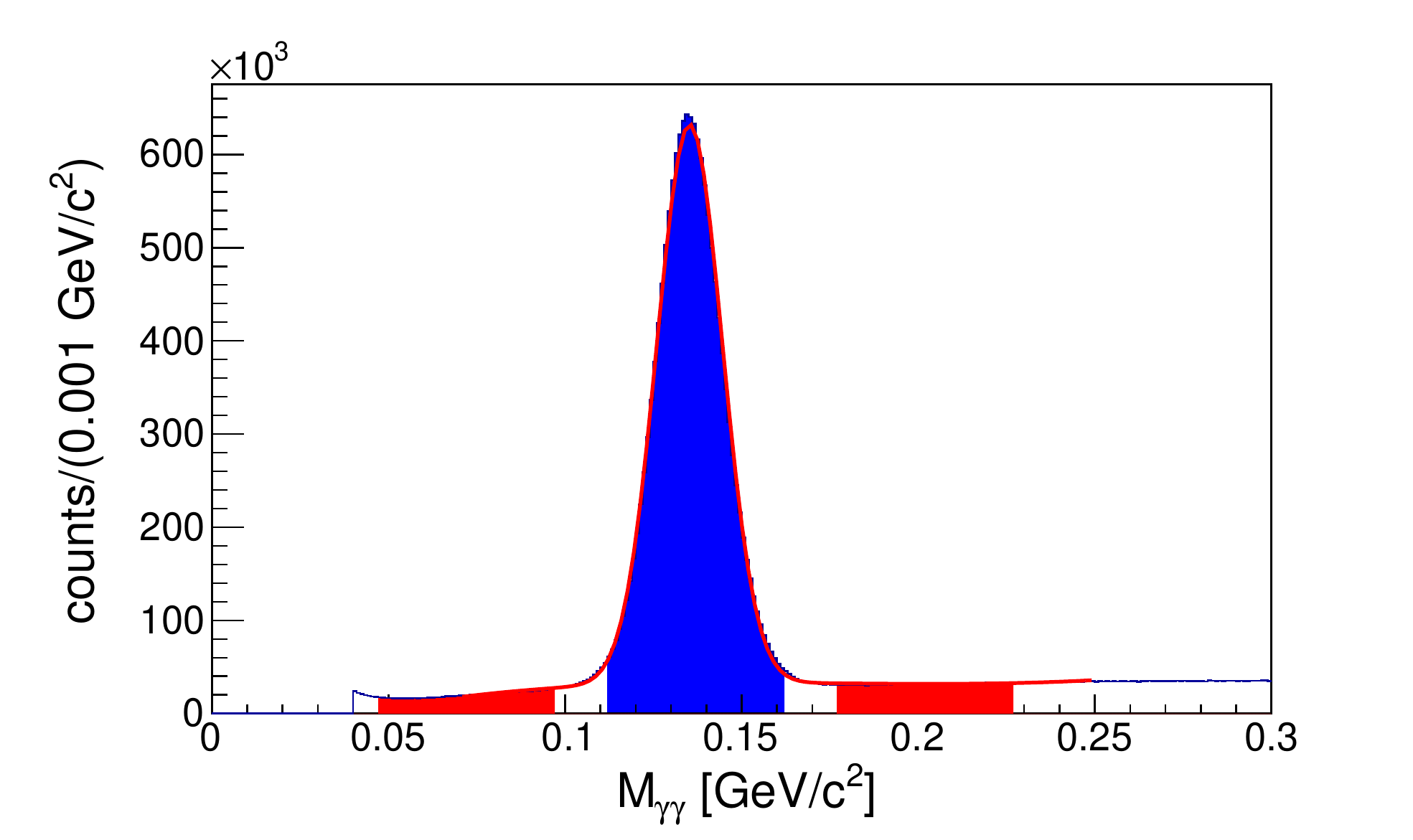} } 
\subfigure[\( \eta \) pairs: \( 2 < p_T^\eta < 3 \) GeV/c\label{Figure:etaInv2to3}]{ \includegraphics[scale = 0.36]{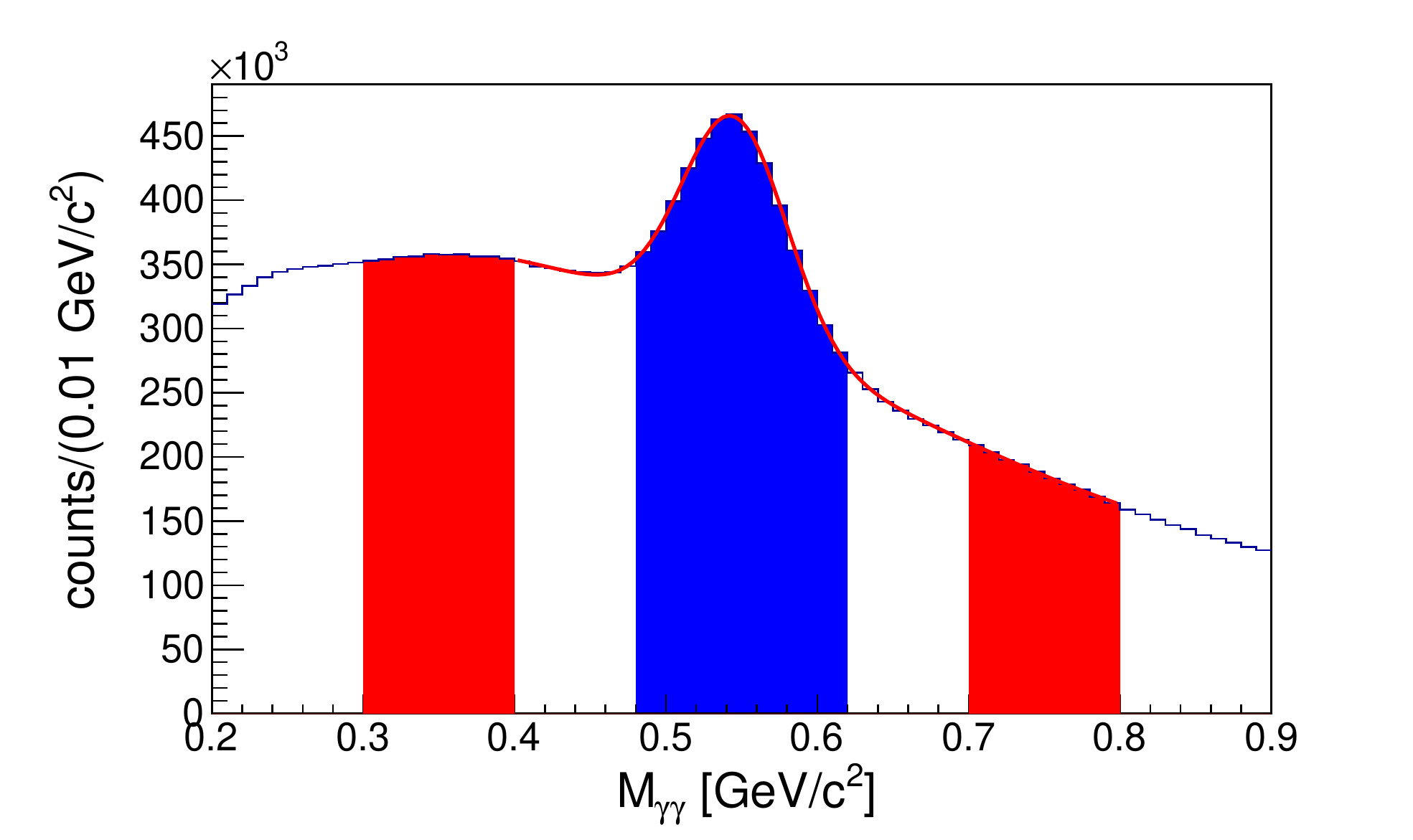} } \\
\subfigure[\( \pi^0 \) pairs: \( 4 < p_T^{\pi^0} < 5 \) GeV/c\label{Figure:pi0Inv4to5}]{ \includegraphics[scale = 0.36]{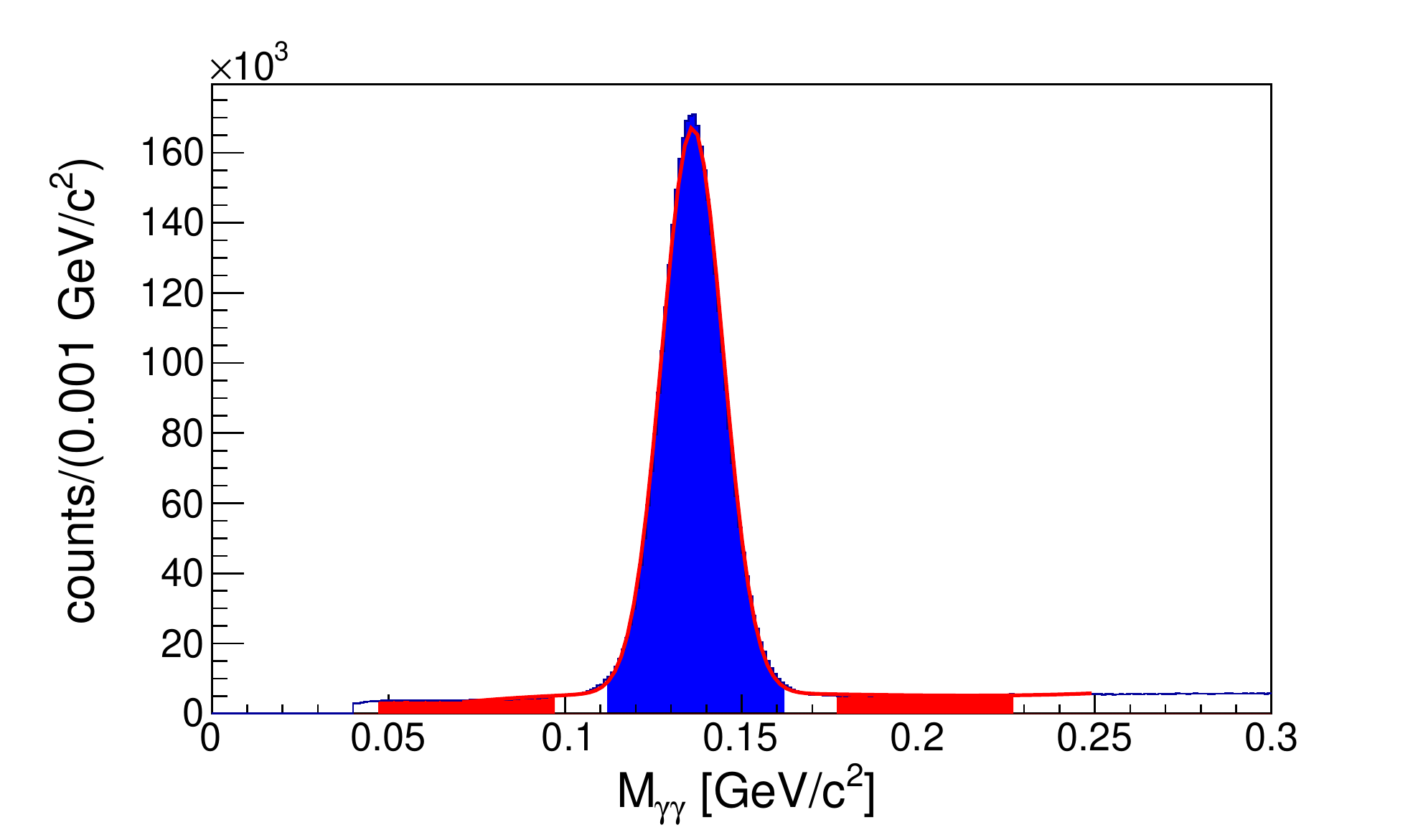} } 
\subfigure[\( \eta \) pairs: \( 4 < p_T^\eta < 5 \) GeV/c\label{Figure:etaInv4to5}]{ \includegraphics[scale = 0.36]{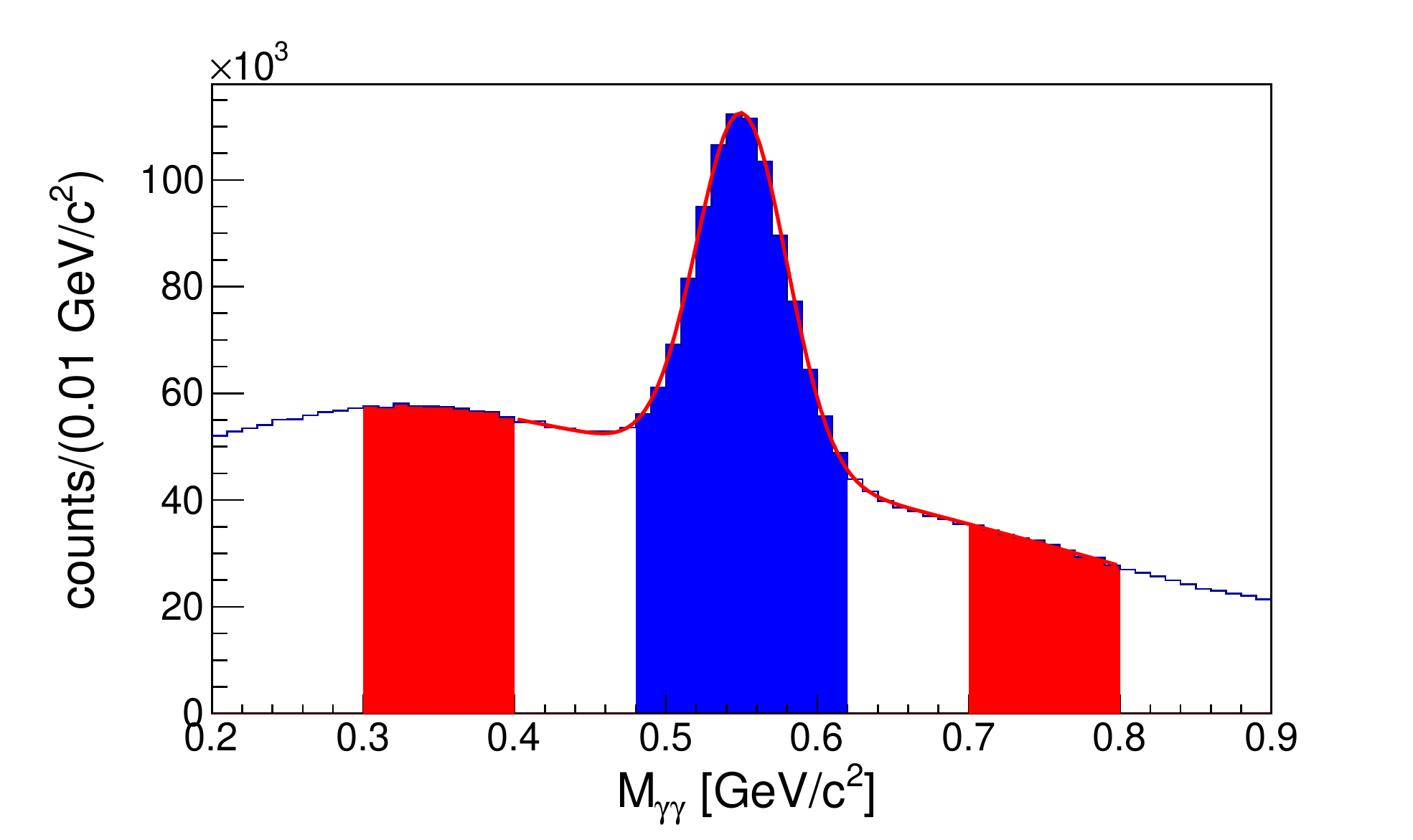} } \\
\subfigure[\( \pi^0 \) pairs: \( 10 < p_T^{\pi^0} < 12 \) GeV/c\label{Figure:pi0Inv10to12}]{ \includegraphics[scale = 0.36]{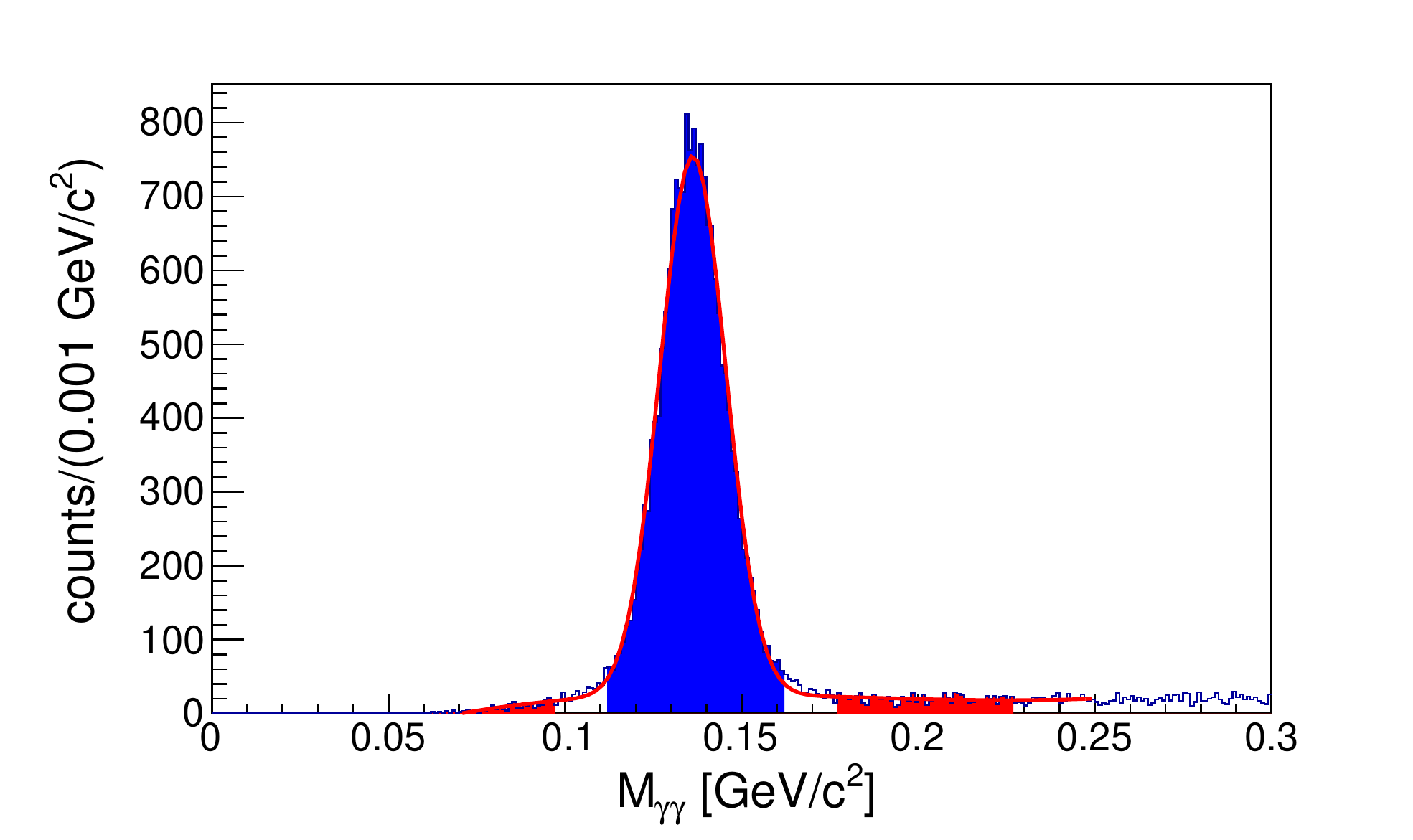} } 
\subfigure[\( \eta \) pairs: \( 10 < p_T^\eta < 20 \) GeV/c\label{Figure:etaInv10to20}]{ \includegraphics[scale = 0.36]{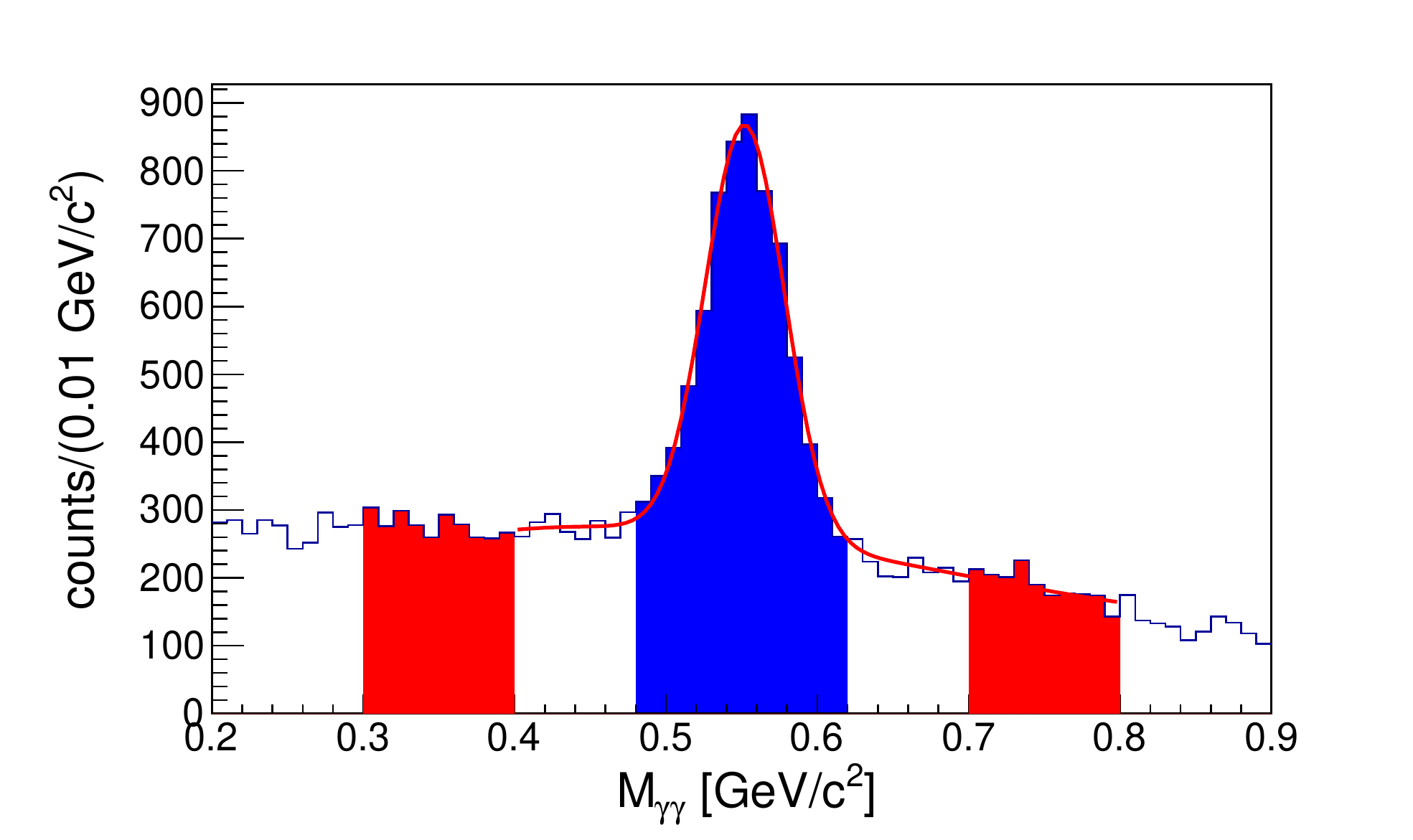} } 
\caption[Example plots of the west arm diphoton invariant mass spectra around the \( \pi^0 \) and \( \eta \) peak as a function of hadron \( p_T \).]{Example plots of the west arm diphoton invariant mass spectra around the \( \pi^0 \) and \( \eta \) peak as a function of hadron \( p_T \).  The middle blue bands represent the invariant mass regions used for the \( A_N^{peak} \) asymmetries and the red regions show the invariant mass regions used to calculate the \( A_N^{bg} \) asymmetries.  The red curves are the fits to these distributions that are used to calculate the background fraction.}
\label{Figure:invPT}
\end{figure}

Since \( \pi^0 \) and \( \eta \) mesons are collected via their diphoton decays, their main source of background comes from combinatorial photon pairs: two photons that just happened to be in the same event, pass all of the pair cuts, and have an invariant mass that fell under either the \( \pi^0 \) or \( \eta \) peak, but are not actually produced in a diphoton decay.  So, for these photon pair analyses the background subtraction formula becomes:
\begin{equation} \label{Equation:pairBackgroundSubtraction}
    A_N^{sig} = \frac{ A_N^{peak} - r A_N^{bg} }{1 - r}
\end{equation}
\noindent where \( A_N^{peak} \) is the asymmetry calculated using photon pairs within this invariant mass peak and \( A_N^{bg} \) is calculated using photon pairs with invariant mass ranges near this peak, but not within it.  These background ranges are stated explicitly in Section~\ref{Section:PhotonPairSelection} and are chosen such that the behavior of the combinatorial background within the peak is well approximated, but the background asymmetry contains none of the physics asymmetry from the actual \( h \rightarrow \gamma \gamma \) decays.  Figure~\ref{Figure:invPT} shows example plots of these invariant mass spectra for both the \( \pi^0 \) and \( \eta \) analyses at different \( p_T \).  The blue region represents the peak photon pairs that are used to calculate \( A_N^{peak} \) and the red regions represent the side band ranges that are used to calculate \( A_N^{bg} \).  By noting that the invariant mass range of the \( \eta \) plots overlap with \( M_{\gamma\gamma} \) range of the \( \pi^0 \) plots, one can see how many more photon pairs are within the \( \pi^0 \) invariant mass peak as compared to the \( \eta \) peak.  

For direct photons, the main source of background comes from isolated photons that are from a diphoton decay, but the second photon was missed.  The vast majority of these photons are from either \( \pi^0 \) or \( \eta \) meson decays and so the direct photon TSSA background subtraction formula becomes: 
\begin{equation} \label{Equation:dpBackgroundSubtraction}
    A_N^{dir} = \frac{A_{N}^{iso}  - r_{\pi^0} A_{N}^{iso, \pi^0} -  r_{\eta} A_{N}^{iso, \eta}}{1 - r_{\pi^0} - r_{\eta}}
\end{equation}
\noindent where \( A_{N}^{iso} \) is the TSSA of the direct photon candidate sample (photons that have passed both the isolation cut and tagging cuts) that were shown previously in Figures~\ref{Figure:dpLumiLeftRight} through ~\ref{Figure:dpCompare}.  \( A_{N}^{iso, \pi^0} \) and \( A_{N}^{iso, \eta} \) are the asymmetries of photons that are tagged as coming from either \( \pi^0 \rightarrow \gamma \gamma \) or \( \eta \rightarrow \gamma \gamma \) decays respectively and pass the photon \textit{pair} isolation cut (Equation~\ref{Equation:PhotonPairIsolationCut}), such that if the second photon from the decay had been missed these decay photons would have been added to the direct photon sample.  

These asymmetries need to be measured as a function of decay photon \( p_T \) or undergo a complicated conversion from hadron to decay photon \( p_T \).  However, this makes these background asymmetries statistically limited especially in the \( p_T \) range of this analysis which is above 5 GeV/c.  But the midrapidity \( A_N^{\pi^0} \) and \( A_N^\eta \) asymmetries have been measured to be consistent with zero, so instead of plugging \( A_{N}^{iso, \pi^0} \) and \( A_{N}^{iso, \eta} \) into the background subtraction formula and vastly increasing the statistical uncertainty of the direct photon results, these background asymmetries are set to zero:
\begin{equation} \label{Equation:dpBackgroundRescaling}
    A_N^{dir} = \frac{ A_{N}^{iso} }{1 - r_{\pi^0} - r_{\eta}}
\end{equation}
\noindent This means that the direct photon background is treated like a dilution: there is some direct photon physical TSSA that is being diluted by photons from \( \pi^0\) and \( \eta \) decays whose TSSAs are zero and so the direct photon asymmetry must be rescaled by the background fraction.  There is a systematic uncertainty assigned due to setting the background asymmetry to zero and it dominates the total systematic uncertainty for the direct photon TSSA, as shown in Table~\ref{Table:dpTSSA}.  

\subsection{Background Fraction}\label{Section:BackgroundFraction}
The background fraction is the estimated contribution from background to the data sample:  \( r = N^B/(N^B + N^S) \).  Because this generally changes as a function of \( p_T \), \( r \) is calculated separately for each \( p_T \) bin.  Also, because the reconstruction efficiencies are different for the two arms, there are two separate background fractions calculated for the west and east arms.  The background correction for the cross-check square root formula result uses the average value between the two arms.  

\subsubsection{Photon Pair Background Fraction}

\begin{table}
\centering
\begin{tabular}{| c | c | c |} 
\hline
\(p_T\) (GeV/c) & West Arm & East Arm\\
\hline
2 - 3     & 0.104   & 0.107   \\
3 - 4     & 0.0838 & 0.0869 \\
4 - 5     & 0.0769 & 0.0803 \\
5 - 6     & 0.0746 & 0.0780 \\
6 - 7     & 0.0733 & 0.0791 \\
7 - 8     & 0.0812 & 0.0811 \\
8 - 9     & 0.0879 & 0.0841 \\
9 - 10   & 0.0834 & 0.0803 \\
10 - 12 & 0.0683 & 0.0791 \\
12 - 20 & 0.0631 & 0.0556 \\
\hline
\end{tabular}
\caption[The fraction contribution of combinatorial background to the \( \pi^0 \) TSSA. ]{The fraction contribution of combinatorial background to the \( \pi^0 \) TSSA. }
\label{Table:rPi0}
\end{table}

\begin{table}
\centering
\begin{tabular}{| c | c | c |} 
\hline
\(p_T\) (GeV/c) & West Arm & East Arm\\
\hline
2 - 3    & 0.715 & 0.712 \\
3 - 4    & 0.606 & 0.601 \\
4 - 5    & 0.554 & 0.552 \\
5 - 6    & 0.520 & 0.519 \\
6 - 7    & 0.501 & 0.497 \\
7 - 8    & 0.483 & 0.502 \\
8 - 10  & 0.466 & 0.482 \\
10 - 20 & 0.452 & 0.487 \\
\hline
\end{tabular}
\caption[The fraction contribution of combinatorial background to the \( \eta \) TSSA.]{The fraction contribution of combinatorial background to the \( \eta \) TSSA. }
\label{Table:rEta}
\end{table}

The photon pair background fraction is calculated by fitting to the invariant mass spectra, as shown in the red curves in Figure~\ref{Figure:invPT}.  A Gaussian is used to model the behavior of the \( h \rightarrow \gamma \gamma \) signal and a third order polynomial is used to model the combinatorial background.  Tables~\ref{Table:rPi0} and ~\ref{Table:rEta} show these background fractions for the \( \pi^0 \) and \( \eta \) analyses respectively.   Consistent with what is shown in Figure~\ref{Figure:invPT}, the \( \eta \) analysis has a much higher fractional background contribution when compared to the \( \pi^0 \) analysis.    The \( M_{\gamma\gamma} \) ranges used for these fits will affect the resulting fit function, which will affect the calculated background fractions, which will in turn cause the overall asymmetry calculated with Equation~\ref{Equation:pairBackgroundSubtraction} to change slightly.  Thus, the uncertainty on these background fractions is estimated by adjusting the \( M_{\gamma\gamma} \) ranges used for these fits slightly and seeing how much that affects the overall background corrected asymmetry.  These uncertainties becomes a component for the final result's systematic error and are listed in Tables~\ref{Table:pi0TSSA} and ~\ref{Table:etaTSSA}.

\subsubsection{Direct Photon Background Fraction}\label{Section:DirectPhotonBackgroundFraction}
The main source of background for isolated direct photons are photons that came from hadronic diphoton decays where the second photon is missed.   Using the photon yield notation detailed in Table~\ref{Table:DPBackgroundNotation}, this background fraction can be expressed as:
\begin{equation}
r_h = \frac{ N^{iso, h}_{miss} }{ N^{iso} }
\end{equation} 
\noindent where \( N^{iso, h}_{miss} \) is a subset of \( N^{iso} \) which cannot be directly measured.  The one-miss ratio, \( R_h \), \cite{PPG136} is a quantity measured in single particle Monte Carlo.  It is defined as the number of photons where only one photon in the simulated \( h \rightarrow \gamma \gamma \) decay is measured divided by the number of photons where both photons are measured. This calculation is done as a function of photon \( p_T \): 
\( R_h = N^h_{miss} (p_T^\gamma) / N^h_{tag} (p_T^\gamma) \).  The one-miss ratio can then be used to convert from the number of tagged decay photons to the numbers of missed photons:
\begin{equation}
r_h = \frac{ N^{iso, h}_{miss} }{ N^{iso} }
     \approx \frac{ N^h_{miss} }{ N^h_{tag} }\frac{ N^{iso, h}_{tag} }{ N^{iso} } 
      = R_h \frac{ N^{iso, h}_{tag} }{ N^{iso} }
\end{equation} 

\noindent where \( N^{iso, h}_{tag} \) are photons that have been tagged as coming from isolated photon \textit{pairs} as determined by Equation~\ref{Equation:PhotonPairIsolationCut}.  So if the second photon had been missed, these photons would have passed the \textit{photon} isolation cut, Equation~\ref{Equation:PhotonIsolationCut}, and been added to the \( N^{iso, h}_{miss} \) sample.  

\begin{table}
\centering
\begin{tabular}{ |m{3em} | m{30em}| } 
\hline
\( N^{iso} \)                   & The direct photon candidate sample, the number of isolated photons (Equation~\ref{Equation:PhotonIsolationCut}) where the photons that 
                                         are tagged as coming from either \( \pi^0 \) or \( \eta \) decays have already been eliminated. \\ 
\hline
\( N^{iso, h}_{tag} \)      & The number of photons that are removed from the direct photon sample because they are tagged as coming from a 
                                        \( h \rightarrow \gamma \gamma \) decay which also happen to be in an isolated photon \textit{pair} 
                                        (Equation~\ref{Equation:PhotonPairIsolationCut}).  This is not a subset of the direct photon candidate sample and can be measured in data. \\ 
\hline
 \( N^{iso, h}_{miss} \)   & The number of isolated photons that came from hadronic decays, but the second photon was missed. These photons are not                                         
                                         eliminated by the tagging cut and are a subset of \( N^{iso} \) and must be estimated using Monte Carlo. \\ 
\hline
\( N^{iso}_{merge} \)     & The number of merged \( \pi^0 \rightarrow \gamma \gamma \) clusters that pass the photon isolation cut.  This a very small subset of 
                                        \( N^{iso} \) and needs to be estimated using Monte Carlo.\\ 
\hline
\end{tabular}
\caption[The notation used for discussing the direct photon background fraction.]{The notation used for discussing the direct photon background fraction.}
\label{Table:DPBackgroundNotation}
\end{table}

These background fractions are calculated separately for \( \pi^0 \) and \( \eta \) decays:  \( r_{\pi^0} = R_{\pi^0}\frac{ N_{tag}^{iso,\pi^0} }{ N_{tag}^{iso} } \) and \( r_{\eta} = R_{\eta}\frac{ N_{tag}^{iso,\eta} }{ N_{tag}^{iso} } \).  This is done partially because there are about four times more \( \pi^0 \rightarrow \gamma \gamma \) photon pairs produced than \( \eta \rightarrow \gamma \gamma \) photon pairs.  This difference will be captured in the \( N^{iso, h}_{tag}/ N^{iso} \) ratios which are calculated using the photon yields measured in data.  However, \( \eta \) mesons also tend to have wider decay angles between their photons as compared to \( \pi^0\)s.  (This is illustrated in Figure~\ref{Figure:pairPhi}.)  This means that for the same hadron \( p_T \), it is slightly more likely to miss one of the photons from an \( \eta \rightarrow \gamma \gamma \) decay than from a \( \pi^0 \rightarrow \gamma \gamma \) decay.  The difference in their decay kinematics is captured with separate single particle Monte Carlo simulations to calculate different one-miss ratios for the different mesons.  To ensure that the simulations capture the decay photon \( p_T \) spectra accurately, these hadrons are simulated to have a power law \( p_T \) spectrum whose exponents are extracted from power law fits to previous PHENIX \( \pi^0 \)\cite{PPG063} and \( \eta \)\cite{PPG107} cross sections from \( p + p \) collisions with \( \sqrt{s} = 200 \) GeV.  The simulated hadrons are then run through a full detector simulation which considered EMCal dead areas and various data cuts.  

\begin{table}
\centering
  \begin{tabular}{|c|c|c|c|c|}
    \hline
         & \multicolumn{2}{c|}{ \( r_{\pi^0} \) } & \multicolumn{2}{c|}{ \( r_\eta \) } \\
	\hline
    \( p_T \) [GeV ] & West Arm & East Arm & West Arm & East Arm  \\
    \hline
    5 to 6   & 0.384 \( \pm \) 0.003 & 0.409 \( \pm \) 0.004 & 0.085 \( \pm \) 0.001 & 0.087  \( \pm \) 0.001 \\
    \hline
    6 to 8   & 0.320 \( \pm \) 0.004 & 0.343 \( \pm \) 0.005 & 0.073 \( \pm \) 0.002 & 0.078  \( \pm \) 0.002 \\
    \hline
    8 to 10  & 0.225 \( \pm \) 0.009 & 0.257 \( \pm \) 0.011 & 0.055 \( \pm \) 0.004 & 0.063  \( \pm \) 0.005 \\
    \hline
    10 to 18 & 0.124 \( \pm \) 0.011 & 0.137 \( \pm \) 0.013 & 0.035 \( \pm \) 0.005 & 0.046  \( \pm \) 0.007 \\
    \hline
  \end{tabular}
  \caption[The fraction contribution of decay photons to the direct photon asymmetry.]{The fraction contribution of decay photons to the direct photon asymmetry.}
\label{Table:rDirectPhoton}
\end{table}

\begin{figure}
\centering
\subfigure[Yellow Beam]{ \includegraphics[scale = 0.36]{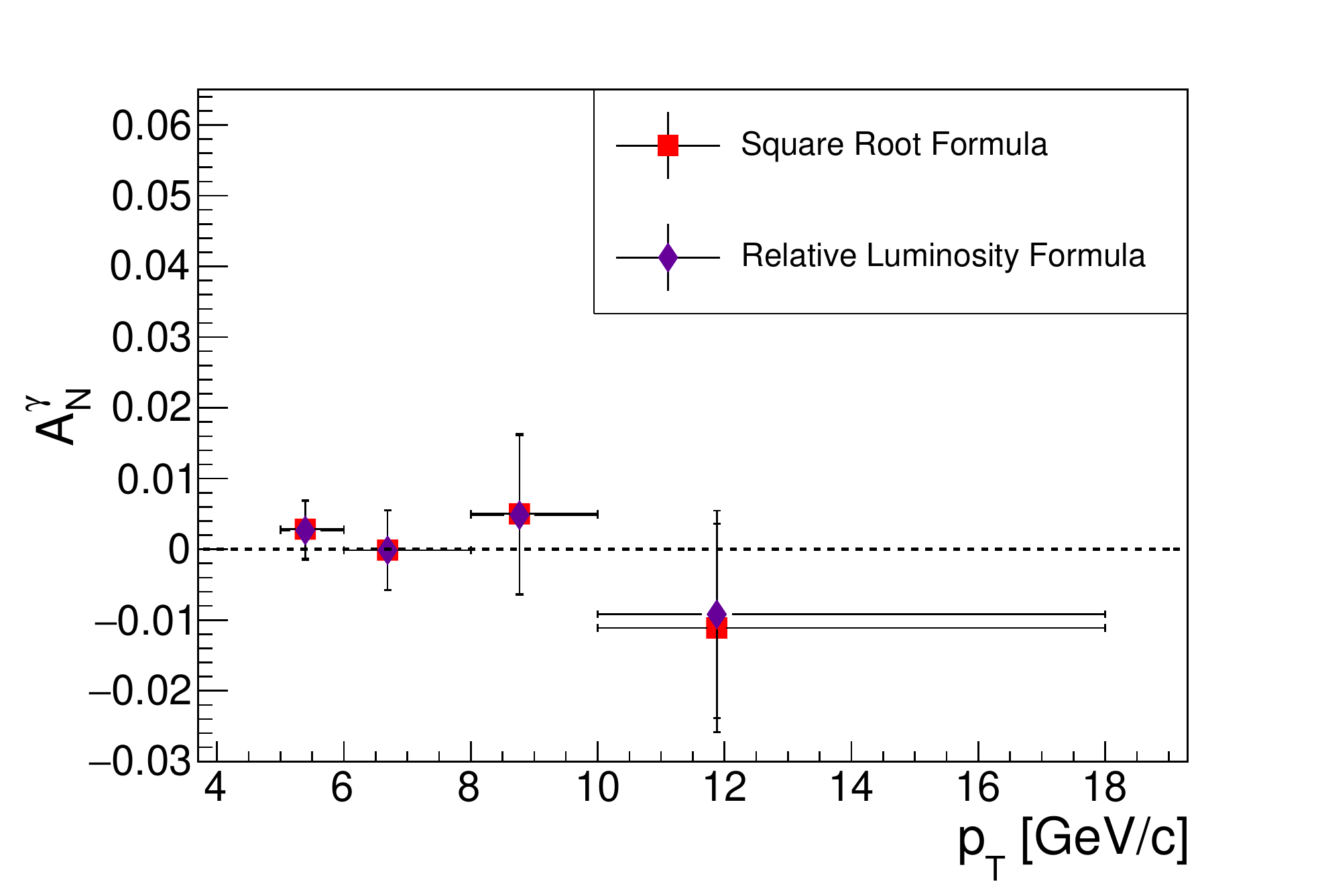} }
\subfigure[T test comparing the yellow beam square root and relative luminosity formula results]{ \includegraphics[scale = 0.36]{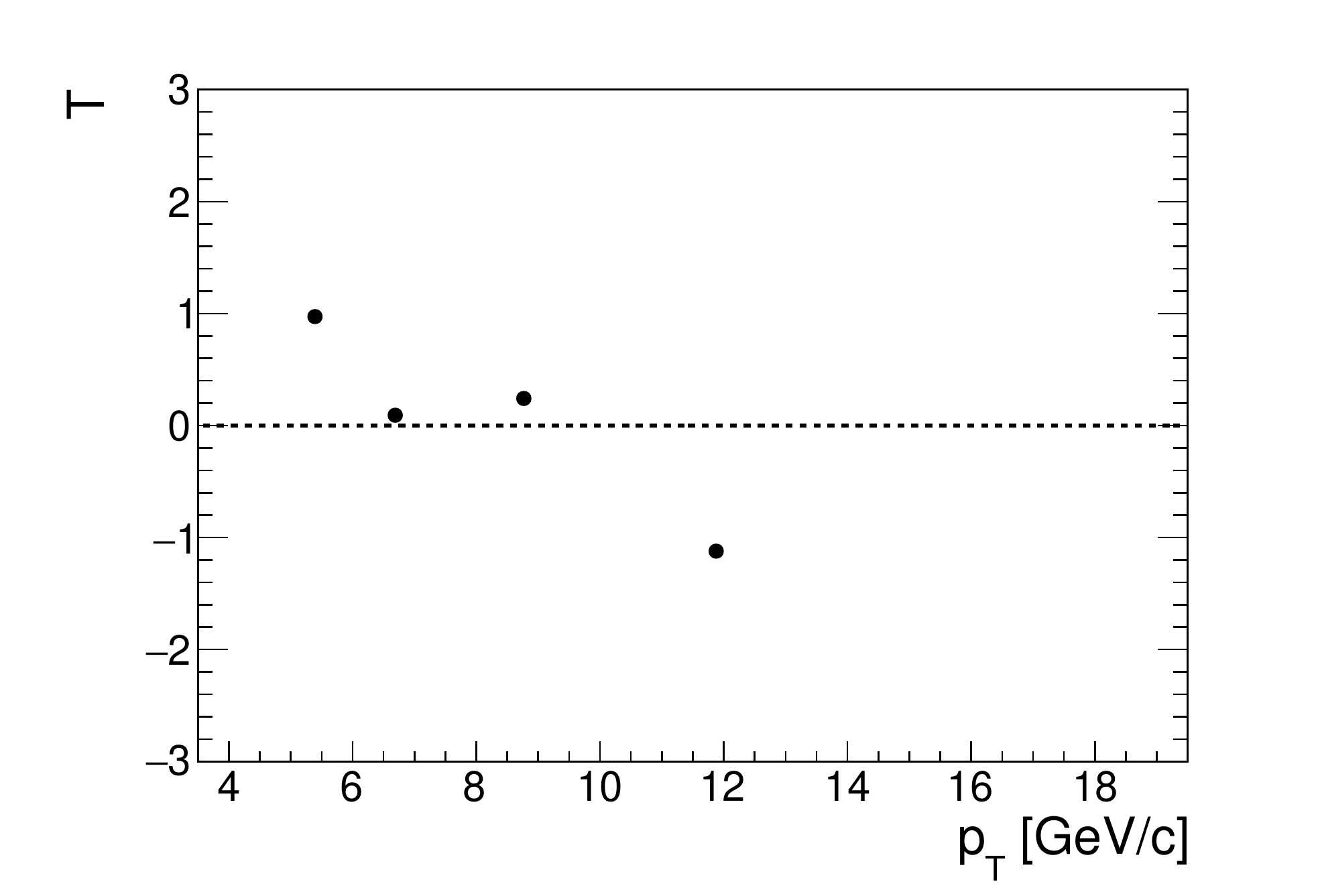} }\\
\subfigure[Blue Beam]{ \includegraphics[scale = 0.36]{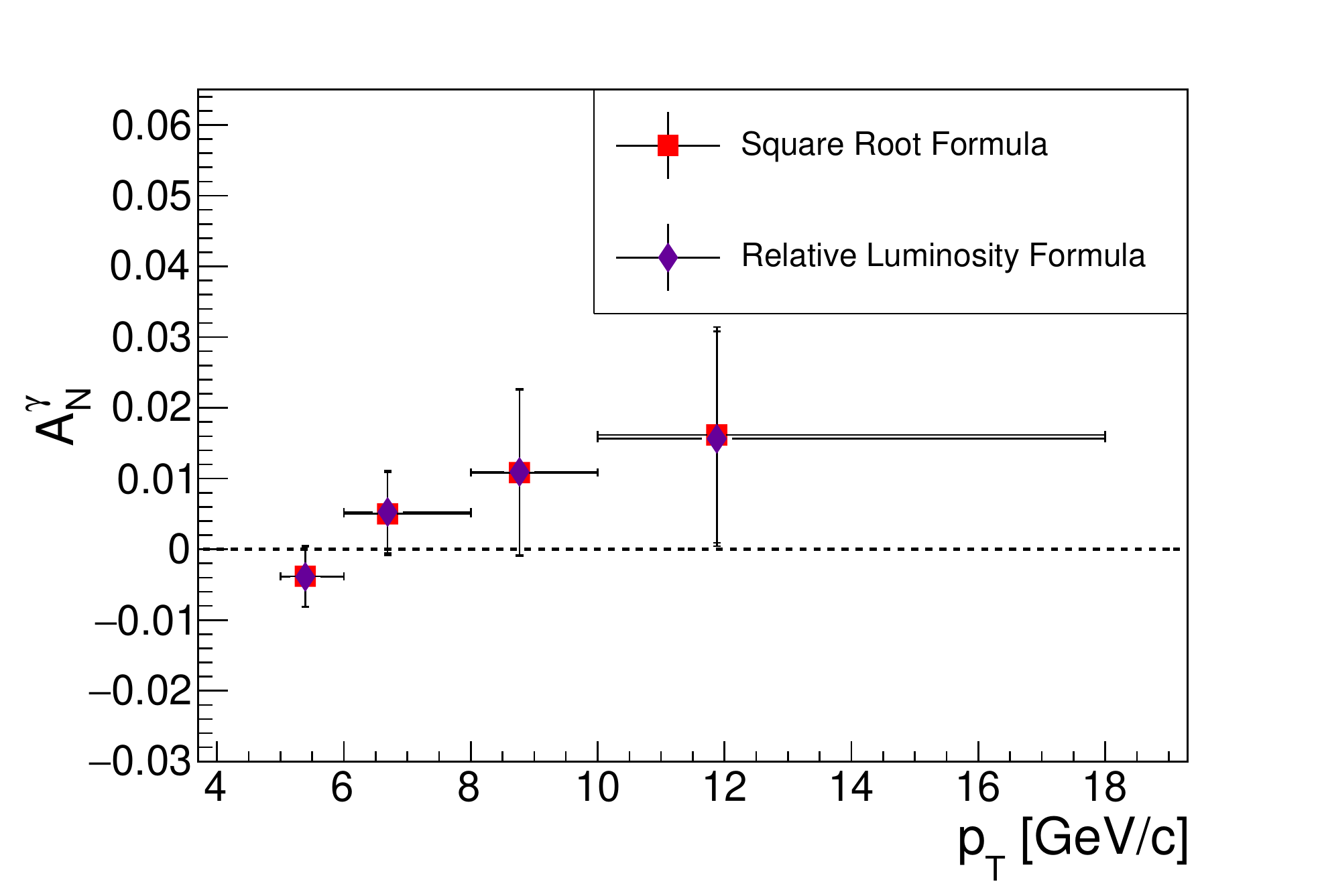} } 
\subfigure[T test comparing the blue beam square root and relative luminosity formula results]{ \includegraphics[scale = 0.36]{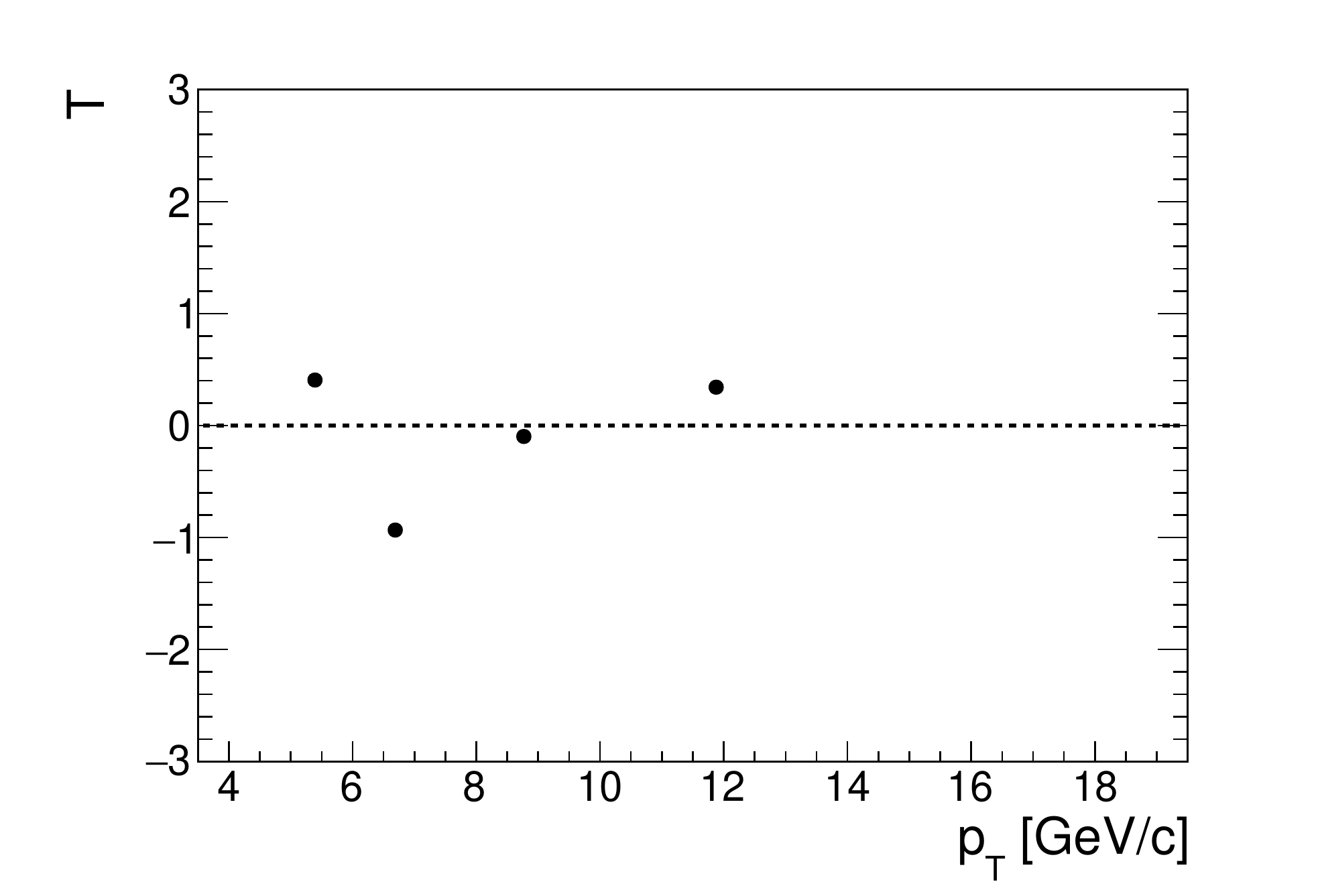} } \\
\subfigure[Final Averaged Asymmetry\label{Figure:finalDPCorrectedCompared}]{ \includegraphics[scale = 0.36]{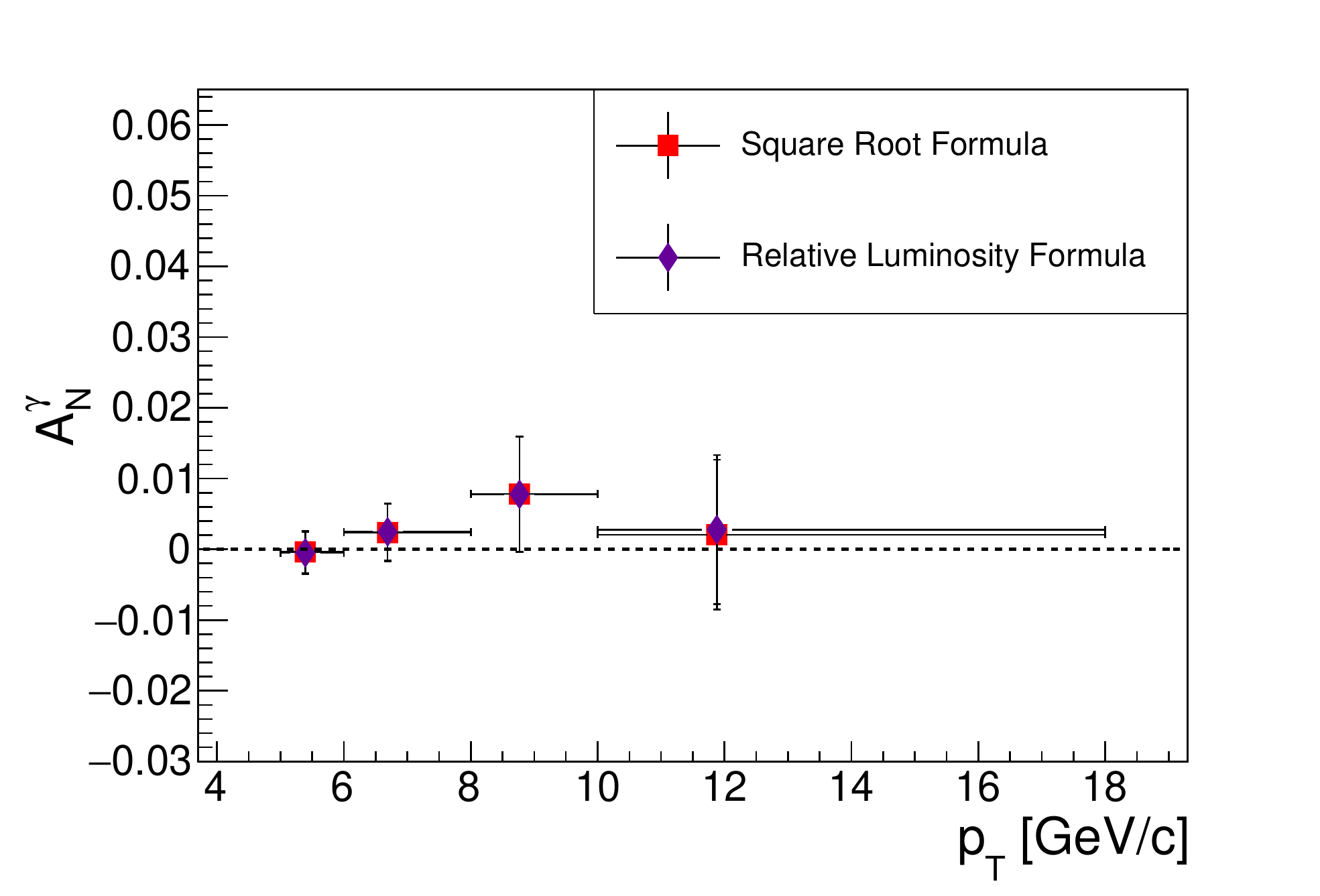} } 
\subfigure[T test comparing the averaged asymmetry results for the square root and relative luminosity results]{ \includegraphics[scale = 0.36]{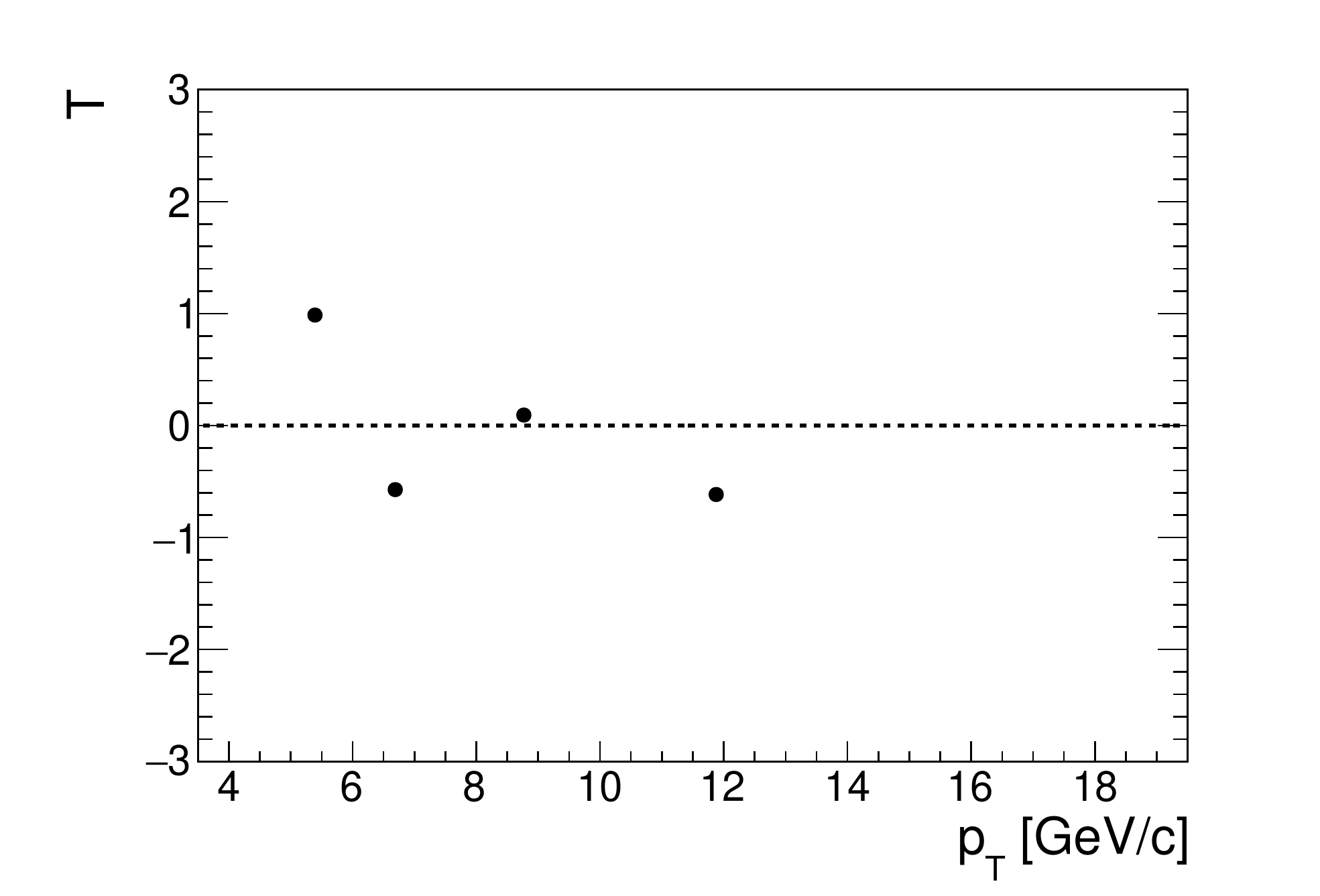} } 
\caption[Comparing the results of the relative luminosity and square root formulas for the background corrected direct photon asymmetry]{Comparing the results of the relative luminosity and square root formulas for the background corrected direct photon asymmetry.}
\label{Figure:dp_correctedCompare}
\end{figure}

As shown in Table~\ref{Table:rDirectPhoton}, the background fractions are up to 5\% larger in the east arm when compared to the west because the PbGl sectors have more dead area  than the PbSc sectors.  The background fraction is lower at higher \( p_T \) partially because the production mechanisms for \( \pi^0 \) and \( \eta \) mesons falls off with \( p_T \) more quickly when compared the direct photon production mechanism.  But the main reason that the background fraction decreases at higher \( p_T \) is that the \( h \rightarrow \gamma \gamma \) decays have smaller decay angles and so it is less likely that only one photon will be measured.  The uncertainties on \( r_h \) are assigned through standard error propagation of the statistical uncertainty for both \( R_h \) and \( N^{iso, h}_{tag}/ N^{iso} \).  These uncertainties are then propagated through Equation~\ref{Equation:dpBackgroundRescaling} and used to assign an additional systematic uncertainty to the direct photon asymmetry, Table~\ref{Table:dpTSSA}.  Figure~\ref{Figure:dp_correctedCompare} shows the results for the direct photon asymmetry after being corrected for background.  When comparing to Figure~\ref{Figure:dpCompare} one can see that many of the T value magnitudes have now been reduced and now there is a more even split between positive and negative values.  The background corrected \( \pi^0 \) and \( \eta \) TSSA results can be seen in Appendix~\ref{Appendix:AsymmetryPloots}. The differences in the relative luminosity and square root formula results (Figure~\ref{Figure:finalDPCorrectedCompared} for the direct photon asymmetry) are assigned as an additional systematic uncertainty for all three TSSA results, which are listed in Tables~\ref{Table:pi0TSSA}, ~\ref{Table:etaTSSA}, and ~\ref{Table:dpTSSA}.  

The background fraction due to \( \pi^0 \) merging is comparatively negligible for the \( p_T \) range of this direct photon analysis.  \( \eta \rightarrow \gamma \gamma \) merging does not contribute at all to this direct photon background because \( \eta \) meson decays tend to have wider decay angles when compared to a \( \pi^0 \) at a similar \( p_T \) and so they do not start to merge until about \( p_T^\eta >50 \) GeV/c at PHENIX, which is well beyond the scope of this measurement.  
The background fraction due to \( \pi^0 \) merging is calculated using a very similar method to the previous direct photon background fractions:
\begin{equation}
r_{merge} = \frac{ N^{iso}_{merge} }{ N^{iso} } \approx \frac{ N_{merge} }{ N_{tag} } \frac{ N^{iso, \pi^0}_{tag} }{ N^{iso} }
\end{equation}
\noindent where \(  N^{iso, \pi^0}_{tag} / N^{iso} \) is the same ratio that is used to calculate \( r_{\pi^0} \).  The \( N_{merge} / N_{tag} \) ratio is calculated using \( \pi^0 \rightarrow \gamma \gamma \) single particle Monte Carlo.  It is the number of merged photon clusters divided by the number of photons that are reconstructed as individual clusters and tagged as coming from a \( \pi^0 \) diphoton decay.  Similar to the one-miss ratio, this value is used to convert from the number of tagged photons to the number of merged clusters.  The simulated \( \pi^0 \) decays are fed through a full detector simulation that took into account the geometry of the EMCal and the clustering algorithm.  Once a shower shape cut is implemented, the \( N_{merge} / N_{tag} \) ratio reduces by several order of magnitude at high \( p_T \).  This makes sense given that this shower shape cut requires that most of the cluster’s energy be concentrated at its center.  \( r_{merge} \) is calculated as a function of cluster \( p_T \) and found to range from about 0.02\% at low \( p_T \) to about 0.1\% at high \( p_T \).  The contribution from merged photons, while still negligible, increases at higher \( p_T \) because this is when the angle between diphoton pairs tends to be the smallest.  Because \( r_{merge} \) is so small, it is added to the uncertainty 

\pagebreak
\noindent of \( r_{\pi^0} \) (which is included in Table~\ref{Table:rDirectPhoton}) and contributes (very slightly) to the overall systematic uncertainty of the direct photon TSSA (Table~\ref{Table:dpTSSA}).  


\section{Systematic Studies}
\subsection{Bunch Shuffling}\label{Section:BunchShuffling}
\begin{figure}
\centering
\subfigure[\( 5 < p_T^\gamma < 6 \) GeV/c]{ \includegraphics[scale = 0.36]{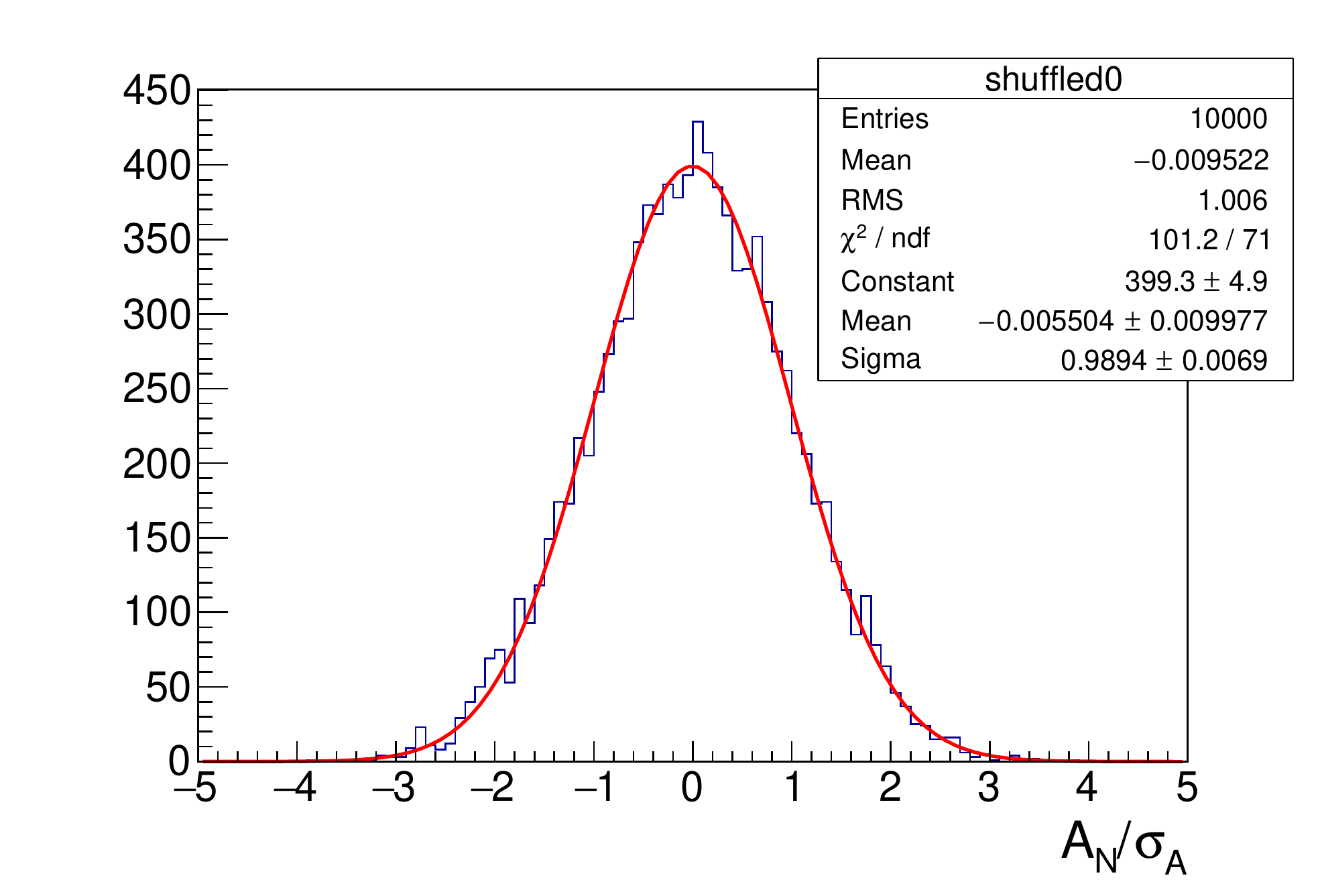} } 
\subfigure[\( 6 < p_T^\gamma < 8 \) GeV/c]{ \includegraphics[scale = 0.36]{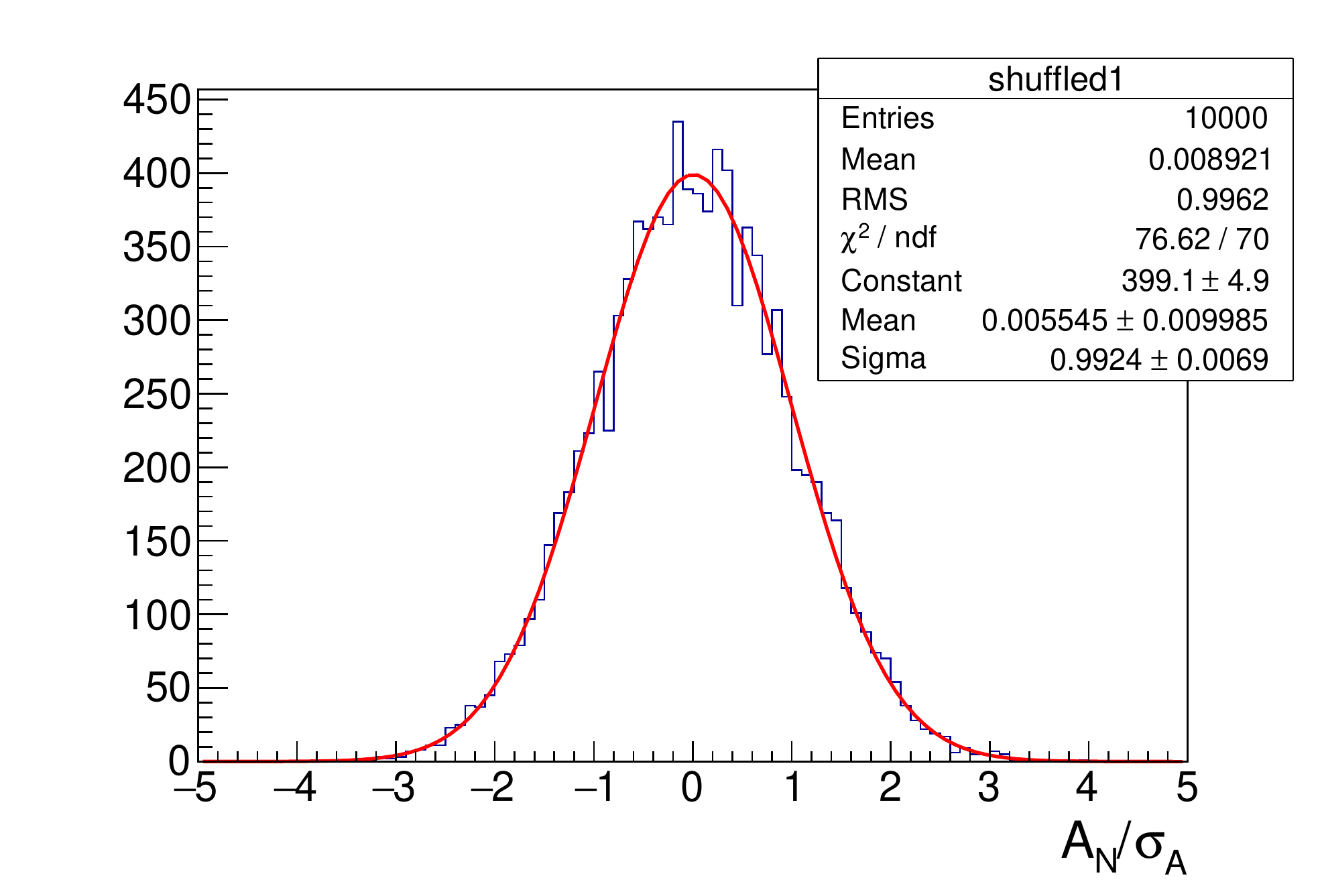} } \\
\subfigure[\( 8 < p_T^\gamma < 10 \) GeV/c]{ \includegraphics[scale = 0.36]{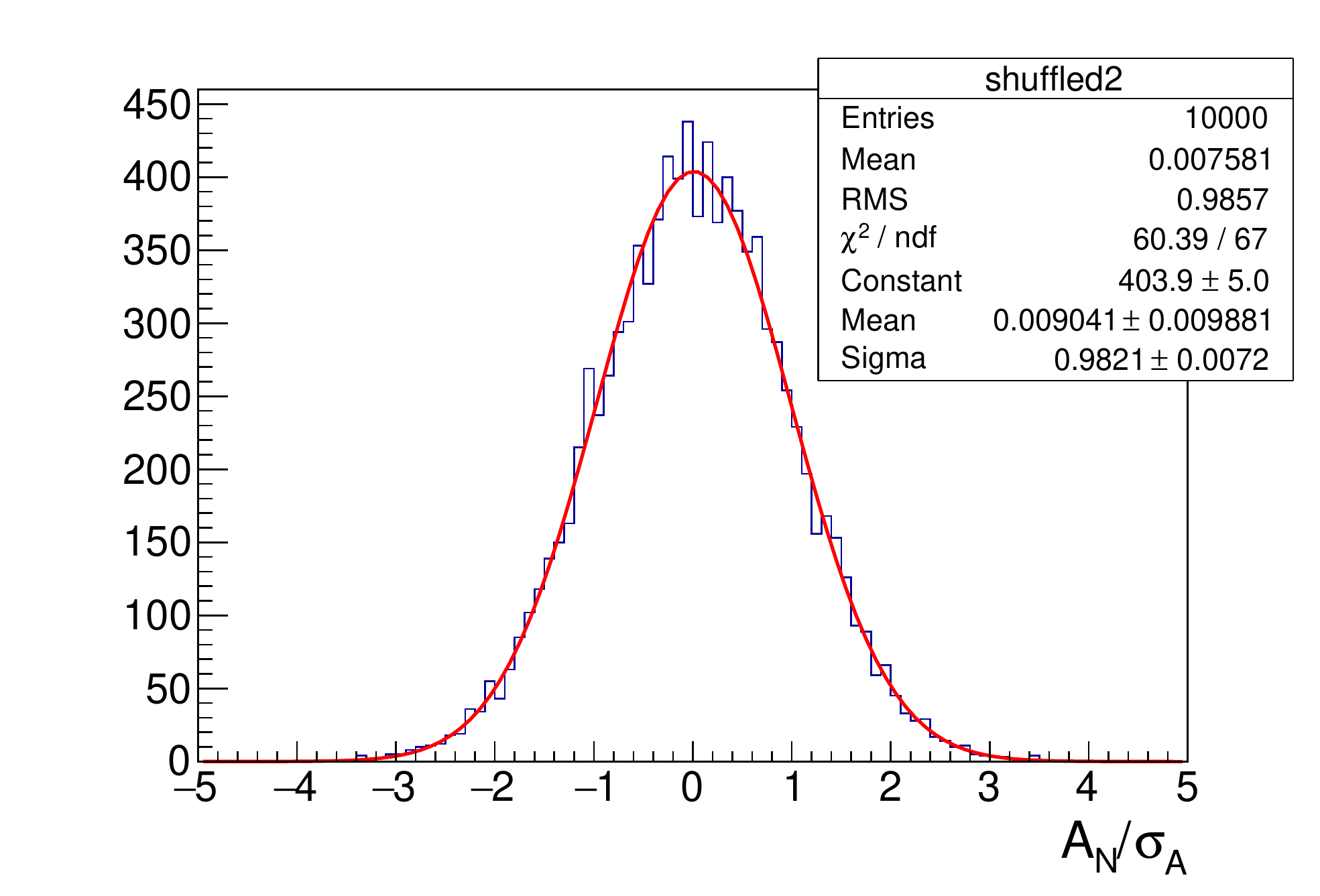} }
\subfigure[\( 10 < p_T^\gamma < 18 \) GeV/c]{ \includegraphics[scale = 0.36]{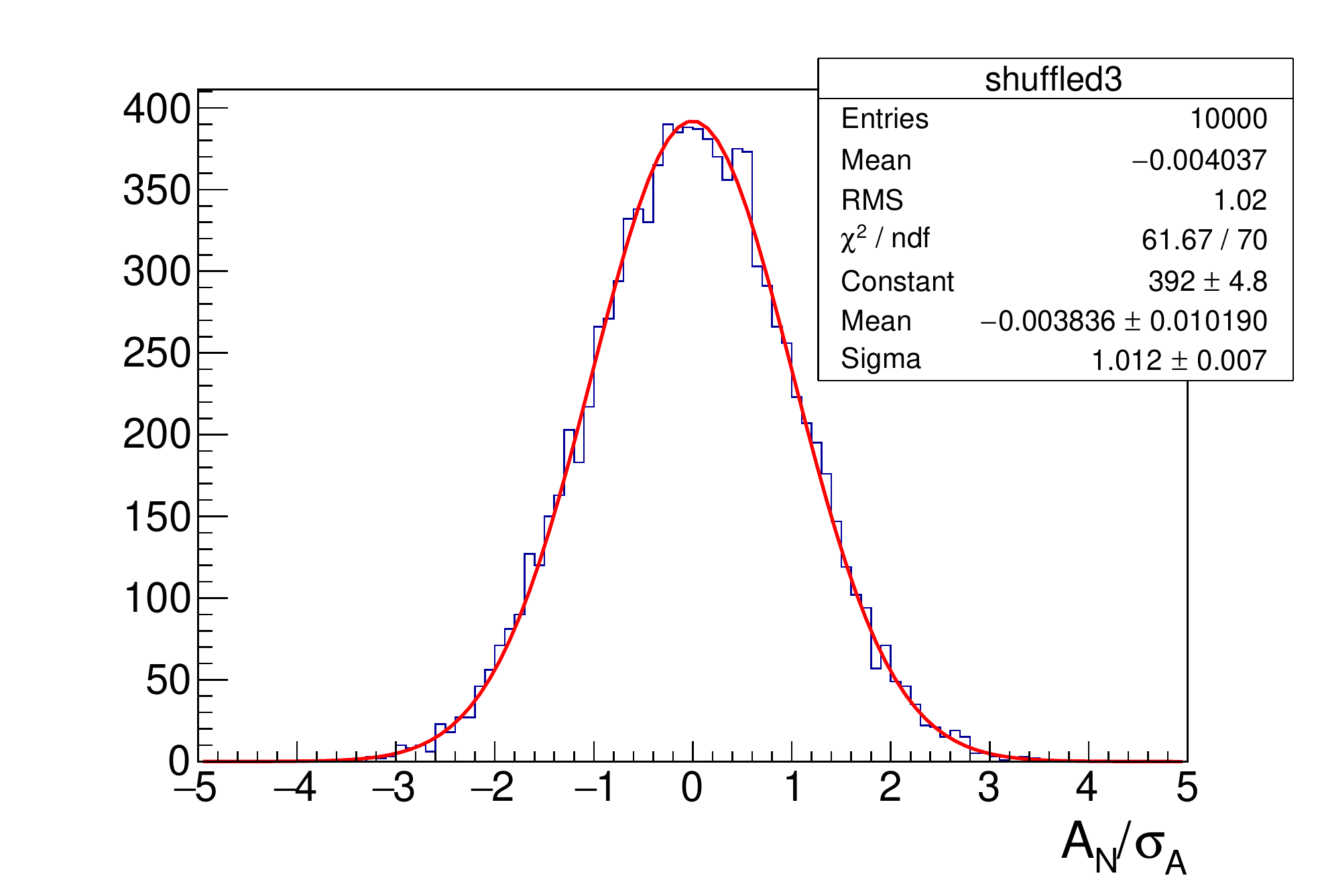} } 
\caption[Results from bunch shuffling for the direct photon asymmetry]{Results from bunch shuffling for the direct photon asymmetry.}
\label{Figure:dpBunchShuffling}
\end{figure}

Bunch shuffling is a technique used to investigate potential sources of systematic uncertainty that could cause the measured results to vary from the true values beyond statistical fluctuations.  It involves randomizing the polarization directions of the beam such that the physics asymmetry disappears and all that is left are the statistical fluctuations present in the data.  For each fill, the polarization directions of each crossing are randomized, and the asymmetry is calculated using the fill group method explained at the beginning of Section~\ref{Section:CalculatingTheAsymmetry}. The square root formula (Section~\ref{Section:SquareRootFormula}) is used to avoid having to recalculate the relative luminosity.  There are 10,000 of these randomized asymmetries  calculated such that there is more than enough statistics to evaluate fluctuations in the data set with Gaussian statistics.  Each of these shuffled asymmetries is divided by its statistical error and added to a histogram.  Figure~\ref{Figure:dpBunchShuffling} shows the bunch shuffling histograms for all four direct photon TSSA \( p_T \) bins.  These distributions are fit to a Gaussian to measure how closely they resembled random noise.  As expected, the means of all of the fits in Figure~\ref{Figure:dpBunchShuffling} are consistent with zero and the widths are all close to one, which means that there is no evidence that there are additional systematic errors present in the data.  This process is repeated for the \( \pi^0 \) and \( \eta \) TSSA analyses which included the asymmetries calculated with photon pairs within the invariant mass peaks and also the background asymmetries.  The Gaussian fits to all of these bunch shuffling histograms also have means that are consistent with zero and widths that are consistent with 1, except for the lowest \( p_T \) bin for both the \( \pi^0 \) and \( \eta \) results which can be seen in Appendix~\ref{Appendix:BunchShuffling}.  The widths of these Gaussians are a few percent larger than 1, which is used to assign an additional systematic uncertainty to this lowest \( p_T \) bin for both of the \( \pi^0 \) and \( \eta \) results.  This value dominates the systematic uncertainty for that bin, which can be seen in Tables~\ref{Table:pi0TSSA} and ~\ref{Table:etaTSSA}.  

\subsection{Direct Photon Cross Section}
The direct photon analysis does not have an invariant mass spectrum to demonstrate that the yields are being extracted correctly.  Instead, the direct photon cross section and isolation ratios are calculated and compared to previously published PHENIX results for \( p + p \) collisions at \( \sqrt{s} = 200 \) GeV. \cite{PPG136}  To simplify both the efficiency and the luminosity calculations needed for this cross check, only the ERTA data is used because this was the only 4 x 4 ERT trigger that was taken in coincidence with the MB trigger that had a 30 cm vertex cut.  The analysis also only uses runs where the ERTA scale down was set to zero, which is about 80\% of the runs.  This invariant cross section is calculated with the following formula:
\begin{equation}\label{Equation:CrossSection}
E \frac{ d^3 \sigma_{dir} }{dp^3} = \frac{1}{ \mathcal{L}} \cdot \frac{1}{2\pi} \cdot \frac{1}{p_T} \cdot  C_{eff}^{trig}(p_T) 
                                    \cdot  C_{eff}^{geo}(p_T) \cdot C_{p + p}^{BBC bias}\cdot  \frac{ N_{dir} }{ \Delta p_T \Delta y }
\end{equation}

Here \( N_{dir} \) is the direct photon yield and quantity of interest.  It is calculated by subtracting off the contribution from background photons:
\begin{equation}
N_{dir} = N_{incl} -( 1 +  R_{ \pi^0 } )\cdot N_{\pi^0} - ( 1 +  R_{ \eta } ) \cdot N_{\eta}
\label{Equation:inclDPYield}
\end{equation}
\noindent where \( N_{incl} \) is the number of inclusive photons before tagging cuts  are applied and \( N_{\pi^0} \) and \( N_{\eta} \) are the number of photons that are tagged as coming from \( \pi^0 \rightarrow \gamma \gamma \) and \( \eta \rightarrow \gamma \gamma \) decays respectively.  No isolation cut is used to allow for an apples-to-apples comparison to the previously published inclusive cross section.  All three of these photon counts (\( N_{incl} \), \( N_{\pi^0} \), and \( N_{\eta} \)), are required to be: in a run where the ERTA scale down was set to zero, in an event that fired the scaled ERTA trigger, and in an EMCal supermodule that fired this ERTA trigger.  \( R_{\pi^0} \) and \( R_\eta \) are the one-miss ratios that are described Section~\ref{Section:DirectPhotonBackgroundFraction} and allow us to estimate the number of photons from \( h \rightarrow \gamma \gamma \) decays but the second photon was missed.  This calculation is done separately for the west and east arms and then the yields are averaged together. 

\begin{figure} 
 \centering
 \centering
 \includegraphics[scale = 0.7]{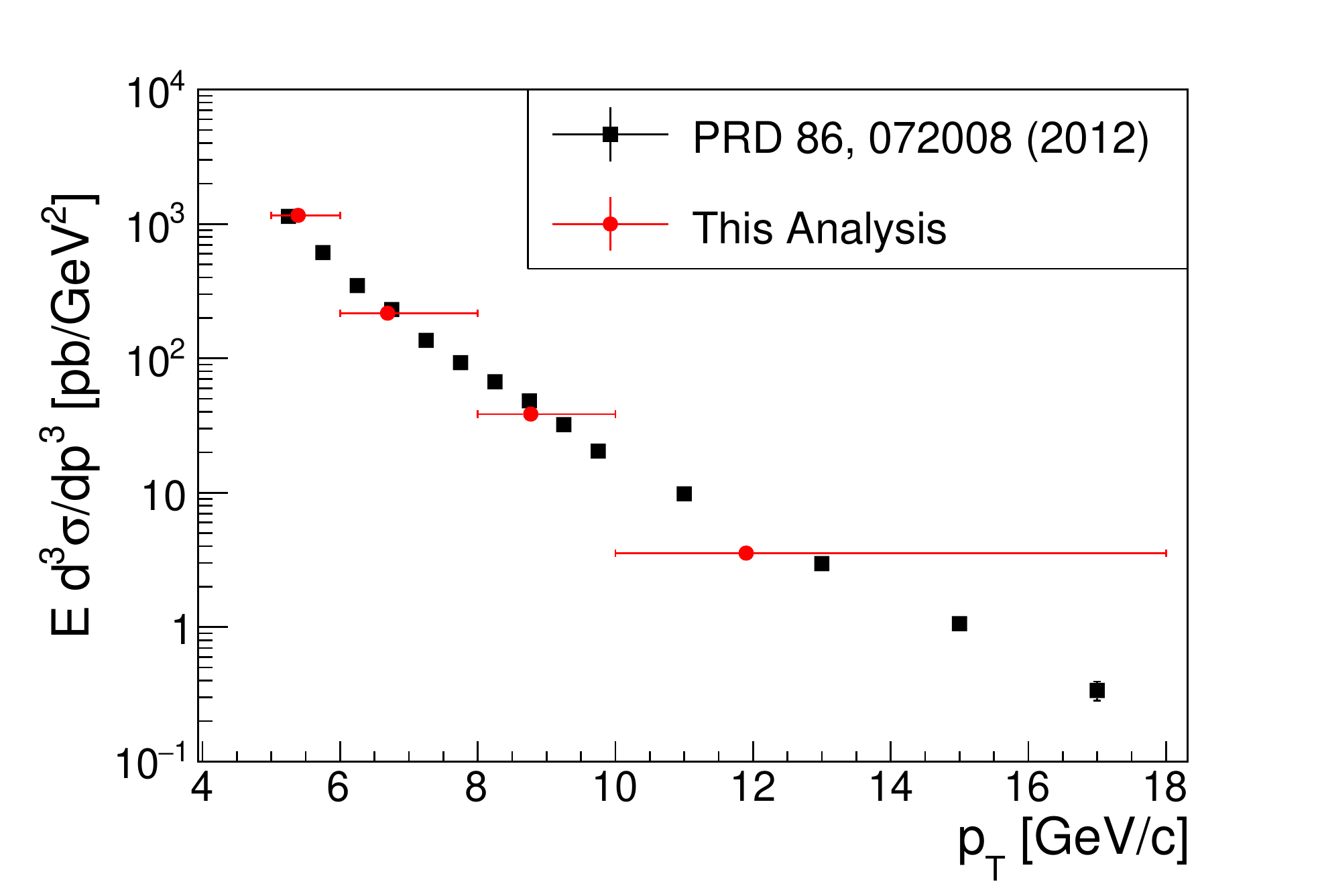}
\caption[The inclusive direct photon cross section cross check plotted with previously published results]{The inclusive direct photon cross section cross check plotted with previously published results, from \cite{PPG136}.}\label{Figure:dpCrossSectionCrossSection}
\end{figure}

In order to be converted to the direct photon invariant cross section, the direct photon yield first needs to be corrected by the width of the \( p_T \) bin, \( \Delta p_T \), and the width of the rapidity range, \( \Delta y \) , which has been set to 1.  \(  C_{p + p}^{BBC bias} \) is the BBC bias factor for measuring ERT events.  Even though the BBC efficiency is about 50\% for all inelastic \( p + p \) collisions at \( \sqrt{s} = 200 \) GeV, PHENIX is optimized such that the BBC has an efficiency of about 75\% for midrapidity ERT events, \cite{PPG136} the reciprocal of which is \(  C_{p + p}^{BBC bias} = 1.337 \).  The geometric efficiency factor, \(  C_{eff}^{geo}(p_T) \), corrects for the EMCal's limited acceptance both from detector geometry and from data cuts.  This is calculated with photon single particle Monte Carlo and includes the conversion from PHENIX's limited central pseudorapidity range of \( \Delta \eta = 0.7 \) to the full unit in rapidity \( \Delta y  = 1\) that is used for this calculation.  \(  C_{eff}^{trig}(p_T) \) is the trigger efficiency factor which is calculated by counting how many ERTA trigger photons were in the minimum bias trigger sample.  The \( p_T \) that is plugged into Equation~\ref{Equation:CrossSection} is  the average \( p_T \) value for that bin and the factor of \( 1 / 2 \pi \) comes from integrating the cross section over \( \phi \).  The integrated luminosity, \( \mathcal{L} \), is
36.62 pb\( ^{-1} \) for the specific data set used for this cross check.   Figure~\ref{Figure:dpCrossSectionCrossSection} shows that this analysis's cross section is consistent with the previous PHENIX cross section from \cite{PPG136}, indicating that the inclusive direct photon yield \( N_{dir} \) is being calculated correctly.

\begin{figure} 
 \centering
 \centering
\subfigure[Isolation ratios from \cite{PPG136}\label{Figure:Fig13_PPG136}]{ \includegraphics[scale = 0.355]{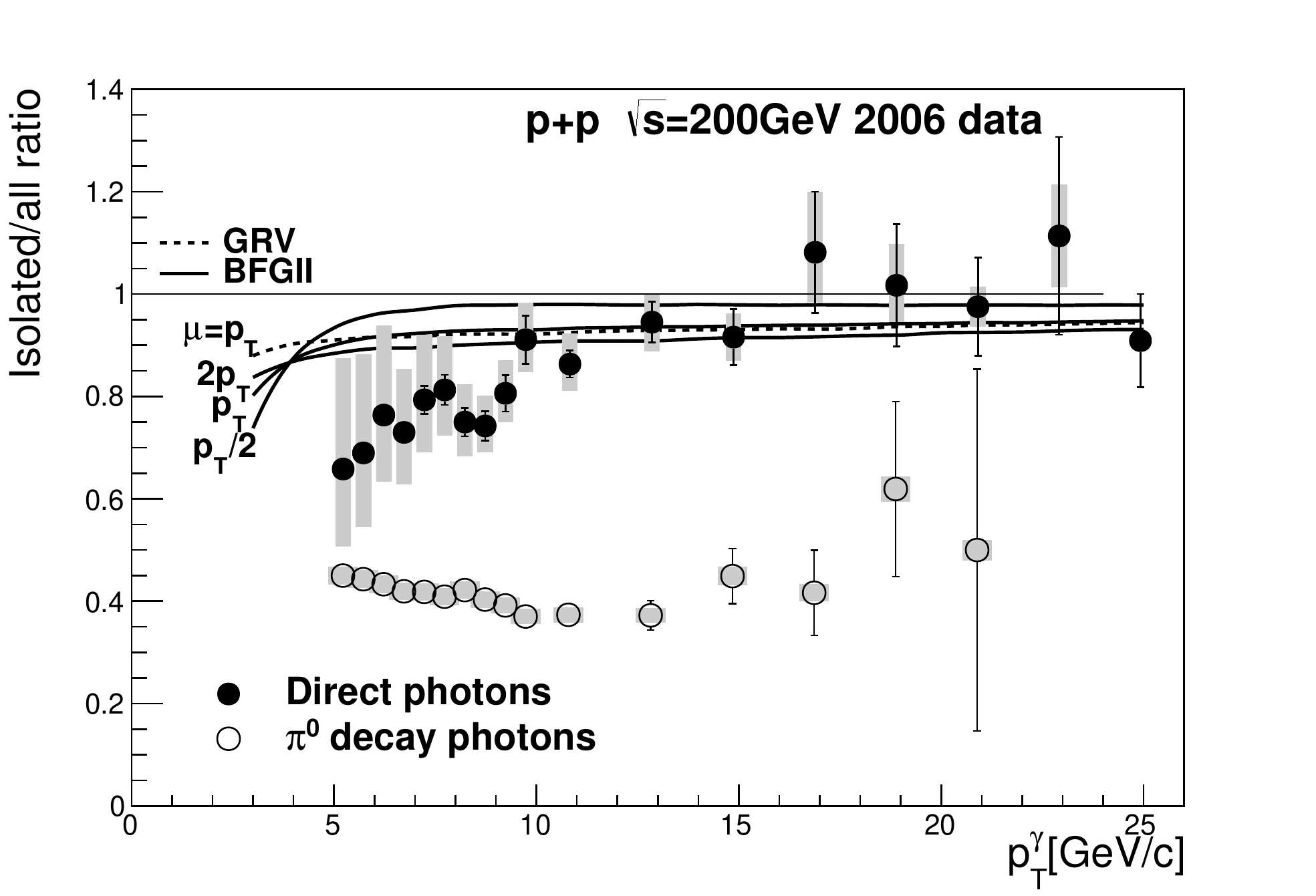} } 
\subfigure[Isolation ratios from this analysis\label{Figure:myIsolatedRatio}]{ \includegraphics[scale = 0.365]{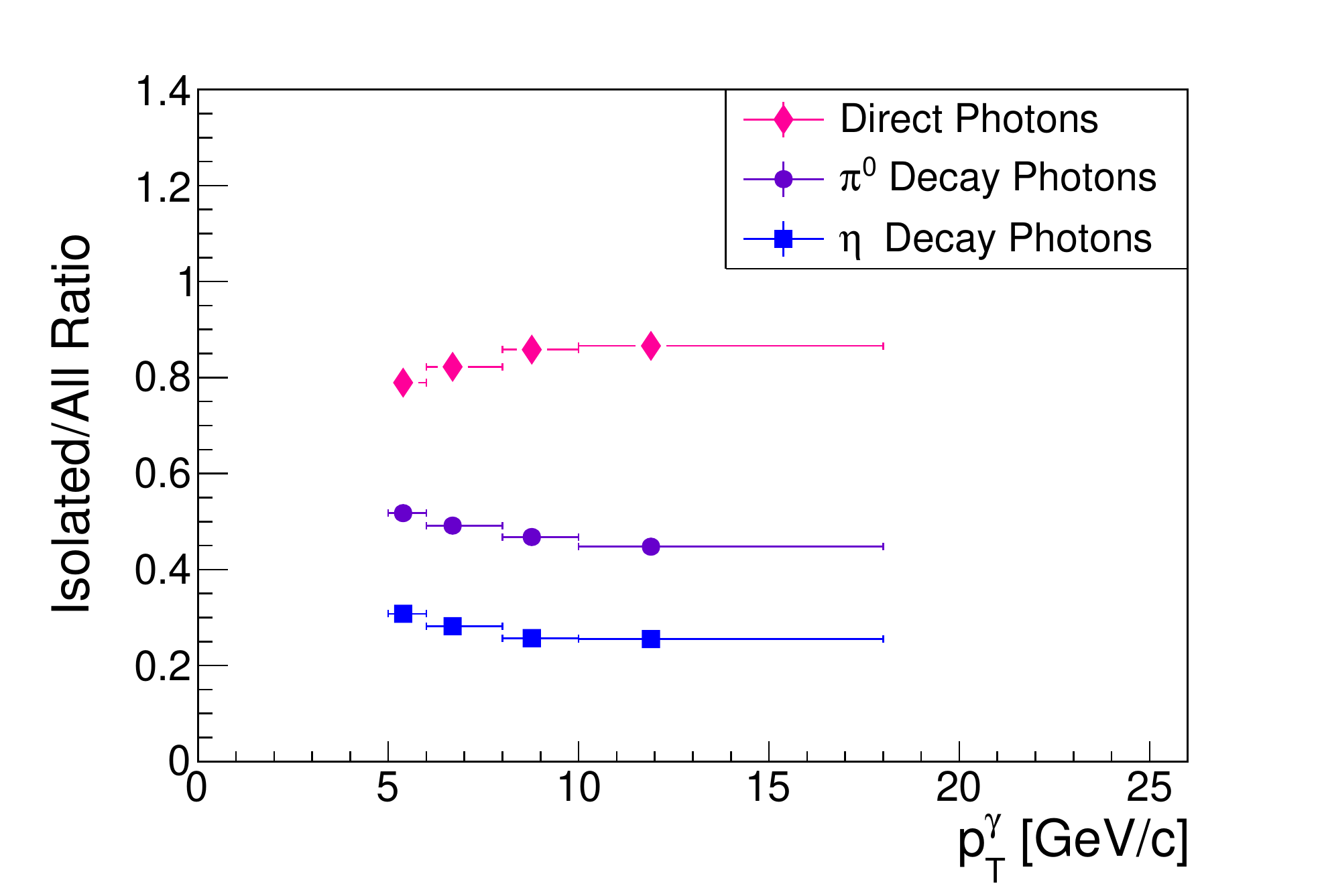} }
\caption[ Isolation Ratios for the direct, \( \pi^0 \), and \( \eta \) meson photons. ]{ Isolation Ratios for the direct, \( \pi^0 \), and \( \eta \) meson photons. }\label{Figure:dpIsolationRatioCrossCheck}
\end{figure}

\pagebreak
In order to verify that the isolation cut is being implemented correctly, the ratio of isolated to inclusive yields is compared to previous results from \cite{PPG136}.  The number of isolated direct photons is calculated using a slightly different formula from the inclusive photon yield:
\begin{equation}
N_{dir}^{iso} = N^{iso} - R_{ \pi^0 } \cdot N^{iso, \pi^0}_{tag} - R_{ \eta } \cdot N^{iso, \eta}_{tag}
\end{equation}
\noindent This expression uses the notation explained in Table~\ref{Table:DPBackgroundNotation}. \( N^{iso} \) is the direct photon sample so the photons that are tagged as coming from either \( \pi^0 \) or \( \eta \) decays have already been eliminated.  \( N^{iso, \pi^0}_{tag} \) and \( N^{iso, \eta}_{tag} \) are some of these tagged decay photons that are also in an isolated photon pair as determined by Equation~\ref{Equation:PhotonPairIsolationCut}.  Figure~\ref{Figure:dpIsolationRatioCrossCheck} shows these ratios compared to what was found in \cite{PPG136}, where the inclusive  direct photon yield is calculated using Equation~\ref{Equation:inclDPYield}.  (The direct photon ratios in Figure~\ref{Figure:Fig13_PPG136} are slightly larger than 1 at high \( p_T \) because \( N_{dir}^{iso} \) and \( N_{dir} \) are calculated using these different formulas.)  These plots also show the ratio of \( N^{iso, \pi^0}_{tag} \) and \( N^{iso, \eta}_{tag} \) photons to the number of inclusive tagged \( \pi^0 \) and \( \eta \) photons respectively.  The figures show the same shape in isolation ratios for the direct photon and \( \pi^0 \) yields as a function of \( p_T \), indicating that the isolation cut used for the direct photon TSSA is being applied correctly.  The magnitude of these values is slightly different because the previously published ratios used an isolation cut with different requirements for what tracks and clusters could be included in the \( E_{cone} \) sum and a larger cone radius.  The fact that the isolated over inclusive yield ratios for both the \( \pi^0 \) and \( \eta \) mesons are less than 1, shows that the isolation cut is doing its job in reducing the direct photon background due to decay photons.

%% file: Chap4/chap4.tex
The midrapidity \( \pi^0 \) and \( \eta \) TSSA were found to be consistent with zero and achieve a factor of three increase in precision when compared to the previously published PHENIX TSSA results for the transversely polarized \( p + p \), \( \sqrt{s} = 200 \) GeV data set that was taken in 2008.  \cite{PPG135} 
Run-15 is PHENIX's last ever polarized proton data set and as such these new asymmetries are the definitive results from PHENIX for these observables.   
Because they measure hadrons, the \( \pi^0 \) and \( \eta \) asymmetries are sensitive to both initial- and final-state effects.  Light hadron TSSA results at forward rapidity sample the polarized proton at higher \( x \) and so are dominated by valence quark spin-momentum correlations.  At midrapidity, the polarized proton is being sampled at more moderate \( x \) and so these midrapidity \( \pi^0 \) and \( \eta \) TSSA results are sensitive at leading order to both quark and gluon dynamics in the polarized proton.  These results have the potential to further constrain the gluon Sivers function as well as the trigluon twist-3 collinear correlation function.  

\begin{figure}
  \centering
  \includegraphics[scale = 0.7]{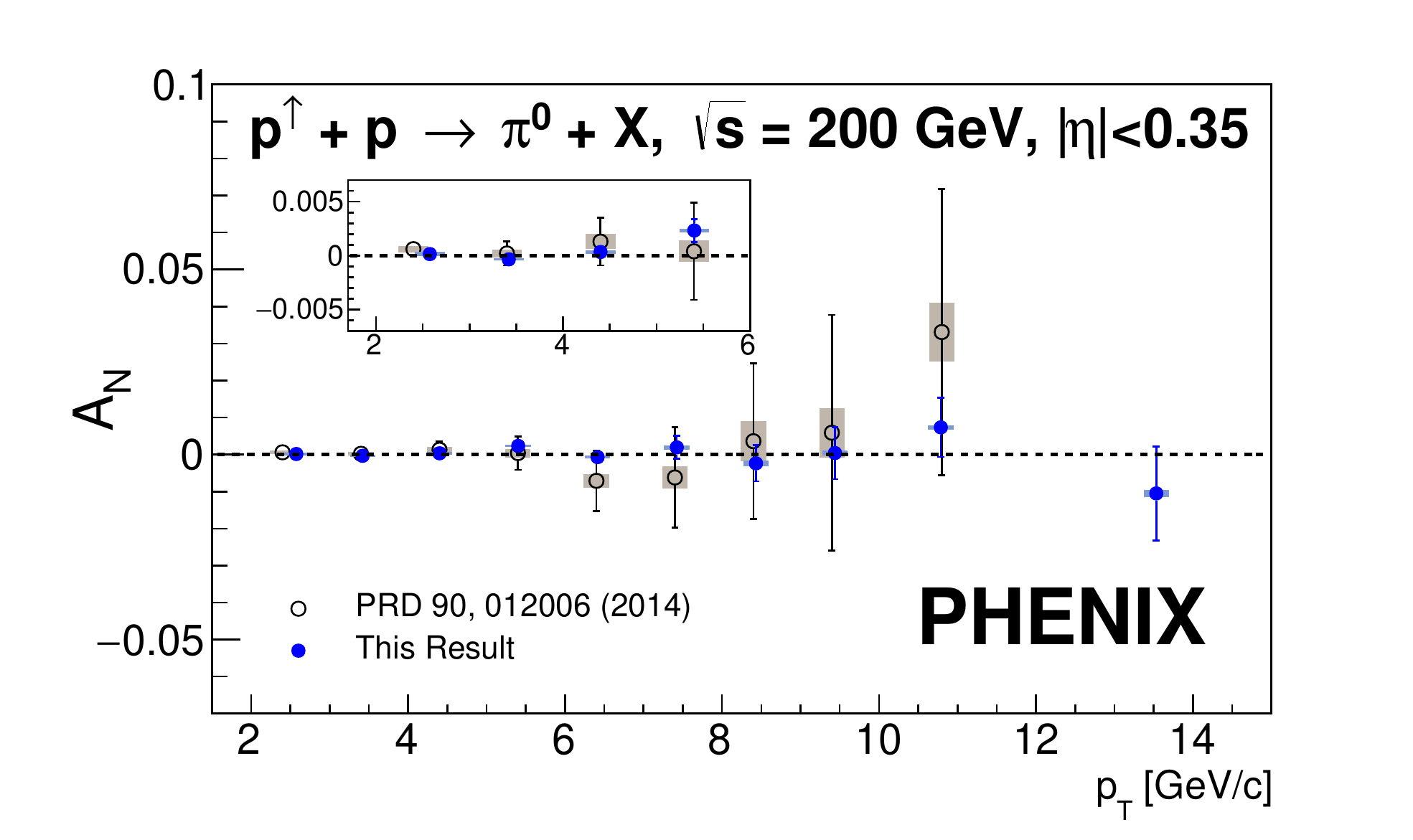}
  \caption[The final midrapidity \( \pi^0 \) TSSA  measured in \(p^\uparrow + p\) collisions with \( \sqrt{s} = 200 \) GeV and plotted with the previously published results]{The final midrapidity \( \pi^0 \) TSSA  measured in \(p^\uparrow + p\) collisions with \( \sqrt{s} = 200 \) GeV and plotted with the previously  published results. \cite{PPG135}}
  \label{Figure:pi0TSSA}
\end{figure}

\begin{table}
\centering
\begin{tabular}{| c | c | c | c | c | c | c | }
\hline
             &                 &                  &               &               & Sys. unc.  &             \\
             &                 &                  &  Sys. unc. & Sys. unc. & (from       &             \\
\( p_T \) &\( \pi^0 \)  &  Statistical  & (rel lumi   & (from bg & bunch       & Sys unc. \\
(GeV/c)  & Asymmetry & uncertainty &  vs. sqrt)   & fraction) & shuffling)  & (total)    \\
\hline  
2.58   &  0.000143 & 0.000281 & 5.71e-05  & 3.92e-07  & 0.000106 & 0.000120 \\
3.42   & -0.000343 & 0.000321 & 1.73e-05  & 3.92e-06  & 0             & 1.77e-05 \\
4.40   &  0.000335 & 0.000571 & 6.56e-05  & 1.91e-06  & 0             & 6.57e-05 \\
5.40   &  0.00233   & 0.00106   & 9.61e-05  & 6.68e-07  & 0             & 9.61e-05 \\
6.41   & -0.000689 & 0.00187   & 0.000112 & 2.11e-05  & 0             & 0.000114 \\
7.42   &  0.00193   & 0.00311   & 0.000341 & 7.61e-05  & 0             & 0.000350 \\
8.43   & -0.00238   & 0.00488   & 0.000245 & 0.000399 & 0             & 0.000469 \\
9.43   & 0.000404  & 0.00703   & 0.000331 & 0.000116 & 0             & 0.000351 \\
10.79 & 0.00734    & 0.00799   & 9.71e-05  & 0.000313 & 0             & 0.000328 \\
13.53 & -0.0105     & 0.0127    & 0.000686 & 1.15e-05  & 0             & 0.000686 \\
\hline
\end{tabular}
\caption[The final midrapidity \( \pi^0 \) TSSA summary table with statistical and systematic errors.]{The final midrapidity \( \pi^0 \) TSSA summary table with statistical and systematic errors.}
\label{Table:pi0TSSA}
\end{table}

\begin{figure}
  \centering
  \includegraphics[scale = 0.7]{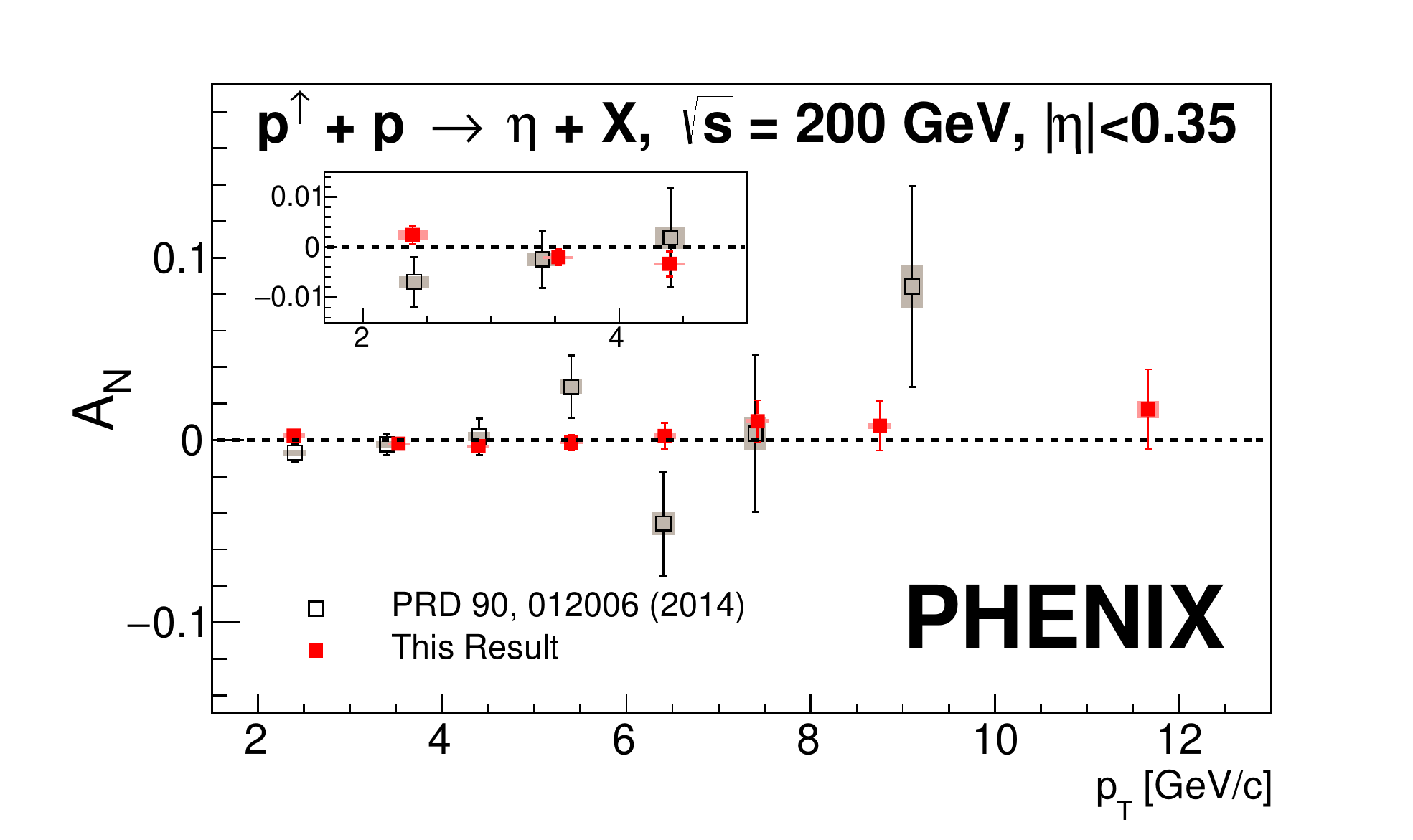}
  \caption[The final midrapidity \( \eta \) TSSA measured in \(p^\uparrow + p\) collisions with \( \sqrt{s} = 200 \) GeV plotted with previously published results.]{The final midrapidity \( \eta \) TSSA measured in \(p^\uparrow + p\) collisions with \( \sqrt{s} = 200 \) GeV plotted with previously published results. \cite{PPG135} }
  \label{Figure:etaTSSA}
\end{figure}

\begin{table}
\centering
\begin{tabular}{| c | c | c | c | c | c | c | }
\hline
             &                 &                  &               &               & Sys. unc.  &             \\
             &                 &                  &  Sys. unc. & Sys. unc. & (from       &             \\
\( p_T \) &\( \eta \)  &  Statistical  & (rel lumi   & (from bg & bunch       & Sys unc. \\
(GeV/c)  & Asymmetry & uncertainty &  vs. sqrt)   & fraction) & shuffling)  & (total)    \\
\hline  
2.39  &  0.00244 & 0.00183 &  0.000518  & 4.58e-05  & 0.000620 & 0.000809 \\
3.53  & -0.00199 & 0.00159 &  8.36e-05  & 3.31e-05  & 0              & 8.99e-05 \\
4.39  & -0.00331 & 0.00248 & 0.000144 & 4.55e-05  & 0              & 0.000151 \\
5.40  & -0.00139 & 0.00421 &  0.000241 & 3.59e-05  & 0              & 0.000244 \\
6.41  &  0.00222 & 0.00709 &  0.00112   & 6.35e-06  & 0              & 0.00112 \\
7.42  &  0.0103   & 0.0115   &  0.000703 & 0.000160 & 0              & 0.000720 \\
8.75  &  0.00790 & 0.0137   &  0.00124   & 0.000188 & 0              & 0.00125 \\
11.76 &  0.0168  & 0.0219   & 0.00425   & 0.000370 & 0              & 0.00426 \\

\hline
\end{tabular}
\caption[The final midrapidity \( \eta \) meson TSSA summary table with statistical and systematic errors.]{The final midrapidity \( \eta \) meson TSSA summary table with statistical and systematic errors.}
\label{Table:etaTSSA}
\end{table}

\begin{figure}
  \centering
  \includegraphics[scale = 0.65]{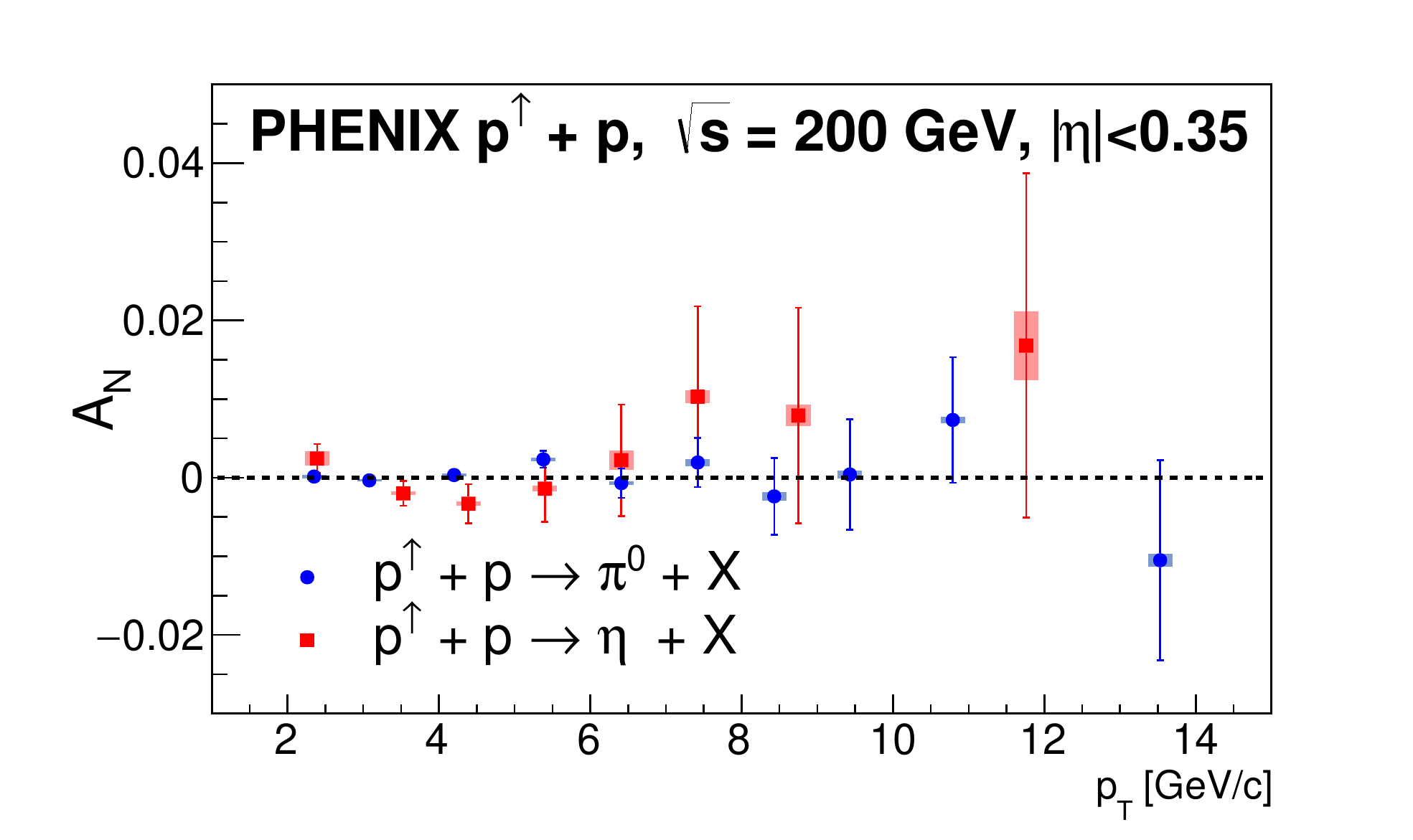}
  \caption[Comparison of the new \( \pi^0 \) and \( \eta \) midrapidity TSSA results  measured in \(p^\uparrow + p\) collisions with \( \sqrt{s} = 200 \) GeV]{Comparison of the new \( \pi^0 \) and \( \eta \) midrapidity TSSA results  measured in \(p^\uparrow + p\) collisions with \( \sqrt{s} = 200 \) GeV}
  \label{Figure:pi0EtaCompare}
\end{figure}

Figure~\ref{Figure:pi0TSSA} shows the final result for this midrapidity \( \pi^0 \) TSSA analysis which extends into higher \( p_T \) when compared to the previously published result from \cite{PPG135}.  There is an additional global uncertainty of 3\% from the polarization normalization on this result and all other new results in this chapter. \cite{PPG203} The final \( A_N^{\pi^0} \) values along with the statistical and systematic errors are listed in Table~\ref{Table:pi0TSSA} where the total systematic uncertainty is calculated by adding the three separate systematic uncertainties listed in the table in quadrature.   This new result is consistent with zero over the entire \( p_T \) range and even consistent with zero to within \( 10^{-4} \) at low \( p_T \).  This analysis's  final midrapidity \( \eta \) TSSA result is plotted in Figure~\ref{Figure:etaTSSA}, which illustrates how much more statistically precise this result is when compared to the previously published asymmetry and shows that this new asymmetry also extends to higher \( p_T \).  This result is also consistent with zero across its entire \( p_T \) range, where Table~\ref{Table:etaTSSA} explicitly lists the final \( A_N^\eta \) values along with the statistical and systematic errors.  Figure~\ref{Figure:pi0EtaCompare} shows these two results plotted together, which shows that in addition to both being consistent with zero, the midrapidity \( \pi^0 \) and \( \eta \) TSSA are consistent with each other.   

\begin{figure}[t]
  \centering
  \includegraphics[scale = 0.7]{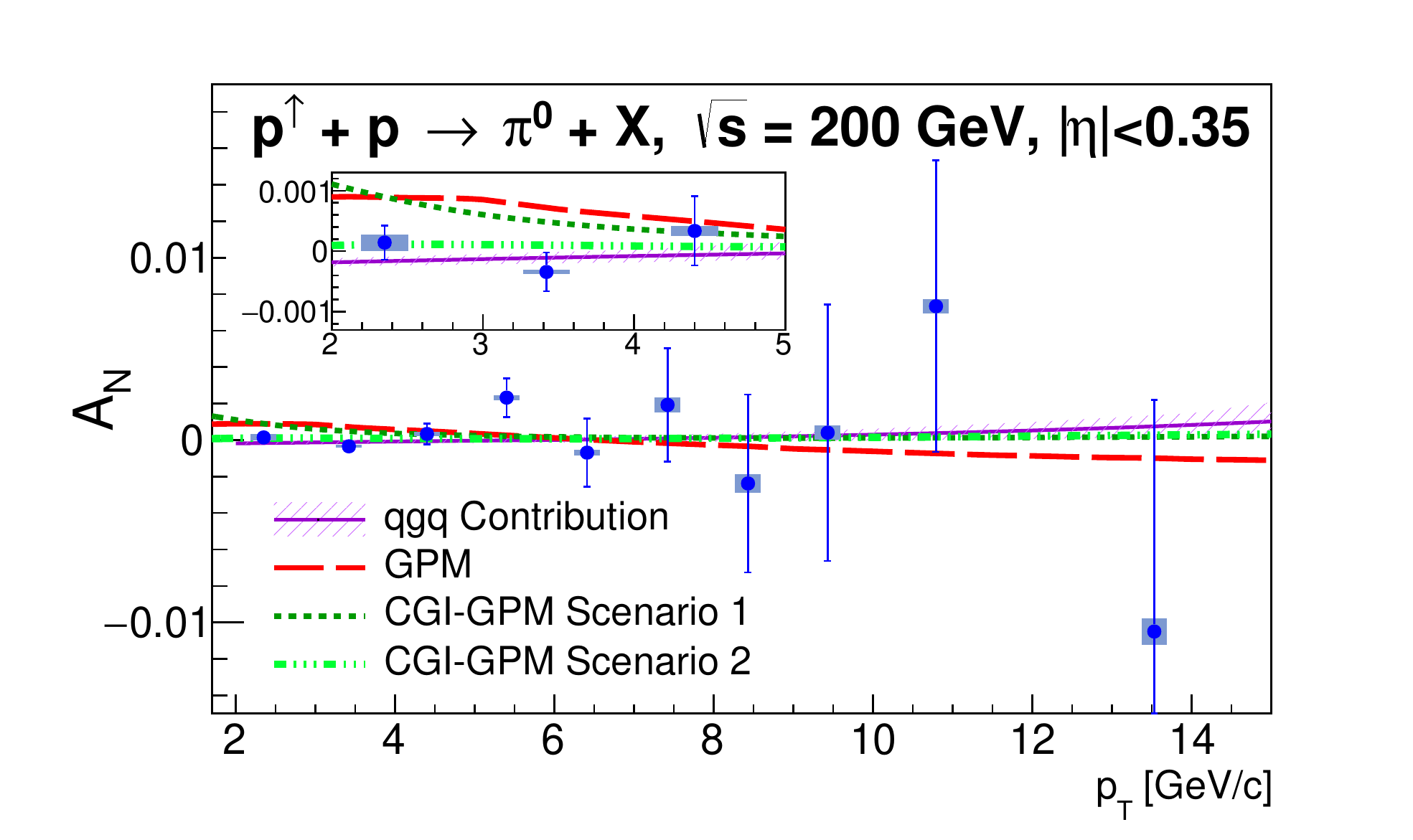}
  \caption[The new \( \pi^0 \) midrapidity TSSA with theoretical predictions in both the TMD and collinear twist-3 frameworks.]{The new \( \pi^0 \) midrapidity TSSA with theoretical predictions in both the TMD and collinear twist-3 frameworks.} 
  \label{Figure:pi0Theory}
\end{figure}

Figure~\ref{Figure:pi0Theory} shows this \( \pi^0 \) TSSA result plotted with theoretical predictions for this asymmetry.  The qgq curve shows the predicted contribution to the \( \pi^0 \) asymmetry from the Qiu-Sterman function and collinear twist-3 fragmentation functions.   This curve was calculated with fits to forward pion asymmetries which were published in Ref. \cite{2020EverythingFits} and have been reevaluated with the \( |\eta| < 0.35 \) pseudorapidity range of this measurement. \cite{DanielPitonyakEmailPi0}  Because midrapidity \( \pi^0 \) production also includes a large fractional contribution from gluon scattering in the proton, a complete collinear twist-3 description of the midrapidity \( \pi^0 \) TSSA would also need to include the contribution from the trigluon correlation function, which can be seen in Figure 11 of Ref. \cite{ForwardPionggg}.   The midrapidity \( \pi^0 \) asymmetry is sensitive to gluon dynamics at leading order and the qgq correlation function's contribution to this asymmetry is predicted to be small, thus this measurement will constrain future extractions of the ggg correlation function.

The rest of the theory curves in Figure~\ref{Figure:pi0Theory} show predictions for the midrapidity \( \pi^0 \) TSSA  generated by the Sivers TMD PDF.  These curves include contributions from both the quark and gluon Sivers functions and use the generalized parton model (GPM).  This framework takes the \( k_T \) moment of these Sivers functions and does not include NLO interactions with the proton fragments.  The ``GPM'' curve in Figure~\ref{Figure:pi0Theory} uses the parameters stated in Equation 32 of Ref.~\cite{CGI-GPM}.  The  color gauge invariant generalized parton model (CGI-GPM) expands on the GPM by including initial- and final-state interactions through the one-gluon exchange approximation.  The CGI-GPM curves plotted in Figure~\ref{Figure:pi0Theory}  show two different scenarios for this model, which can be found in Equation 34 of Ref.~\cite{CGI-GPM}.  
The values that are used for the Scenario 1 curve are chosen to maximize the open heavy flavor TSSA generated by the gluon Sivers function while still keeping this asymmetry within the statistical error bars of the published result in Ref.~\cite{PPG196} and simultaneously describing the previously published midrapidity \( \pi^0 \) TSSA from Ref. \cite{PPG135}.  The values used in the Scenario 2 curve do the same, except that they minimize the open heavy flavor TSSA within the range of the published statistical error bars.   
All three of these curves have been evaluated for \( x_F =0 \).\cite{UmbertoEmail}  These TMD calculations do not include the Collins effects because it has been calculated \cite{CollinsEffect, SiversEffect} and also measured \cite{STAR500GeVpreliminary} to be small when compared to the Sivers effect.  As shown in the zoomed in top panel of Figure~\ref{Figure:pi0Theory}, this \( \pi^0 \) TSSA result has the statistical precision at low \( p_T \) to distinguish between the GPM and CGI-GPM models, preferring CGI-GPM Scenario 2.  
 
\begin{figure}
\centering
\includegraphics[scale = 0.7]{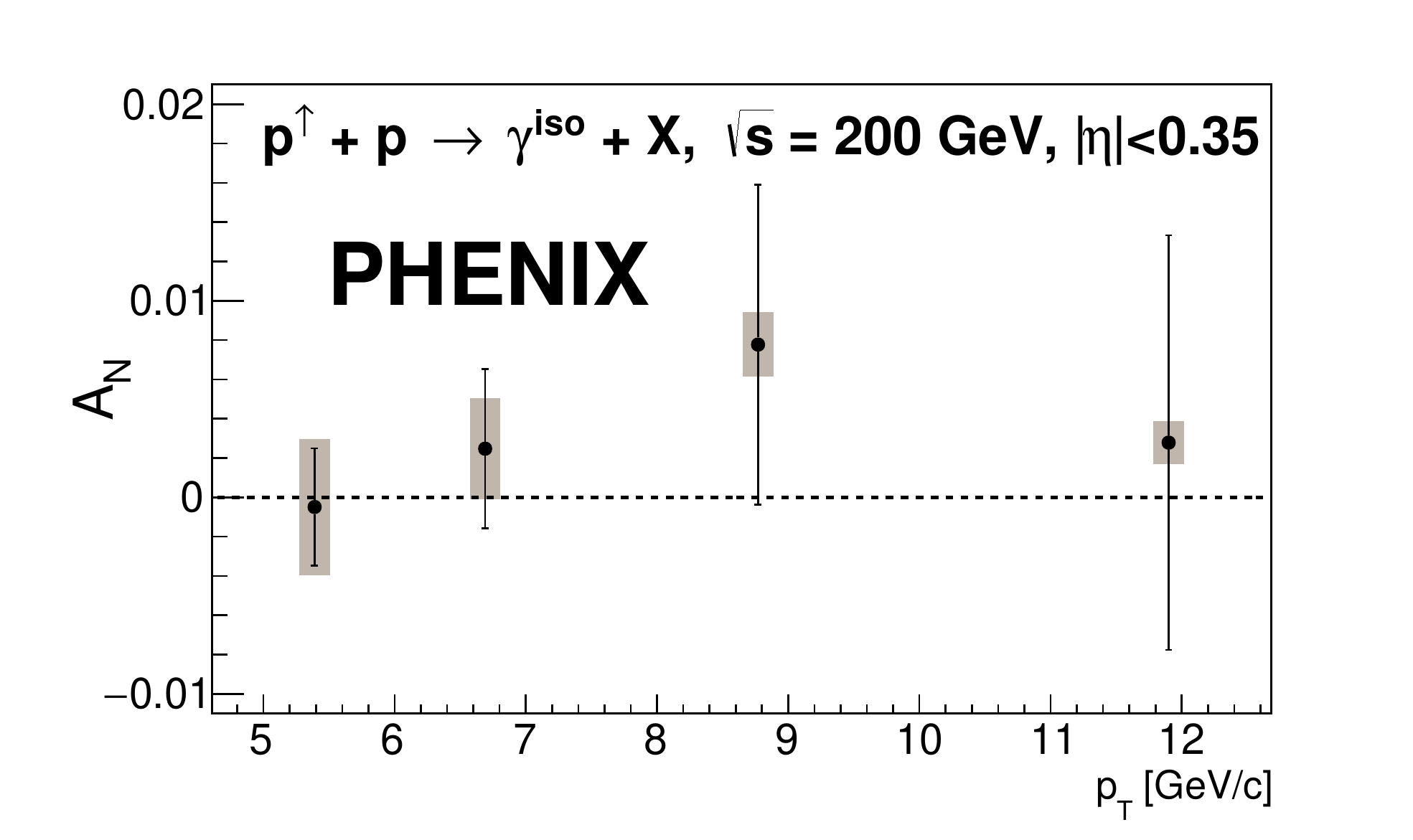}
\caption[The midrapidity isolated direct photon TSSA that was measured using \( \sqrt{s} = 200 \) GeV  \(p^\uparrow + p\) collisions]{The midrapidity isolated direct photon TSSA that was measured using \( \sqrt{s} = 200 \) GeV  \(p^\uparrow + p\) collisions}
\label{Figure:dpTSSA}
\end{figure}

The direct photon TSSA was measured for the first time at RHIC.  It was found to be consistent with zero to within about 2\% as shown in Figure~\ref{Figure:dpTSSA} and listed in detail in Table~\ref{Table:dpTSSA}.  Direct photons do not undergo any hadronization and so are only sensitive to initial-state effects.  Midrapidity direct photon production is also dominated at leading order by \( g + q \rightarrow \gamma + q \) scattering and so provides a uniquely clean way of studying gluon dynamics in the proton.  

\begin{table}
\centering
\begin{tabular}{| c | c | c | c | c | c | c | }
\hline
             &    Direct    &                   & Sys. unc. & Sys. unc.  & Sys. unc. &             \\
\( p_T \) &    Photon   & Statistical   & (rel lumi  & (from bg  & (from bg  & Sys unc. \\
(GeV/c)  & Asymmetry  & Uncertainty &  vs. sqrt) & fraction)  & asymm.)  &  (total)   \\
\hline 
5.39  & -0.000492 & 0.00299 & 0.000131 & 1.27e-05  & 0.00341  & 0.00341 \\
6.69  &  0.00247   & 0.00404 & 0.000112 & 2.69e-05  & 0.00252  & 0.00252 \\
8.77  &  0.00777   & 0.00814 & 5.49e-05  & 0.000160 & 0.00159  & 0.00159 \\
11.88 & 0.00278   & 0.0105   & 0.000715 & 4.39e-05  & 0.000775 & 0.00106 \\
\hline
\end{tabular}
\caption[The final midrapidity isolated direct photon TSSA summary table with statistical and systematic errors]{The final midrapidity isolated direct photon TSSA summary table with statistical and systematic errors.}
\label{Table:dpTSSA}
\end{table}

\begin{figure}
  \centering
  \includegraphics[scale = 0.7]{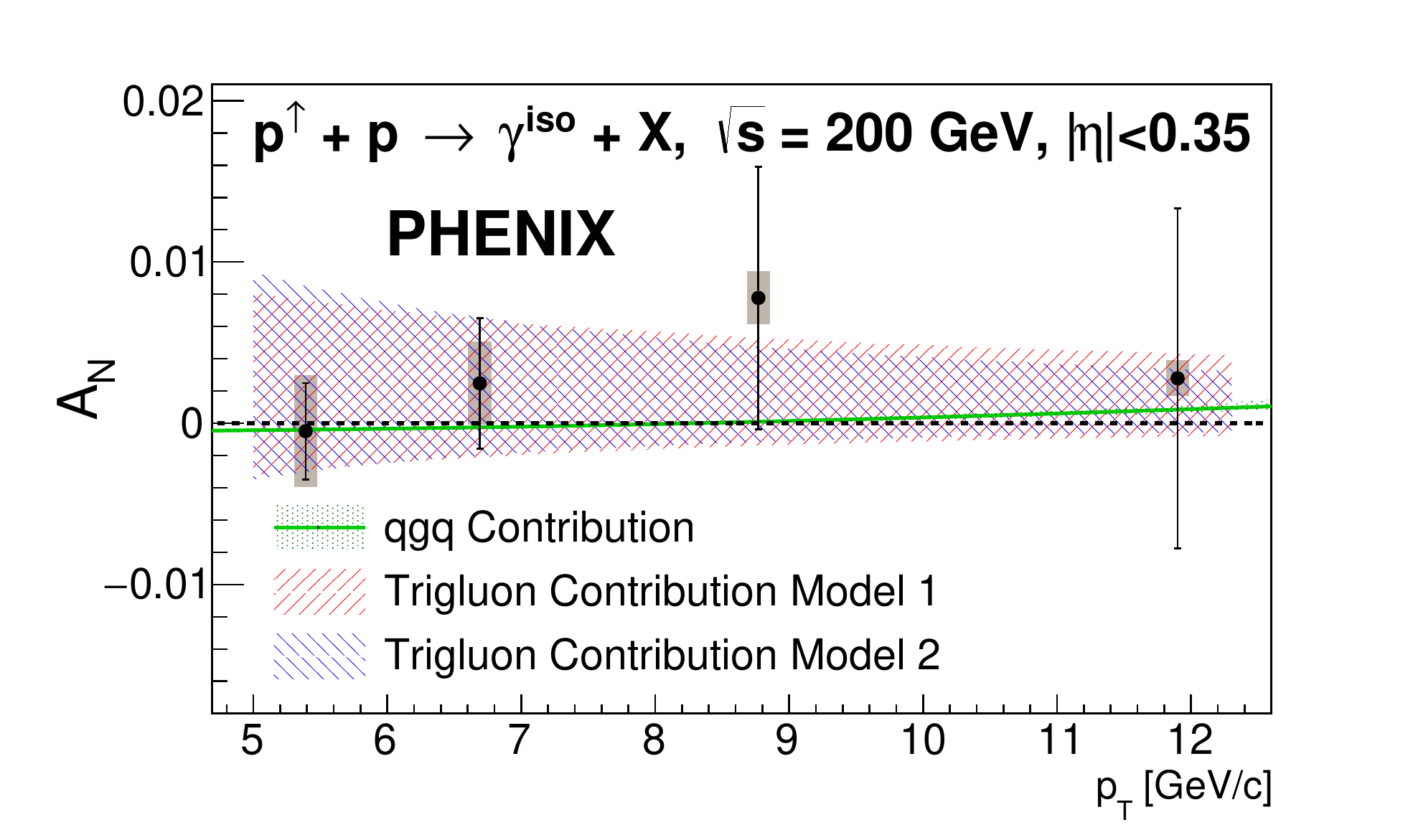}
  \caption[The direct photon asymmetry plotted with predictions from twist-3 collinear correlation functions.]{The direct photon asymmetry plotted with predictions from twist-3 collinear correlation functions.}
  \label{Figure:dpTheory}
\end{figure}

Figure~\ref{Figure:dpTheory} shows this same isolated \( A_N^\gamma \) result plotted with predictions from collinear twist-3 correlation functions.  The green curve plots the contribution of qgq correlation functions to the direct photon asymmetry which is calculated using functions that were published in Ref. \cite{dpTSSAqgq} that are integrated over the \( |\eta| < 0.35 \) pseudorapidity range of this analysis. \cite{DanielPitonyakEmail}  This calculation includes contributions from both the polarized and unpolarized proton.   At forward rapidity,  \( A_N^\gamma \) is dominated by the Qiu-Sterman function (the SGP term of the qgq function of the polarized proton), while contributions from the SFP of the polarized proton and the SGP of the unpolarized proton are comparatively small and the contribution from the SFP of the unpolarized proton is exactly zero.~\cite{dpTSSAqgq} All of this carries over to the midrapidity direct photon asymmetry, except that term from the SFP of the polarized proton is the same size as the contribution from the Qiu-Sterman function.  For this calculation, the Qiu-Sterman function is calculated as the \( k_T \) moment of the Sivers function that is extracted from the global fit in Ref. \cite{2020EverythingFits}.  It is assumed that the SFP of the polarized proton could also be expressed as a \( k_T \) moment of this Sivers function because very little is known about these SFP functions. \cite{DanielPitonyakEmail}  Because the contributions from the Qiu-Sterman function and the SFP of the polarized proton are of the same order and have the opposite sign, the total contribution of qgq correlation functions to the midrapidity asymmetry is predicted to be small. 

This means that the midrapidity direct photon TSSA is can be used to extract the trigluon correlation function.  
The predicted range of this function's contribution to this asymmetry measurement is also plotted in Figure~\ref{Figure:dpTheory}.  This theoretical calculation uses results that were published in Ref. \cite{dpTSSAggg} and were reevaluated as a function of photon \( p_T \) for \( \eta = 0 \). \cite{ShinsukeYoshidaEmail}  As this plot clearly shows, this \( A_N^\gamma \) measurement has the statistical precision at low \( p_T \) to constrain this trigluon correlation function.


\pagebreak

The fact that all three of these asymmetry results are all consistent with zero agrees with previous measurements that were sensitive to transverse gluon spin-momentum correlations.  The previously published midrapidity \( \pi^0 \) and \( \eta \) TSSA PHENIX results were also consistent with zero. \cite{PPG135} Heavy flavor production at RHIC energies is dominated by gluon-gluon fusion and PHENIX forward heavy flavor asymmetry measurements were found to be consistent with zero both for the open heavy flavor \cite{PPG196} and \( J/\psi \)  TSSA. \cite{PPG211}.  Midrapidity jets that are produced in \( p + p \) collisions at RHIC also have a large contribution from gluon scattering (similar to the direct photon).  The STAR inclusive jet TSSA measurements have also been found to be consistent with zero at both \( \sqrt{s} = 200 \) GeV\cite{STARJetTSSA200GeV} and \( \sqrt{s} = 500 \) GeV.  \cite{STARJetTSSA500GeV}  RHIC TSSA measurements that are sensitive at leading order to gluon kinematics have consistently determined that transverse gluon spin-momentum correlations in a transversely polarized protons are small.  And these new results do so with the highest level of precision of any other TSSA measured at RHIC.

%% file: Chap5/chap5.tex


\section{Future Measurements}

Measurements at RHIC have played a vital role in increasing our knowledge of how spin behaves in QCD, but there remains much to learn.  The sPHENIX experiment is currently being built and is scheduled to start taking data in 2023. It will be housed in what was the PHENIX experimental hall.  sPHENIX was designed to be able to probe the quark-gluon plasma (QGP) at different length and temperature scales by measuring jets, jet correlations, and bottomonium states in \( p + p \), \( p \) \( + \) Au, and Au \( + \) Au collisions.\cite{sPHENIXproposal}  sPHENIX will be the first dedicated high-rate jet detector at RHIC and so it will include precision tracking along with high rate and large acceptance hadronic and electromagnetic calorimetry.  It will incorporate the magnetic solenoid that was originally used in the BaBar experiment at SLAC \cite{BaBarMagnet} and so the central barrel was designed to fit around this magnet.  The EMCal layer will fit around the central tracking system and it will be surrounded by other detector layers which in the direction radially away from the beam will be: the inner hadronic calorimeter, the BaBar solenoid, and the outer hadronic calorimeter layer, which also serves as a magnetic flux return.  This calorimetry system will have a pseudorapidity range of  \( | \eta | < 1 \) where at larger rapidities the towers will be tilted towards the interaction point to increase the detector's angular acceptance.  A segment of sPHENIX's future EMCal which will be centered at pseudorapidity \( \eta \sim 1 \) was tested in 2018 at the T-1044 experiment located in the Fermilab Test Beam Facility, with participation by the University of Michigan.  It was found to have an energy resolution well beyond what will be needed to measure jets.  \cite{2018TestBeam}  

sPHENIX will be able to measure particle-in-jet asymmetries which are sensitive to a nonperturbative transverse momentum scale and can be used to extract TMD FFs.  These measurements will be similar to the Collins asymmetry measurements from STAR \cite{STARJetTSSA500GeV, STAR500GeVpreliminary} except with smaller systematic uncertainties because the jets will include input from hadronic calorimetry data.  sPHENIX's excellent secondary vertex tagging will mean that the open heavy flavor TSSA can be measured with smaller systematic errors.  sPHENIX will also be capable of measuring direct photons at midrapidity and so will be able to follow up the direct photon TSSA measurement with higher statistical precision.  Both of these asymmetries are sensitive to the trigluon correlation function.  sPHENIX will furthermore be able to measure angular correlations between direct photons and jets which are sensitive to nonperturbative transverse momentum scales and are an ideal probe of factorization breaking. \cite{sPHENIXcoldQCD}

A high rate EMCal with large acceptance and high energy and spatial resolution is essential to accomplishing these photon and jet measurements.  The readouts of all of the calorimeters will use silicon photomultipliers (SiPMs) which have the ability to provide high gain even in a large magnetic field all while requiring minimal space.  sPHENIX will be using S12572-015P SiPMs which are made by Hamamatsu and have an active area of \( 3 \times 3 \) mm\textsuperscript{2} that contains 40,000  pixels which are about 15 \( \mu \)m in length.  The SiPM circuit boards will be attached to light guides that are placed behind the calorimeter towers. \cite{2018TestBeam} SiPMs are powered by what is referred to as a bias voltage or operating voltage, which is around 70 V for these devices.  Adjusting the bias voltage within a range of about \( \pm 5 \) V will cause the gain to change linearly.  The gain also depends on the temperature of the SiPM as well as material characteristics that vary device to device.  To conserve the space needed for the cables and circuit boards, the sPHENIX calorimeter electronics were designed to power multiple SiPMs with the same bias voltage.  Thus sPHENIX got Hamamatsu to agree to sort these SiPMs such that for an operating voltage within a specified 40 mV range, each SiPM in that group would have a gain of about \( 2.4 \times 10^5 \) when the SiPM is at \( 25^\circ \)C.  The University of Michigan group was responsible for operating a tester designed by the University of Debrecen in Hungary to verify that the sorting matched these specifications.  

The forward rapidity direct photon TSSA has been shown to be a clean method for extracting the Qiu-Sterman function. \cite{dpTSSAqgq}  The backward rapidity direct photon TSSA has been shown to be sensitive to the magnitude of the trigluon correlation function. \cite{dpTSSAggg}  STAR is currently working on a measurement of both the forward and backward direct photon asymmetries for both \( \sqrt{s} = 200 \) GeV and \( \sqrt{s} = 500 \) GeV.  \cite{STARdirectPhotonMeasurement}  The STAR forward upgrade, which is scheduled to start taking data around 2021, will include both tracking and a hadronic calorimeter, in addition to a high resolution EMCal.  This means it will be able to measure forward \textit{isolated} direct photon asymmetries and so include a smaller contribution from fragmentation photons.  The STAR forward upgrade will also be able to measure forward rapidity particle-in-jet asymmetries.  \cite{STARforwardUpgrade}  

The E1039 experiment or SpinQuest is an upcoming polarized fixed target experiment at Fermilab.  It will use \( \mu^+\mu^- \) pairs to measure the Sivers asymmetry in Drell-Yan and verify the Sivers sign change.  Due to the detector acceptance and collision kinematics, SpinQuest will be particularly sensitive to antiquark distributions in the polarized target.  It will be able to measure the antiup and antidown Sivers functions for the first time by comparing asymmetries measured using the transversely polarized hydrogen and deuterium targets.  Additionally, SpinQuest will be able to measure the \( J/\psi \) TSSA which is sensitive to the trigluon correlation function.  \cite{spinquest} There is also a proposal for a transversely polarized target at the LHCb which would start taking data around 2027.  It would be able to measure backward rapidity TSSAs using the high trigger rates and particle identification capabilities of the LHCb experiment. \cite{LHCSpin}

The Electron-Ion Collider (EIC) is the next planned high energy nuclear physics facility in the US and will offer an unprecedented way of studying TMD physics.   \cite{nuclearPhysicsLongRangePlan} The EIC will be built at BNL and is scheduled to start taking data in 2030 and will collide polarized beams of electrons and ions.  Through SIDIS, collisions at the EIC will be able to measure TMD functions for a wide range of collision energies and so will be able to constrain TMD evolution.  Due to its high luminosity, the EIC will also be able to measure exclusive processes in electron-proton collisions such as deeply virtual Compton scattering (DVCS), where the scattered electron is measured along with the intact proton and an additional radiated photon.  DVCS is sensitive to the spatial distribution of partons within the proton.  Longitudinally polarized collisions will allow the EIC to measure the spin structure of the proton.  Excellent forward detection combined with high luminosity will allow the EIC to probe the polarized proton at lower \( x \) than has ever been measured, constraining gluon distributions in the polarized proton.  Additionally the EIC will be the first ever \( eA \) collider, giving clean access to gluon saturation effects. \cite{gluonSaturationEIC}  The EIC will able to access parton distributions within the neutron through electron-deuterium collisions.  Furthermore, polarized \( eA \) collisions will help constrain TMD functions for light nuclei beyond the proton.

\section{Summary}
The parton model was first proposed in 1969 as a method of analyzing high energy hadronic collisions, \cite{FeynmanPartonModel} but it was always known to be an approximation.  Collinear functions in the context of the parton model assume that the partons only move longitudinally within the proton and completely integrate over their internal dynamics.  Leading twist calculations only allow for a single parton within the proton to participate in a scattering event.  The assumption of universality tells us that when this parton is exiting the proton, it will behave exactly the same regardless of the types of color fields that are produced by different types of scattering processes.  Closely coupled, the assumption of factorization states that initial- and final-state effects can be expressed in separate functions.  pQCD calculations that incorporate all of these assumptions have been able to successfully interpret and predict spin-averaged hadronic cross sections at high momentum transfer.  

However, when it came to interpreting spin-momentum correlations like the spontaneous polarization of baryons or TSSAs, these assumptions started to break down.  Collinear twist-3 and TMD formalisms are two different methods of relaxing the assumptions of the parton model and allowing quarks and gluons to interact with their surrounding color fields.  Collinear twist-3 functions describe the quantum mechanical interference between interacting with one parton versus interacting with two and TMD functions depend explicitly on the parton's nonperturbative transverse momentum.  Both have been shown to be able to generate the large TSSAs that have been measured in proton-proton collisions at forward rapidity.  Nonzero PT-odd TMD functions lead to the prediction of color entanglement effects from  soft gluon exchanges with proton remnants both before and after the partonic scattering event.   Some theories have even expanded to include observable effects from quantum entanglement \textit{within} the proton. When undergoing a high energy collision, the proton can be split into probed and unprobed regions which remain quantum mechanically entangled throughout the whole collision process. \cite{protonColorEntanglement}  QCD research is beginning to reach the stage where it is ready to consider color-dynamics within strong force bound states.  

This dissertation presented the TSSAs of midrapidity direct photons, neutral pions, and eta mesons at PHENIX.  The \( \pi^0 \) and \( \eta \) TSSAs were found to be consistent with zero and a factor of three increase in precision when compared with previous results.  These asymmetries are sensitive to both initial- and final state-effects for a mix of parton flavors.  The direct photon TSSA was measured for the first time at RHIC and was also found to be consistent with zero.  Direct photons do not undergo hadronization and so can be used as a clean probe of proton structure.  These asymmetry results will help constrain the trigluon correlation function in the transversely polarized proton as well as the gluons Sivers function, both of which are steps towards creating a more complete, three dimensional picture of proton structure.

%% file: Appendices/pi0_eta_Asymmetry_Ploots.tex
This section shows various asymmetry cross checks for both the \( \pi^0 \) and \( \eta \) TSSA results.  All of the asymmetries plotted in this section have been corrected for combinatorial background using the method explained in Section~\ref{Section:BackgroundSubraction}.  The \( \pi^0 \) background fraction can be found in Table~\ref{Table:rPi0} and the \( \eta \) background fraction can be found in Table~\ref{Table:rEta}.  The relative luminosity formula is described in detail in Section~\ref{Section:RelativeLuminosityFormula} and the square root formula is explained in Section~\ref{Section:SquareRootFormula}.  The T values comparing the left versus the right relative luminosity formula results are calculated with Equation~\ref{Equation:TTestLeftRight}.  The T values comparing the blue and yellow beam results are calculated using Equation~\ref{Equation:TTestYellowBlue}. And T values that are comparing the differences in the relative luminosity and square root formula results are calculated with Equation~\ref{Equation:TTestSqrtLumi}.


\begin{figure}
\centering
\subfigure[Yellow Beam]{ \includegraphics[scale = 0.32]{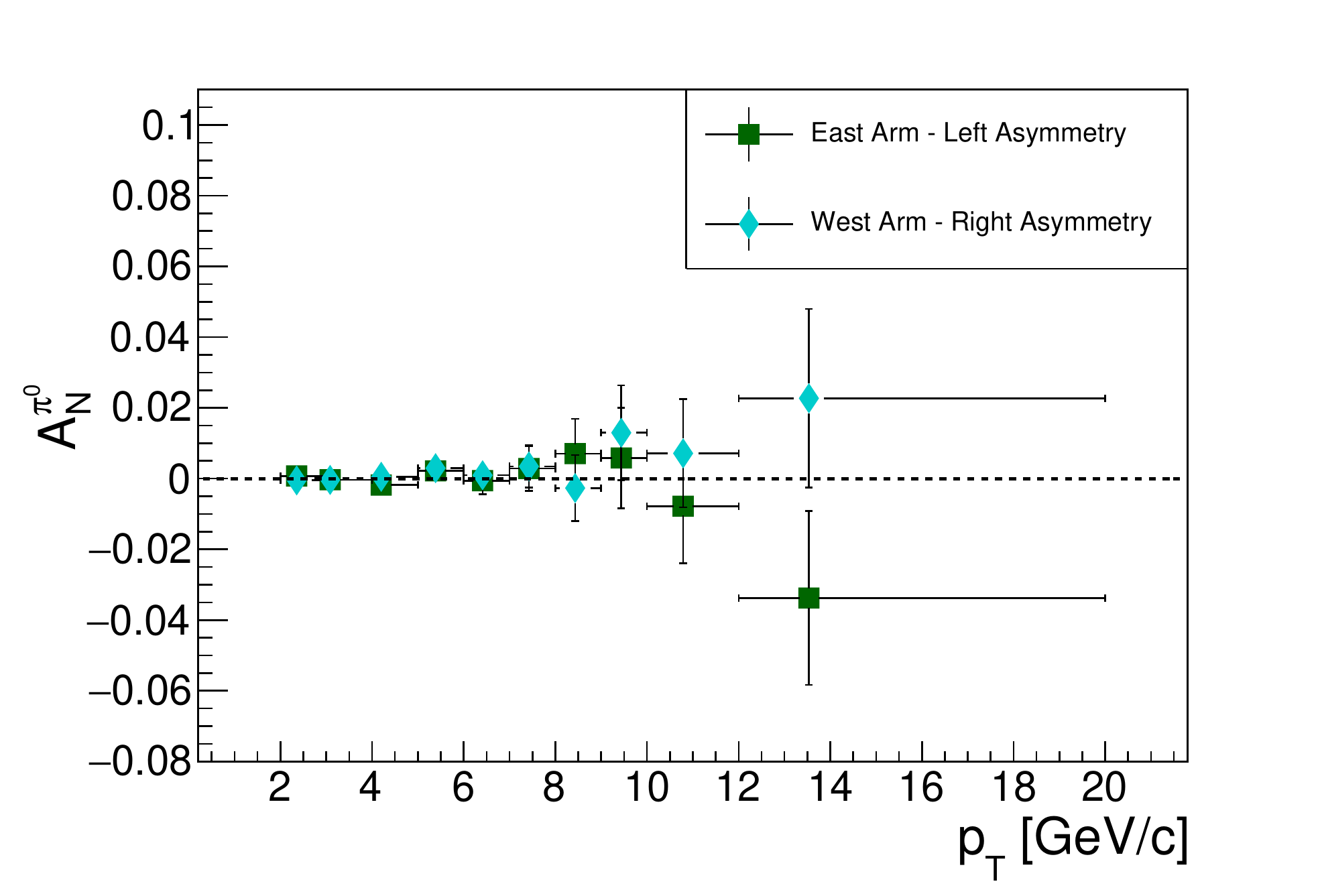} }
\subfigure[T test of the yellow beam left and right asymmetry results]{ \includegraphics[scale = 0.32]{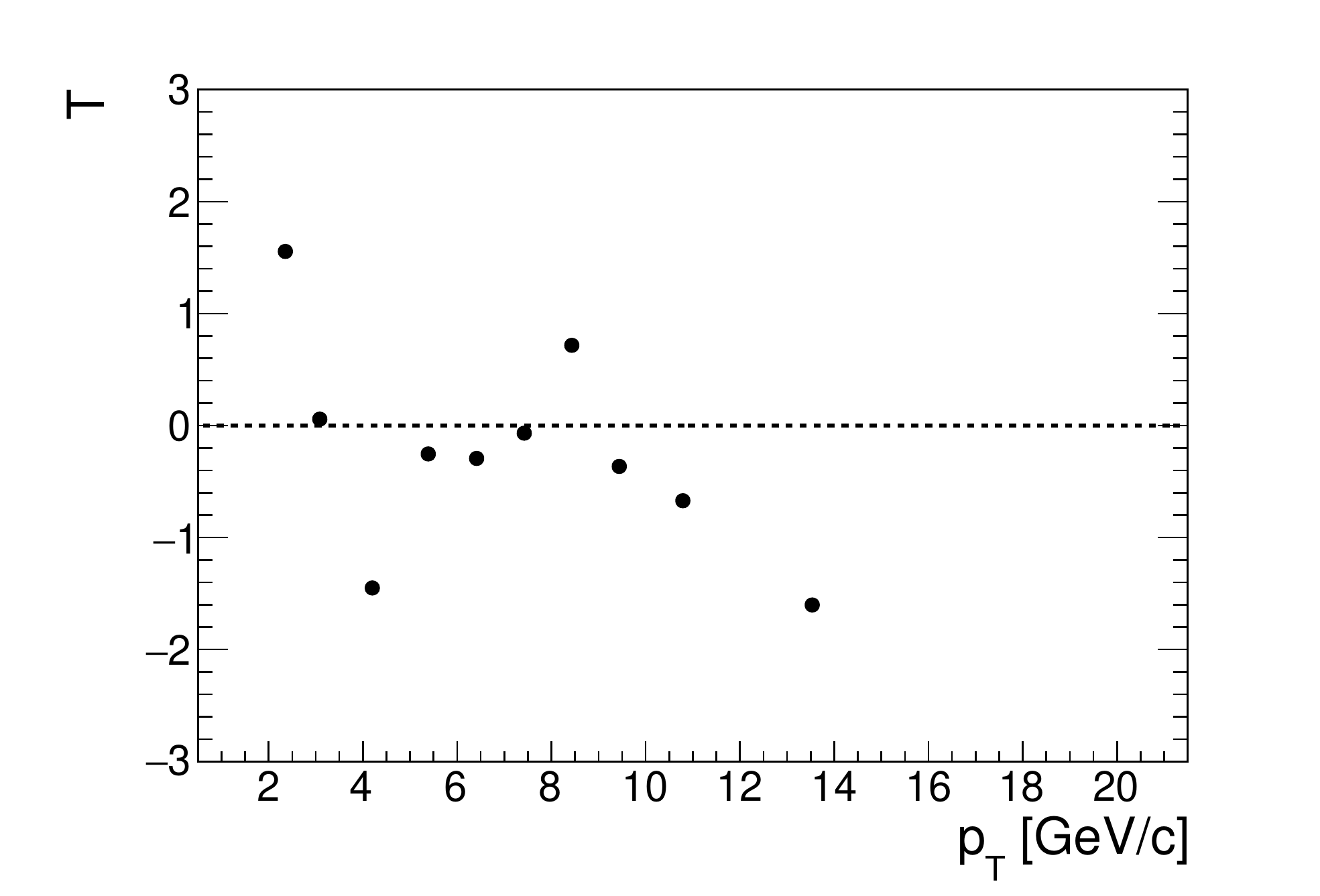} }\\
\subfigure[Blue Beam]{ \includegraphics[scale = 0.32]{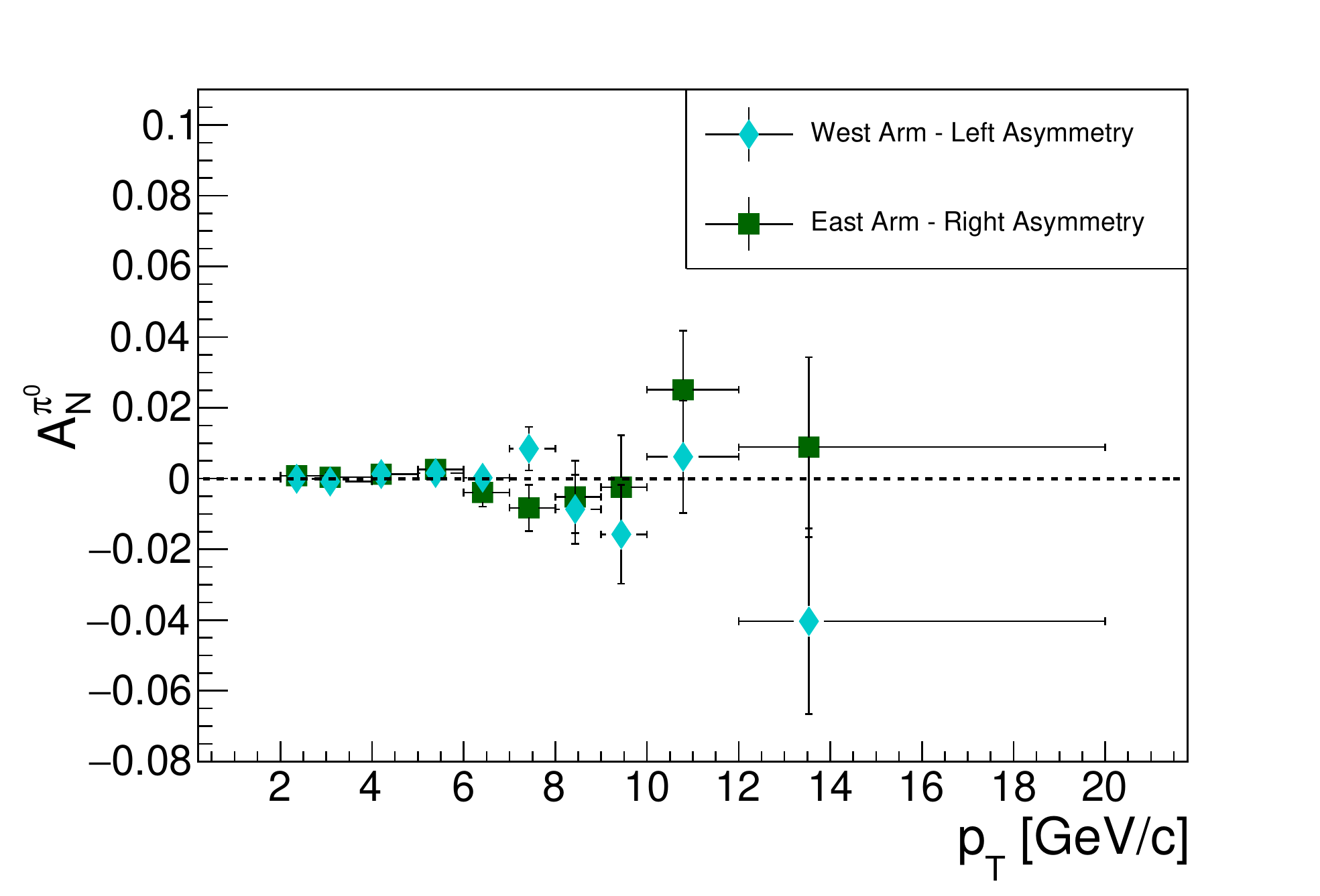} } 
\subfigure[T test of the blue beam left and right asymmetry results]{ \includegraphics[scale = 0.32]{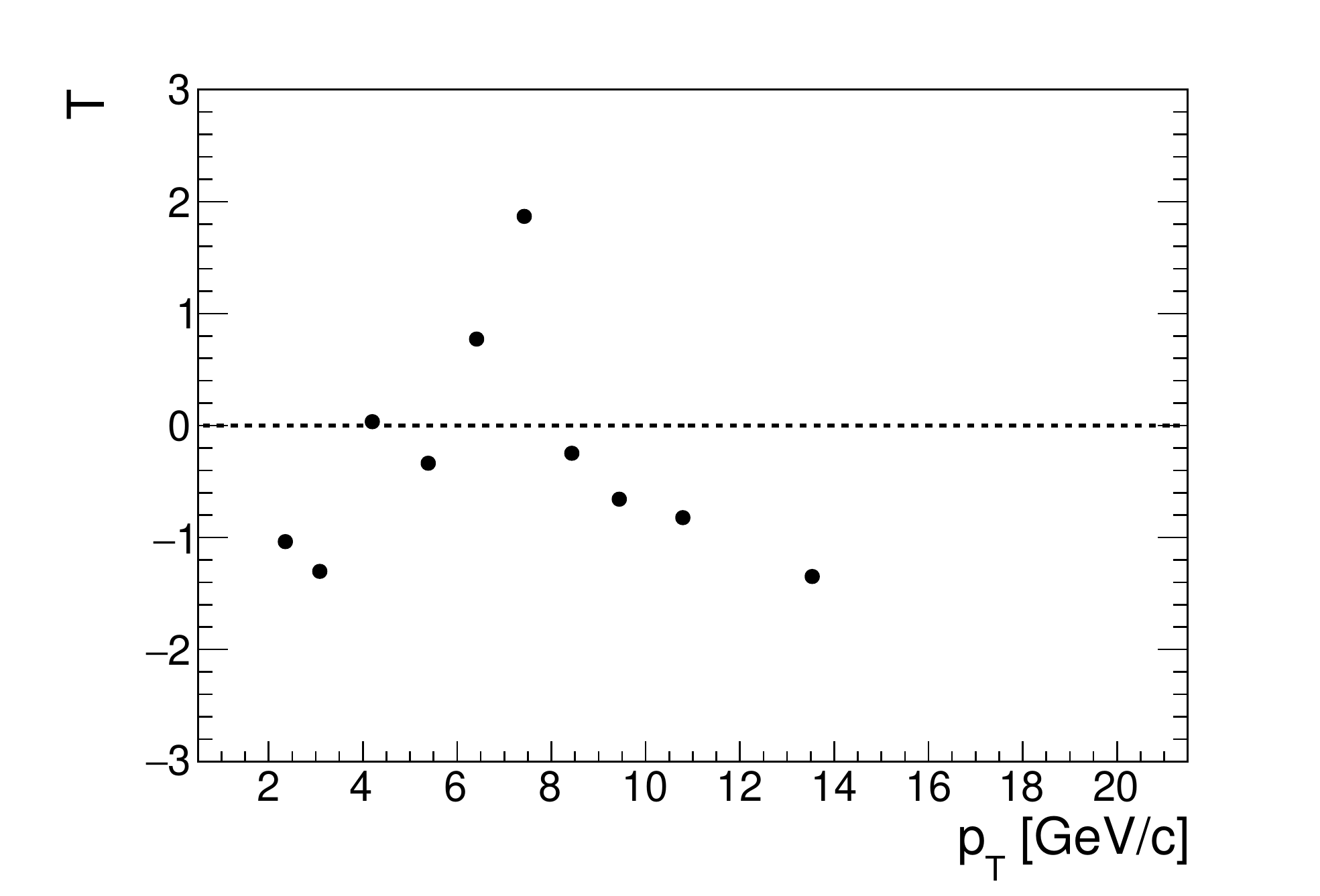} } 
\caption[The background corrected \( \pi^0 \) asymmetry calculated using the relative luminosity formula]
{The background corrected \( \pi^0 \) asymmetry calculated using the relative luminosity formula.  }
\label{Figure:pi0_correctedLumiLeftRight}
\end{figure}

\begin{figure}
\centering
\subfigure[Yellow and Blue Beam Asymmetries]{\includegraphics[scale = 0.32]{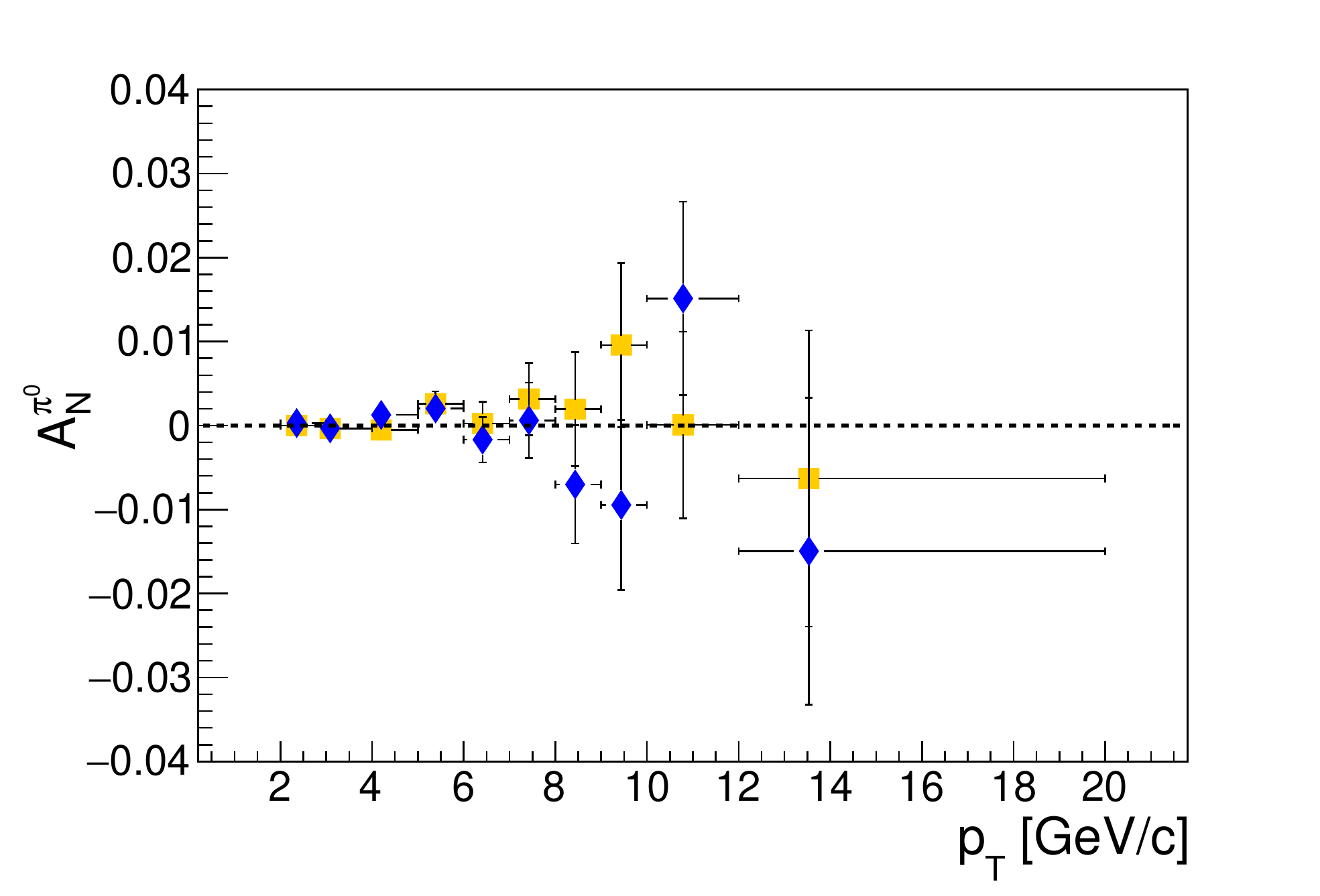} } 
\subfigure[T test comparing the result of the blue and yellow beam asymmetries]{\includegraphics[scale = 0.32]{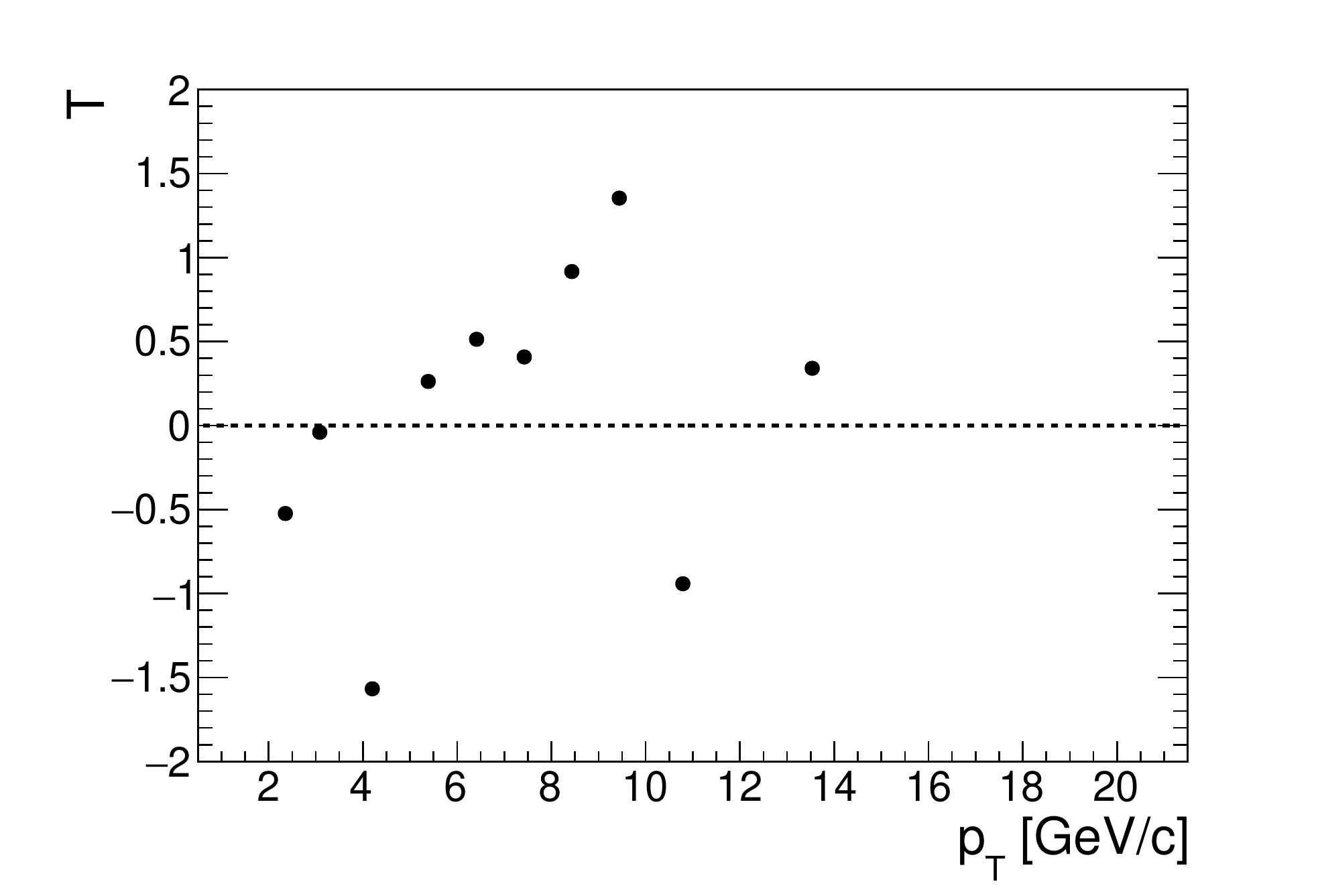} }
\caption[Relative luminosity yellow and blue beam asymmetries for background corrected \( \pi^0 \) asymmetry.]
{Relative luminosity yellow and blue beam asymmetries for background corrected \( \pi^0 \) asymmetry.  }
\label{Figure:pi0_correctedLumiYellowBlue}
\end{figure}

\begin{figure}
\centering
\subfigure[Yellow Beam]{ \includegraphics[scale = 0.32]{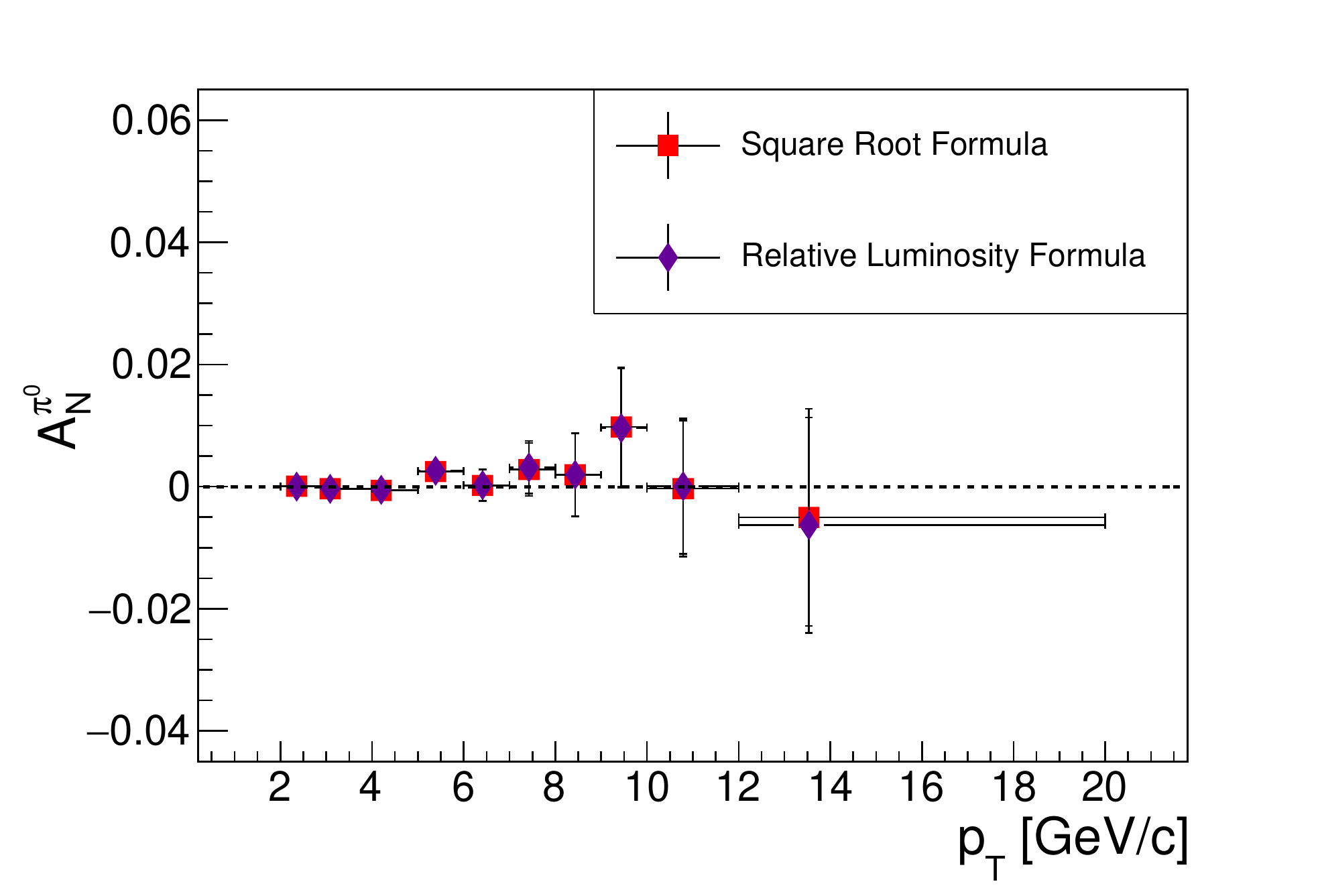} }
\subfigure[T test comparing the yellow beam square root and relative luminosity formula results]{ \includegraphics[scale = 0.32]{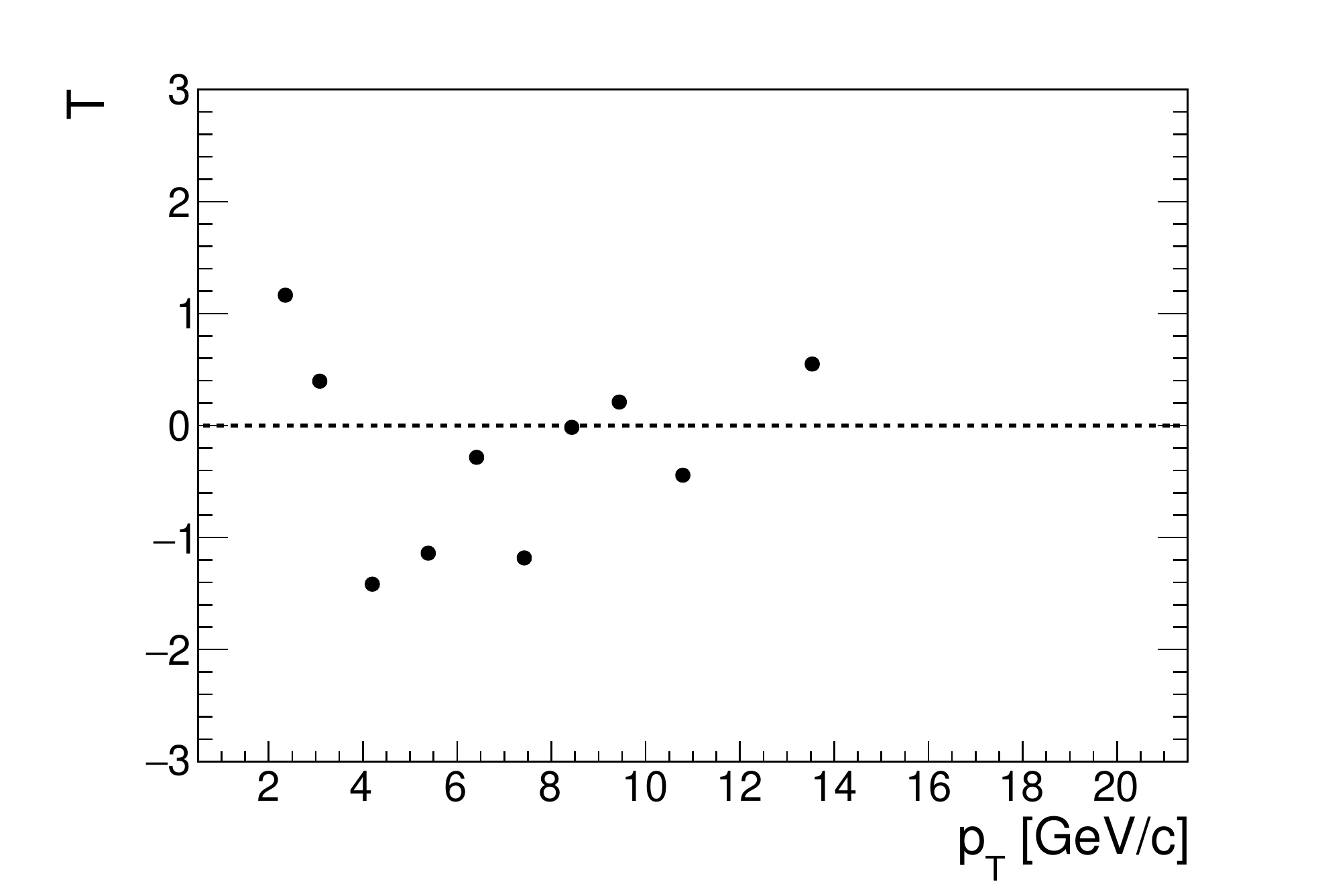} }\\
\subfigure[Blue Beam]{ \includegraphics[scale = 0.32]{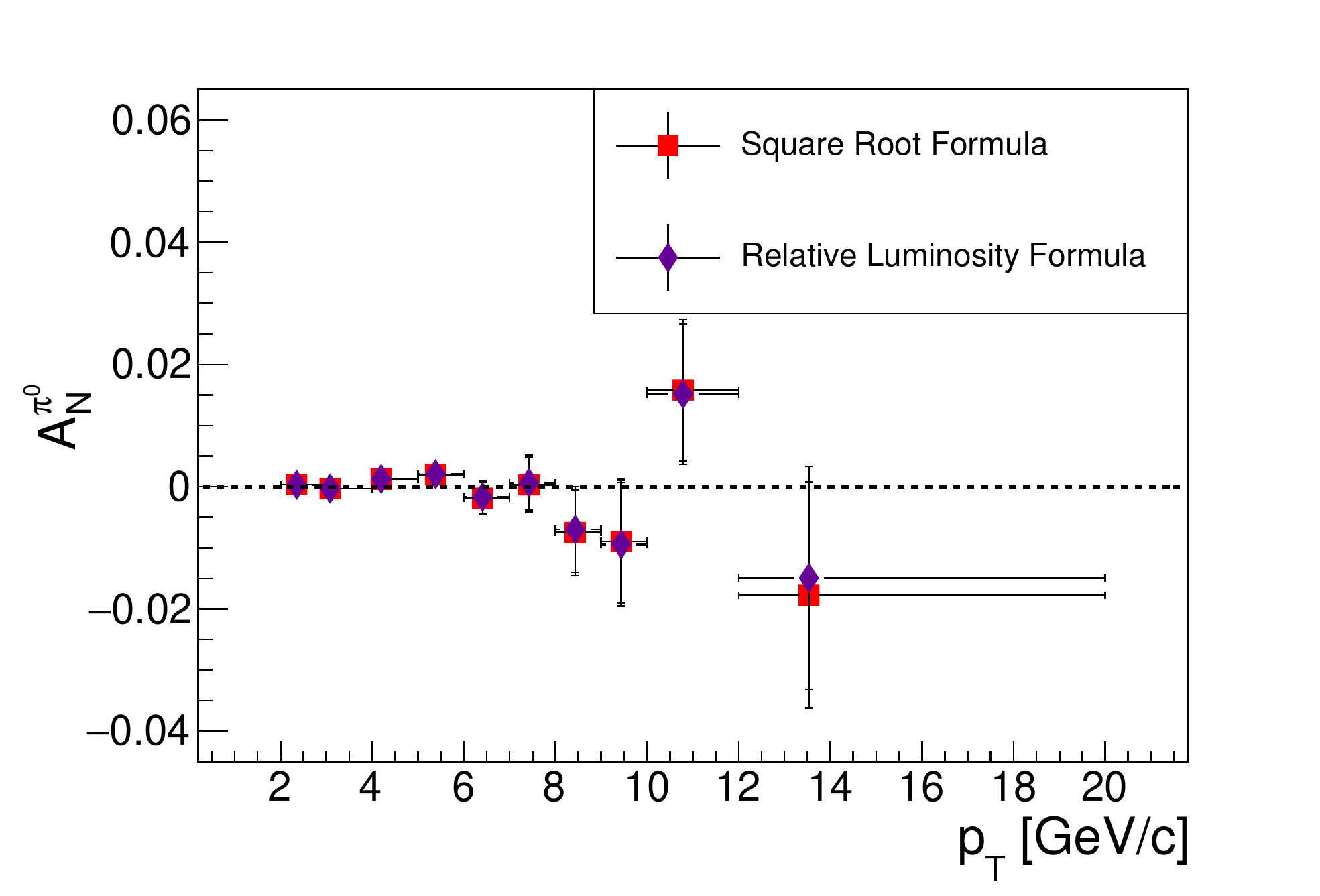} } 
\subfigure[T test comparing the blue beam square root and relative luminosity formula results]{ \includegraphics[scale = 0.32]{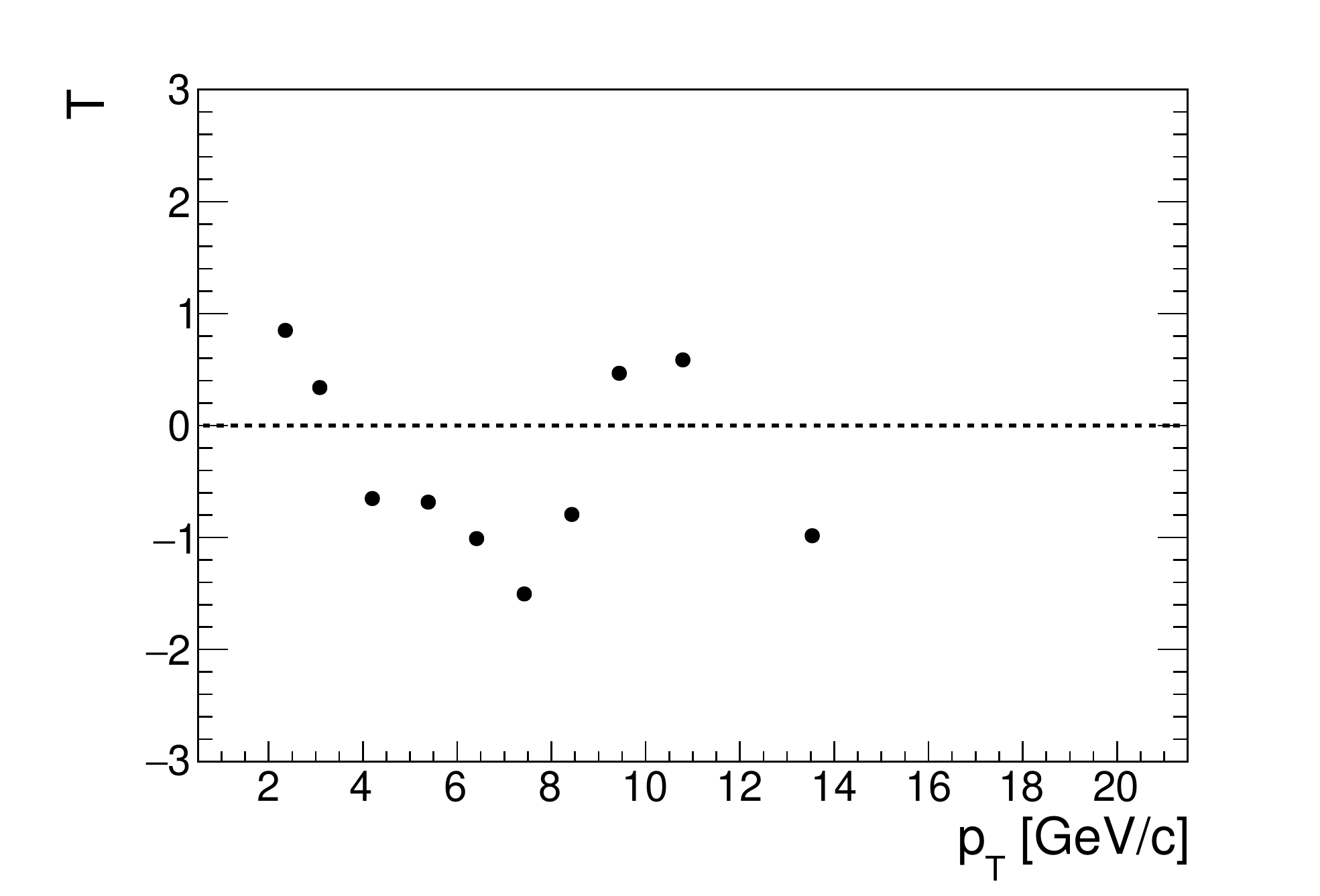} } \\
\subfigure[Final Averaged Asymmetry]{ \includegraphics[scale = 0.32]{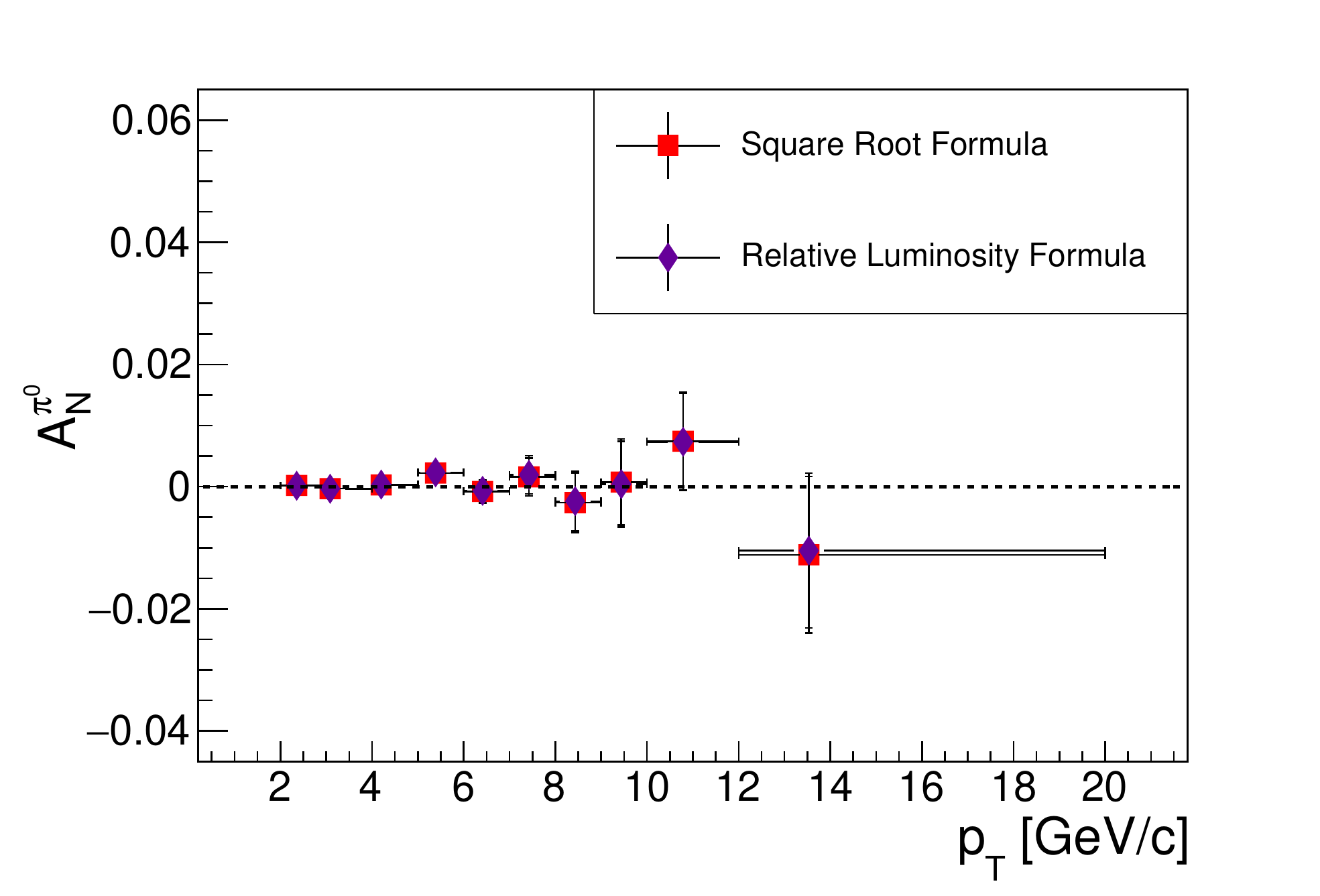} } 
\subfigure[T test comparing the averaged asymmetry results for the square root and relative luminosity formulas]{ \includegraphics[scale = 0.32]{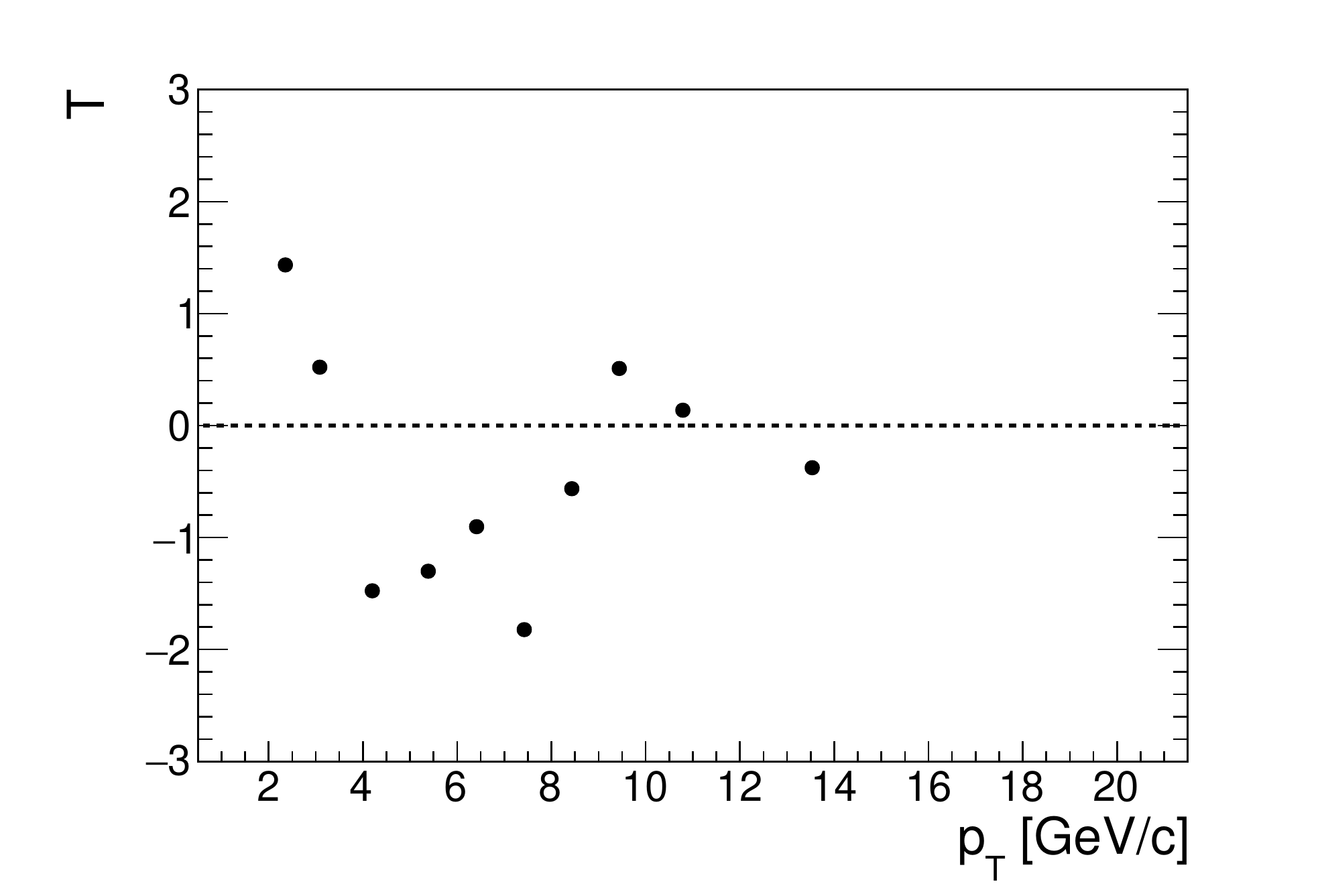} } 
\caption[Comparing the results of the relative luminosity and square root formulas for the background corrected \( \pi^0 \) asymmetry]
{Comparing the results of the relative luminosity and square root formulas for the background corrected \( \pi^0 \) asymmetry.   }
\label{Figure:pi0_correctedCompare}
\end{figure}


\begin{figure}
\centering
\subfigure[Yellow Beam]{ \includegraphics[scale = 0.32]{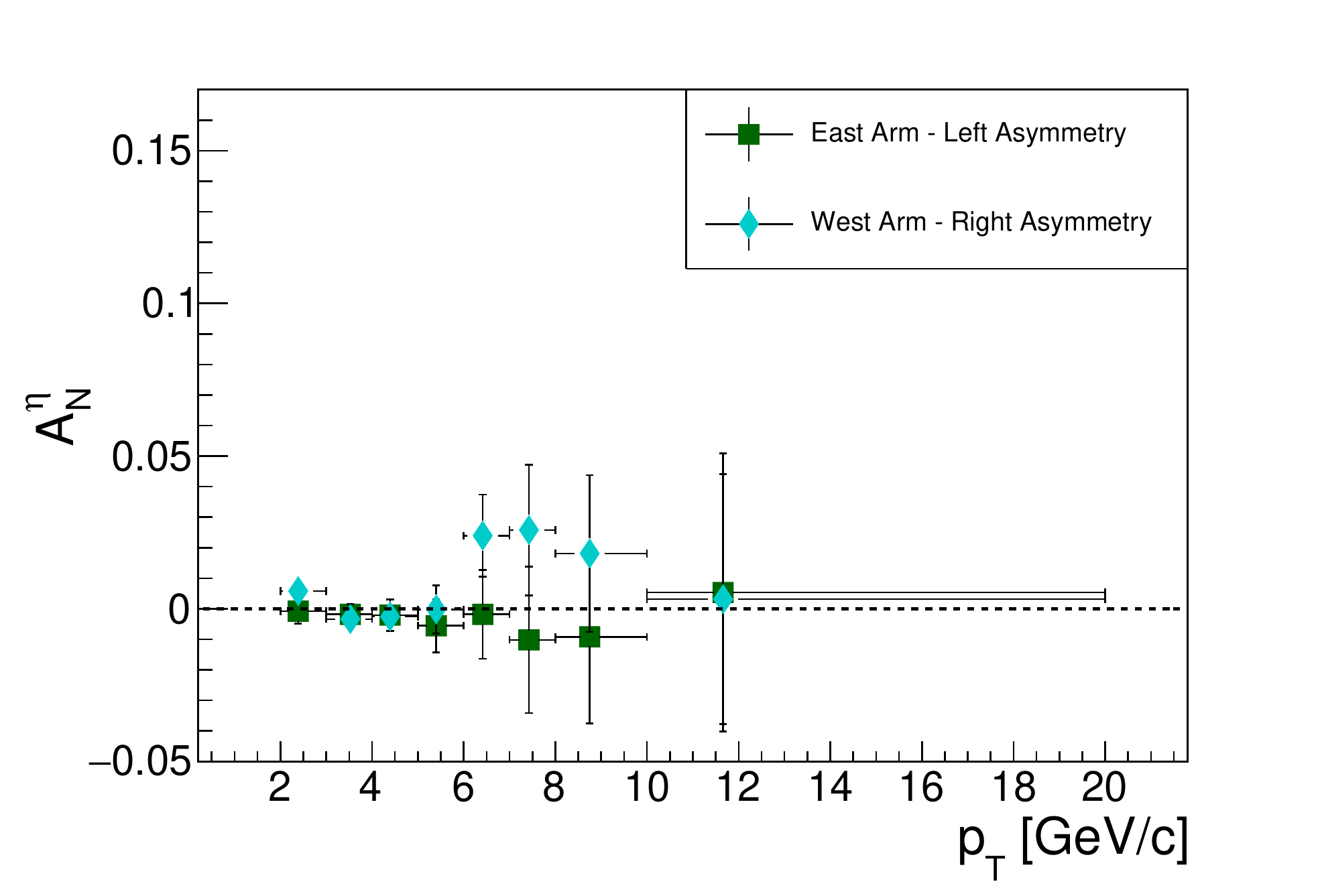} }
\subfigure[T test of the yellow beam left and right asymmetry results]{ \includegraphics[scale = 0.32]{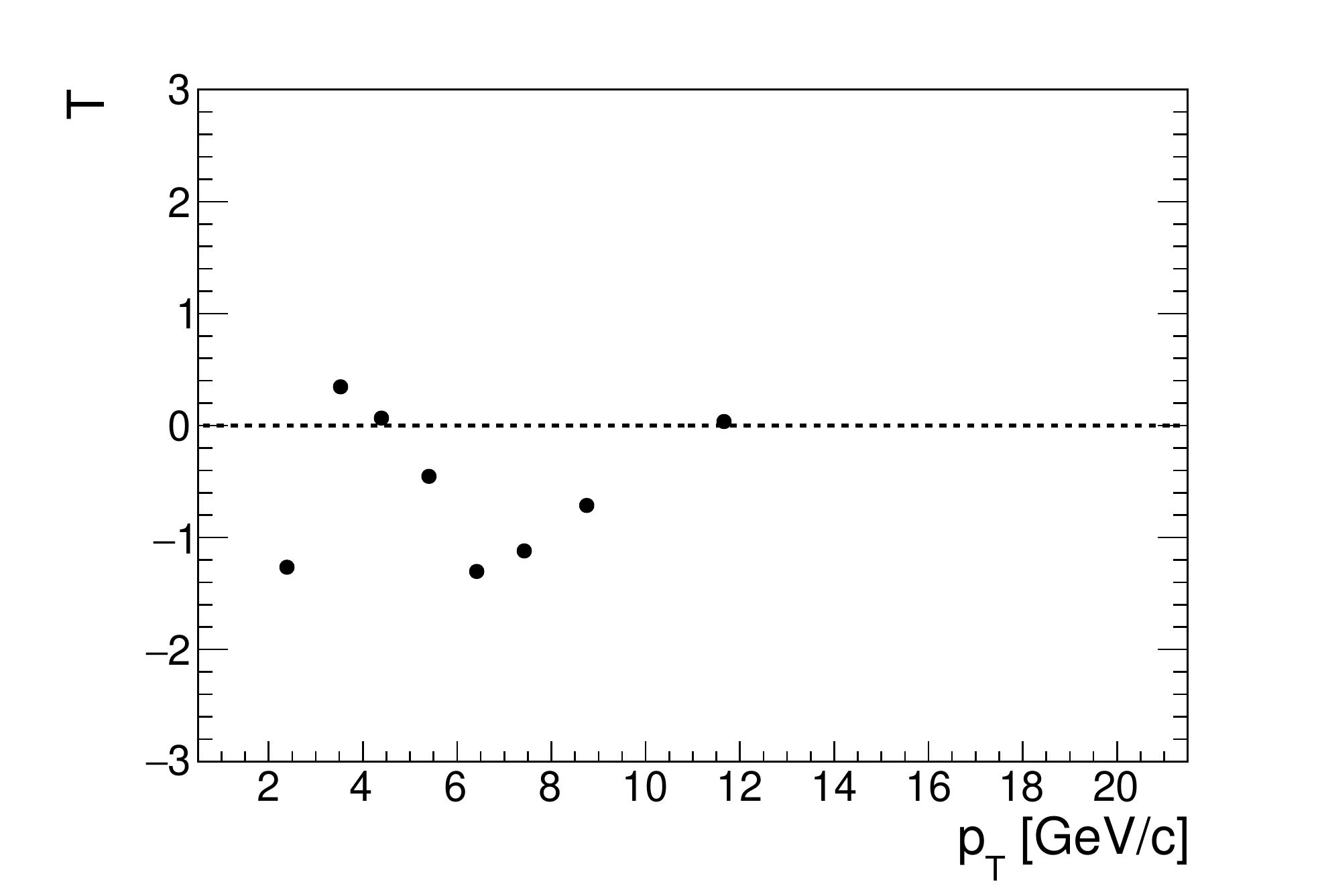} }\\
\subfigure[Blue Beam]{ \includegraphics[scale = 0.32]{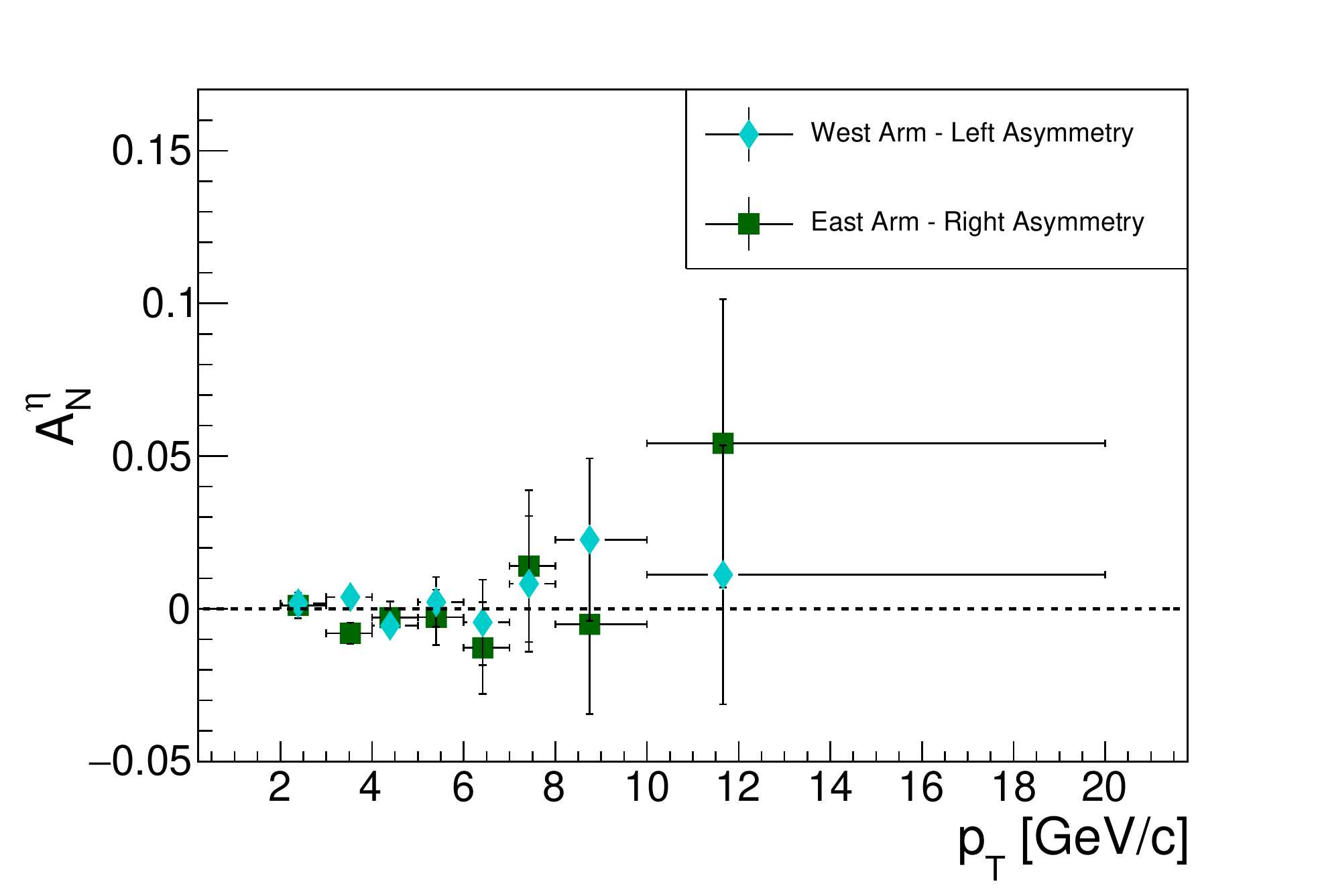} } 
\subfigure[T test of the blue beam left and right asymmetry results]{ \includegraphics[scale = 0.32]{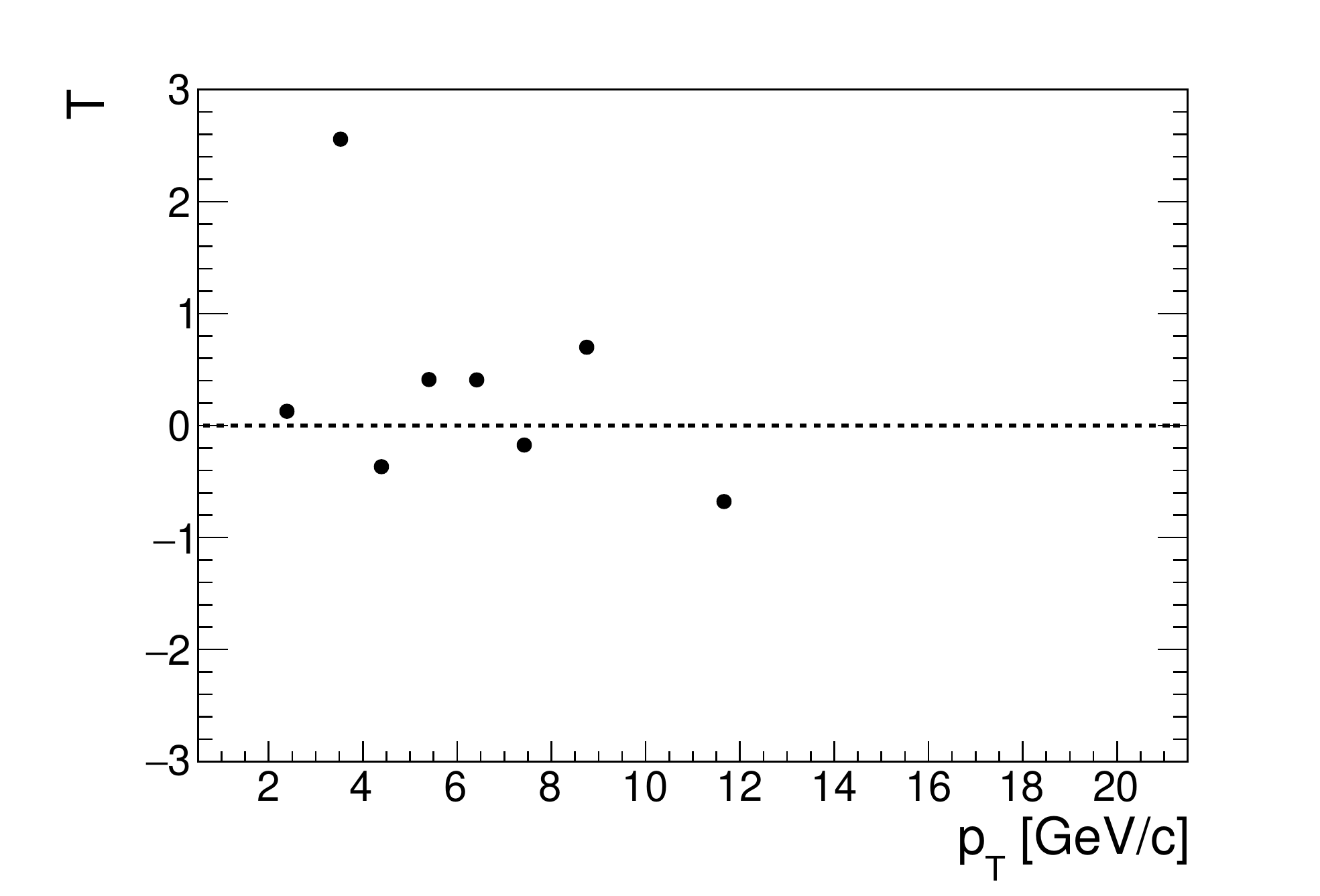} } 
\caption[The background corrected \( \eta \) asymmetry calculated using the relative luminosity formula.  ]
{The background corrected \( \eta \) asymmetry calculated using the relative luminosity formula.  }
\label{Figure:eta_correctedLumiLeftRight}
\end{figure}

\begin{figure}
\centering
\subfigure[Yellow and Blue Beam Asymmetries]{\includegraphics[scale = 0.32]{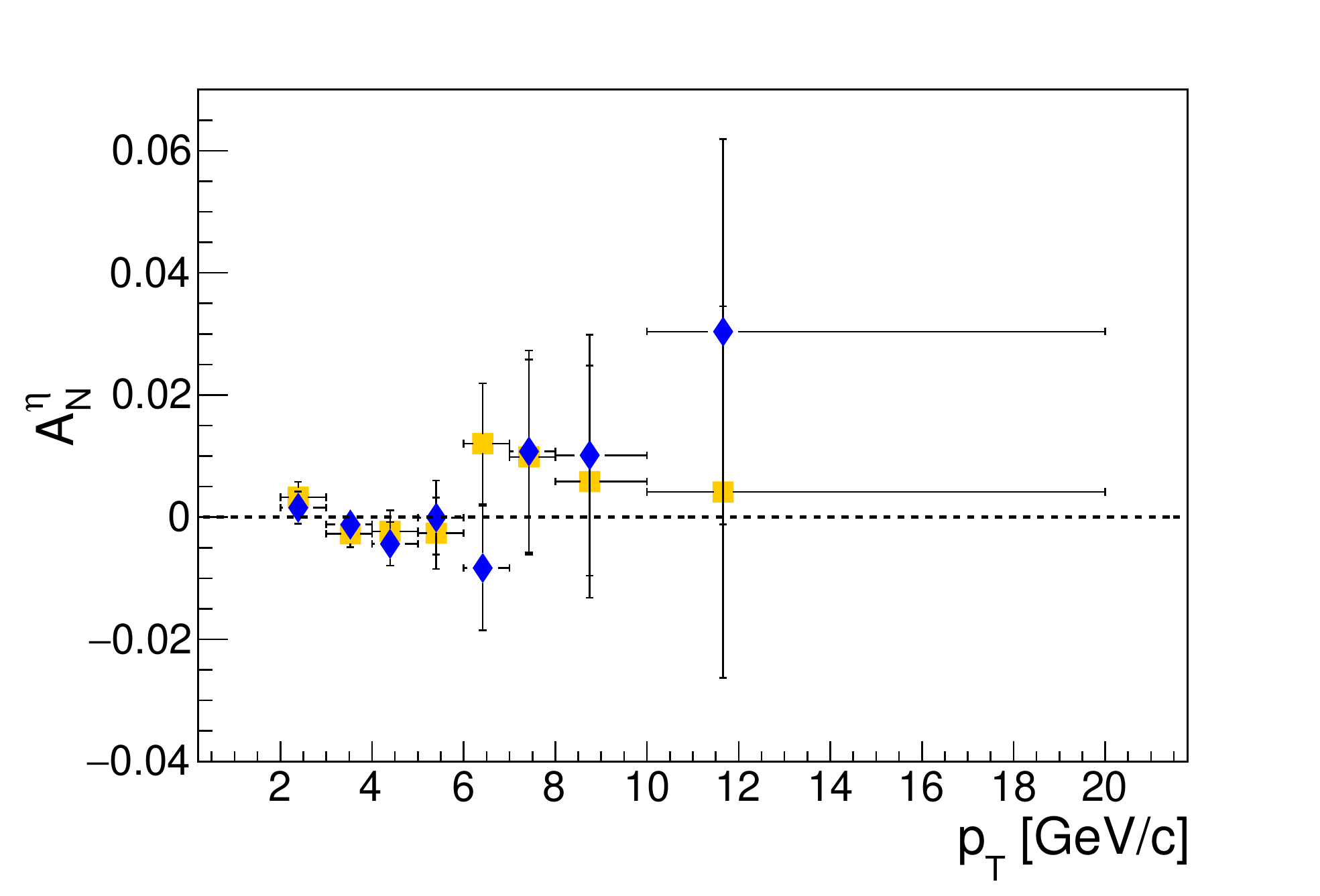} } 
\subfigure[T test comparing the result of the blue and yellow beam asymmetries]{\includegraphics[scale = 0.32]{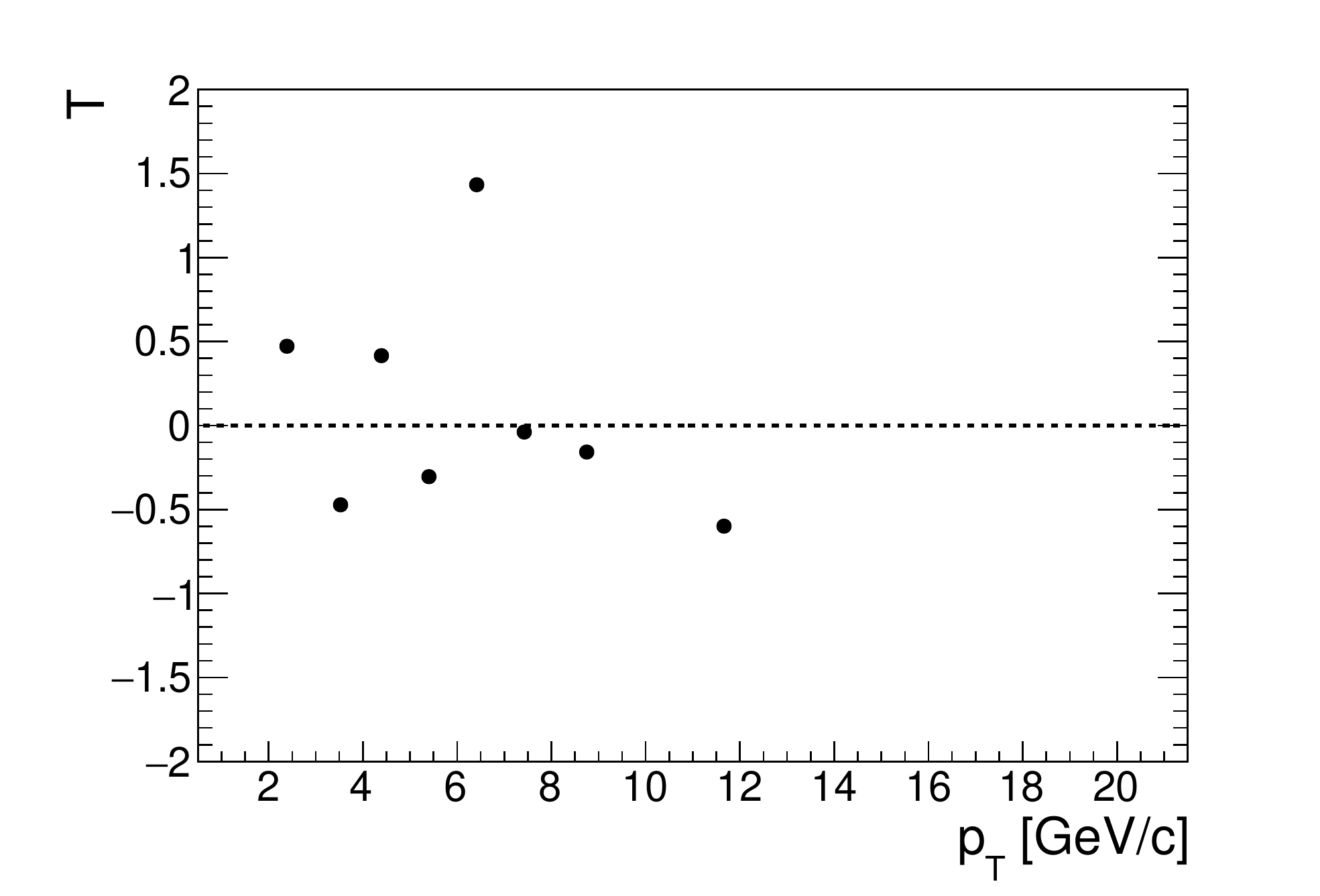} }
\caption[Relative luminosity yellow and blue beam asymmetries for background corrected \( \eta \) asymmetry]
{Relative luminosity yellow and blue beam asymmetries for background corrected \( \eta \) asymmetry.  }
\label{Figure:eta_correctedLumiYellowBlue}
\end{figure}

\begin{figure}
\centering
\subfigure[Yellow Beam]{ \includegraphics[scale = 0.32]{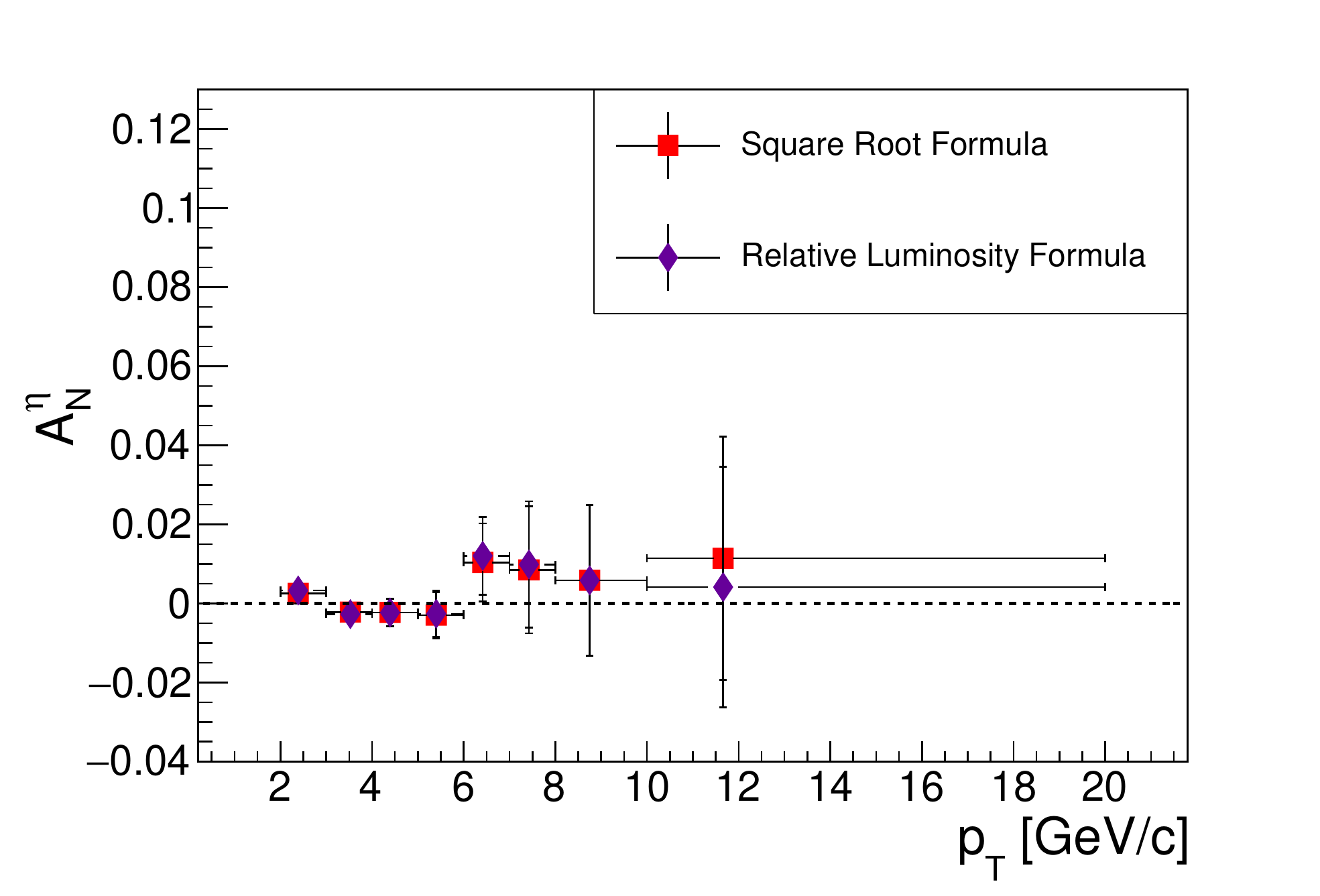} }
\subfigure[T test comparing the yellow beam square root and relative luminosity formula results]{ \includegraphics[scale = 0.32]{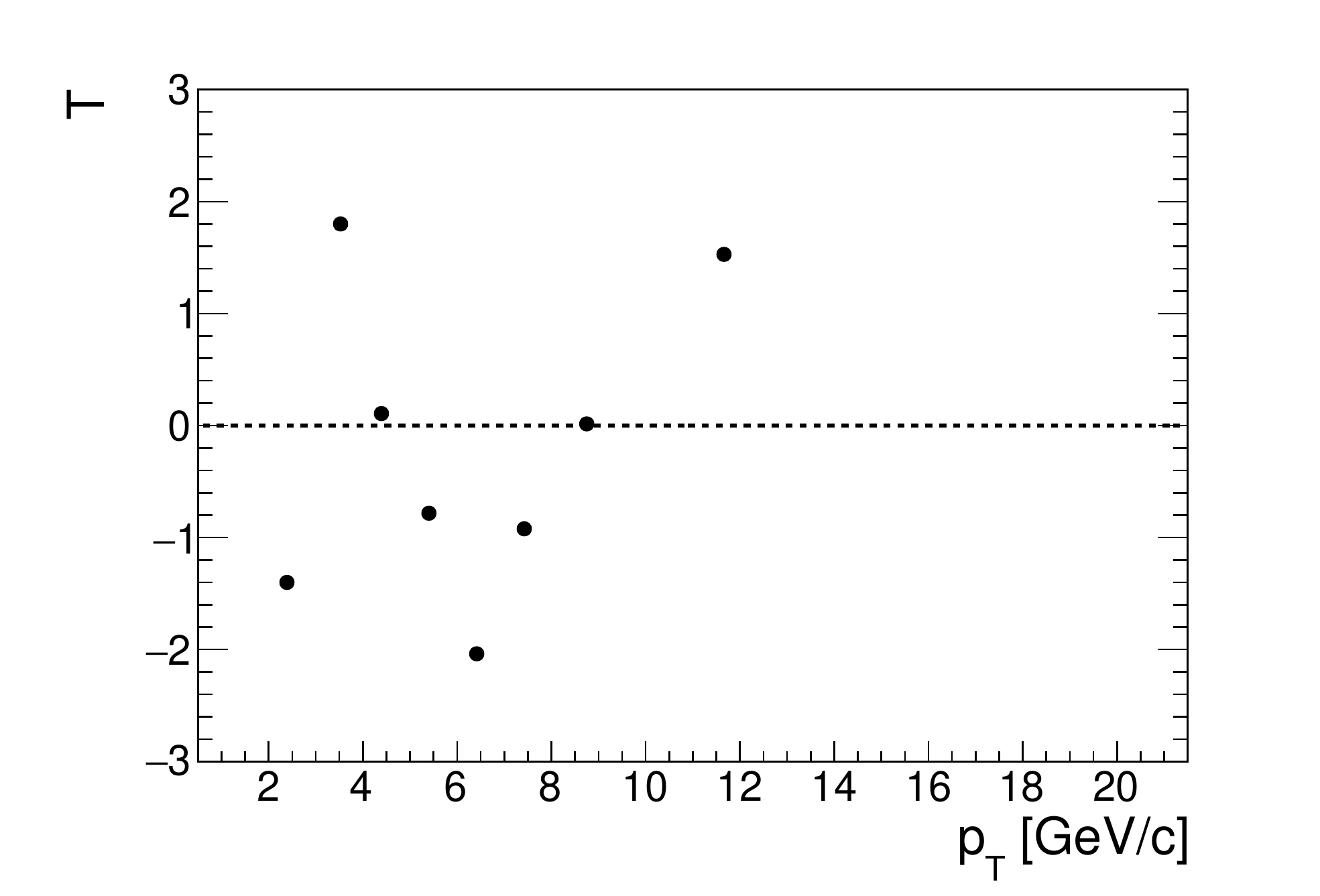} }\\
\subfigure[Blue Beam]{ \includegraphics[scale = 0.32]{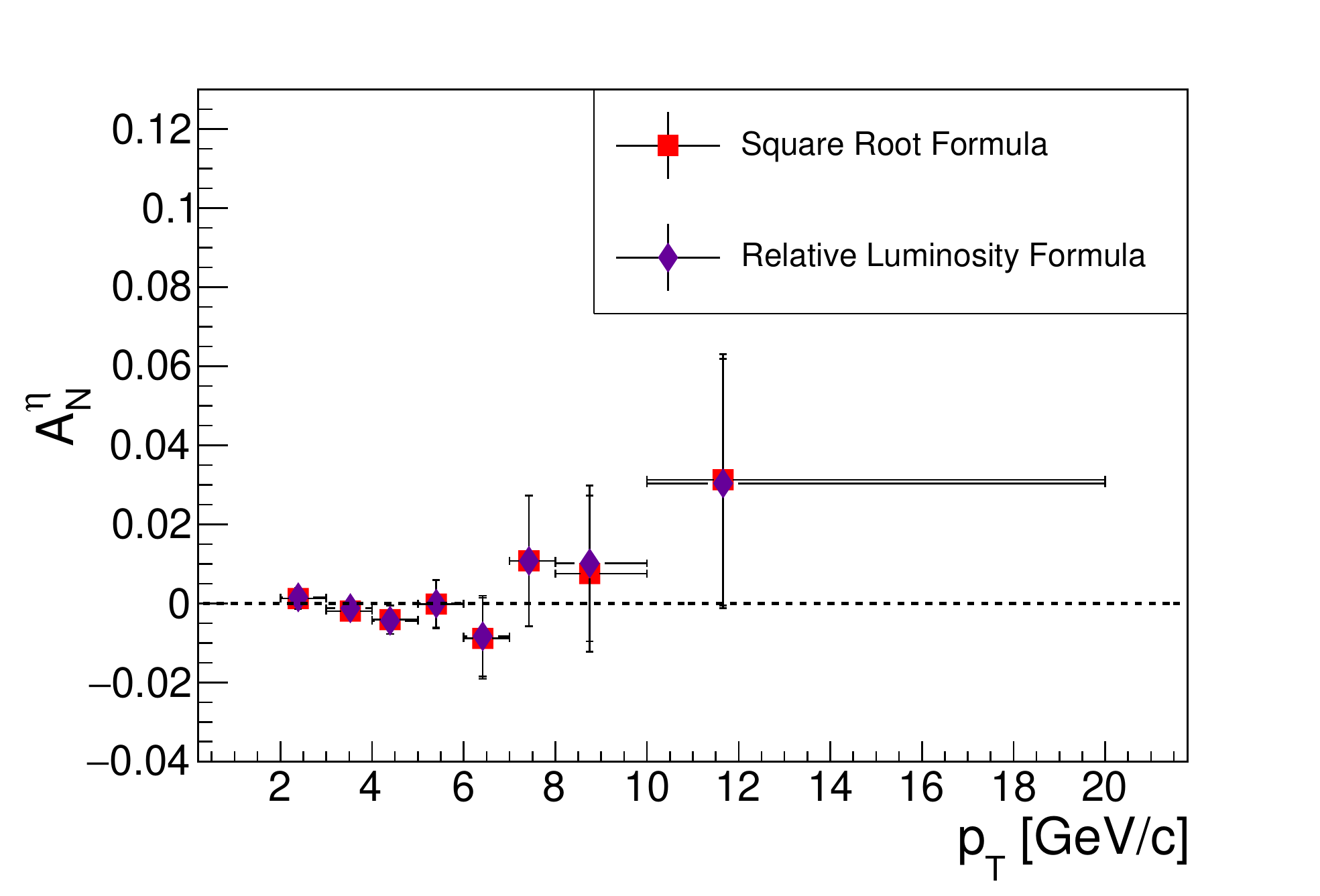} } 
\subfigure[T test comparing the blue beam square root and relative luminosity formula results]{ \includegraphics[scale = 0.32]{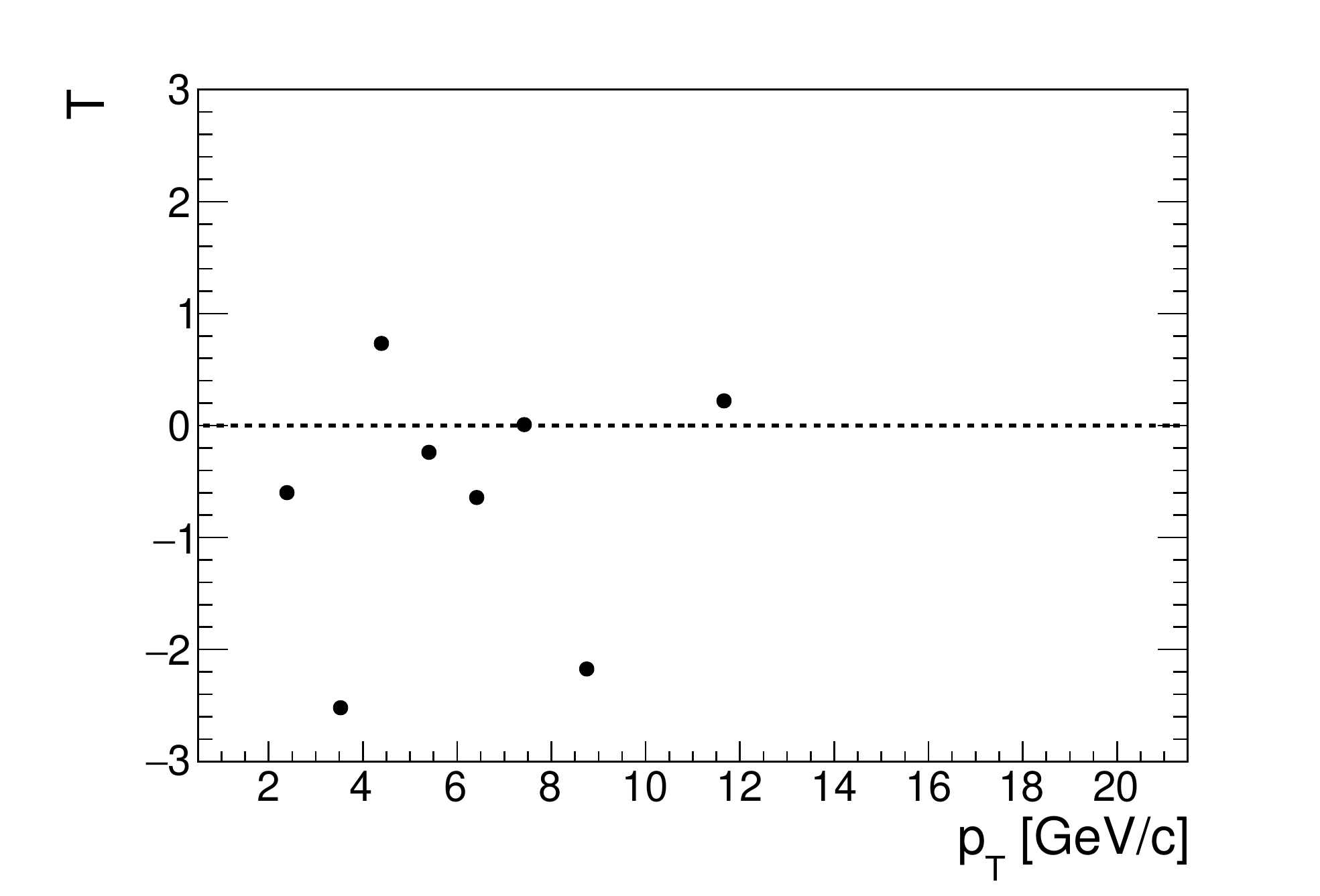} } \\
\subfigure[Final Averaged Asymmetry]{ \includegraphics[scale = 0.32]{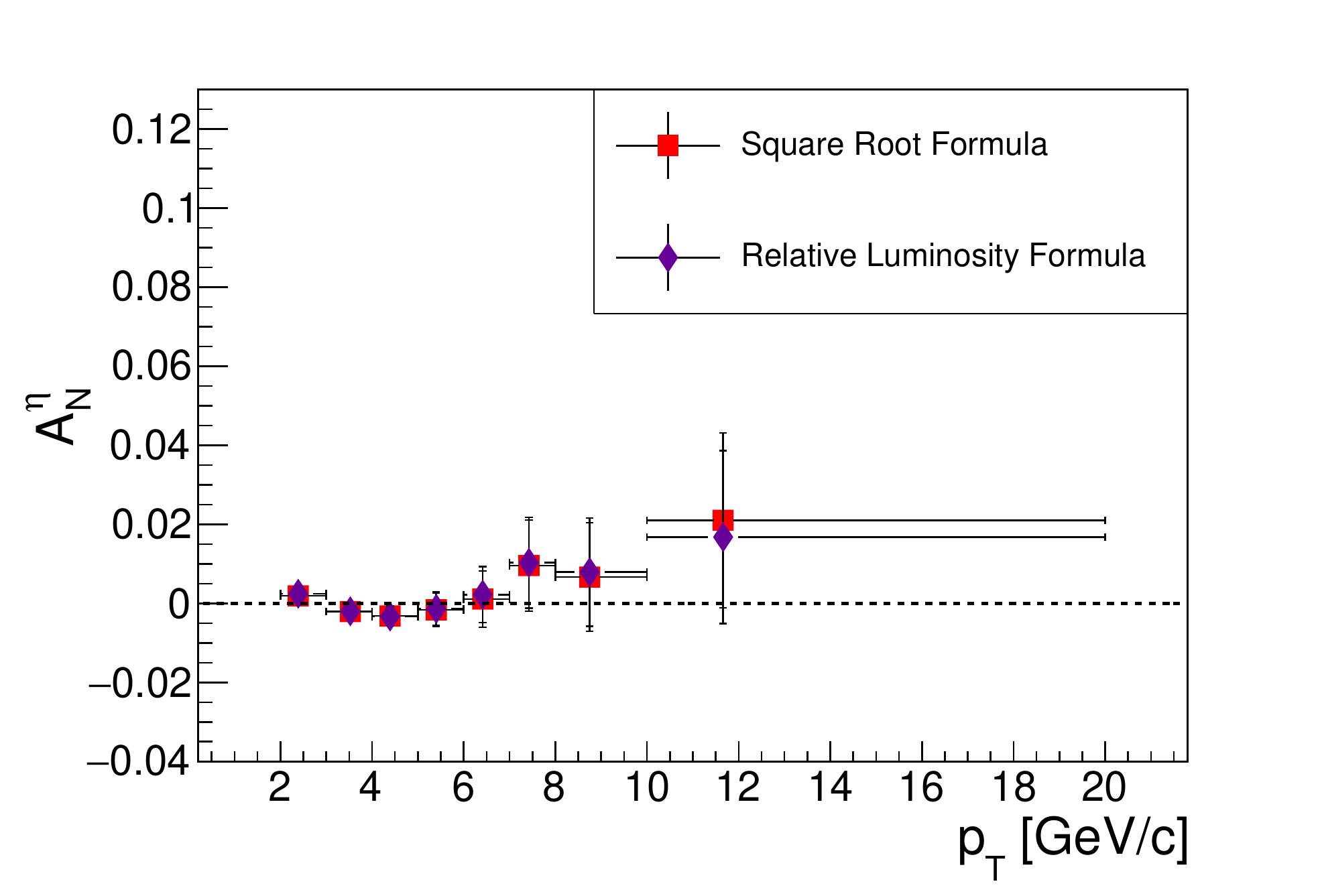} } 
\subfigure[T test comparing the averaged asymmetry results for the square root and relative luminosity formulas]{ \includegraphics[scale = 0.32]{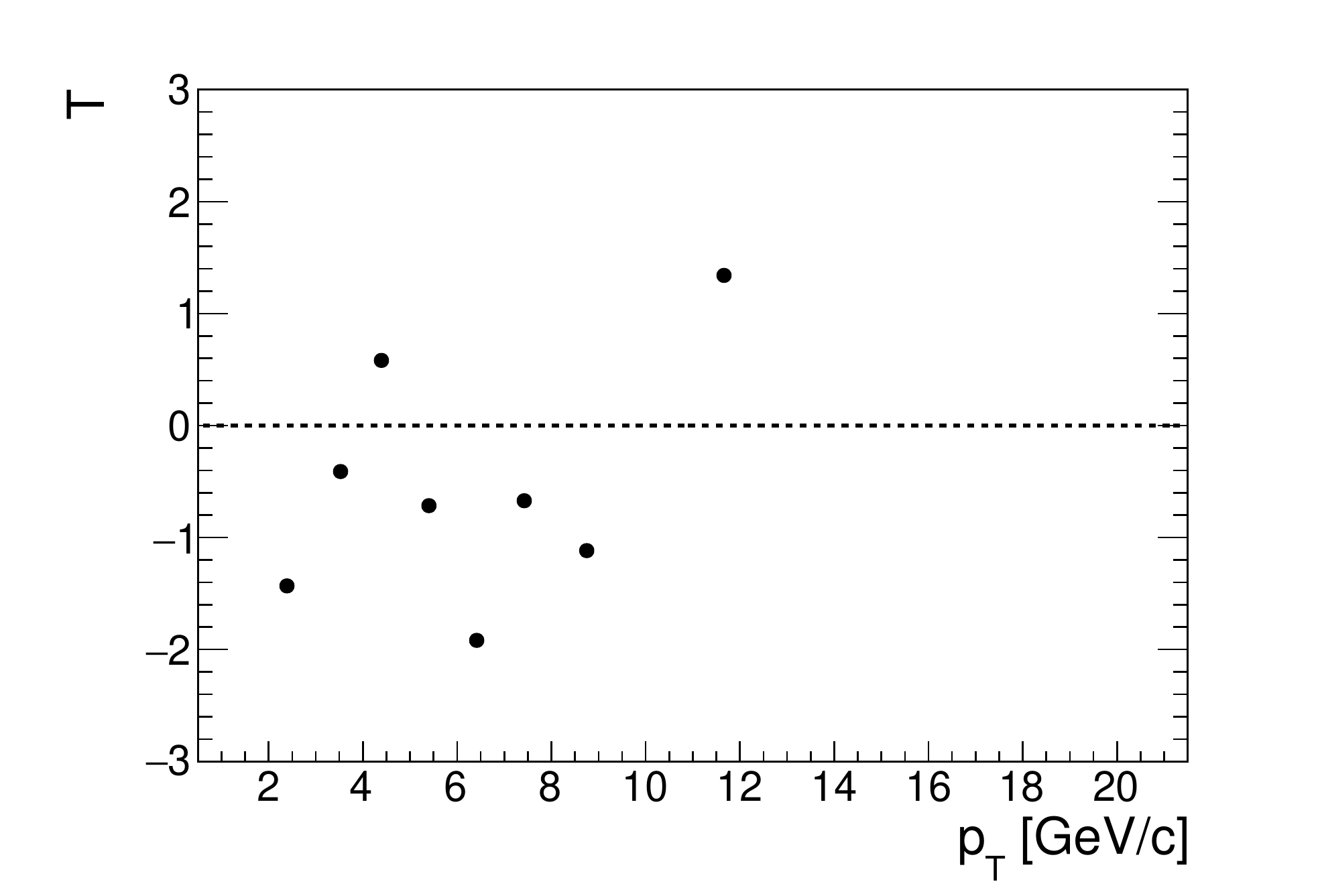} } 
\caption[Comparing the results of the relative luminosity and square root formulas for the background corrected \( \eta \) asymmetry]
{Comparing the results of the relative luminosity and square root formulas for the background corrected \( \eta \) asymmetry.   }
\label{Figure:eta_correctedCompare}
\end{figure}

%% file: Appendices/Bunch_Shuffling.tex
Bunch shuffling is a method used in asymmetry analyses to investigate the fluctuations present in data and test whether or not those fluctuations are consistent with statistical variations. This method of bunch shuffling is described in detail in Section~\ref{Section:BunchShuffling}, but essentially it involves randomizing the assigned polarization directions of the beam and recalculating the asymmetry.  This is done 10000 times and each asymmetry is divided by its statistical error and added to a histogram.  To test whether or not the variations in the asymmetry are consistent with statistical fluctuations, these histograms are fit with a Gaussian to verify that the mean is consistent with zero and the width is consistent with 1.  This is true for all bunch shuffling results for the direct photon (Figure~\ref{Figure:dpBunchShuffling}), \( \pi^0 \) and \( \eta \) asymmetries except for the lowest \( p_T \) bin of both the \( \pi^0 \) and \( \eta \) TSSAs.  This can be seen in the plots of this section where the width of the Gaussian fit for the lowest \( \pi^0 \) asymmetry \( p_T \) bin  is 1.068 and 1.056 for the lowest \( p_T \) bin of \( \eta \) asymmetry.  This indicates that there is some systematic error that was not eliminated by data cuts and so these numbers are used to assign an additional systematic uncertainty to the lowest \( p_T \) bin of both the \( \pi^0 \) and \( \eta \) TSSA results, which can been seen in Tables~\ref{Table:pi0TSSA} and ~\ref{Table:etaTSSA}.

\begin{figure}
\centering
\subfigure[\( 2 < p_T^{\pi^0} < 3 \) GeV/c \label{Figure:pi02to3}]{\includegraphics[scale = 0.32]{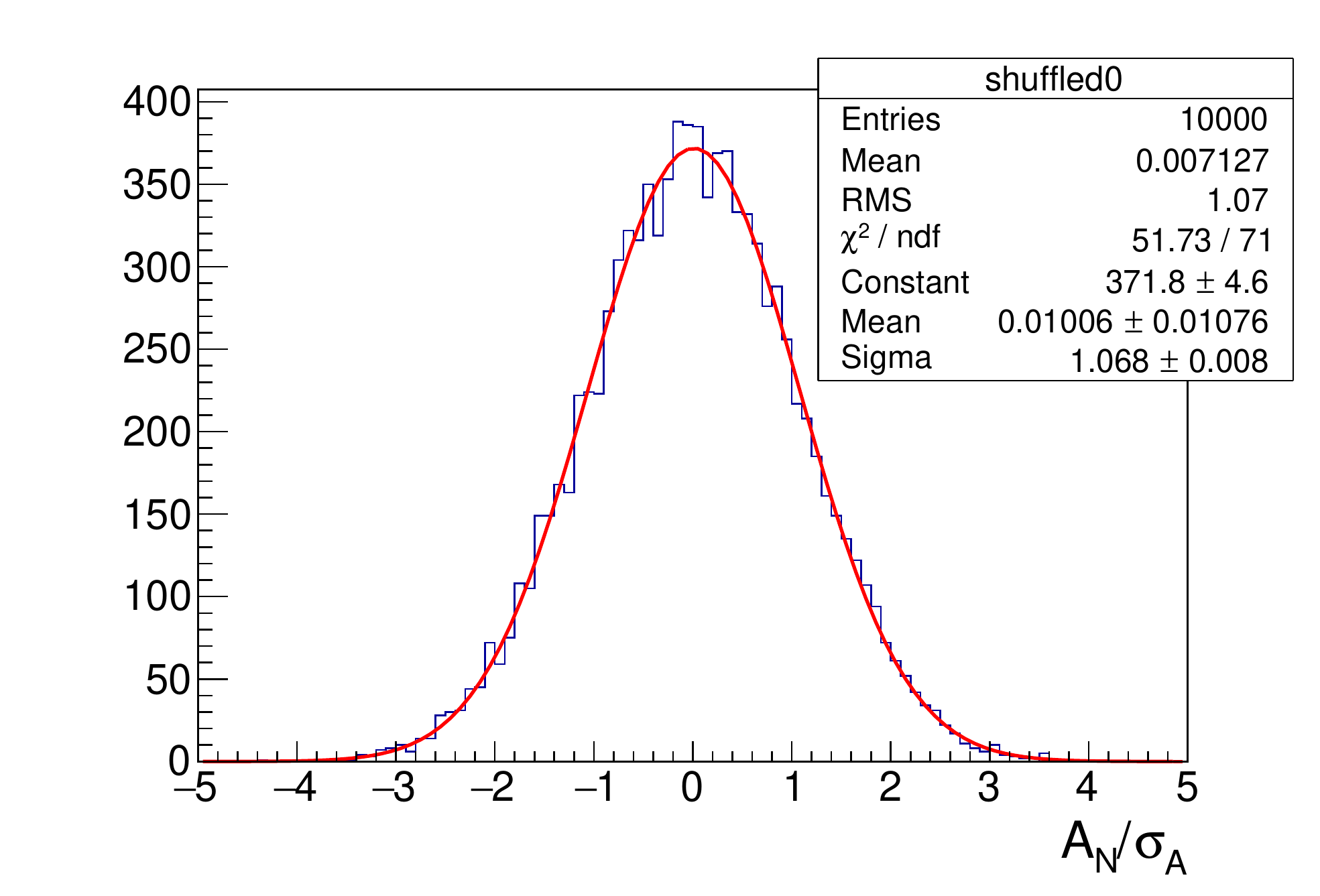} } 
\subfigure[\( 3 < p_T^{\pi^0} < 4 \) GeV/c]{\includegraphics[scale = 0.32]{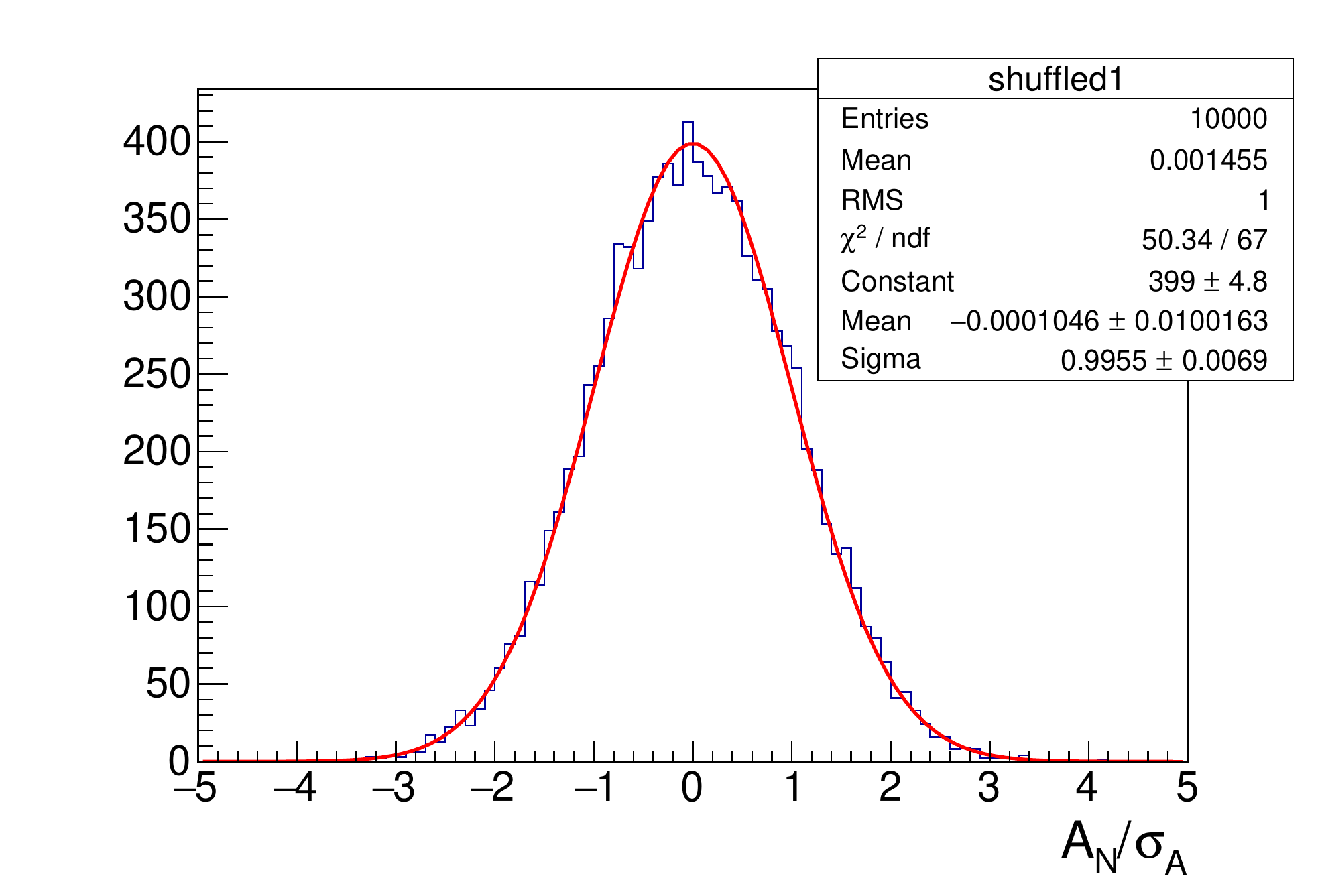} }
\caption[Bunch shuffling results for the lowest two \( p_T \) bins for the \( \pi^0 \) TSSA.  These asymmetries were calculated with photon pairs with \( 112 < M_{\gamma\gamma} < 162 \) MeV/c\textsuperscript{2} ]
{Bunch shuffling results for the lowest two \( p_T \) bins for the \( \pi^0 \) TSSA.  These asymmetries were calculated with photon pairs with \( 112 < M_{\gamma\gamma} < 162 \) MeV/c\textsuperscript{2} }
\label{Figure:pi0BunchShuffle}
\end{figure}

\begin{figure}
\centering
\subfigure[\( 2 < p_T^\eta < 3 \) GeV/c\label{Figure:eta2to3}]{\includegraphics[scale = 0.32]{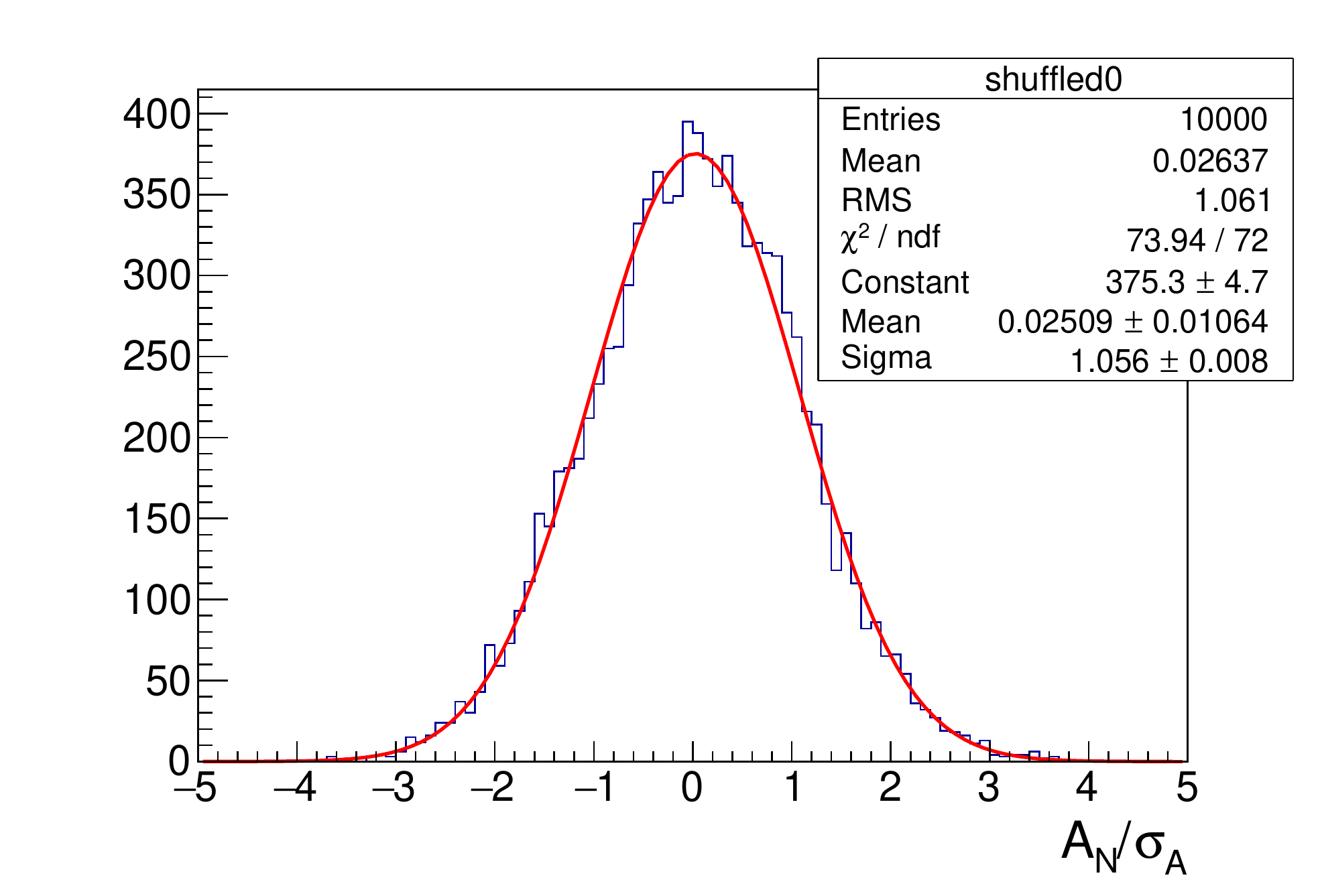} } 
\subfigure[\( 3 < p_T^\eta < 4 \) GeV/c]{\includegraphics[scale = 0.32]{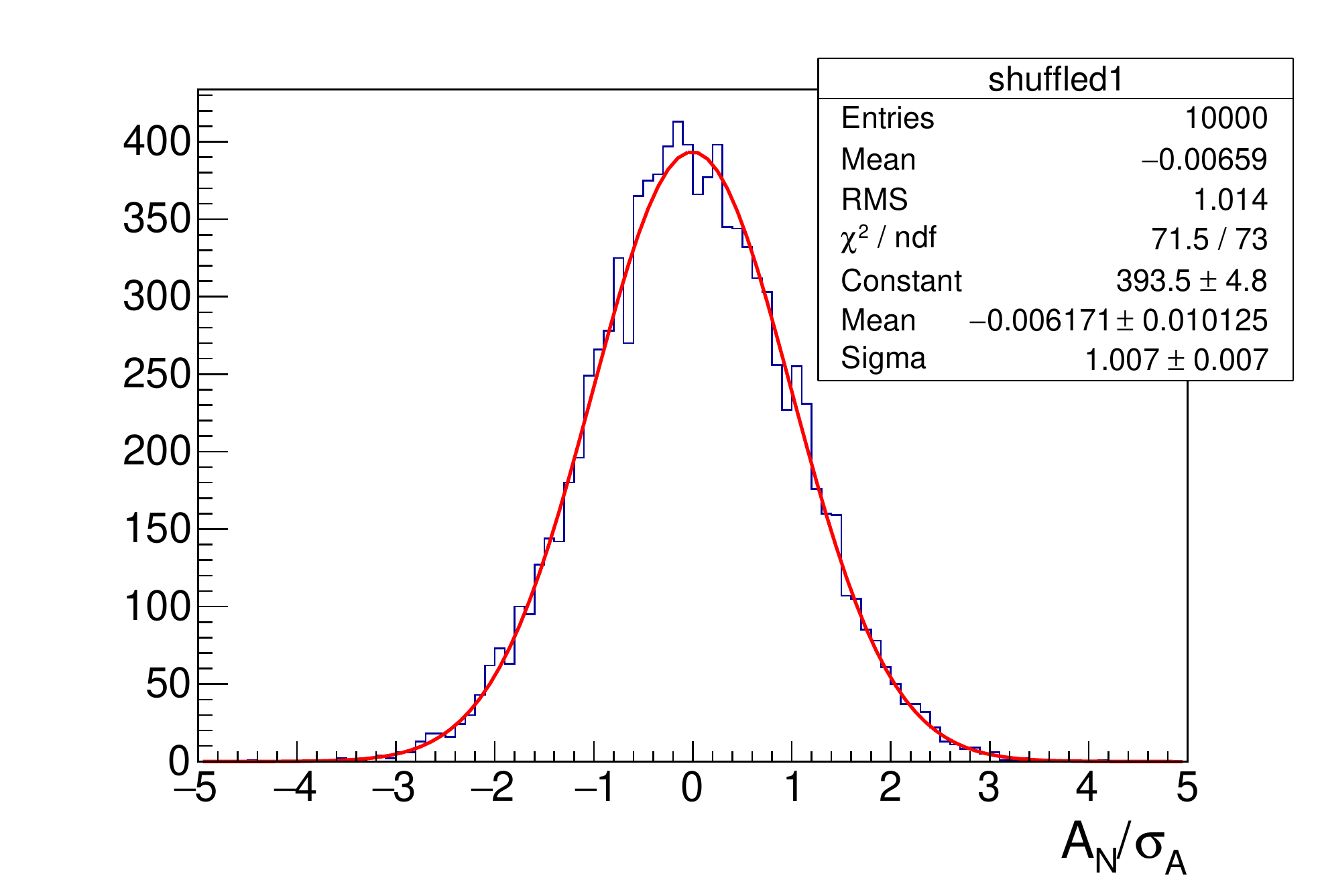} }
\caption[Bunch shuffling results for the lowest two \( p_T \) bins for the \( \eta \) TSSA.  These asymmetries were calculated with photon pairs with  \( 480 < M_{\gamma\gamma} < 620 \) MeV/c\textsuperscript{2}]
{Bunch shuffling results for the lowest two \( p_T \) bins for the \( \eta \) TSSA.  These asymmetries were calculated with photon pairs with  \( 480 < M_{\gamma\gamma} < 620 \) MeV/c\textsuperscript{2} }
\label{Figure:etaBunchShuffle}
\end{figure}